\documentclass[a4paper,11pt]{article}
\pdfoutput=1 

\usepackage{jheppub} 
\usepackage{hyperref}
\hypersetup{
    bookmarks=true,         
    unicode=false,          
    pdftoolbar=true,        
    pdfmenubar=true,        
    pdffitwindow=false,     
    pdfstartview={FitH},    
    pdftitle={My title},    
    pdfauthor={Author},     
    pdfsubject={Subject},   
    pdfcreator={Creator},   
    pdfproducer={Producer}, 
    pdfkeywords={keyword1} {key2} {key3}, 
    pdfnewwindow=true,      
    colorlinks=true,       
    linkcolor=red,          
    citecolor=purple,        
    filecolor=magenta,      
    urlcolor=blue,           
    linktocpage=true
}
\usepackage{graphics}
\usepackage{rotating}
\usepackage{subfigure}
\usepackage{pstricks}
\usepackage{color}
\usepackage{amsfonts}
\usepackage{mathrsfs,amsmath}
\usepackage{epsfig}
\usepackage{amsmath,amssymb,amsthm,graphicx,latexsym}
\usepackage{ulem}

\newcommand{\be}{\begin{equation}}
\newcommand{\ee}{\end{equation}}
\newcommand{\bea}{\begin{eqnarray}}
\newcommand{\eea}{\end{eqnarray}}
\newcommand{\bml}{\begin{subequations}}
\newcommand{\eml}{\end{subequations}}
\newcommand{\bfig}{\begin{figure}}
\newcommand{\efig}{\end{figure}}

\newcommand{\slashed}{\hspace{-1.1ex}/}

\newcommand{\bmat}{\begin{pmatrix}}
\newcommand{\emat}{\end{pmatrix}}
\begin{document}	
	\title{\textsc{\fontsize{25}{17}\selectfont \sffamily \bfseries \textcolor{violet}{Cosmological spectrum of two-point correlation function from vacuum fluctuation of Stringy Axion field in De Sitter space: A study of the role of Quantum Entanglement}}}

	\author[a]{Sayantan Choudhury,
		\footnote{\textcolor{purple}{\bf Alternative
				E-mail: sayanphysicsisi@gmail.com}. ${}^{}$}}
\author[b,c]{Sudhakar Panda
}
	\affiliation[a]{Quantum Gravity and Unified Theory and Theoretical Cosmology Group, Max Planck Institute for Gravitational Physics (Albert Einstein Institute),
	   Am M$\ddot{u}$hlenberg 1,
	   14476 Potsdam-Golm, Germany.}
		\affiliation[b]{	
		National Institute of Science Education and Research,
		Jatni, Bhubaneswar, Odisha - 752050, India.}
		\affiliation[c]{Homi Bhabha National Institute, Training School Complex,
		Anushakti Nagar, Mumbai-400085, India.
		}
	\emailAdd{sayantan.choudhury@aei.mpg.de,panda@niser.ac.in}

	\abstract{  In this work, we study the impact of quantum entanglement on the two-point correlation function and the associated primordial power spectrum of mean square vacuum fluctuation in a bipartite quantum field theoretic system. The field theory that we consider is the effective theorry of axion  field arising from {\bf Type IIB} string theory compactified to four dimensions. We compute the expression for the power spectrum of vacuum fluctuation in three  different approaches, namely (1)  field operator expansion (FOE) technique with the quantum entangled state, (2) reduced density matrix (RDM) formalism with mixed quantum state and (3) the method of non-entangled state (NES).  For massless  axion field, in all these three formalism, we reproduce, at the leading order, the exact scale invariant power spectrum which is well known in the literature.  We observe that due to quantum entanglement, the sub-leading terms for these thee formalisms are different. Thus, such correction terms break the degeneracy among the analysis of the FOE, RDM and NES formalisms in the super-horizon limit.  On the other hand, for massive  axion field we get a slight deviation from scale invariance and exactly quantify the spectral tilt of the power spectrum in small scales. Apart from that, for massless  and massive  axion field, we find distinguishable features of the power spectrum for the FOE, RDM, and NES on the large scales, which is the result of quantum entanglement. We also find that such large-scale effects are comparable to or greater than the curvature radius of the de Sitter space. Most importantly, in near future if experiments probe for early universe  phenomena, one can detect such small quantum effects. In such a scenario, it is possible to test the implications of quantum entanglement in primordial cosmology. 
	}

	\maketitle
	\flushbottom
	\section{\textcolor{blue}{Introduction}}
	
			    \begin{figure*}[htb]
			    \centering
			    {
			        \includegraphics[width=18.5cm,height=8cm] {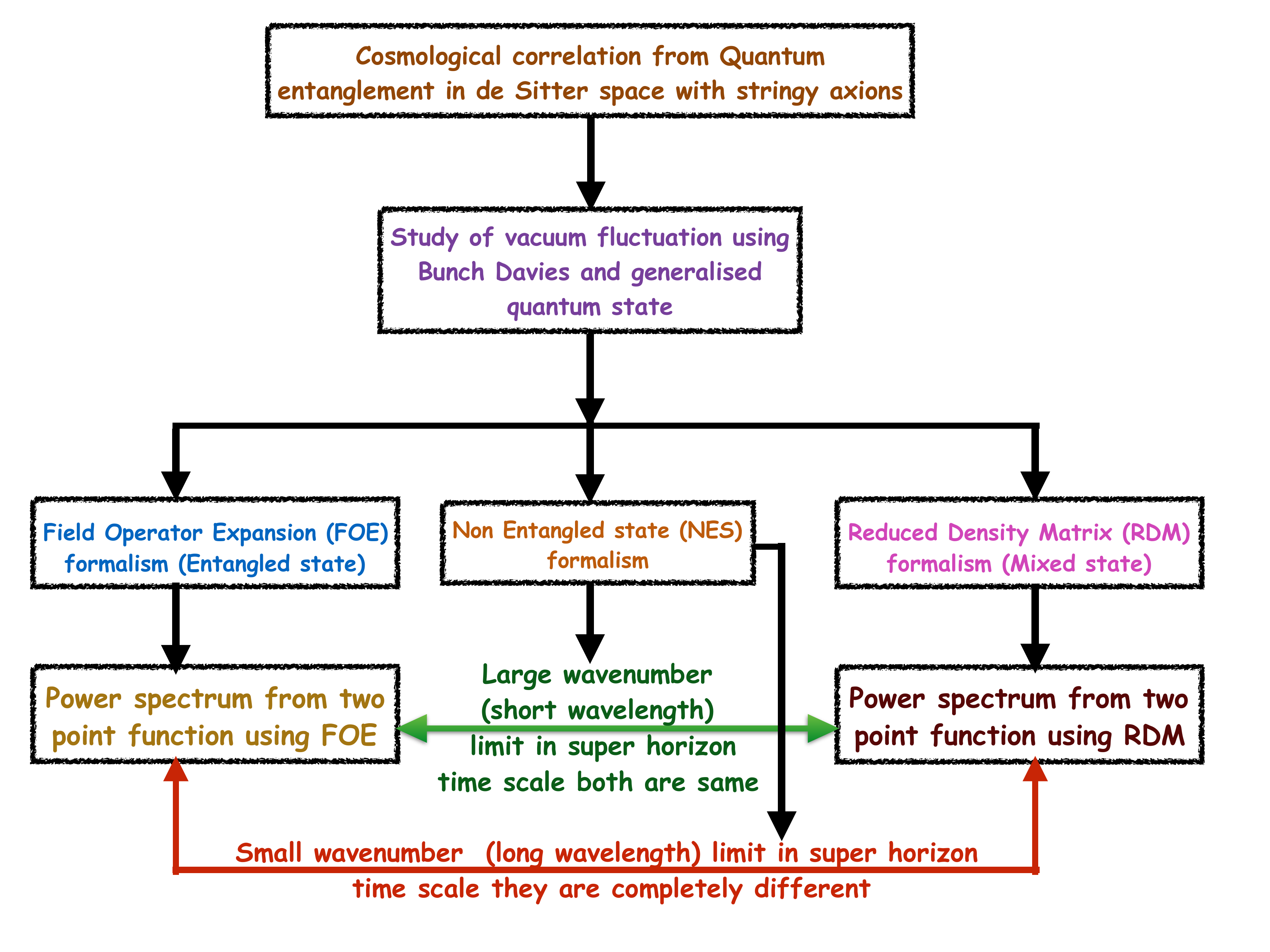}
			    }
			    \caption[Optional caption for list of figures]{Schematic diagram for the computation algorithm of long range effect of cosmological correlation function from quantum entanglement of axion in de Sitter open hyperbolic chart.} 
			    \label{fzaa}
			    \end{figure*}
       The concept of quantum entanglement is one of the most interesting features that one can study in the context of quantum mechanics. Using such idea one can study the instantaneous physical implication of local measurements \cite{Horodecki:2009zz,MartinMartinez:2012sg,Nambu:2008my}. There are several applications in the framework of quantum field theory in which the quantum entanglement play a significant role. For example,  particle creation (EPR Bell pair \cite{Bell:1964kc}) through the bubble nucleation procedure has been  explained using the idea of quantum entanglement where the quantum system is strongly correlated \cite{Coleman:1980aw,Garriga:2012qp,Garriga:2013pga,Frob:2014zka}. Also using the concept of quantum entanglement in QFT one successfully explains many phenomena like entropy bounds,  phase transitions, anomalies, confinement, thermalization and quantum critical quenches, localization in quantum gravity and description of interior of black holes. Apart from that quantum entanglement has huge application in the context of quantum information theory, quantum cryptography and interferometry.
       
       The von-Neumann entropy and  R$\acute{e}$nyi entropy are the  appropriate measures of quantum entanglement the framework of condensed matter theory \cite{Laflorencie:2015eck}, in quantum information theory and in theoretical high energy physics. The idea of entanglement entropy in the context of quantum field theory is the best possible computational tool to quantify and study the nature of the long range effects of quantum correlation. However, the computation of entanglement entropy for a specific class of quantum field theories were not easy  before the method proposed by Ryu and Takayanagi \cite{Ryu:2006bv}. In this work, the authors have  computed the entanglement entropy for a strongly coupled field theory set up with a gravity dual using the techniques of holography and the results are remarkable  as it is in agreement with various expectations from the quantum field theory side \cite{Takayanagi:2012kg}.
       
       Following this success, Maldacena and Pimentel in ref.~\cite{Maldacena:2012xp} further proposed an explicit technique to compute the entanglement entropy in the framework of quantum field theory of de Sitter space with Bunch Davies quantum initial vacuum state~\footnote{It is important to note that, by the term Bunch-Davies vacuum here we actually pointing towards the well known Euclidean vacuum state which is actually a false vacuum state in quantum field theory and commonly used to fix the initial quantum condition of our universe in terms of quantum mechanical state or the wave function of the universe.}. It is important to note that, particularly that the Green functions which verify a condition (commonly known as the {\it Hadamard condition}) consisting to behave on the light-cone as in flat space for Bunch Davies or the Euclidean false vacuum state. On the other hand, the Bunch Davies or the Euclidean false vacuum can also be physically interpreted as being generated by an infinite time tracing operation from the condition that the energy scale of the quantum mechanical fluctuations is much smaller than the characteristic scale in cosmology, which is the Hubble scale. This quantum vacuum state possesses actually no quanta at the limiting asymptotic past infinity. However, in the the framework of quantum field theory of curved space time, there exist a huge class of quantum mechanical vacuum states over the background De Sitter space time which are invariant under all the $SO(1,4)$ isometries and commonly known as the $\alpha$-vacua.  Here $\alpha$ is a real parameter which forms a real parameter family of continuous numbers to describe the issometric classes of invariant quantum vacuum state in De Sitter space. In a more technical sense, sometimes the $\alpha$ vacua is characterized as the squeezed quantum vacuum state in the context of quantum field theory of curved space time. Also it is important to note that, in the original version something called, $\alpha, \beta$ or Motta-Allen (MA) vacua is appearing which is CPT violating and here an additional real parameter $\beta$ is appearing in the phases in the definition of the quantum mechanical vacuum state. This phase factor is responsible for the CPT violation. Once we switch off this phase factor by fixing $\beta=0$, the one can get back the CPT symmetry preserving quantum vacuum state in the present context. The $\alpha$ vacua and the Bunch Davies or Euclidean false vacuum are connected to each other via Bogoliubov transformation. Specially $\alpha=0$ case corresponds to the Bunch Davies or Euclidean vacuum state in which the Hadamard condition in the Green's functions is satisfied. Additionally, the point to be noted here that, the Bunch-Davies or the Euclidean quantum vacuum state is actually representing a zero-particle quantum mechanical state which is observed by a geodesic observer, which implies that, an observer who is in free fall in the expanding state is characterized by this vacuum state. Because of this reason to explain the origin of quantum mechanical fluctuations appearing in the context of cosmological perturbation theory in the inflationary models or during the particle production phenomena the concept of Euclidean false quantum vacuum state is commonly used in primordial cosmology literature. Here, the authors have studied the gravitational dual of the quantum field theory of De Sitter space using holographic techniques in detail. Further in ref. \cite{Kanno:2014lma} the authors have extended this computation in the context of $\alpha$ vacua \cite{Allen:1985ux} in the same context.  In ref. \cite{Choudhury:2017bou} and \cite{Choudhury:2017qyl} the computation of quantum entanglement entropy and the formation of EPR Bell pair from stringy Axion~\footnote{Here we want to point few works, refs.~\cite{Capolupo:2019peg,Patrascu:2018sia}, where the authors have studied quantum field theory of axion fields and its relation with quantum entanglement.} were discussed with Bunch Davies and $\alpha$ vacua respectively.  
       
    Based on the physical set up used in our previous works \cite{Choudhury:2017bou} and \cite{Choudhury:2017qyl}, in this paper we have studied the cosmological implications of  quantum entanglement  by focussing on the long range effects of the two point correlation function computed from the mean square vacuum fluctuation of stringy Axion field with Bunch Davies and $\alpha$ quantum states as initial choice of vacua . We expect from this analysis that the signature and impact of quantum entanglement could be manifest  in the correlation function even  beyond the Hubble horizon scale. Our expectation is mainly due to the fact that de Sitter expansion of universe distinguish between a pair of Axions \cite{Maldacena:2015bha,Choudhury:2016cso,Choudhury:2016pfr,Kanno:2017dci}, known as EPR Bell pair which is created within causally connected Hubble region. For this purpose, we use three different techniques:
    \begin{enumerate}
   \item  Field operator expansion (FOE) method with entangled state, 
   \item Reduced density matrix formalism (RDM) with mixed state and
   \item Non-entangled state (NES) method. 
\end{enumerate}   
Here one can ask the following sets of questions regarding the implementation of three different techniques in the present context:
\begin{itemize}
\item Q1.~~Why we have used three different formalisms to compute the cosmological two point correlation function? 
\item Q2.~~What is the correct physics they believe that happens in the setup of the space time?
\item  Q3.~~In those three formalisms, the physics is completely different. So which one is correct? 
\item Q4.~~We finally could only observe one possible observational consequence. So which one is correct? 
\end{itemize}
  The appropriate answers to above mentioned questions are appended below point wise:
  \begin{itemize}
  \item A1.~~We have used three different formalisms to compute the cosmological two point correlation function to check the explicit role of quantum mechanical entanglement in the primordial cosmology. In these three formalisms the leading order expressions become same. But the difference only can be found once we look into the small quantum corrections appearing in these formalisms. If the signature of quantum entanglement will be detected in near future in the observational probes of early universe, then one can explicitly rule out the possibility of appearing of NES method in the  context of quantum field theory of primordial cosmology. On the other hand, if the signatures of quantum entanglement cannot be confirmed then one can strongly rely on the result obtained in the NES method. Additionally it is important to note that, these three frameworks provide us the quantum mechanical origin of quantum field theory of early universe cosmology.
  
  \item A2 \& A3.~~From the theoretical perspective these three different formalisms have their own merit on the physical ground. If the quantum mechanical origin of the quantum correction of the primordial fluctuation is coming from the non entangled state then NES formalism is the only single option which can take care of the correct physics. On the other hand, if the quantum mechanical origin of the quantum correction of the primordial fluctuation is coming from the entangled mixed state then RDM formalism applicable to the subsystem is the most promising option which supports correct physical explanation. The last option is FOE formalism which is applicable when the quantum mechanical origin of the quantum correction of the primordial fluctuation is guided by the total entangled state (not the subsystem) then FOE formalism is useful to describe the correct physics. 
  
  \item A4.~~It is very well known fact that at late time scale all the large scale structure is formed due to long range persistent correlation originated from the primordial quantum mechanical fluctuation in the early universe. This can only be consistently theoretically established by using FOE and RDM formalisms which supports the concept of quantum entanglement in early universe cosmology. Now RDM formalism is more theoretically consistent than the FOE method as it is based on the quantum description of the reduced subsystem. Now as far as the detection in the observation is concerned, if we can detect the quantum mechanical origin of the sub leading  quantum correction in near future probes then one can explicitly very the explicit role of quantum entanglement, precisely test FOE or RDM formalism is correct. If we cannot detect the role of quantum entanglement then NES formalism will provide the correct physical explanation of the quantum origin of the sub leading correction term in the two point primordial correlation function.

  \end{itemize}
We implement the RDM formalism using the previous work done by Maldacena and Pimentel in ref.~\cite{Maldacena:2012xp} in the context of de Sitter cosmology.  In our computation we have explicitly included the effect of Stringy Axion in the small field regime  and as a result we get perturbatively corrected contributions in the expression for the power spectrum derived using FOE, RDM and NES formalisms. Such correction terms can be interpreted as quantum effects which are appearing from the UV complete theory, such as a specific type of bipartite quantum field theory driven by axion. We note that the axion field which is being considered here, is actually originating from {\bf Type IIB} string theory compactified on a Calabi-Yau three fold (${\bf CY^3}$), in presence of a ${\bf NS5}$ brane sitting at the bottom of a long throat \cite{Panda:2010uq}.  Most importantly, in the large wave number~\footnote{Here the wave number $p$ mimics the role of ${\bf SO(3,1)}$ principal quantum number in the de Sitter hyperbolic open chart which is continuous and lying within the range $0<p<\infty$. The other ${\bf SO(3,1)}$ quantum numbers $m$ (azimuthal) and $l$ (orbital) play no significant role in the final result as the expression for the power spectrum for mean square vacuum fluctuation only depends on the quantum number  $p$. } limit  (small scale or small wave length approximation \cite{Kanno:2014ifa}) we have shown the results for the power spectrum derived from these three formalism perfectly match with each other if we consider only the leading order contribution.  However, the results are different for these three formalisms if we we include the contributions from next and next to next leading order.
In a way one can say that such additional small perturbative correction terms play a pivotal role to distinguish between the FOE, RDM and NES formalisms. This is obviously an important information because using the present observational data on early universe cosmology \cite{Kolevatov:2017dze,Kopeikin:2001uk,Choudhury:2015hvr,Maharana:1997cz} one can further constrain the present model and also test the appropriateness of these formalisms.  Apart from this, for completeness, we have also analysed  the behaviour of the power spectrum in the small wave number limit (large scale or large wave length approximation). We find that all these three formalisms yield distinctive results in terms of the momentum (quantum number) dependence of the power spectrum in  order by order. But the lack of observational data on this particular regime  does not allow us to test the appropriateness and correctness of the proposed methods. We hope that  in near future when the observational data for this regime  will be available, our results  can further constrain the model and rule out two of the possibilities between the three formalisms discussed here. We would like  to mention here that in our computation of the power spectrum for mean square vacuum fluctuation we have not considered the quantum fluctuation of the pseudo scalar Axion field as a classical back ground field,  the approach which is mostly used in the context of the  cosmological correlations from early universe. Instead , we have chosen the field operator of the Axion field itself as quantum operator whose fluctuation with respect to a quantum mechanical vacuum state (Bunch Davies and $\alpha$ vacua).  Thus, in this paper, we have followed:
\begin{enumerate}
\item A complete quantum approach to compute the primordial power spectrum of mean square vacuum fluctuation, which is not usually followed in the context of cosmology. 
\item For the specific structure of the axion effective potential , we have computed the explicit form of the corrections which are due to  quantum effects.
\item For our calculation, we have used three different approaches at super horizon time scale hoping  that the quantum corrections,  at small and large wave number limits when confronted with observations, can select the most effective approach and the nature of quantum corrections.. From the cosmological perspective we believe this is a very important step forward.
\end{enumerate}
   
    The plan of the paper is as follows: In section \ref{3a}, we  begin our discussion with the computation of the wave function of the Axion field in a de Sitter hyperbolic open chart. For this purpose we have discussed the details of the background de Sitter geometrical set up in subsection \ref{3av1}. Further in subsection \ref{3aq} and \ref{3aq1}, we have solved the total wave function for Axion for Bunch Davies vacuum and generalised $\alpha$-  vacua respectively.  
Using these solutions we have derived the cosmological power spectrum of mean square quantum vacuum fluctuation in section \ref{fg1}. In subsections \ref{x1a} and \ref{x1b} we have discussed the quantum vacuum fluctuation using field operator expansion (FOE) formalism with entangled state for Axion. field.  We have also derived the explicit form of the wave function in this formalism.  This solution is used to derive the power spectrum by computing the two point quantum correlation function from mean square vacuum fluctuation.  In subsection \ref{x2a}and \ref{x2b} we have discussed the quantum vacuum fluctuation using reduced density matrix (RDM) formalism using mixed state for Axion field and we have derived the explicit form of the reduced density matrix in the de Sitter hyperbolic open chart. Further, this result is used to derive the power spectrum by computing the two point quantum correlation function from mean square vacuum fluctuation in large and small wave number limits for both massless and massve Axion fields.  In subsection \ref{x3a}and \ref{x3b} we have studied the quantum vacuum fluctuation using non entangled state (NES) formalism for Axion field and  have discussed the NES formalism in detail.  This result has been used to  derive the power spectrum by computing the two point quantum correlation function from mean square vacuum fluctuation. Finally, section \ref{x4} has been devoted to summery and conclusion and future prospects . In Figure~(\ref{fzaa}), we have presented a schematic diagram for the computation algorithm of long range effect of cosmological correlation function from quantum entanglement of axion in de Sitter open hyperbolic chart.
        \section{\textcolor{blue}{Wave function of axion in open chart}}
        \label{3a}
 We briefly review here, for sake of completeness, the background geometry and the results for wave function of the axion field. 
        \subsection{Background geometry}
         \label{3av1}
  		    \begin{figure*}[htb]
  		    \centering
  		    {
  		        \includegraphics[width=15.5cm,height=18cm] {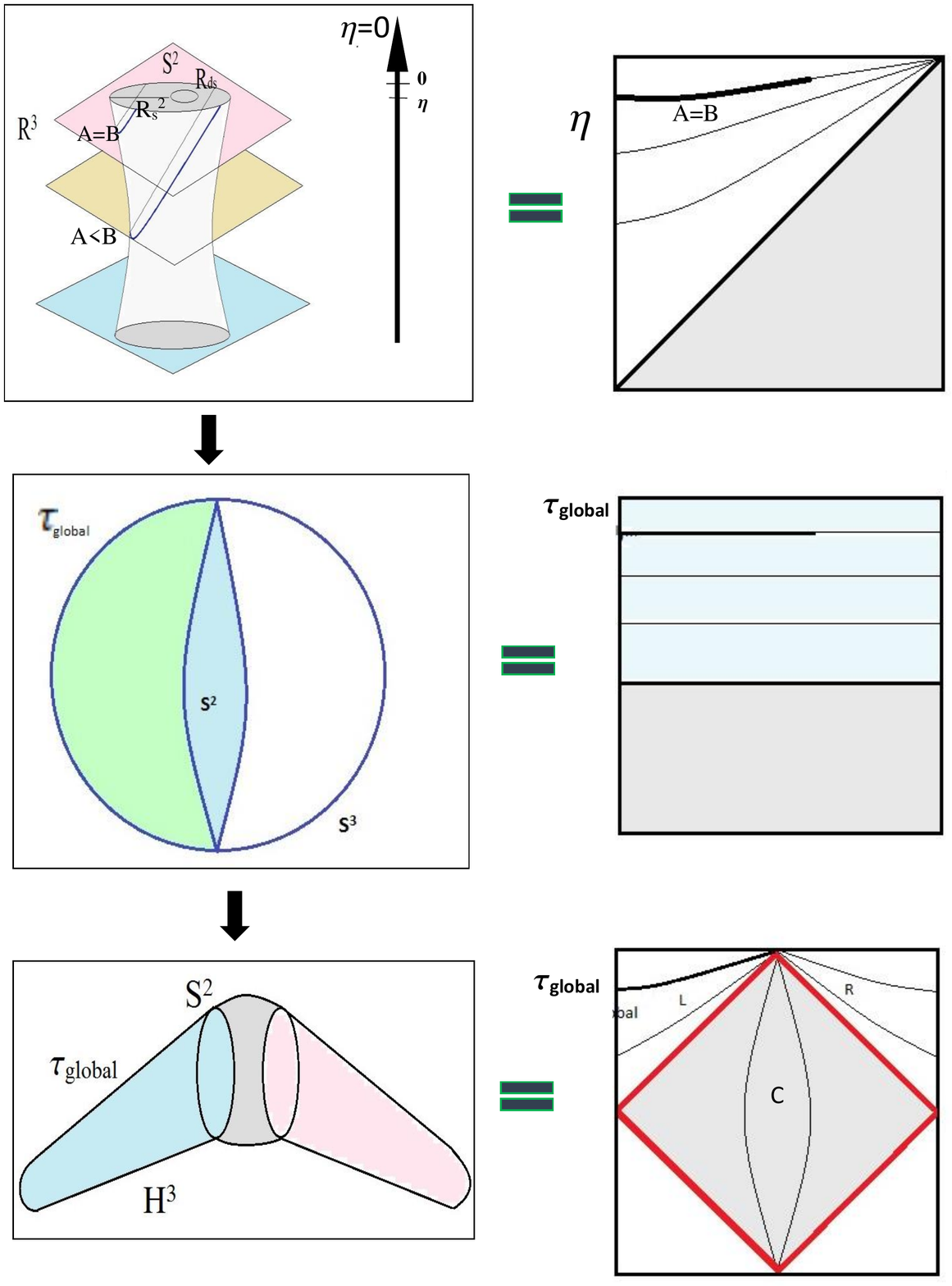}
  		    }
  		    \caption[Optional caption for list of figures]{Schematic diagram for the geometrical construction and underlying symmetries of the bipartite quantum field theoretic system of de Sitter hyperbolic open chart. Corresponding Penrose
  		    diagrams are also drawn for completeness.}
  		    \label{fbn}
  		    \end{figure*}
                 We consider a time preserving space-like hypersurface ${\bf S^2}$ in the open hyperbolic chart of the de Sitter space.. As a result ${\bf S^2}$ is divided into two sub regions-interior and exterior which are identified  by \textcolor{purple}{\bf RI} ($\equiv$\textcolor{purple}{\bf L})/ \textcolor{red}{\bf RII} ($\equiv$\textcolor{red}{\bf R}). In terms of the Lorentzian signature an open chart in de Sitter space is described by three different subregions : 
                \bea
                                \label{x2xx}
                                \displaystyle \textcolor{red}{\bf R(=RII)}/\textcolor{purple}{\bf L(=RI)}&:&\displaystyle\small\left\{\begin{array}{ll}
                                	\displaystyle \tau_{\rm E}=\pm \frac{\pi}{2}\mp it_{\bf R/L}~~~~~ &
                                	\mbox{\small { \textcolor{red}{\bf  $t_{\bf R}\geq 0$/$t_{\bf L}\geq 0$}}}  
                                	\\ 
                                	\displaystyle \rho_{\rm E}=-ir_{\bf R/L} & \mbox{\small {  \textcolor{red}{\bf  $r_{\bf R}\geq 0$/ $r_{\bf L}\geq 0$}}}~~~~~~~~\\
    \displaystyle ds^2_{\bf R/L}=\frac{1}{H^2}\left[-dt^2_{\bf R/L}+\sinh^2t_{\bf R/L}\left(dr^2_{\bf R/L}+\sinh^2r_{\bf R//L}~d\Omega^2_{\bf 2}\right)\right]                            	
                                \end{array}
                                \right.\\
                                \label{x3}
                                \displaystyle \textcolor{blue}{\bf C}&:&\displaystyle\small\left\{\begin{array}{ll}
                                	\displaystyle \tau_{\rm E}=t_{\bf C}~~~~ &
                                	\mbox{\small {\textcolor{blue}{\bf  $-\frac{\pi}{2}\leq t_{\bf C}\leq \frac{\pi}{2}$}}}  
                                	\\ 
                                	\displaystyle \rho_{\rm E}=\frac{\pi}{2}-ir_{\bf C} & \mbox{\small { \textcolor{blue}{\bf  $-\infty<r_{\bf c}< \infty$}}}.~~~~~~~~~~~~~\\ \displaystyle ds^2_{\bf C}=\frac{1}{H^2}\left[dt^2_{\bf C}+\cos^2t_{\bf C}\left(-dr^2_{\bf C}+\cosh^2r_{\bf C}~d\Omega^2_{\bf 2}\right)\right] \end{array} 
                                \right.
                                                  \eea                                                            
 where $H=\dot{a}/a$ is the Hubble parameter and $d\Omega^2_{\bf 2}$ represents angular part of the metric on ${\bf S}^2$. Now let us assume that the total Hilbert space of the local quantum mechanical system is described by ${\bf \cal H}$, which can be written using bipartite decomposition in a direct product space as,
                ${\bf \cal H}={\bf \cal H}_{\bf INT}\otimes {\bf \cal H}_{\bf EXT}$.
                Here ${\bf \cal H}_{\bf INT}$ and ${\bf \cal H}_{\bf EXT}$ are the Hilbert space associated with interior and exterior region and describe the localised modes in \textcolor{purple}{\bf RI}/ \textcolor{red}{\bf RII} respectively. 
                
      In Figure~(\ref{fbn}) we have shown the schematic diagram for the geometrical construction and underlying symmetries of the bipartite quantum field theoretic system of de Sitter hyperbolic open chart. Corresponding Penrose
  		    diagrams are also drawn for completeness.         
         \subsection{Wave function for Axion using Bunch Davies vacuum}
         \label{3aq}
    \begin{figure*}[htb]
    \centering
        \includegraphics[width=13.2cm,height=7.5cm] {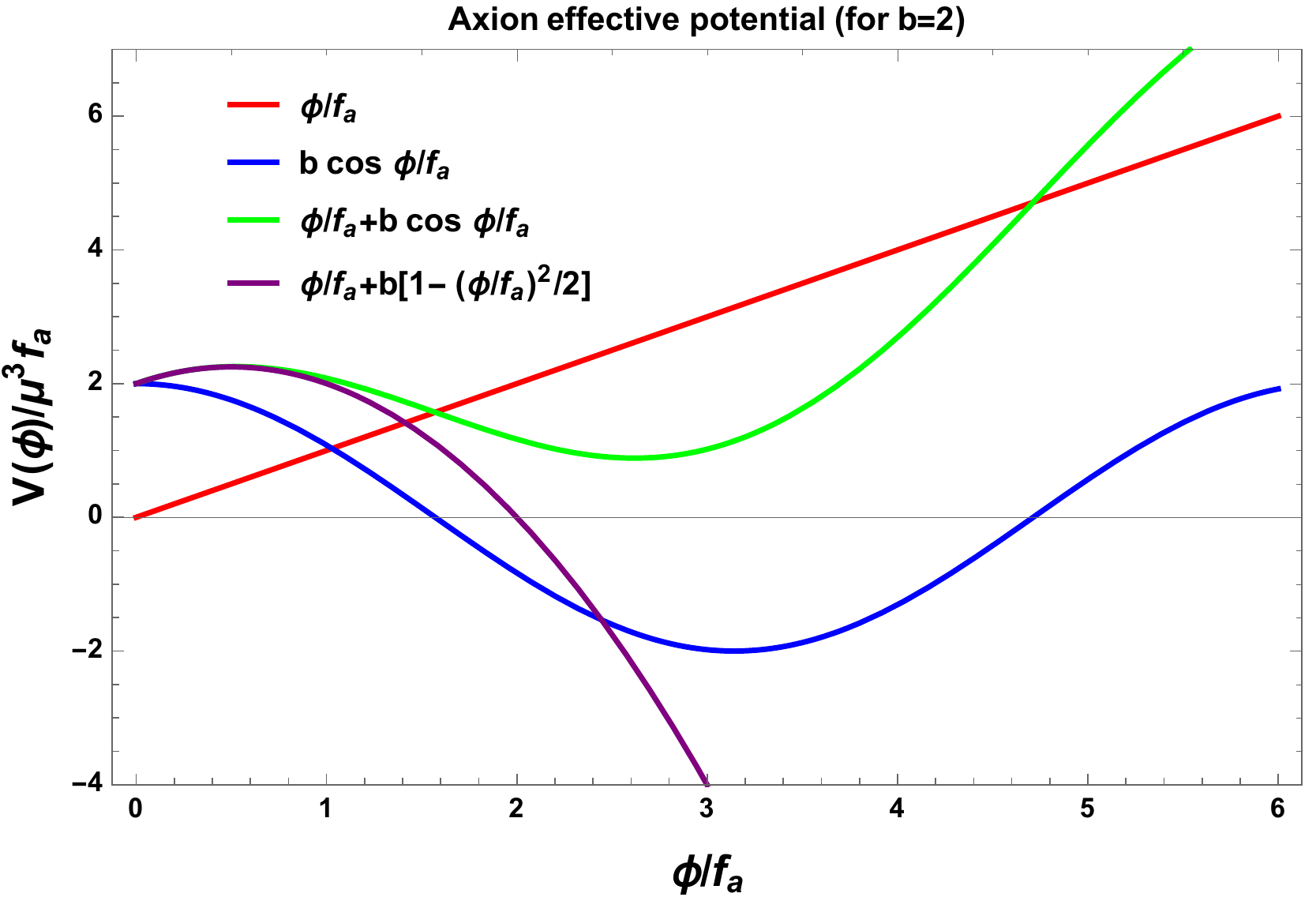}
    \caption[Optional caption for list of figures]{Behaviour of the axion effective potential obtained from {\bf Type IIB} String Theory with respect to the dimensionless field value $\phi/f_a$, where $f_a$ is the axion decay constant.} 
    \label{dfs}
    \end{figure*}
        Though our prime objective is to compute the cosmological correlation functions 
        for axion field in de Sitter space, we need the results for the wave function of the axion field in the just mentioned geometrical set up.  Note that he axion field under consideration is coming from RR sector of {\bf Type IIB} string theory compactified on ${\bf CY^3}$ in presence of ${\bf NS~5}$ brane \cite{Svrcek:2006yi,Panda:2010uq}. The effective action for the axion field is given by \cite{Panda:2010uq}:
        \bea\label{axi}  S_{axion}&=& \int d^{4}x \sqrt{-g}\left[-\frac{1}{2}(\partial \phi)^2 +\mu^3\left\{\phi+bf_{a}\cos\left(\frac{\phi}{f_{a}}\right)\right\}\right],\eea
        where $\mu^3$ is the mass scale, $f_a$ is axion decay constant and the parameter $b$ is defined as,  
        $b= \Lambda^4_{G}/\mu^3 f_{a}.$ Here $\Lambda_{G}$ depend on the string coupling $g_s$, slope parameter $\alpha^{'}$ and details of SUSY breaking parameter. For $\phi<<f_a$, effective potential for axion can be expressed as:
        	\bea\label{axiz2} V(\phi)&\approx&\mu^3\left(bf_{a}+\phi\right)-\frac{m^2_{axion}}{2}\phi^2,\eea
        	where we introduce the effective mass of the axion as,
        	$m^2_{axion}=\mu^3 b/f_{a}=\Lambda^4_{G}/f^2_{a}.$
        	Here axion decay constant follow a (conformal) time dependent profile, which is explicitly mentioned in refs.~\cite{Maldacena:2015bha,Choudhury:2016cso,Choudhury:2016pfr}.
        	
        	In Figure~(\ref{dfs}) we have explicitly presented the behaviour of the above axion potential with respect to the dimensionless field value $\phi/f_a$.
			    \begin{figure*}[htb]
			    \centering
			    {
			        \includegraphics[width=18.5cm,height=10cm] {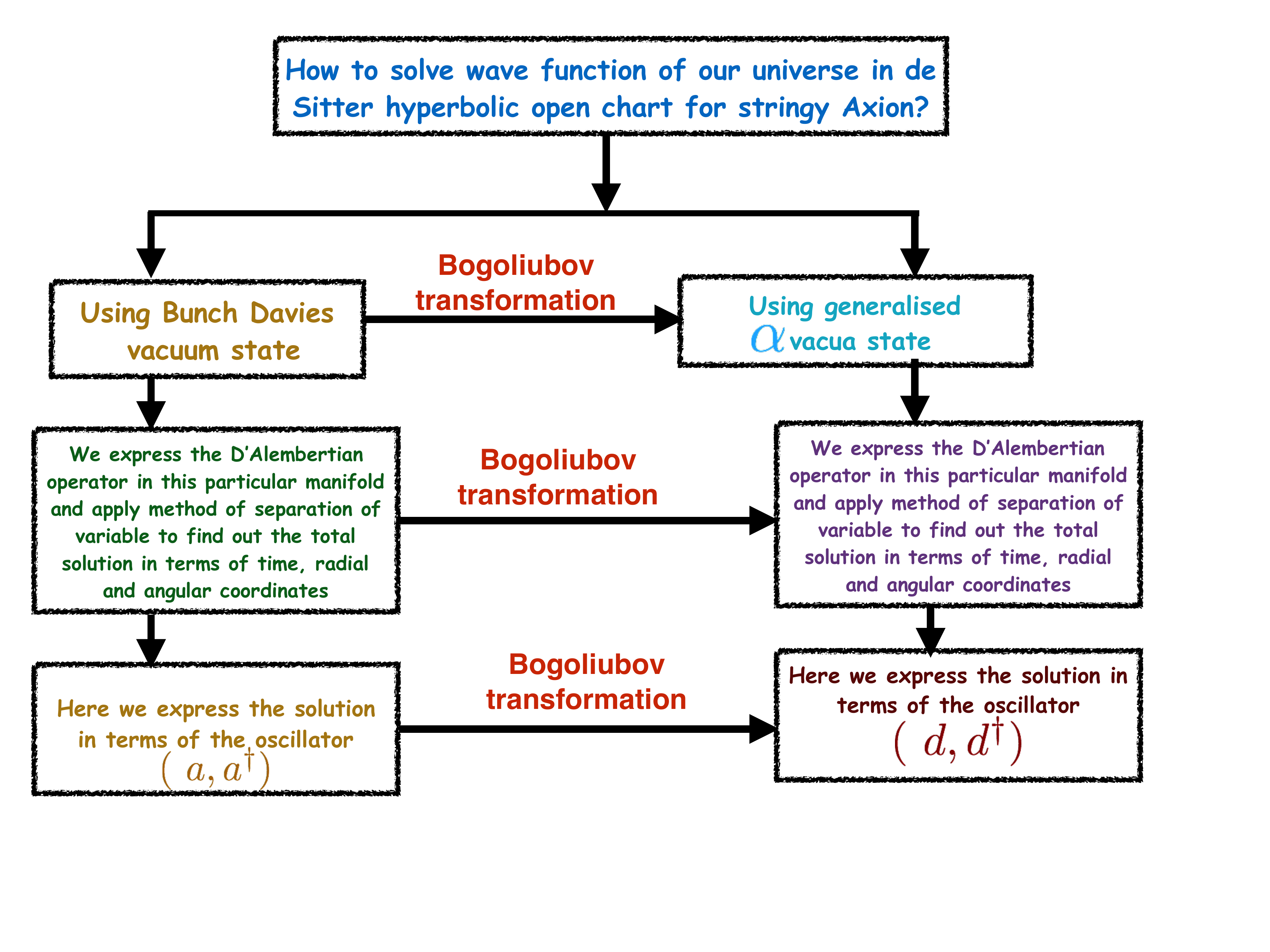}
			    }
			    \caption[Optional caption for list of figures]{Schematic diagram for the computation algorithm of solving the wave function of our universe in de Sitter hyperbolic open chart for stringy axion.} 
			    \label{fzaax}
			    \end{figure*}
        
        Further using Eqn~(\ref{axi}) the field equation of motion for the axion can be written as:
        \bea \left[\frac{1}{a^3(t)}\partial_{t}\left(a^3(t)\partial_{t}\right)-\frac{1}{H^2a^2(t)}\widetilde{\bf L^2}_{\bf H^3}+m^2_{axion}\right]\phi&=&\mu^3,\eea
       where the scale factor $a(t)$ in de Sitter open chart is given by, $a(t)=\sinh t/H$ and $H$ is the Hubble parameter, which is in principle can be time-dependent. But in the global patch of De Sitter space, it appears that the Hubble parameter $H$ can be treated as approximately a constant throughout the evolutionary time scale of our universe and the value is fixed at very high energy scale, $10^{16}~GeV$ at which the inflation and cosmological particle production (in the present context axion production) at very early universe are occurring. This value of the Hubble parameter is appearing from the observational constraint from Planck 2018 on the tensor-to-scalar ratio, which is actually a very important quantity in cosmology and determines the existence of primordial gravitational wave fluctuations at the very early time scale of our universe.
       
       Additionally, it is important to note that, the Laplacian operator $\widetilde{\bf L^2}_{\bf H^3}$, which is defined in the ${\bf H^3}$ geometry can be written as:
              \bea \widetilde{\bf L^2}_{\bf H^3}&=&\frac{1}{\sinh^2r}\left[\partial_{r}\left(\sinh^2r~\partial_{r}\right)+\frac{1}{\sin\theta}\partial_{\theta}\left(\sin\theta~\partial_{\theta}\right)+\frac{1}{\sin^2\theta}\partial^2_{\phi}\right],\eea     
                                    which satisfy the following eigenvalue equation:
              \bea \widetilde{\bf L^2}_{\bf H^3}{\rm\cal Y}_{plm}(r,\theta,\phi)&=&-(1+p^2){\rm\cal Y}_{plm}(r,\theta,\phi).     \eea
              Here ${\cal Y}_{plm}(r,\theta,\phi)$ represents orthonormal eigenfunctions which can be written in terms of a radial and angular part as:
                      \bea  {\cal Y}_{plm}(r,\theta,\phi)&=&\frac{\Gamma\left(ip+l+1\right)}{\Gamma\left(ip+1\right)}~\frac{p}{\sqrt{\sinh r}}~{\cal P}^{-\left(l+\frac{1}{2}\right)}_{\left(ip-\frac{1}{2}\right)}\left(\cosh r\right)Y_{lm}(\theta,\phi),\eea          
              where $Y_{lm}(\theta,\phi)$ is the spherical harmonics. 
             Consequently, the total solution of the equations of motion can be written as:
              \bea \Phi(t,r,\theta,\phi)&=&\sum_{\sigma=\pm 1}\sum_{Q=p,l,m}\left[a_{Q}{\cal V}_{Q}(t,r,\theta,\phi)+a^{\dagger}_{Q}{\cal V}^{*}_{Q}(t,r,\theta,\phi)\right]~~~~,\eea  
              Here the total solution ${\cal V}_{Q}(t,r,\theta,\phi)$ for Bunch Davies vacuum can be expressed as:                                    \bea {\cal V}_{Q}(t,r,\theta,\phi)&=&\frac{1}{a(t)}\chi_{p,\sigma}(t){\cal Y}_{plm}(r,\theta,\phi)=\frac{H}{\sinh t}\chi_{p,\sigma}(t){\cal Y}_{plm}(r,\theta,\phi),~~~~~~~\eea 
              where $\chi_{p,\sigma}(t)$ forms a complete set of positive frequency function. Also this can be written as a sum of complementary ($\chi^{(c)}_{p,\sigma}(t)$) and particular integral  ($\chi^{(p)}_{p,\sigma}(t)$) part, as given by:
              \bea \chi_{p,\sigma}(t)&=&\chi^{(c)}_{p,\sigma}(t)+\chi^{(p)}_{p,\sigma}(t).\eea
             Explicitly the solution for the 
       complementary part and the particular integral part can be expressed as:
     \bea
    \label{ccdzzz}
                             \displaystyle  \chi^{(c)}_{p,\sigma}(t)&=&\footnotesize\displaystyle\left\{\begin{array}{ll}
                            \displaystyle \frac{1}{2\sinh\pi p}\left[\frac{\left(e^{\pi p}-i\sigma~e^{-i\pi\nu}\right)}{\Gamma\left(\nu+\frac{1}{2}+ip\right)}{\cal P}^{ip}_{\left(\nu-\frac{1}{2}\right)}(\cosh t_{\bf R}) 
                            -\frac{\left(e^{-\pi p}-i\sigma~e^{-i\pi\nu}\right)}{\Gamma\left(\nu+\frac{1}{2}-ip\right)}{\cal P}^{-ip}_{\left(\nu-\frac{1}{2}\right)}(\cosh t_{\bf R})\right]~~ &
                                                                                      \mbox{\small {\textcolor{red}{\bf for R}}}  
                                                                                     \\ 
                                     \displaystyle \frac{\sigma}{2\sinh\pi p}\left[\frac{\left(e^{\pi p}-i\sigma~e^{-i\pi\nu}\right)}{\Gamma\left(\nu+\frac{1}{2}+ip\right)}{\cal P}^{ip}_{\left(\nu-\frac{1}{2}\right)}(\cosh t_{\bf L}) 
                                     -\frac{\left(e^{-\pi p}-i\sigma~e^{-i\pi\nu}\right)}{\Gamma\left(\nu+\frac{1}{2}-ip\right)}{\cal P}^{-ip}_{\left(\nu-\frac{1}{2}\right)}(\cosh t_{\bf L})\right] & \mbox{\small {\textcolor{red}{\bf for L}}},~~
                                                                                               \end{array}                          \right.                                                                                       \\  \chi^{(p)}_{p,\sigma}(t)&=&\sinh^2 t\sum^{\infty}_{n=0}\frac{1}{\left(p^2-p^2_{n}\right)}\chi^{(c)}_{p_{n},\sigma}(t)\int dt^{'}~\chi^{(c)}_{p_{n},\sigma}(t^{'})~\mu^3~.    \eea                                                                                            
    where 
    the parameter $\nu$ is defined as:
    \bea \nu&=&\sqrt{\frac{9}{4}-\frac{m^2_{axion}}{H^2}}=\sqrt{\frac{9}{4}-\frac{\mu^3 b}{f_a H^2}}=\sqrt{\frac{9}{4}-\frac{\Lambda^4_G}{f^2_a H^2}}.\eea
  	In Figure~(\ref{fzaax}) we have given a schematic diagram for the computation algorithm of solving the wave function of our universe in de Sitter hyperbolic open chart for stringy axion.

     \subsection{Wave function for Axion using $\alpha$ vacua}
     \label{3aq1}                               
    Here we use two subspaces in CPT invariant ${\bf SO(1,4)}$ isometric de Sitter space, which is identified as \textcolor{red}{\bf RI} and \textcolor{red}{\bf RII} respectively. Use the result obtained for Bunch Davies vacuum, and performing a  Bogoliubov transformation the mode functions for the $\alpha$-vacua can be expressed as:
\bea \label{ass5} \Phi(r,t,\theta,\phi)&=&\int^{\infty}_{0} dp \sum_{\sigma=\pm 1}\sum^{\infty}_{l=0}\sum^{+l}_{m=-l}\left[d_{\sigma plm}{\cal F}^{(\alpha)}_{\sigma plm}(r,t,\theta,\phi)+d^{\dagger}_{\sigma plm}({\cal F}^{(\alpha)}_{\sigma plm})^{*}(r,t,\theta,\phi)\right],~~~~~~ \eea
where the $\alpha$-vacua state are defined as:
\bea  d_{\sigma p l m}|\alpha\rangle&=&0~~~~~~~~~~ \forall \sigma=(+1,-1);0<p<\infty;l=0,\cdots,\infty,m=-l,\cdots,+l.~~~~~~~~~~~ \eea
In this context, the $\alpha$-vacua mode function ${\cal F}^{(\alpha)}_{\sigma plm}$ can be expressed in terms of Bunch Davies mode function ${\cal V}_{\sigma plm}(r,t,\theta,\phi)$ using Bogoliubov transformation as:
\bea \label{ass3} {\cal F}^{(\alpha)}_{\sigma plm}&=&\left[\cosh\alpha~{\cal V}_{\sigma plm}(r,t,\theta,\phi)+\sinh\alpha~{\cal V}^{*}_{\sigma plm}(r,t,\theta,\phi)\right].\eea 
Here ${\cal V}_{\sigma plm}(r,t,\theta,\phi)$ is the Bunch Davies vacuum states, which is defined as:
\bea \label{ass4} {\cal V}_{\sigma plm}(r,t,\theta,\phi)&=&\frac{H}{\sinh t}\chi_{p,\sigma}(t){\cal Y}_{plm}(r,\theta,\phi).\eea 
After substituting Eq~(\ref{ass3}) and Eq~(\ref{ass4}) in Eq~(\ref{ass5}) we get the following expression for the wave function:
\bea \Phi(r,t,\theta,\phi)&=&\frac{H}{\sinh t}\int^{\infty}_{0} dp \sum_{\sigma=\pm 1}\sum^{p-1}_{l=0}\sum^{+l}_{m=-l}\left[d_{\sigma plm}\cosh\alpha~\chi_{p,\sigma}(t)+d^{\dagger}_{\sigma plm}\sinh\alpha~\chi^{*}_{p,\sigma}(t)\right]{\cal Y}_{plm}(r,\theta,\phi),~~~~~~~~~~ \eea
Finally, the solution of the time dependent part of the wave function can be recast as:
     \bea
                                             \label{cvvcxcvb}
                                    \displaystyle \boxed{\chi_{p,\sigma}(t)=\sum_{q={\bf R},{\bf L}}\left\{\underbrace{\frac{1}{{\cal N}_{p}}\left[\alpha^{\sigma}_{q}~{\cal P}^{q}+\beta^{\sigma}_{q}~{\cal P}^{q*}\right]}_{\textcolor{red}{\bf Complementary~solution}}+\underbrace{\sum^{\infty}_{n=0}\frac{1}{{\cal N}_{p_n}\left(p^2_n-p^2\right)}\left[\bar{\alpha}^{\sigma}_{q,n}~\bar{\cal P}^{q,n}+\bar{\beta}^{\sigma}_{q,n}~\bar{\cal P}^{q*,n}\right]}_{\textcolor{red}{\bf Particular~solution}}\right\}\forall \sigma=\pm 1}~~~~~~~\eea 
                                    where we use the following shorthand notation:
                                     \bea\bar{\cal P}^{q,n}&=& \sinh^2t~ \int dt^{'}~\chi^{(c)}_{p_n,\sigma,q}(t^{'})~\mu^3~ {\cal P}^{q,n}.\eea
   Here we also use the shorthand notations ${\cal P}^{q}$, ${\cal P}^{q,n}$, for the Legendre polynomial. Also the coefficient functions $(\alpha^{\sigma}_{q}, \beta^{\sigma}_{q})$ and $(\alpha^{\sigma}_{q,n}, \beta^{\sigma}_{q,n})$, normalization constants ${\cal N}_{p}$, ${\cal N}_{p_n}$  for the complementary and particular part of the solution which are defined as:
   \bea {\cal N}_{p}&=& \frac{4\sinh \pi p}{\sqrt{\pi}}\frac{\sqrt{\cosh \pi p-\sigma\sin \pi \nu}}{|\Gamma\left(\nu+ip+\frac{1}{2}\right)|},\\
  {\cal N}_{p,(n)}&=& \frac{4\sinh \pi p_n}{\sqrt{\pi}}\frac{\sqrt{\cosh \pi p_n-\sigma\sin \pi \nu}}{|\Gamma\left(\nu+ip_n+\frac{1}{2}\right)|}. \eea

 \section{\textcolor{blue}{Cosmological spectrum of quantum vacuum fluctuation}}   
  \label{fg1}     
   In this section, we present our computation of the spectrum of Bunch Davies vacuum and $\alpha$ vacua fluctuation from two point correlation function . We will be discussing  the computation of two point correlation function and their associated cosmological spectra from three completely different formalisms:-
\begin{enumerate}

\item \underline{\textcolor{red}{\bf Field operator expansion (FOE) method}}:\\
This method is useful for  entangled quantum states with the wave function of the de Sitter universe for   Bunch Davies and most generalised $\alpha$ vacua. Technically this formalism is based on the wave function $\chi^{\cal I}$ which we will explicitly derive .  The cosmological spectrum is characterised by the two point correlation function and their associated power spectrum. Using such entangled state in this formalism one can construct the usual density matrix for Bunch Davies and most generalised $\alpha$ vacua. 

\item \underline{\textcolor{red}{\bf Reduced density matrix (RDM) formalism}}:\\
 This formalism is helpful for mixed quantum states and is useful for the construction of reduced density matrix in a diagonalised representation of Bunch Davies and  $\alpha$ vacua by tracing over the all possible degrees of freedom from the region \textcolor{red}{\bf R}. Technically the formalism is based on the wave function $\psi^{\cal I}$ which we explicitly derive.
 
 \item  \underline{\textcolor{red}{\bf Non entangled state (NES) formalism}}:\\
 This formalism in presence of non entangled quantum state which deals with the construction of wave function in the region \textcolor{red}{\bf L} in which the total universe is described. Here we also use Bunch Davies and most generalised $\alpha$ vacua in the region \textcolor{red}{\bf L}. Technically this formalism is based on the wave function ${\cal \phi}^{\cal I}$ which we explicitly derive in this paper.

 \end{enumerate}      
  We will now derive the expression for the mean square fluctuation considering both Bunch Davies vacuum and $\alpha$ vacua using the results presented in the previous section. For this computation we will follow the steps which are  outlined below:
   \begin{enumerate}
   	\item First of all, we trace out all contributions which belong to the \textcolor{red}{\bf R} region. As a result the required field operator is only defined in the \textcolor{red}{\bf L} region. This method we use in FOE formalism where the  quantum states for \textcolor{red}{\bf L} and \textcolor{red}{\bf R} region are entangled with each other. On the other hand, doing a partial trace over region \textcolor{red}{\bf R} one can construct reduced density matrix  which leads to RDM formalism. Instead, if we use the non entangled quantum state and compute the wave function solely in \textcolor{red}{\bf L} region we will be lead to the NES formalism.  Note that all of these three methods are used to compute mean square vacuum fluctuation or more precisely the quantum mechanical computation of two point correlation function for axion and the associated  power spectrum.
   	
   	\item Instead of doing the computation in $|{\bf L}\rangle$ basis we use a new basis $|{\bf L}^{'}\rangle$,  obtained by applying Bogoliubov transformation in $|{\bf L}\rangle$. Consequently the field operators will act on $|{\bf L}^{'}\rangle$ and the FOE method is developed in this transformed basis. On the other hand, as mentioned earlier it will appear in the expression for the reduced density matrix to be  used in the RDM formalism. But in the NES formalism this  transformation is not very useful  since in this case the total wave function  is solely described by the quantum mechanical state appearing in the \textcolor{red}{\bf L} region and the corresponding Hilbert space is spanned by only $|{\bf L}\rangle$ which forms a complete basis.
   	
   	\item Further, we will compute the expressions for the mean square quantum vacuum fluctuation and  the corresponding cosmological power spectrum after horizon exit using all the three formalisms i.e. FOE, RDM and NES. We will finally consider two limiting situations : long wave length and short wave length approximation for the computation of the power spectrum. 
   \end{enumerate}
 
   \subsection{Quantum vacuum fluctuation using field operator expansion (FOE)  (with entangled state)} 
   \subsubsection{Wave function in field operator expansion (FOE)}
   \label{x1a}
   
			    \begin{figure*}[htb]
			    \centering
			    {
			        \includegraphics[width=18.5cm,height=10cm] {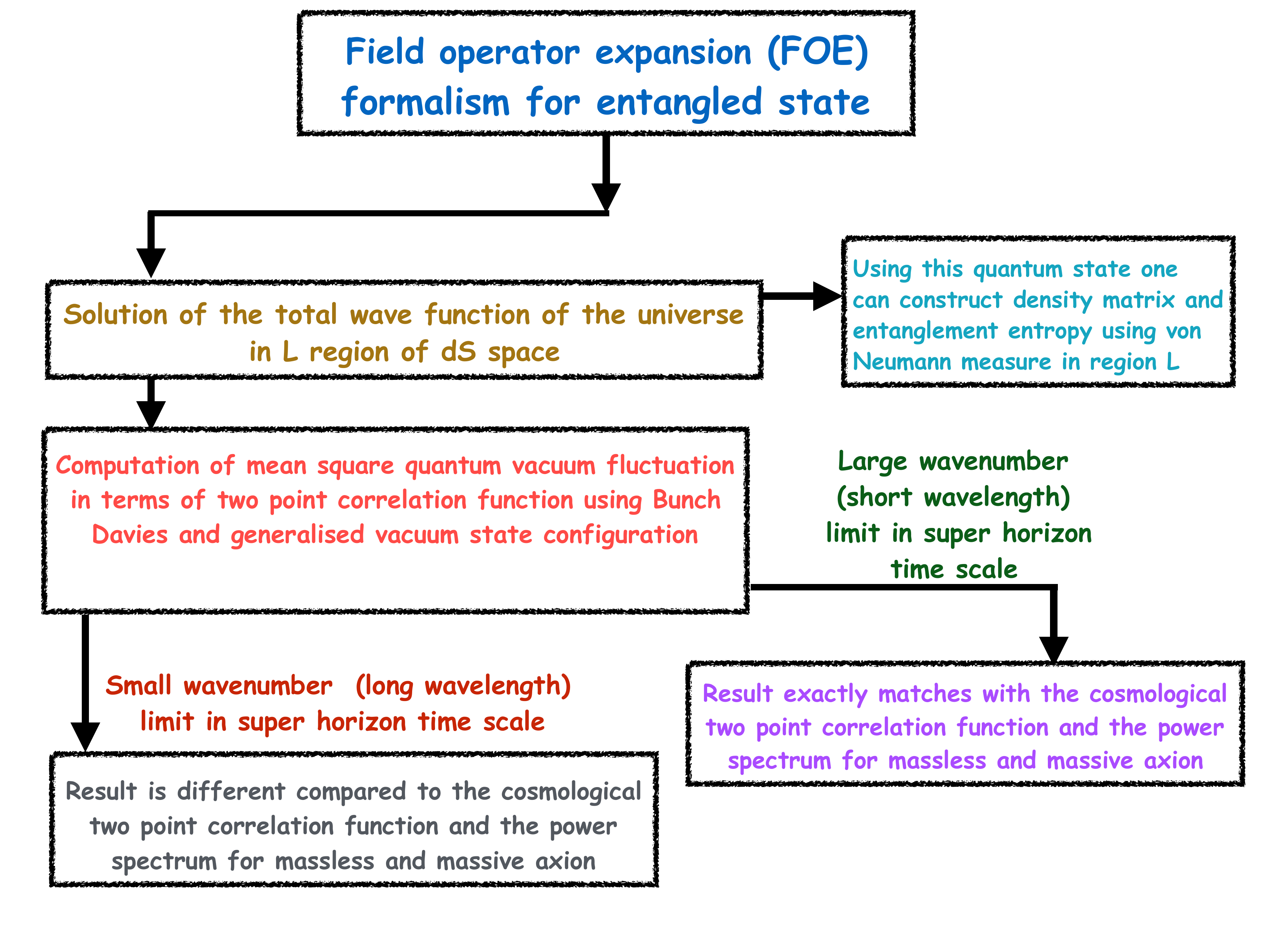}
			    }
			    \caption[Optional caption for list of figures]{Schematic diagram for the computation algorithm of field operator expansion method for entangled state of axion in de Sitter hyperbolic open chart.} 
			    \label{fzaa4}
			    \end{figure*}
			    
   Let us first compute the spectrum of vacuum fluctuation using field operator expansion (FOE). In figure~(\ref{fzaa4}) we have presented a schematic diagram for the computation algorithm of field operator expansion method for entangled state of axion in de Sitter hyperbolic open chart. To compute the vacuum fluctuation using FOE, we focus only with the left region \textcolor{red}{\bf L} as it is completely symmetric to the right region \textcolor{red}{\bf R}.  We use the time dependent mode function for the left region \textcolor{red}{\bf L} which we have presented in section 2. Thus instead of getting a $(4\times 4)$ square matrix  (when both sectors are considered)  we have a $(4\times 2)$ matrix which appears in the solution of the field equation as:
   \bea  \widetilde{\chi^{I}}=\frac{1}{{\cal N}_p}\widetilde{\cal  M}^{I}_{\cal J}\widetilde{\cal P}^{\cal J}+\sum^{\infty}_{n=0}\frac{1}{{\cal N}_{p,(n)}}\widetilde{\left({\cal  M}_{(n)}\right)^{I}_{\cal J}}\widetilde{\cal P}^{\cal J}_{(n)},\eea
   where the index ${\cal J}=1,2$ is appearing for the contribution from region \textcolor{red}{\bf L}.  To write down the total solution in region \textcolor{red}{\bf L} we define the following matrices:
   \bea \widetilde{\cal M}^{I}_{\cal J}&=&\left(\begin{array}{ccc} \alpha^{\sigma}_{\bf L} &~~~ \beta^{\sigma}_{\bf L} \\ \beta^{\sigma^{*}}_{\bf L} &~~~ \alpha^{\sigma^{*}}_{\bf L}  \end{array}\right),~~~~~~\widetilde{\left({\cal M}_{(n)}\right)^{I}_{\cal J}}=\left(\begin{array}{ccc} \bar{\alpha}^{\sigma}_{{\bf L},n} &~~~ \bar{\beta}^{\sigma}_{{\bf L},n} \\ \bar{\beta}^{\sigma^{*}}_{{\bf L},n} &~~~ \bar{\alpha}^{\sigma^{*}}_{{\bf L},n}  \end{array}\right),~~~~~~\\
   \widetilde{\chi^{I}}&=&\left(\begin{array}{ccc} \chi^{\sigma}(t) \\ \chi^{\sigma*}(t),
   \end{array}\right),~~~~~~
   \widetilde{\cal P}^{\cal J}=\left(\begin{array}{ccc} \widetilde{\cal P}^{{\bf L}} \\ \widetilde{\cal P}^{{{\bf L}^*}},\\
   \end{array}\right),~~~~~~~ 
   \widetilde{\cal P}^{\cal J}_{(n)}=\left(\begin{array}{ccc} \widetilde{\cal P}^{{\bf L},n} \\ \widetilde{\cal P}^{{{\bf L}^*},n}\\
   \end{array}\right),\eea                where $\sigma=\pm 1$, $I=1,2,3,4$ and ${\cal J}=1,2$. 
   The Fourier mode of the field operator, which is also the total solution of the field equation for axion (in presence of source contribution) can be expressed as:
    \bea \widetilde{\Phi(t_{\bf L})}=\frac{H}{\sinh t_{\bf L}}Q_{I}\widetilde{\chi^{I}}=\frac{H}{\sinh t_{\bf L}}Q_{I}\left[\frac{1}{{\cal N}_p}\widetilde{\cal  M}^{I}_{\cal J}\widetilde{\cal P}^{J}+\sum^{\infty}_{n=0}\frac{1}{{\cal N}_{p,(n)}}\widetilde{\left({\cal  M}_{(n)}\right)^{I}_{\cal J}}\widetilde{\cal P}^{\cal J}_{(n)} \right],~~~~~~
    \eea                                   where the operator $Q_{I}$ represent a set of creation and annihilation operators which are defined (in section 2) for Bunch Davies vacuum ($\alpha=0$)  and $\alpha$ vacua ($\alpha\neq 0$) as:
    \bea Q_{I}\equiv \displaystyle\left\{\begin{array}{ll}
    	\displaystyle a_{I}=(a_{\sigma},
    		a^{\dagger}_{\sigma})=\left[a^{(c)}_{I}+\sum^{\infty}_{n=0}a^{(p)}_{I(n)}\right]~~~~~~~~~~~~~~~~~ &
    	\mbox{\small {\textcolor{red}{\bf for Bunch Davies vacuum}}}  
    	\\ 
    	\displaystyle d_{I}=(d_{\sigma},
    	d^{\dagger}_{\sigma})=\left[d^{(c)}_{I}+\sum^{\infty}_{n=0}d^{(p)}_{I(n)}\right]  ~~~~~~~ & \mbox{\small {\textcolor{red}{\bf for $\alpha$ vacua}}}.~~
    \end{array}
    \right.\eea
     Here we have labeled  the time coordinate $t$ by $t_{\bf L}$ since we are considering the left region \textcolor{red}{\bf L} only.
    
    To explicitly write down the expression for the amplitude of the normalized power spectrum, we start with the column matrix representation of the time dependent part of the solution of the wave function,  given by:
    \bea\label{gh1} \widetilde{\chi^{I}}&=&\left(\begin{array}{ccc} \chi^{\sigma}(t) \\ \chi^{\sigma*}(t)
    \end{array}\right)=\left(\begin{array}{ccc} {\cal A}^{\sigma}_{\bf L}\widetilde{\cal P}^{\bf L}+ {\cal B}^{\sigma}_{\bf L}\widetilde{\cal P}^{\bf L *}  \\ {\cal B}^{\sigma *}_{\bf L}\widetilde{\cal P}^{\bf L}+ {\cal A}^{\sigma *}_{\bf L}\widetilde{\cal P}^{\bf L *}
\end{array}\right)+\sum^{\infty}_{n=0}\left(\begin{array}{ccc} {\cal A}^{\sigma}_{{\bf L},(n)}\widetilde{\cal P}^{\bf L}_{(n)}+ {\cal B}^{\sigma}_{{\bf L},(n)}\widetilde{\cal P}^{\bf L *}_{(n)}  \\ {\cal B}^{\sigma *}_{{\bf L},(n)}\widetilde{\cal P}^{\bf L}_{(n)}+ {\cal A}^{\sigma *}_{{\bf L},(n)}\widetilde{\cal P}^{\bf L *}_{(n)}
\end{array}\right),\eea
where the entries of the column matrix for the complementary and particular integral part of the solution are given by the following expressions:
\bea {\cal A}^{\sigma}_{\bf L}&=& \frac{\alpha^{\sigma}_{\bf L}}{{\cal N}_p}=\sigma\frac{e^{\pi p}-i\sigma~e^{-i\pi\nu}}{{\cal N}_p\Gamma\left(\nu+ip+\frac{1}{2}\right)},\\ {\cal B}^{\sigma}_{\bf L}&=& \frac{\beta^{\sigma}_{\bf L}}{{\cal N}_p}=-\sigma\frac{e^{-\pi p}-i\sigma~e^{-i\pi\nu}}{{\cal N}_p\Gamma\left(\nu-ip+\frac{1}{2}\right)},~~~~~~~\\{\cal A}^{\sigma}_{{\bf L},(n)}&=& \frac{\alpha^{\sigma}_{\bf {\bf L},(n)}}{{\cal N}_{p,(n)}}=\sigma\frac{e^{\pi p_n}-i\sigma~e^{-i\pi\nu}}{{\cal N}_{p,(n)}\Gamma\left(\nu+ip_n+\frac{1}{2}\right)},\\{\cal B}^{\sigma}_{{\bf L},(n)}&=& \frac{\beta^{\sigma}_{{\bf L},(n)}}{{\cal N}_{p,(n)}}=-\sigma\frac{e^{-\pi p_n}-i\sigma~e^{-i\pi\nu}}{{\cal N}_{p,(n)}\Gamma\left(\nu-ip_n+\frac{1}{2}\right)}.\eea
 ${\cal N}_p$ and ${\cal N}_{p,(n)}$ in the above equations are  the normalization constants for the complementary part and particular integral part of the solution as defined section 2.
  
   \subsubsection{Two point correlation function}
   \label{x1b}
   To compute the expression for the two point correlation function for the vacuum fluctuation let us now concentrate on a single mode with fixed value of the ${\bf SO(3,1)}$ quantum numbers $p$, $l$ and $m$. As a result the mean square vacuum fluctuation of axion for any generalized arbitrary vacuum state ($|{\bf \Omega}\rangle $) can be expressed as:
    \bea \langle {\bf \Omega}|\widetilde{\Phi_{plm}(t_{\bf L})}\left(\widetilde{\Phi_{p^{'}l^{'}m^{'}}(t_{\bf L})}\right)^{\dagger}|{\bf \Omega}\rangle &=& \frac{H^2}{\sinh^2 t_{\bf L}}\langle {\bf \Omega}|\left[Q_{I}\widetilde{\chi^{I}}\right]_{plm}\left(\left[Q_{I}\widetilde{\chi^{I}}\right]_{p^{'}l^{'}m^{'}}\right)^{\dagger}|{\bf \Omega}\rangle.\eea
   Further explicitly writing the expression for the mean square vacuum fluctuation of axion for Bunch Davies vacuum we get the following simplified expressions:
   \bea \langle {\bf BD}|\widetilde{\Phi_{plm}(t_{\bf L})}\left(\widetilde{\Phi_{p^{'}l^{'}m^{'}}(t_{\bf L})}\right)^{\dagger}|{\bf BD}\rangle &=& \frac{H^2}{\sinh^2 t_{\bf L}}\langle {\bf BD}|\left[a_{I}\widetilde{\chi^{I}}\right]_{plm}\left(\left[a_{I}\widetilde{\chi^{I}}\right]_{p^{'}l^{'}m^{'}}\right)^{\dagger}|{\bf BD}\rangle\nonumber\\
   &=&\frac{H^2}{\sinh^2 t_{\bf L}}\sum_{\sigma=\pm 1}|\widetilde{\chi^{\sigma}}|^2 ~\delta(p-p^{'})~\delta_{l l^{'}}~\delta_{m m^{'}}\nonumber\\
   &\equiv&P_{\bf BD}(p,t_{\bf L}) ~\delta(p-p^{'})~\delta_{l l^{'}}~\delta_{m m^{'}},\eea
   where we define the amplitude of the normalized power spectrum of axion as:
   \bea {\cal P}_{\bf BD}(p,t_{\bf L})=\frac{p^3}{2\pi^2}~P_{\bf BD}(p,t_{\bf L})=\frac{p^3}{2\pi^2}~\frac{H^2}{\sinh^2 t_{\bf L}}\sum_{\sigma=\pm 1}|\widetilde{\chi^{\sigma}}|^2.\eea

Further using Eq~(\ref{gh1}) we compute the following expression, which is appearing in the expression for the amplitude of the normalized power spectrum:
\bea \label{df1} \sum_{\sigma=\pm 1}|\widetilde{\chi^{\sigma}}|^2=\sum_{\sigma=\pm 1}\left(\widetilde{\chi^{\sigma}}\right)^{\dagger}\widetilde{\chi^{\sigma}}&=&\left[\left(|{\cal A}^{\sigma}_{\bf L}|^2+|{\cal B}^{\sigma}_{\bf L}|^2\right)\widetilde{\cal P}^{\bf L}\widetilde{\cal P}^{{\bf L}*}+{\cal A}^{\sigma}_{\bf L}{\cal B}^{\sigma *}_{\bf L}\left(\widetilde{\cal P}^{\bf L}\right)^2+{\cal A}^{\sigma *}_{\bf L}{\cal B}^{\sigma}_{\bf L}\left(\widetilde{\cal P}^{{\bf L}*}\right)^2\right.\nonumber\\ && \left.
~+\sum^{\infty}_{n=0}\left\{\left({\cal A}^{\sigma}_{{\bf L}}{\cal A}^{\sigma*}_{{\bf L},(n)}+{\cal B}^{\sigma}_{{\bf L}}{\cal B}^{\sigma*}_{{\bf L},(n)}\right)\widetilde{\cal P}^{\bf L}\widetilde{\cal P}^{{\bf L}*}_{(n)}\right.\right.\nonumber\\&&\left.\left.~~~~~~~~~~~~~~~~+\left({\cal A}^{\sigma}_{{\bf L}}{\cal B}^{\sigma*}_{{\bf L},(n)}+{\cal A}^{\sigma}_{{\bf L},(n)}{\cal B}^{\sigma*}_{{\bf L}}\right)\widetilde{\cal P}^{\bf L}\widetilde{\cal P}^{\bf L}_{(n)}\right.\right.\nonumber\\&&\left.\left.~~~~~~~~~~~~~~~~+\left({\cal A}^{\sigma*}_{{\bf L},(n)}{\cal B}^{\sigma}_{{\bf L}}+{\cal A}^{\sigma*}_{{\bf L}}{\cal B}^{\sigma}_{{\bf L},(n)}\right)\widetilde{\cal P}^{{\bf L}*}_{(n)}\widetilde{\cal P}^{{\bf L}*}\right\}\right.\nonumber\\ && \left.
~+\sum^{\infty}_{n=0}\sum^{\infty}_{m=0}\left\{\left({\cal A}^{\sigma}_{{\bf L},(n)}{\cal A}^{\sigma*}_{{\bf L},(m)}+{\cal B}^{\sigma}_{{\bf L},(n)}{\cal B}^{\sigma*}_{{\bf L},(m)}\right)\widetilde{\cal P}^{\bf L}_{(n)}\widetilde{\cal P}^{{\bf L}*}_{(m)}\right.\right.\nonumber\\&&\left.\left.~~~~~~~~~~~~~~~~+{\cal A}^{\sigma}_{{\bf L},(n)}{\cal B}^{\sigma*}_{{\bf L},(m)}\widetilde{\cal P}^{\bf L}_{(n)}\widetilde{\cal P}^{\bf L}_{(m)}+{\cal A}^{\sigma*}_{{\bf L},(n)}{\cal B}^{\sigma}_{{\bf L},(m)}\widetilde{\cal P}^{{\bf L}*}_{(n)}\widetilde{\cal P}^{{\bf L}*}_{(m)}\right\}\right].~~~~~~~~~~~\eea
Using Eq~(\ref{df1}), the amplitude of the normalized power spectrum of axion from Bunch Davies vacuum can be expressed in all time scales of region \textcolor{red}{\bf L} as:
\bea \label{po1} {\cal P}_{\bf BD}(p,t_{\bf L})&=&\frac{p^3}{2\pi^2}~\frac{H^2}{\sinh^2 t_{\bf L}}\sum_{\sigma=\pm 1}|\widetilde{\chi^{\sigma}}|^2\nonumber\\
&=&\frac{p^3}{2\pi^2}~\frac{H^2}{\sinh^2 t_{\bf L}}\left[\left(|{\cal A}^{\sigma}_{\bf L}|^2+|{\cal B}^{\sigma}_{\bf L}|^2\right)\widetilde{\cal P}^{\bf L}\widetilde{\cal P}^{{\bf L}*}+{\cal A}^{\sigma}_{\bf L}{\cal B}^{\sigma *}_{\bf L}\left(\widetilde{\cal P}^{\bf L}\right)^2+{\cal A}^{\sigma *}_{\bf L}{\cal B}^{\sigma}_{\bf L}\left(\widetilde{\cal P}^{{\bf L}*}\right)^2\right.\nonumber\\ && \left.
~+\sum^{\infty}_{n=0}\left\{\left({\cal A}^{\sigma}_{{\bf L}}{\cal A}^{\sigma*}_{{\bf L},(n)}+{\cal B}^{\sigma}_{{\bf L}}{\cal B}^{\sigma*}_{{\bf L},(n)}\right)\widetilde{\cal P}^{\bf L}\widetilde{\cal P}^{{\bf L}*}_{(n)}\right.\right.\nonumber\\&&\left.\left.~~~~~~~~~~~~~~~~+\left({\cal A}^{\sigma}_{{\bf L}}{\cal B}^{\sigma*}_{{\bf L},(n)}+{\cal A}^{\sigma}_{{\bf L},(n)}{\cal B}^{\sigma*}_{{\bf L}}\right)\widetilde{\cal P}^{\bf L}\widetilde{\cal P}^{\bf L}_{(n)}\right.\right.\nonumber\\&&\left.\left.~~~~~~~~~~~~~~~~+\left({\cal A}^{\sigma*}_{{\bf L},(n)}{\cal B}^{\sigma}_{{\bf L}}+{\cal A}^{\sigma*}_{{\bf L}}{\cal B}^{\sigma}_{{\bf L},(n)}\right)\widetilde{\cal P}^{{\bf L}*}_{(n)}\widetilde{\cal P}^{{\bf L}*}\right\}\right.\nonumber\\ && \left.
~+\sum^{\infty}_{n=0}\sum^{\infty}_{m=0}\left\{\left({\cal A}^{\sigma}_{{\bf L},(n)}{\cal A}^{\sigma*}_{{\bf L},(m)}+{\cal B}^{\sigma}_{{\bf L},(n)}{\cal B}^{\sigma*}_{{\bf L},(m)}\right)\widetilde{\cal P}^{\bf L}_{(n)}\widetilde{\cal P}^{{\bf L}*}_{(m)}\right.\right.\nonumber\\&&\left.\left.~~~~~~~~~~~~~~~~+{\cal A}^{\sigma}_{{\bf L},(n)}{\cal B}^{\sigma*}_{{\bf L},(m)}\widetilde{\cal P}^{\bf L}_{(n)}\widetilde{\cal P}^{\bf L}_{(m)}+{\cal A}^{\sigma*}_{{\bf L},(n)}{\cal B}^{\sigma}_{{\bf L},(m)}\widetilde{\cal P}^{{\bf L}*}_{(n)}\widetilde{\cal P}^{{\bf L}*}_{(m)}\right\}\right].~~~~~~~~~~~\eea
However, it is not easy to extract any information from Eqn~(\ref{po1}) for  cosmological predictions. Hence, we consider the superhorizon time scales ($t_{\bf L}>>1$) of region \textcolor{red}{\bf L}. In such a case, the Legendre functions,  appearing in the complementary part and the particular integral part of the time dependent solution, can be approximated as :
\bea \left(\widetilde{\cal P}^{\bf L},\widetilde{\cal P}^{\bf L*}\right)&\equiv& P^{\pm ip}_{\nu-\frac{1}{2}}\left(\cosh t_{\bf L}\right)~\underrightarrow{t_{\bf L}>>1}~\frac{2^{\nu-\frac{1}{2}}\left(\cosh t_{\bf L}\right)^{\nu-\frac{1}{2}}\Gamma(\nu)}{\sqrt{\pi}\Gamma\left(\nu\mp ip +\frac{1}{2}\right)},\\
\left(\widetilde{\cal P}^{\bf L}_{(n)},\widetilde{\cal P}^{\bf L*}_{(n)}\right)&\equiv& P^{\pm ip_n}_{\nu-\frac{1}{2}}\left(\cosh t_{\bf L}\right)~\underrightarrow{t_{\bf L}>>1}~\frac{2^{\nu-\frac{1}{2}}\left(\cosh t_{\bf L}\right)^{\nu-\frac{1}{2}}\Gamma(\nu)}{\sqrt{\pi}\Gamma\left(\nu\mp ip_n +\frac{1}{2}\right)}.\eea
Consequently, in the superhorizon time scales ($t_{\bf L}>>1$) of region \textcolor{red}{\bf L} eqn~(\ref{df1}) can be further simplified as:
\bea \label{df2} \sum_{\sigma=\pm 1}|\widetilde{\chi^{\sigma}}|^2=\sum_{\sigma=\pm 1}\left(\widetilde{\chi^{\sigma}}\right)^{\dagger}\widetilde{\chi^{\sigma}}&~\underrightarrow{t_{\bf L}>>1}~& \widetilde{{\cal M}(p,\nu)}\left(\cosh t_{\bf L}\right)^{2\nu-1}\eea
where the time independent function $\widetilde{{\cal M}(p,\nu)}$ is defined as:
\bea\widetilde{{\cal M}(p,\nu)}&=&\frac{2^{2\nu-1}\left(\Gamma(\nu)\right)^2}{\pi}\times \sum_{\sigma=\pm 1}\left[\frac{\left(|{\cal A}^{\sigma}_{\bf L}|^2+|{\cal B}^{\sigma}_{\bf L}|^2\right)}{\left|\Gamma\left(\nu+ ip +\frac{1}{2}\right)\right|^2}+\frac{{\cal A}^{\sigma}_{\bf L}{\cal B}^{\sigma *}_{\bf L}}{\left(\Gamma\left(\nu- ip +\frac{1}{2}\right)\right)^2}+\frac{{\cal A}^{\sigma *}_{\bf L}{\cal B}^{\sigma}_{\bf L}}{\left(\Gamma\left(\nu+ ip +\frac{1}{2}\right)\right)^2}\right.\nonumber\\ && \left.
~+\sum^{\infty}_{n=0}\left\{\frac{\left({\cal A}^{\sigma}_{{\bf L}}{\cal A}^{\sigma*}_{{\bf L},(n)}+{\cal B}^{\sigma}_{{\bf L}}{\cal B}^{\sigma*}_{{\bf L},(n)}\right)}{\Gamma\left(\nu- ip +\frac{1}{2}\right)\Gamma\left(\nu+ ip_n +\frac{1}{2}\right)}+\frac{\left({\cal A}^{\sigma}_{{\bf L}}{\cal B}^{\sigma*}_{{\bf L},(n)}+{\cal A}^{\sigma}_{{\bf L},(n)}{\cal B}^{\sigma*}_{{\bf L}}\right)}{\Gamma\left(\nu- ip +\frac{1}{2}\right)\Gamma\left(\nu- ip_n +\frac{1}{2}\right)}\right.\right.\nonumber\\&&\left.\left.~~~~~~~~~~~~~~~~~~~~~~~~~~~~~~~~~~~~~~~~~~~~~~~~+\frac{\left({\cal A}^{\sigma*}_{{\bf L},(n)}{\cal B}^{\sigma}_{{\bf L}}+{\cal A}^{\sigma*}_{{\bf L}}{\cal B}^{\sigma}_{{\bf L},(n)}\right)}{\Gamma\left(\nu+ ip_n +\frac{1}{2}\right)\Gamma\left(\nu+ ip +\frac{1}{2}\right)}\right\}\right.\nonumber\\ && \left.
~+\sum^{\infty}_{n=0}\sum^{\infty}_{m=0}\left\{\frac{\left({\cal A}^{\sigma}_{{\bf L},(n)}{\cal A}^{\sigma*}_{{\bf L},(m)}+{\cal B}^{\sigma}_{{\bf L},(n)}{\cal B}^{\sigma*}_{{\bf L},(m)}\right)}{\Gamma\left(\nu- ip_n +\frac{1}{2}\right)\Gamma\left(\nu+ ip_m +\frac{1}{2}\right)}\right.\right.\nonumber\\&&\left.\left.~~~~~~~~~~~~~~~~~+\frac{{\cal A}^{\sigma}_{{\bf L},(n)}{\cal B}^{\sigma*}_{{\bf L},(m)}}{\Gamma\left(\nu- ip_n +\frac{1}{2}\right)\Gamma\left(\nu- ip_m +\frac{1}{2}\right)}+\frac{{\cal A}^{\sigma*}_{{\bf L},(n)}{\cal B}^{\sigma}_{{\bf L},(m)}}{\Gamma\left(\nu+ ip_n +\frac{1}{2}\right)\Gamma\left(\nu+ ip_m +\frac{1}{2}\right)}\right\}\right].~~~~~~~~~~~\eea
As a result, in the superhorizon time scales ($t_{\bf L}>>1$) of region \textcolor{red}{\bf L} the amplitude of the normalized power spectrum of axion from Bunch Davies vacuum can be expressed as:
\bea \label{po2} {\cal P}_{\bf BD}(p,t_{\bf L})&=&\frac{p^3}{2\pi^2}~\frac{H^2}{\sinh^2 t_{\bf L}}\sum_{\sigma=\pm 1}|\widetilde{\chi^{\sigma}}|^2~\underrightarrow{t_{\bf L}>>1}~\frac{p^3}{2\pi^2}~\left(\cosh t_{\bf L}\right)^{2\nu-3}~H^2\widetilde{{\cal M}(p,\nu)}.~~~~~~~~~~~\eea
Here, it is important to note that in the superhorizon time scales ($t_{\bf L}>>1$) of region \textcolor{red}{\bf L} if we consider the massless case where we fix the mass parameter to be $\nu=3/2$, then the time dependent contribution can be approximated as:
\bea \left(\frac{\left(\cosh t_{\bf L}\right)^{2\nu-1}}{\sinh^2 t_{\bf L}}\right)_{\nu=3/2}~\underrightarrow{t_{\bf L}>>1}~~~1.\eea 
Consequently, in the superhorizon time scales of region \textcolor{red}{\bf L} and for the massless axion case, the amplitude of the normalized power spectrum of axion from Bunch Davies vacuum can be expressed as:
\bea \label{po3} {\cal P}_{\bf BD}(p,t_{\bf L})&=&\frac{p^3}{2\pi^2}~\frac{H^2}{\sinh^2 t_{\bf L}}\sum_{\sigma=\pm 1}|\widetilde{\chi^{\sigma}}|^2~\underrightarrow{t_{\bf L}>>1,\nu=3/2}~\frac{p^3}{2\pi^2}~H^2\widetilde{{\cal M}(p,\nu=3/2)}.~~~~~~~~~~~\eea
This implies that in the massless case, the amplitude of the vacuum fluctuation gets frozen with respect to the time scale when the associated modes exit the horizon.

Further to infer the exact wave number dependence of the amplitude of the normalized power spectrum from Bunch Davies vacuum we need to know the behaviour of the power spectrum at very short wavelengths ($p,p_n>>1$). In this limit it is expected that the power spectrum  should match  the result obtained for spatially flat universe. Note that in the short wave length approximation the time independent function $\widetilde{{\cal M}(p>>1,\nu)}$ for any arbitrary mass parameter $\nu$ can be expressed as:
\bea\widetilde{{\cal M}(p>>1,\nu)}&=&\frac{2^{2(\nu-1)}\left(\Gamma(\nu)\right)^2}{p^3\pi}\widetilde{{\cal G}(p>>1)},~~~~~~~~~~~\eea
where we have defined a new function $\widetilde{{\cal G}(p>>1)}$ in the short wave length limit as :
\bea\widetilde{{\cal G}(p)}&=&\frac{1}{\left(1+\frac{1}{82944p^4}\right)}\times \nonumber\\
&&\left[\left(1+e^{-2\pi p}\right)^2+\sum^{\infty}_{n=0}\left(\frac{p}{p_n}\right)^{\frac{3}{2}}\frac{\sqrt{1+\frac{1}{82944p^4}}}{\sqrt{1+\frac{1}{82944p^4_n}}}\left(1+2\left(e^{-2\pi p}+e^{-2\pi p_n}\right)+e^{-2\pi (p+p_n)}\right)\right.\nonumber\\ && \left.
~+\sum^{\infty}_{n=0}\sum^{\infty}_{m=0}\frac{p^3}{\left(p_n p_m\right)^{3/2}}\frac{\left(1+\frac{1}{82944p^4}\right)}{\sqrt{1+\frac{1}{82944p^4_n}}\sqrt{1+\frac{1}{82944p^4_m}}}\left(1+e^{-\pi (p_m+p_n)}\right)^2\right].~~~~~~~~~~~\eea
            \begin{figure*}[htb]
            \centering
           \subfigure[Large wave number dependence of FOE power spectrum for $\alpha=0$.]{
               \includegraphics[width=7.7cm,height=7.5cm] {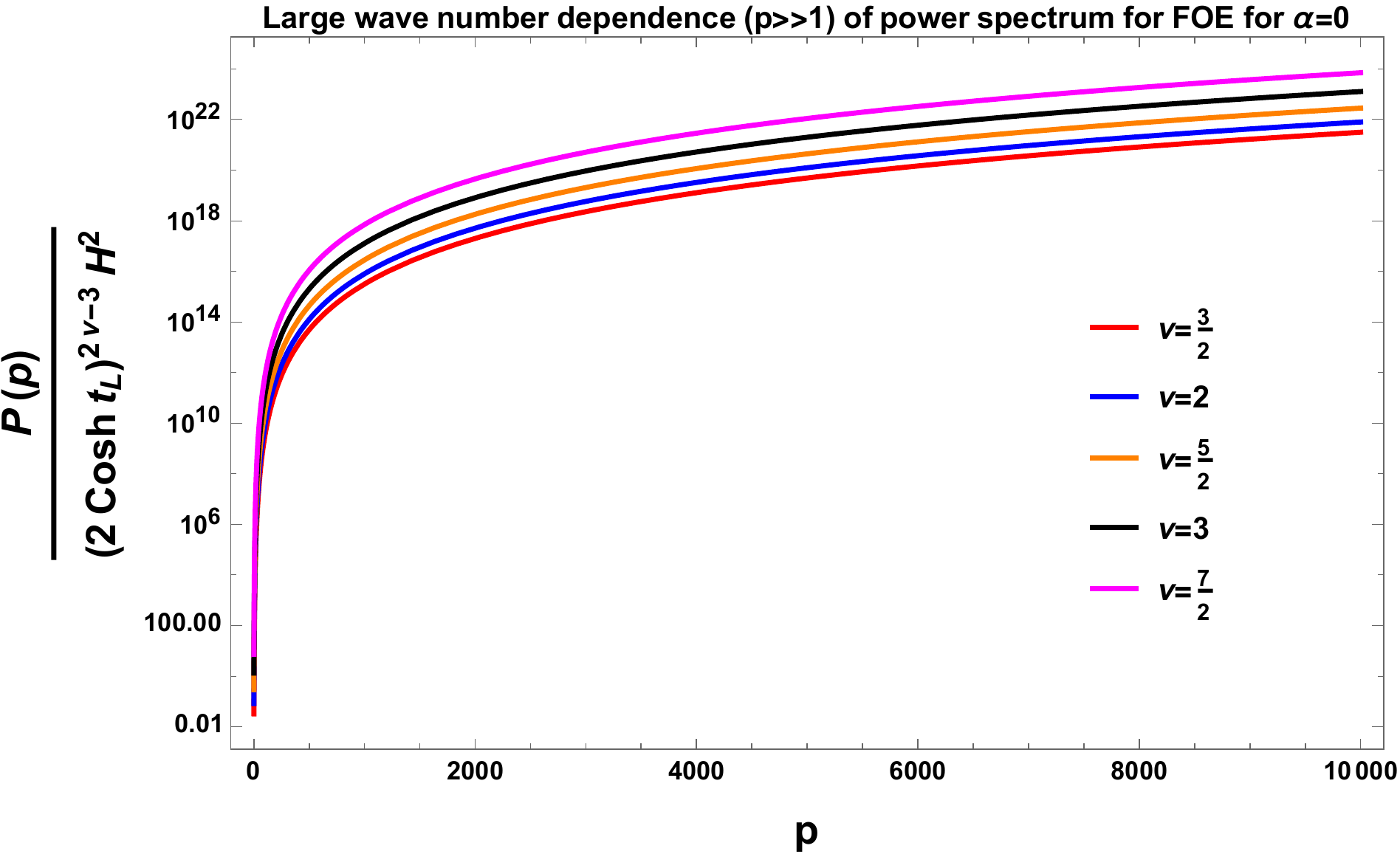}
               \label{fig1a}
            }
            \subfigure[Large wave number dependence of FOE power spectrum for $\alpha=0.1$.]{
                \includegraphics[width=7.7cm,height=7.5cm] {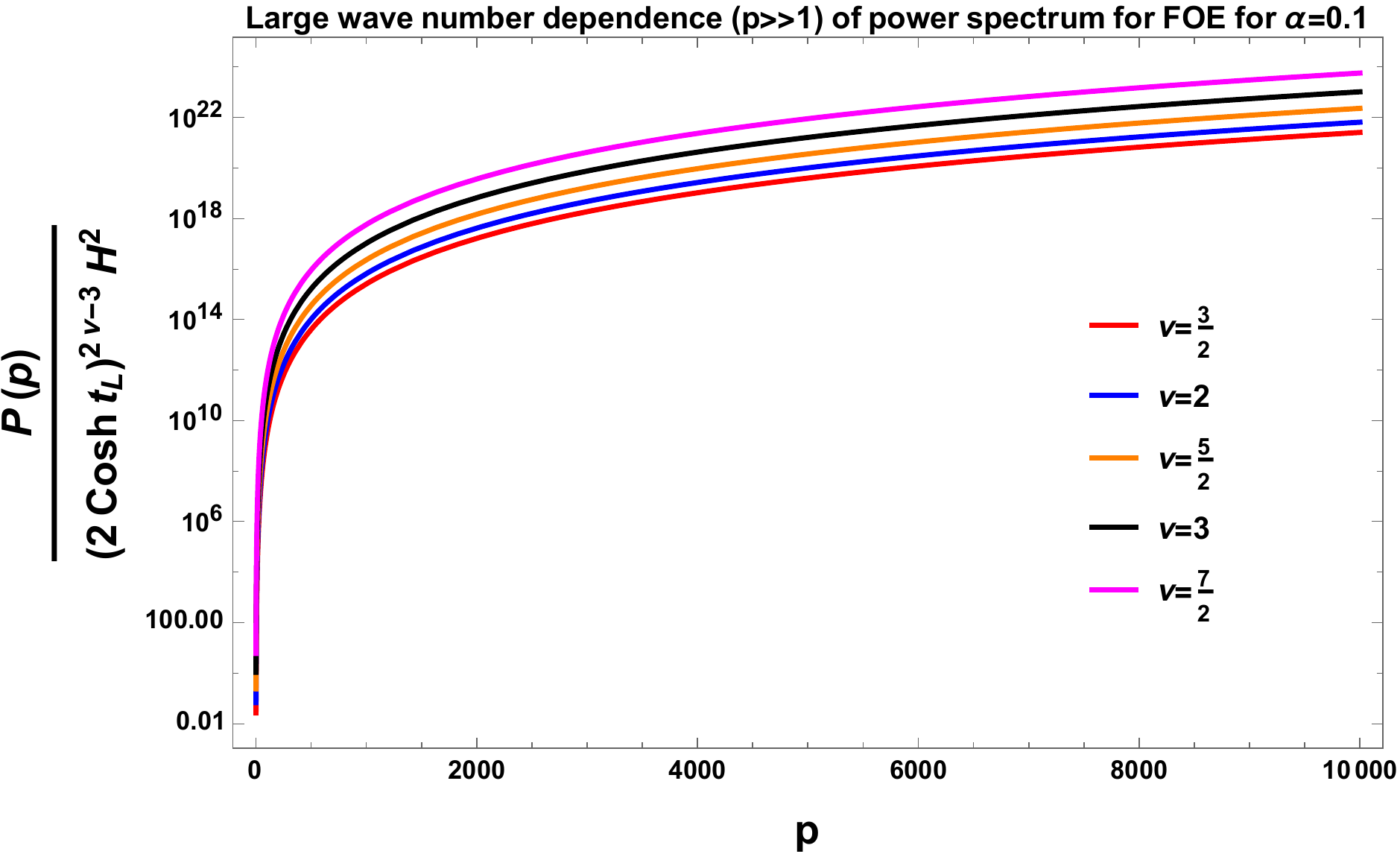}
               \label{fig1b}
              }
             \subfigure[Mass parameter dependence of FOE power spectrum for $p>>1$.]{
                    \includegraphics[width=10.7cm,height=7.5cm] {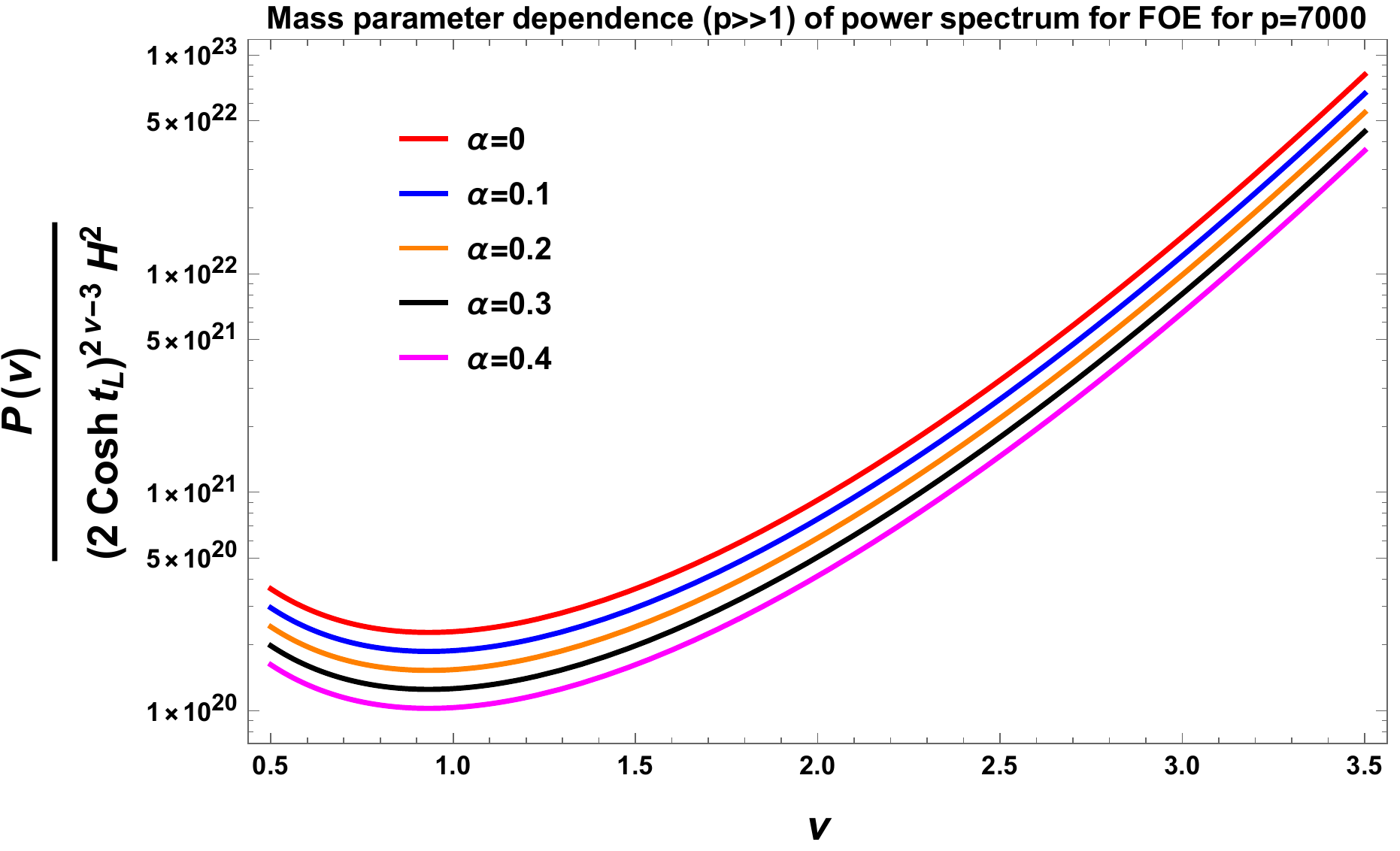}
                    \label{fig1c}   
           }
           \caption[Optional caption for list of figures]{Features of FOE power spectrum in large wave number region.} 
            \label{fig1x1}
            \end{figure*}
The above equation implies that for very large  $p,p_n>>1$ one can rewrite this as, $\widetilde{{\cal G}(p)}\sim 1+\cdots$, and all  the $\cdots$  terms can be considered as small correction terms.
Also for the mass less case ($\nu=3/2$) and  in the short wave length approximation, the time independent function $\widetilde{{\cal M}(p,\nu=3/2)}$ can be further simplified as:
\bea\widetilde{{\cal M}(p>>1,\nu=3/2)}&=&\frac{\widetilde{{\cal G}(p>>1)}}{2p^3}.~~~~~~~~~~~\eea
Finally, in the superhorizon time scales ($t_{\bf L}>>1$) of region \textcolor{red}{\bf L}, the amplitude of the normalized power spectrum of axion from Bunch Davies vacuum in the short wave length limit can be expressed as:
\bea \label{po2v} {\cal P}_{\bf BD}(p>>1,t_{\bf L}>>1)&=&\frac{p^3}{2\pi^2}~\left(\cosh t_{\bf L}\right)^{2\nu-3}~H^2\widetilde{{\cal M}(p,\nu)}\nonumber\\
&=&\left(2\cosh t_{\bf L}\right)^{2\nu-3}~\left(\frac{H}{2\pi}\right)^2~\left(\frac{\Gamma(\nu)}{\Gamma\left(\frac{3}{2}\right)}\right)^2\widetilde{{\cal G}(p>>1)}.~~~~~~~~~~~\eea
Also for the massless case ($\nu=3/2$) in the superhorizon time scales ($t_{\bf L}>>1$) of region \textcolor{red}{\bf L} the amplitude of the normalized power spectrum of axion from Bunch Davies vacuum in the short wave length limit can be simplified as:
\bea \label{po3z} {\cal P}_{\bf BD}(p>>1,t_{\bf L}>>1)&=&\frac{p^3}{2\pi^2}~H^2\widetilde{{\cal M}(p>>1,\nu=3/2)}=\left(\frac{H}{2\pi}\right)^2~\widetilde{{\cal G}(p>>1)}.~~~~~~~~~~~\eea

Now, we generalize the above results for the two point correlation function and the associated power spectrum for $\alpha$ vacua. For $\alpha$ vacua the mean square vacuum fluctuation of axion in the short wave length limit can be expressed as:
  \bea \langle \alpha |\widetilde{\Phi_{plm}(t_{\bf L})}\left(\widetilde{\Phi_{p^{'}l^{'}m^{'}}(t_{\bf L})}\right)^{\dagger}|\alpha \rangle &=& \frac{H^2}{\sinh^2 t_{\bf L}}\langle \alpha |\left[d_{I}\widetilde{\chi^{I}}\right]_{plm}\left(\left[d_{I}\widetilde{\chi^{I}}\right]_{p^{'}l^{'}m^{'}}\right)^{\dagger}|\alpha \rangle\nonumber\\
  &=&\frac{H^2}{\sinh^2 t_{\bf L}}\sum_{\sigma=\pm 1}|\widetilde{\chi^{\sigma}}|^2 ~\delta(p-p^{'})~\delta_{l l^{'}}~\delta_{m m^{'}}\nonumber\\
  &\equiv&P(p>>1,\alpha ,t_{\bf L}) ~\delta(p-p^{'})~\delta_{l l^{'}}~\delta_{m m^{'}}.\eea
  where we have defined the amplitude of the normalized power spectrum of axion in the short wave length limit as:
  \bea {\cal P}(p>>1,\alpha,t_{\bf L})&=&\frac{p^3}{2\pi^2}~P(p>>1,\alpha,t_{\bf L})\nonumber\\
  &=&{ P}_{\bf BD}(p>>1,t_{\bf L})~\left(\cosh 2\alpha-\sinh 2\alpha\right)\nonumber\\&=&\exp(-2\alpha)~{ P}_{\bf BD}(p>>1,t_{\bf L}).~~~~~~~\eea
  In the above equation,  ${ P}_{\bf BD}(p,t_{\bf L})$ is defined as:
  \bea { P}_{\bf BD}(p>>1,t_{\bf L})&=&\frac{p^3}{2\pi^2}~\frac{H^2}{\sinh^2 t_{\bf L}}\sum_{\sigma=\pm 1}|\widetilde{\chi^{\sigma}}|^2.\eea
 We carry out the same approximations as earlier and we note that  in the superhorizon time scales ($t_{\bf L}>>1$) of region \textcolor{red}{\bf L} the amplitude of the normalized power spectrum of axion in the short wave length limit from $\alpha$ vacua can be expressed as:
  \bea \label{po2vv} {\cal P}(p>>1,\alpha,t_{\bf L}>>1)&=&{\cal P}_{\bf BD}(p>>1,t_{\bf L}>>1)~\left(\cosh 2\alpha-\sinh 2\alpha\right)=\exp(-2\alpha)~{\cal P}_{\bf BD}(p,t_{\bf L}>>1),~~~~~~~~~~~\eea
  where the normalized power spectrum in superhorizon scale for Bunch Davies vacuum ${\cal P}_{\bf BD}(p>>1,t_{\bf L}>>1)$ is defined in Equation (3.26). Here it is important to note that, with $\alpha=0$ then we can reproduce the results obtained for Bunch Davies vacuum.
  
  In figure~(\ref{fig1a}) and figure~(\ref{fig1b}) we have shown the behaviour of the power spectrum of the mean square vacuum fluctuation computed from FOE formalism in the short wave length regime for $\alpha=0$ and $\alpha=0.1$  and for fixed values of the mass parameter $\nu (=3/2,2,5/2,3,7/2)$ respectively. In both the cases we have found almost similar behaviour. Additionally, in figure~(\ref{fig1c}) we have depicted the behaviour of the power spectrum with respect to the mass parameter $\nu$ with fixed values of the parameter $\alpha (=0,0.1,0.2,0.3,0.4)$. It is clear from this figure that the power spectrum shows two distinct behaviour in $1/2<\nu<1$ and $\nu>1$ region. For $1/2<\nu<1$ region, the amplitude of the normalized power spectrum decreases to a certain value but just after $\nu=1$ it increases.

 On the other hand, to know the exact wavenumber dependence of the amplitude of the normalised power spectrum from Bunch Davies vacuum in the long wavelength limit we need to know the behaviour of the power spectrum at $p,p_n<<1$. In this limit it is expected that the power spectrum of axion match with the result obtained for spatially flat universe. Here the time independent function $\widetilde{{\cal M}(p<<1,\nu)}$ for any arbitrary mass parameter $\nu$ can be expressed as:
\bea\widetilde{{\cal M}(p<<1,\nu)}&=&\frac{2^{2(\nu-1)}\left(\Gamma(\nu)\right)^2}{\pi}\widetilde{{\cal G}(p<<1)},~~~~~~~~~~~\eea
where we have defined a new function $\widetilde{{\cal G}(p<<1)}$ in the long wave length limit as :
\bea\widetilde{{\cal G}(p<<1)}&=&\frac{\pi}{|\Gamma\left(\nu+\frac{1}{2}\right)|^2}\left[1+\frac{|\Gamma\left(\nu+\frac{1}{2}\right)|^2}{\left(\Gamma\left(\nu+\frac{1}{2}\right)\right)^2}\left\{1+3e^{-\pi p}\sum^{\infty}_{n=0}e^{-\pi p_n}+2\sum^{\infty}_{n=0}\sum^{\infty}_{m=0}e^{-\pi (p_n+p_m)}\right\}\right].~~~~~~~~~~~\eea
This implies that for very small wave numbers $p,p_n<<1$, one can write, $\widetilde{{\cal G}(p<<1)}\sim \frac{\pi}{|\Gamma\left(\nu+\frac{1}{2}\right)|^2}\left[1+\cdots\right]$, where all  the$\cdots$  terms are small correction terms.

            \begin{figure*}[htb]
            \centering
           \subfigure[Small wave number dependence of FOE power spectrum for $\alpha=0$.]{
               \includegraphics[width=7.7cm,height=7.5cm] {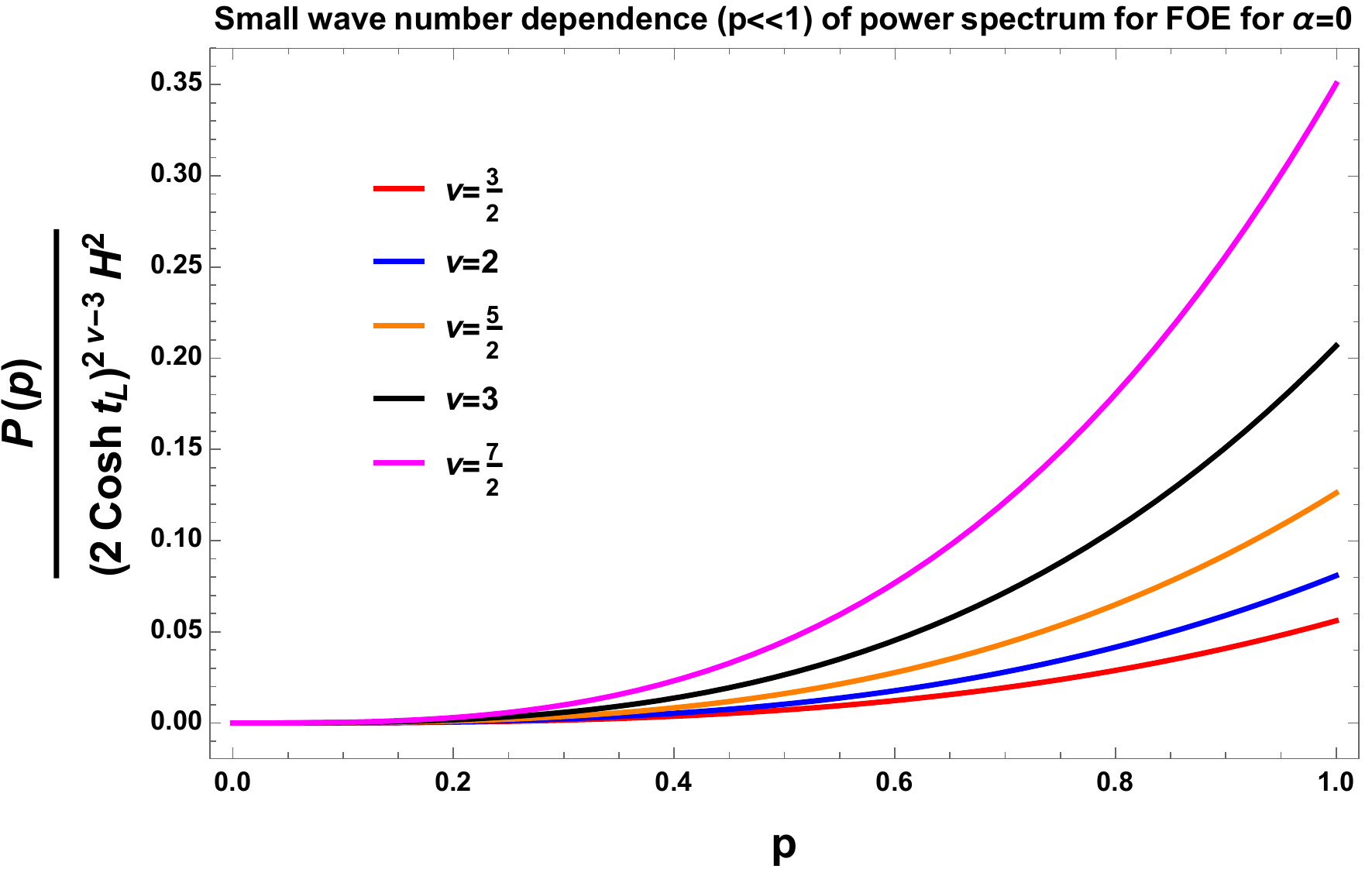}
               \label{fig2a}
            }
            \subfigure[Small wave number dependence of FOE power spectrum for $\alpha=0.1$.]{
                \includegraphics[width=7.7cm,height=7.5cm] {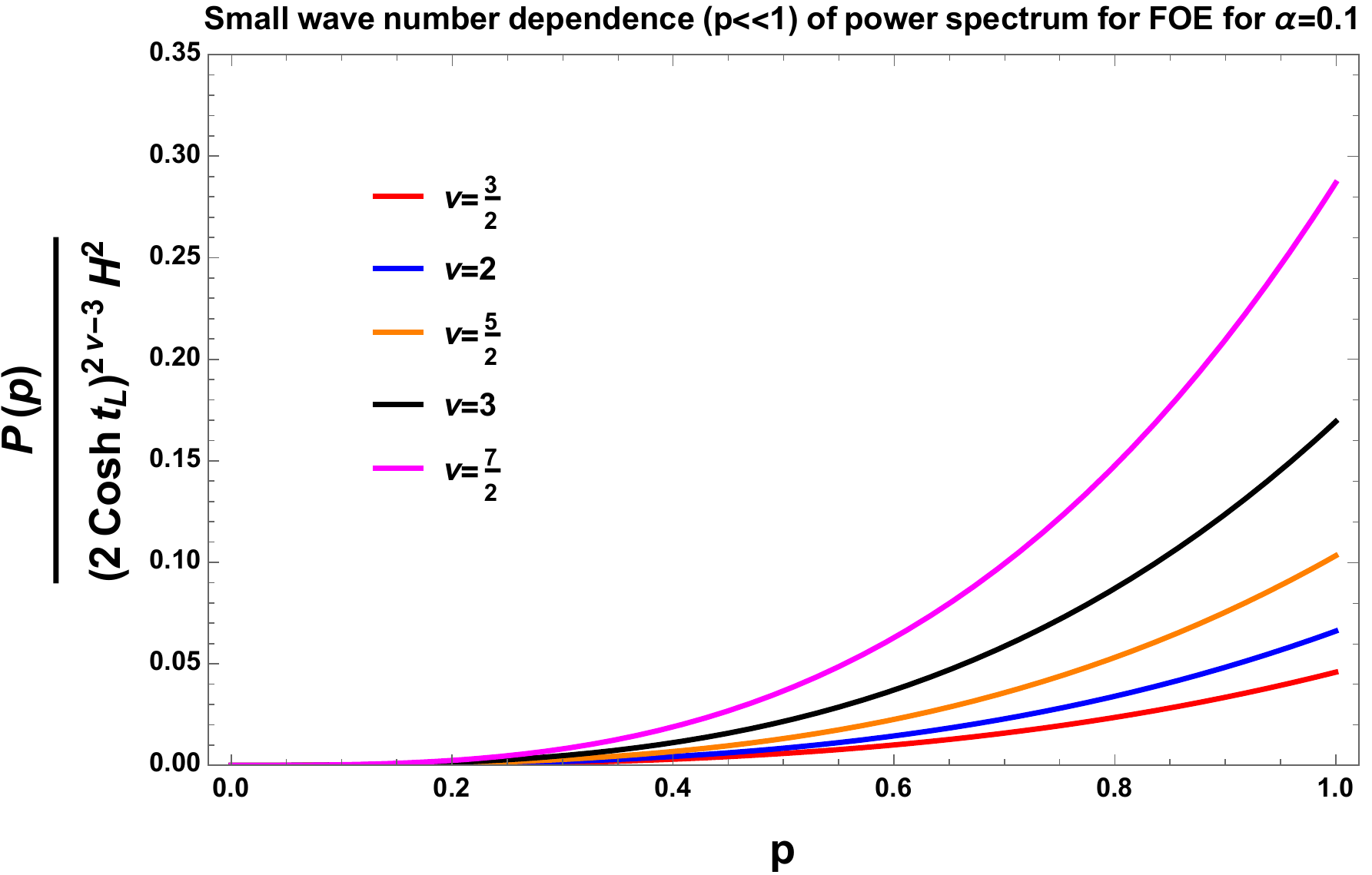}
               \label{fig2b}
              }
             \subfigure[Mass parameter dependence of FOE power spectrum for $p<<1$.]{
                    \includegraphics[width=10.7cm,height=7.5cm] {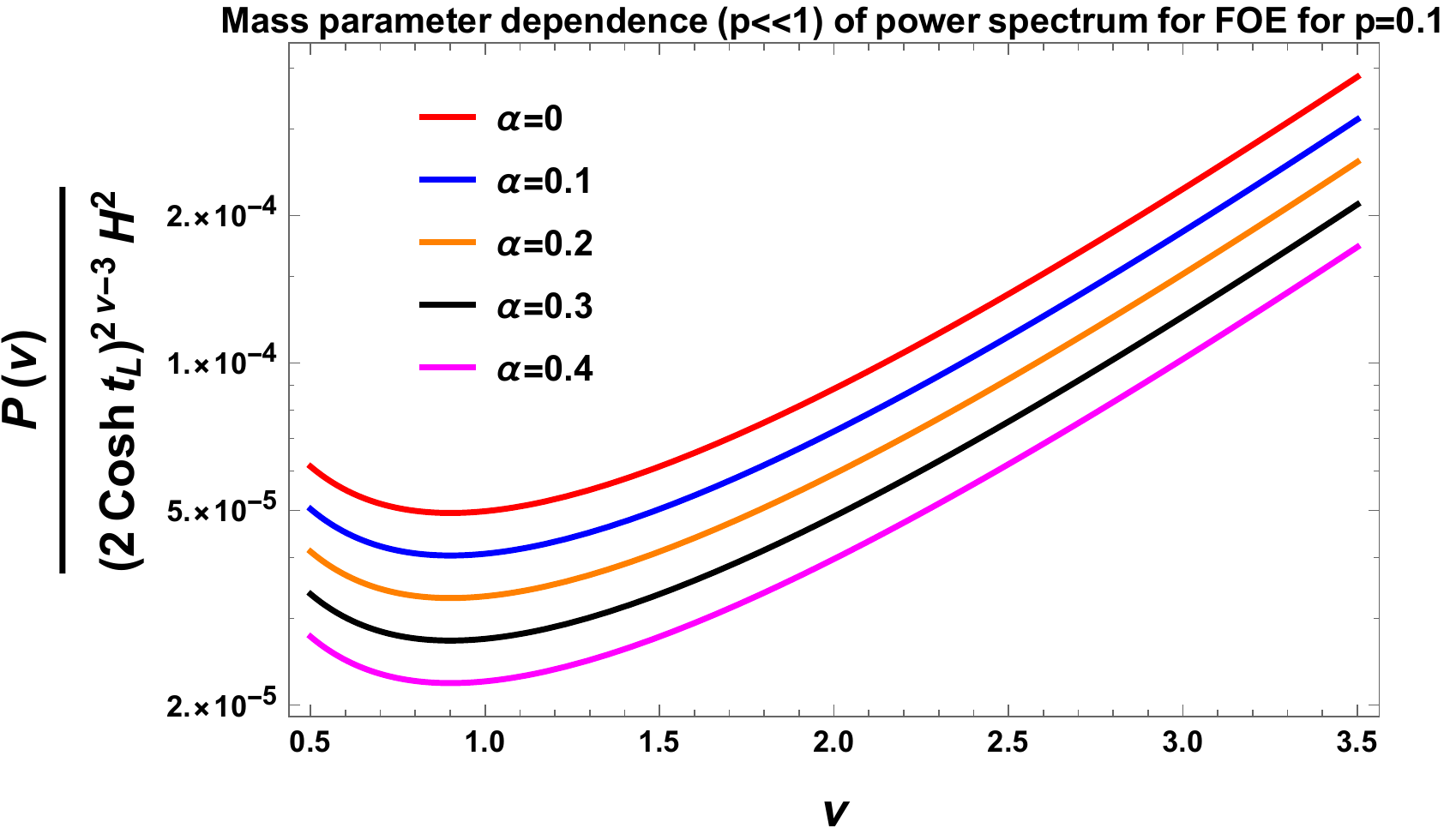}
                    \label{fig2c}   
           }
           \caption[Optional caption for list of figures]{Features of FOE power spectrum in small wave number region.} 
            \label{fig1x2}
            \end{figure*}

Also for the massless case ($\nu=3/2$) and in the long wave length approximation,  the time independent function $\widetilde{{\cal M}(p<<1,\nu=3/2)}$ can  further be simplified as:
\bea\widetilde{{\cal M}(p<<1,\nu=3/2)}&=&\frac{\widetilde{{\cal G}(p<<1)}}{2}.~~~~~~~~~~~\eea
Finally, in the super horizon time scales ($t_{\bf L}>>1$) of region \textcolor{red}{\bf L} the amplitude of the normalized power spectrum of axion from Bunch Davies vacuum, in the long wave length limit, can be expressed as:
\bea \label{po2v} {\cal P}_{\bf BD}(p<<1,t_{\bf L}>>1)&=&\frac{p^3}{2\pi^2}~\left(\cosh t_{\bf L}\right)^{2\nu-3}~H^2\widetilde{{\cal M}(p<<1,\nu)}\nonumber\\
&=&\left(2\cosh t_{\bf L}\right)^{2\nu-3}~\left(\frac{H}{2\pi}\right)^2~p^3~\left(\frac{\Gamma(\nu)}{\Gamma\left(\frac{3}{2}\right)}\right)^2\widetilde{{\cal G}(p<<1
)},~~~~~~~~~~~\eea
and for the massless case ($\nu=3/2$) this simplifies to:
\bea \label{po3z} {\cal P}_{\bf BD}(p<<1,t_{\bf L}>>1)&=&\frac{p^3}{2\pi^2}~H^2\widetilde{{\cal M}(p<<1,\nu=3/2)}=\left(\frac{H}{2\pi}\right)^2~p^3~\widetilde{{\cal G}(p<<1)}.~~~~~~~~~~~\eea
Here it is important to note that both of Eq~(\ref{po2v}) and Eq~(\ref{po3z}) are valid after horizon exit. 

Next, we generalize the result for the two point correlation function and the associated power spectrum for $\alpha$ vacua. For $\alpha$ vacua the mean square vacuum fluctuation of axion in the long wave length limit can be expressed as:
  \bea \langle \alpha |\widetilde{\Phi_{plm}(t_{\bf L})}\left(\widetilde{\Phi_{p^{'}l^{'}m^{'}}(t_{\bf L})}\right)^{\dagger}|\alpha \rangle &=& \frac{H^2}{\sinh^2 t_{\bf L}}\langle \alpha |\left[d_{I}\widetilde{\chi^{I}}\right]_{plm}\left(\left[d_{I}\widetilde{\chi^{I}}\right]_{p^{'}l^{'}m^{'}}\right)^{\dagger}|\alpha \rangle\nonumber\\
  &=&\frac{H^2}{\sinh^2 t_{\bf L}}\sum_{\sigma=\pm 1}|\widetilde{\chi^{\sigma}}|^2 ~\delta(p-p^{'})~\delta_{l l^{'}}~\delta_{m m^{'}}\nonumber\\
  &\equiv&P(p<<1,\alpha ,t_{\bf L}) ~\delta(p-p^{'})~\delta_{l l^{'}}~\delta_{m m^{'}},\eea
  where  the amplitude of the normalized power spectrum of axion at long wave length limit is defined as:
  \bea {\cal P}(p<<1,\alpha,t_{\bf L})&=&\frac{p^3}{2\pi^2}~P(p<<1,\alpha,t_{\bf L})\nonumber\\&=&{ P}_{\bf BD}(p,t_{\bf L})~\left(\cosh 2\alpha-\sinh 2\alpha\right)\nonumber\\&=&\exp(-2\alpha)~{ P}_{\bf BD}(p<<1,t_{\bf L}),~~~~~~~\eea
  with ${ P}_{\bf BD}(p<<1,t_{\bf L})$ as defined  earlier.

  In the super horizon time scales ($t_{\bf L}>>1$) of region \textcolor{red}{\bf L} the amplitude of the normalized power spectrum of axion in the long wave length approximation from $\alpha$ vacua can be expressed as:
  \bea \label{po2vv} {\cal P}(p<<1,\alpha,t_{\bf L}>>1)&=&{\cal P}_{\bf BD}(p<<1,t_{\bf L}>>1)~\left(\cosh 2\alpha-\sinh 2\alpha\right)\nonumber\\&=&\exp(-2\alpha)~{\cal P}_{\bf BD}(p<<1,t_{\bf L}>>1),~~~~~~~~~~~\eea
  where  ${\cal P}_{\bf BD}(p<<1,t_{\bf L}>>1)$ is defined in Eq~(\ref{po2v}). It may be noted that, for $\alpha=0$ we get back the results obtained for Bunch Davies vacuum.
  
  In figure~(\ref{fig2a}) , figure~(\ref{fig2b})  and in figure~(\ref{fig2c}) we have shown the behaviour of the power spectrum of the mean square vacuum fluctuation computed from FOE formalism in the small wave number regime. The values of $\alpha$  and the values of the mass parameter $\nu$  used here are  same as those taken for large wave number regime. As expected, the behaviour for the the two limiting cases are distinct. However, the characteristics observed for $\alpha$ and $\nu$ dependences for both the cases are almost similar.

 \subsection{Quantum vacuum fluctuation using reduced density matrix (RDM) formalism (with mixed state)} 
\begin{figure*}[htb]
			    \centering
			    {
			        \includegraphics[width=18.5cm,height=10cm] {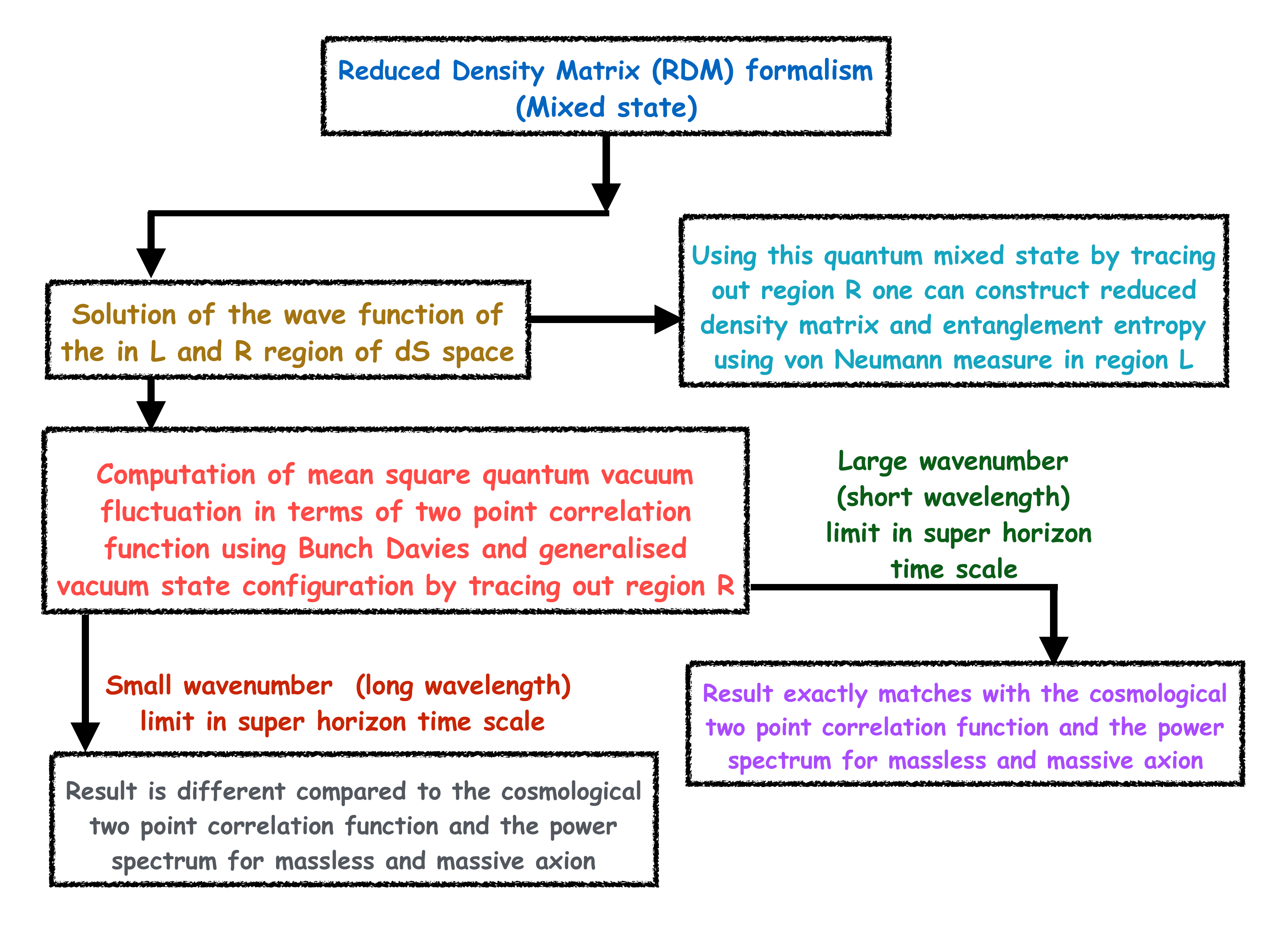}
			    }
			    \caption[Optional caption for list of figures]{Schematic diagram for the computation algorithm of reduced density matrix formalism for mixed quantum state of axion in de Sitter hyperbolic open chart. } 
			    \label{fzaaqrt}
			    \end{figure*}

  In this section, we study the features of the two point correlation function of the quantum vacuum fluctuations and the associated primordial power spectrum using the reduced density matrix formalism. In figure~(\ref{fzaaqrt}) we have presented a schematic diagram for the computation algorithm of reduced density matrix formalism for mixed quantum state of axion in de Sitter hyperbolic open chart. 
 \subsubsection{Reduced density matrix (RDM) formalism}
 \label{x2a}
  We first write down the Fourier mode of the field operator, which is also the total solution of the field equation for axion in presence of source contribution.  We start directly from the solution  obtained in Eqn~(\ref{cvvcxcvb}) and rewrite it in terms of the following matrix equation:
   \bea\label{xxxc} \boxed{{\bf \chi}^{I}=\frac{1}{{\cal N}_p}{\cal  M}^{I}_{J}{\cal P}^{J}+\sum^{\infty}_{n=0}\frac{1}{{\cal N}_{p,(n)}}\left({\cal  M}_{(n)}\right)^{I}_{J}{\cal P}^{J}_{(n)}} \eea 
   where for the complementary part of the solution we have defined the following matrices:
   \bea {\cal M}^{I}_{J}&=&\left(\begin{array}{ccc} \alpha^{\sigma}_{q} &~~~ \beta^{\sigma}_{q} \\ \beta^{\sigma^{*}}_{q} &~~~ \alpha^{\sigma^{*}}_{q}  \end{array}\right),~~~~~~
   \chi^{I}=\left(\begin{array}{ccc} \chi_{\sigma}(t) \\ \chi^{*}_{\sigma}(t),
   \end{array}\right),~~~~~~
   {\cal P}^{J}=\left(\begin{array}{ccc} {\cal P}^{q} \\ {\cal P}^{{q^*}},\\
   \end{array}\right).\eea
   Similarly for the particular solution, we define the following matrices:
   \bea \left({\cal M}_{(n)}\right)^{I}_{J}&=&\left(\begin{array}{ccc} \bar{\alpha}^{\sigma}_{q,n} &~~~ \bar{\beta}^{\sigma}_{q,n} \\ \bar{\beta}^{\sigma^{*}}_{q,n} &~~~ \bar{\alpha}^{\sigma^{*}}_{q,n}  \end{array}\right),~~~~~~
   {\cal P}^{J}_{(n)}=\left(\begin{array}{ccc} {\cal P}^{q,n} \\ {\cal P}^{{q^*},n}\\
   \end{array}\right),\eea                 where $\sigma=\pm 1$, $q={\bf R},{\bf L}$ and $I,J=1,2,3,4 $.
   
    The redefined normalization constant for the particular part of the solution ${\cal N}_{p,(n)}$ can be expressed as, 
   ${\cal N}_{p,(n)}=2\sinh \pi p_n ~\sqrt{{\cal N}_{p_n\sigma}} ~\left(p^2-p^2_n\right)$. Further using Eqn~(\ref{xxxc}) the Bunch-Davies mode function can be written as:  
   \bea   \frac{H}{\sinh t}a_{I}\chi^{I}=\frac{H}{\sinh t}a_{I}\left[\frac{1}{{\cal N}_p}{\cal  M}^{I}_{J}{\cal P}^{J}+\sum^{\infty}_{n=0}\frac{1}{{\cal N}_{p,(n)}}\left({\cal  M}_{(n)}\right)^{I}_{J}{\cal P}^{J}_{(n)} \right],~~~~~~
   \eea                                   where $a_{I}=(a_{\sigma},
   a^{\dagger}_{\sigma})$ represents a set of creation and annihilation operators.  
   
   We also define the following operators:
   \bea\label{oq1} b_{J} &=& a^{(c)}_{I}{\cal M}^{I}_{J},~~~ b_{J(n)} = a^{(p)}_{I(n)}\left({{\cal M}_{(n)}}\right)^{I}_{J},
   \eea
   where $a^{(c)}_{I}=(a^{(c)}_{\sigma},
   a^{(c)\dagger}_{\sigma})$ and $a^{(p)}_{I(n)}=(a^{(p)}_{\sigma,n},
   a^{(p)\dagger}_{\sigma,n})$  are 
   the set of creation and annihilation operators which act on the complementary and particular part respectively. Thus, the operator contribution for the total solution is:
   \bea a_{I}&=& \left[a^{(c)}_{I}+\sum^{\infty}_{n=0}a^{(p)}_{I(n)}\right],\eea 
   where by inverting Eqn~(\ref{oq1}) we have expressed:
   \bea\label{oq1a} a^{(c)}_{I} &=& b_{J}\left({\cal M}^{-1}\right)^{I}_{J},~~~ a^{(p)}_{I(n)} = b_{J(n)}\left({\cal M}^{-1}_{(n)}\right)^{I}_{J}.
   \eea
   The inverse matrices are defined as:
   \bea \left({\cal M}^{-1}\right)^{I}_{J}&=&\left(\begin{array}{ccc} \gamma_{\sigma q} &~~~ \delta_{\sigma q} \\ \delta^{*}_{\sigma q} &~~~ \gamma^{*}_{\sigma q}  \end{array}\right),
   ~~~~\left({\cal M}^{-1}_{(n)}\right)^{I}_{J}=\left(\begin{array}{ccc} \bar{\gamma}_{\sigma q,n} &~~~ \bar{\delta}_{\sigma q,n} \\ \bar{\delta}^{*}_{\sigma q,n} &~~~ \bar{\gamma}^{*}_{\sigma q,n}  \end{array}\right),\eea
   where $\sigma=\pm 1$, $q={\bf R},{\bf L}$ and $I,J=1,2,3,4 $.   
    
For further computation, $\alpha$-vacua are defined in terms of Bunch Davies vacuum state as:
\bea |\alpha\rangle &=& \exp\left(\displaystyle\frac{1}{2} \tanh\alpha~\sum_{\sigma=\pm 1}a^{\dagger}_{\sigma}a_{\sigma}\right)|{\bf BD}\rangle.\eea
It is to be noted that for $\alpha=0$ we get, 
$|\alpha=0\rangle =|0\rangle =|{\bf BD}\rangle.$
Moreover, we can also write the ${\bf R}$ and ${\bf L}$ vacua as:
 \bea   |{\bf R}\rangle&=& |{\bf R}\rangle_{(c)}+\sum^{\infty}_{n=0}|{\bf R}\rangle_{(p),n},~~~  |{\bf L}\rangle= |{\bf L}\rangle_{(c)}+\sum^{\infty}_{n=0}|{\bf L}\rangle_{(p),n},\eea 
  with subscripts $(c)$ and $(p)$ representing the complementary and particular part respectively.
  
Further assuming the bipartite Hilbert space (${\cal H}_{\alpha}:={\cal H}_{\bf R}\otimes {\cal H}_{\bf L}$) one can also write the $\alpha$-vacua in terms of the ${\bf R}$ and ${\bf L}$ vacuum as:
 \be \label{sdss} |\alpha\rangle = \exp\left(\displaystyle\frac{1}{2} \tanh\alpha~\sum_{\sigma=\pm 1}a^{\dagger}_{\sigma}a_{\sigma}\right)\underbrace{\exp\left(\frac{1}{2}\sum_{i,j={\bf R},{\bf L}}m_{ij}~b^{\dagger}_{i}~b^{\dagger}_{j}+\frac{1}{2}\sum_{i,j={\bf R},{\bf L}}\sum^{\infty}_{n=0}\bar{m}_{ij,n}~\bar{b}^{\dagger}_{i,n}~\bar{b}^{\dagger}_{j,n}\right)(|{\bf R}\rangle \otimes |{\bf L}\rangle)}_{\textcolor{red}{\bf Bunch-Davies~contribution}},\ee
 where the matrices $m_{ij}$ and $\bar{m}_{ij,n}$ are defined for the complementary and particular part of the solution obtained for Bunch Davies vacuum state. In other words by setting $\alpha=0$ we get the following expression for the Bunch Davies quantum state:
 \be \label{sdss} |{\bf BD}\rangle = \exp\left(\frac{1}{2}\sum_{i,j={\bf R},{\bf L}}m_{ij}~b^{\dagger}_{i}~b^{\dagger}_{j}+\frac{1}{2}\sum_{i,j={\bf R},{\bf L}}\sum^{\infty}_{n=0}\bar{m}_{ij,n}~\bar{b}^{\dagger}_{i,n}~\bar{b}^{\dagger}_{j,n}\right)(|{\bf R}\rangle \otimes |{\bf L}\rangle).\ee
 Also the creation and annihilation operators for the ${\bf R}$ and ${\bf L}$ vacuum are defined in terms of new $b$ type of oscillators using Bogoliubov transformation as:
         \bea\label{xxxq1} a_{\sigma}&=&\sum_{q={\bf R},{\bf L}}\left\{\left[\gamma_{q\sigma}b_{q}+\delta^{*}_{q\sigma}b^{\dagger}_{q}\right]+\sum^{\infty}_{n=0}\left[\bar{\gamma}_{q\sigma,n}\bar{b}_{q,n}+\bar{\delta}^{*}_{q\sigma,n}\bar{b}^{\dagger}_{q,n}\right]\right\}~~~\forall \sigma=\pm 1,\\ \label{xxxq2} a^{\dagger}_{\sigma}&=&\sum_{q={\bf R},{\bf L}}\left\{\left[\gamma^{*}_{q\sigma}b^{\dagger}_{q}+\delta_{q\sigma}b_{q}\right]+\sum^{\infty}_{n=0}\left[\bar{\gamma}^{*}_{q\sigma,n}\bar{b}^{\dagger}_{q,n}+\bar{\delta}_{q\sigma,n}\bar{b}_{q,n}\right]\right\}~~~\forall \sigma=\pm 1.\eea   
         Here $\gamma_{q\sigma}$, $\delta_{q\sigma}$, $\bar{\gamma}_{q\sigma,n}$ and $\bar{\delta}_{q\sigma,n}$ are the coefficient matrices. 
          For our further computation we use the definition of $\alpha$-vacuum state (and Bunch Davies vacuum state), which is very useful to compute long range cosmological correlation functions in de Sitter space. In the context of $\alpha$-vacua the creation and annihilation operators are defined in terms of the constituents of ${\bf R}$ or ${\bf L}$ vacuum state as:
 \bea d_{\sigma}&=&\sum_{q={\bf R},{\bf L}}\left\{\left[\left(\cosh\alpha~\gamma_{q\sigma}-\sinh\alpha~\delta_{q\sigma}\right)b_{q}+\left(\cosh\alpha~\delta^{*}_{q\sigma}-\sinh\alpha~\gamma^{*}_{q\sigma}\right)b^{\dagger}_{q}\right]\right.\nonumber\\ && \left.~~~~~~~~~~~~~~~ +\left[\left(\cosh\alpha~\sum^{\infty}_{n=0}\bar{\gamma}_{q\sigma,n}\bar{b}_{q,n}-\sinh\alpha~\sum^{\infty}_{n=0}\bar{\delta}_{q\sigma,n}\bar{b}_{q,n}\right)\right.\right.\nonumber\\ && \left.\left.~~~~~~~~~~~~~~~~~~~~+\left(\cosh\alpha~\sum^{\infty}_{n=0}\bar{\delta}^{*}_{q\sigma,n}\bar{b}^{\dagger}_{q,n}-\sinh\alpha~\sum^{\infty}_{n=0}\bar{\gamma}^{*}_{q\sigma,n}\bar{b}^{\dagger}_{q,n}\right)\right]\right\}~~~\forall \sigma=\pm 1,\eea\bea
 d^{\dagger}_{\sigma}&=&\sum_{q={\bf R},{\bf L}}\left\{\left[\left(\cosh\alpha~\gamma^{*}_{q\sigma}-\sinh\alpha~\delta^{*}_{q\sigma}\right)b^{\dagger}_{q}+\left(\cosh\alpha~\delta_{q\sigma}-\sinh\alpha~\gamma_{q\sigma}\right)b_{q}\right]\right.\nonumber\\ && \left.~~~~~~~~~~~~~~~ +\left[\left(\cosh\alpha~\sum^{\infty}_{n=0}\bar{\gamma}^{*}_{q\sigma,n}\bar{b}^{\dagger}_{q,n}-\sinh\alpha~\sum^{\infty}_{n=0}\bar{\delta}^{*}_{q\sigma,n}\bar{b}^{\dagger}_{q,n}\right)\right.\right.\nonumber\\ && \left.\left.~~~~~~~~~~~~~~~~~~~~+\left(\cosh\alpha~\sum^{\infty}_{n=0}\bar{\delta}_{q\sigma,n}\bar{b}_{q,n}-\sinh\alpha~\sum^{\infty}_{n=0}\bar{\gamma}_{q\sigma,n}\bar{b}_{q,n}\right)\right]\right\}~~~\forall \sigma=\pm 1,\eea
 where we use the definition of creation and annihilation operators in Bunch Davies vacuum as mentioned in Eq~(\ref{xxxq2}) and Eq~(\ref{xxxq1}). In this computation it is important to note that, under Bogoliubov transformation the original matrix $\gamma_{q\sigma}$, $\delta_{q\sigma}$, $\bar{\gamma}_{q\sigma,n}$ and $\bar{\delta}_{q\sigma,n}$ used for Bunch Davies vacuum transform ( for $\alpha$-vacua) as:
 \bea \gamma_{q\sigma}&\longrightarrow& \left(\cosh\alpha~\gamma_{q\sigma}-\sinh\alpha~\delta_{q\sigma}\right),~~
 \delta_{q\sigma}\longrightarrow\left(\cosh\alpha~\delta_{q\sigma}-\sinh\alpha~\gamma_{q\sigma}\right),\\
 \bar{\gamma}_{q\sigma,n}&\longrightarrow&\left(\cosh\alpha~\bar{\gamma}_{q\sigma,n}-\sinh\alpha~\bar{\delta}_{q\sigma,n}\right),~~
 \bar{\delta}_{q\sigma,n}\longrightarrow\left(\cosh\alpha~\bar{\delta}_{q\sigma,n}-\sinh\alpha~\bar{\gamma}_{q\sigma,n}\right).\nonumber\eea
 Thus, after the Bogoliubov transformation, $\alpha$-vacua state can be written in terms of ${\bf R}$ and ${\bf L}$ vacua as:
 \bea |\alpha\rangle &=&\exp\left(\frac{1}{2}\sum_{i,j={\bf R},{\bf L}}\tilde{m}_{ij}~b^{\dagger}_{i}~b^{\dagger}_{j}+\frac{1}{2}\sum_{i,j={\bf R},{\bf L}}\sum^{\infty}_{n=0}\bar{\tilde{m}}_{ij,n}~\bar{b}^{\dagger}_{i,n}~\bar{b}^{\dagger}_{j,n}\right)(|{\bf R}\rangle \otimes |{\bf L}\rangle),\eea
   Here $\tilde{m}_{ij}$ and $\bar{\tilde{m}}_{ij,n}$ represent the entries of the  matrices corresponding to the complementary and particular solution respectively and  we will compute them by demanding 
  $ d_{\sigma}|\alpha\rangle=0,$  and keeping only linear terms of  creation operators.
    This directly yields the following:
          \bea \label{qz1} \left[\tilde{m}_{ij}\left(\cosh\alpha~\gamma_{j\sigma}-\sinh\alpha~\delta_{j\sigma}\right)+\left(\cosh\alpha~\delta^{*}_{i\sigma}-\sinh\alpha~\gamma^{*}_{i\sigma}\right)\right]&=& 0,\\
          \label{qz2} \left[\left(\cosh\alpha~\bar{\tilde{m}}_{ij,n}\bar{\gamma}_{j\sigma,n}-\sinh\alpha~\bar{m}_{ij,n}\bar{\delta}_{j\sigma,n}\right) 
          +\left(\cosh\alpha~\bar{\delta}^{*}_{i\sigma,n}-\sinh\alpha~\bar{\gamma}^{*}_{i\sigma,n}\right)\right]&=& 0\forall ~n.\eea
            From these two equations,  the matrices corresponding to the complementary and particular part of the solution can be expressed as:
            \bea \label{qz1a} \tilde{m}_{ij}&=& -\left(\cosh\alpha~\delta^{*}_{i\sigma}-\sinh\alpha~\gamma^{*}_{i\sigma}\right)\left(\cosh\alpha~\gamma-\sinh\alpha~\delta\right)^{-1}_{\sigma j}=\left(\begin{array}{ccc} \tilde{m}_{\bf RR} &~~~ \tilde{m}_{\bf RL} \\ \tilde{m}_{\bf LR} &~~~ \tilde{m}_{\bf LL}  \end{array}\right),~~~~~\\ \label{qz22a} \bar{\tilde{m}}_{ij,n}&=&-\left(\cosh\alpha~\bar{\delta}^{*}_{i\sigma,n}-\sinh\alpha~\bar{\gamma}^{*}_{i\sigma,n}\right)\left(\cosh\alpha~\bar{\gamma}-\sinh\alpha~\bar{\delta}\right)^{-1}_{\sigma j,n}=\left(\begin{array}{ccc} \bar{m}_{{\bf RR},n} &~~~ \bar{m}_{{\bf RL},n} \\ \bar{m}_{{\bf LR},n} &~~~ \bar{m}_{{\bf LL},n}  \end{array}\right).~~~~~~~~~~\eea 
             Substituting the expressions for $\gamma$, $\delta$, $\gamma_n$ and $\delta_n$ we finally 
obtain the entries of the mass matrices for $i,j={\bf R},{\bf L}$ as:
   \bea 
              \tilde{m}_{ij}=e^{i\theta}~\frac{\sqrt{2}~e^{-p\pi}~{\cal T}^{(\nu)}_{ij}}{\sqrt{\cosh 2\pi p+\cos 2\pi \nu}\left(\cosh^2\alpha+\sinh^2\alpha~e^{-2\pi( p+i\nu)}\right)}~~~~~ \\ \bar{\tilde{m}}_{ij,n}=e^{i\theta}~ \frac{\sqrt{2}~e^{-p_n\pi}~{\cal T}^{(\nu,n)}_{ij}}{\sqrt{\cosh 2\pi p_n+\cos 2\pi \nu}\left(\cosh^2\alpha+\sinh^2\alpha~e^{-2\pi( p_n+i\nu)}\right)}~~~~~\eea
where we defined the ${\cal T}$ matrices as:
\bea ~~~~~~{\cal T}^{(\nu)}_{ij}&=&\left(\begin{array}{ccc} {\cal T}^{(\nu)}_{\bf RR} &~~~ {\cal T}^{(\nu)}_{\bf RL} \\ {\cal T}^{(\nu)}_{\bf LR} &~~~ {\cal T}^{(\nu)}_{\bf LL}  \end{array}\right),~~~~~~~~~~{\cal T}^{(\nu,n)}_{ij}=\left(\begin{array}{ccc} {\cal T}^{(\nu,n)}_{\bf RR} &~~~ {\cal T}^{(\nu,n)}_{\bf RL} \\ {\cal T}^{(\nu,n)}_{\bf LR} &~~~ {\cal T}^{(\nu,n)}_{\bf LL}  \end{array}\right).~~~~~\eea
and the corresponding entries of the ${\cal T}$ matrices are given by: 
\bea 
{\cal T}^{(\nu)}_{\bf RR}&=&{\cal T}^{(\nu)}_{\bf LL}=\left[\left(\cosh^2\alpha+\sinh^2\alpha~e^{-2i\pi\nu}\right)-\sinh 2\alpha~ \sinh^2 \pi p~e^{-i\pi\nu}\sec\pi\nu\right]\cos\pi\nu,\\
{\cal T}^{(\nu)}_{\bf RL}&=&{\cal T}^{(\nu)}_{\bf LR}=i\left[\cosh^2\alpha+\sinh^2\alpha~e^{-2i\pi\nu}+\sinh 2\alpha~\cos\pi\nu~ e^{-i\pi\nu}\right]\sinh \pi p,\\
{\cal T}^{(\nu,n)}_{\bf RR}&=&{\cal T}^{(\nu,n)}_{\bf LL}=\left[\left(\cosh^2\alpha+\sinh^2\alpha~e^{-2i\pi\nu}\right)-\sinh 2\alpha~ \sinh^2 \pi p_n~e^{-i\pi\nu}\sec\pi\nu\right]\cos\pi\nu,\\
{\cal T}^{(\nu,n)}_{\bf RL}&=&{\cal T}^{(\nu,n)}_{\bf LR}=i\left[\cosh^2\alpha+\sinh^2\alpha~e^{-2i\pi\nu}+\sinh 2\alpha~\cos\pi\nu~ e^{-i\pi\nu}\right]\sinh \pi p_n.~~~~~~~~~~~~~~~~\eea 
For the  massless ($\nu=3/2$) axion  case, we obtain the following simplified expressions:  
\bea 
              \tilde{m}_{ij}=e^{i\theta}~\frac{\sqrt{2}~e^{-p\pi}~{\cal T}^{(3/2)}_{ij}}{\sqrt{\cosh 2\pi p-1}\left(\cosh^2\alpha-\sinh^2\alpha~e^{-2\pi p}\right)}~~~~~ \\ \bar{\tilde{m}}_{ij,n}=e^{i\theta}~ \frac{\sqrt{2}~e^{-p_n\pi}~{\cal T}^{(3/2,n)}_{ij}}{\sqrt{\cosh 2\pi p_n-1}\left(\cosh^2\alpha-\sinh^2\alpha~e^{-2\pi p_n}\right)}~~~~~\eea
where we have defined the ${\cal T}^{(3/2)}$ matrices as:
\bea ~~~~~~{\cal T}^{(3/2)}_{ij}&=&\left(\begin{array}{ccc} {\cal T}^{(3/2)}_{\bf RR} &~~~ {\cal T}^{(3/2)}_{\bf RL} \\ {\cal T}^{(3/2)}_{\bf LR} &~~~ {\cal T}^{(3/2)}_{\bf LL}  \end{array}\right),~~~~~~~~~~{\cal T}^{(3/2,n)}_{ij}=\left(\begin{array}{ccc} {\cal T}^{(3/2,n)}_{\bf RR} &~~~ {\cal T}^{(3/2,n)}_{\bf RL} \\ {\cal T}^{(3/2,n)}_{\bf LR} &~~~ {\cal T}^{(3/2,n)}_{\bf LL}  \end{array}\right).~~~~~\eea
and the corresponding entries of the ${\cal T}^{(3/2)}$ matrices are given by: 
\bea 
{\cal T}^{(3/2)}_{\bf RR}&=&{\cal T}^{(3/2)}_{\bf LL}=0,\\
{\cal T}^{(3/2)}_{\bf RL}&=&{\cal T}^{(3/2)}_{\bf LR}=i\sinh \pi p,\\
{\cal T}^{(3/2,n)}_{\bf RR}&=&{\cal T}^{(3/2,n)}_{\bf LL}=0,\\
{\cal T}^{(3/2,n)}_{\bf RL}&=&{\cal T}^{(3/2,n)}_{\bf LR}=i\sinh \pi p_n.~~~~~~~~~~~~~~~~\eea                      
In the above analysis, we have considered small axion mass ($\nu^2>0$) limiting situations with an arbitrary parameter $\alpha$, which corresponds to  Bunch Davies vacuum state with the choice $\alpha=0$.  For completeness, we also consider the large axion mass ($\nu^2<0$ where $\nu\rightarrow -i|\nu|$) limiting situation which is very important to study the imprints of quantum entanglement in cosmological correlation functions. In this large axion mass limiting situation, we actually consider a specific window of ${\bf SO(1,3)}$ principal quantum number, which is bounded within the range $0<p<|\nu|$. Consequently, the entries of the coefficient matrix $\tilde{m}$ can be  approximated as:
 \bea 
                 \label{arx2}
        \displaystyle \tilde{m}_{\bf RR} &=& -\sqrt{\frac{\cosh(|\nu|-p)}{\cosh(|\nu|+p)}} 
                                        \frac{2~\left[\cosh2\alpha\cosh^2\pi |\nu|-\sinh2\alpha\sinh^2\pi p+\frac{1}{2}\sinh 2\pi|\nu|\right]}{(e^{2\pi p}+e^{2\pi |\nu|})\cosh^2\alpha+(e^{2\pi p}+e^{2\pi |\nu|})\sinh^2\alpha},~~~~~~~~~~ \\
                   \label{arx3}
          \displaystyle \tilde{m}_{\bf RL} &=&-\sqrt{\frac{\cosh(|\nu|-p)}{\cosh(|\nu|+p)}} 
                                                  \frac{2~i~\left[\left(\cosh2\alpha+\sinh2\alpha\right)\cosh\pi |\nu|+\sinh\pi|\nu|\right]}{(e^{2\pi p}+e^{2\pi |\nu|})\cosh^2\alpha+(e^{2\pi p}+e^{2\pi |\nu|})\sinh^2\alpha},~~~~~~~~~\eea  
 which for  $\alpha=0$ yield a simplified expression for the  $\tilde{m}$  with Bunch Davies vacuum state.                                                    
We note that for general value of  $\alpha$ and for  large axion mass ($\nu^2<0$ where $\nu\rightarrow -i|\nu|$) , we  always get real value for $\tilde{m}_{\bf RR}$ and  imaginary value for $\tilde{m}_{\bf RL}$. This is an important observation for our further analysis.

From the perspective of cosmological observation in the superhorizon time scale, we again consider two further limiting situations: (a) large wave number ($p>>1$) or small wave length limit and (b)small wave number ($p<<1$) or large wave length limit. 

Using these two limiting situations we can  simplify the expression for the entries of the coefficient matrix $\tilde{m}$ considering both small and large axion mass.  We start with the expressions for small axion mass limit in large wave number ($p>>1$) approximation:
\bea 
              \tilde{m}_{ij}\approx 2~e^{i\theta}~e^{-2p\pi}~\widetilde{\cal T}^{(\nu)}_{ij}~{\rm sech}^2\alpha~~~~~ \\ \bar{\tilde{m}}_{ij,n}\approx 2~e^{i\theta}~ e^{-2p_n\pi}~\widetilde{\cal T}^{(\nu,n)}_{ij}~{\rm sech}^2\alpha~~~~~\eea
where we have defined the $\widetilde{\cal T}$ matrices for $p>>1$ limit as:
\bea ~~~~~~\widetilde{\cal T}^{(\nu)}_{ij}&=&\left(\begin{array}{ccc} \widetilde{\cal T}^{(\nu)}_{\bf RR} &~~~ \widetilde{\cal T}^{(\nu)}_{\bf RL} \\ \widetilde{\cal T}^{(\nu)}_{\bf LR} &~~~ \widetilde{\cal T}^{(\nu)}_{\bf LL}  \end{array}\right),~~~~~~~~~~\widetilde{\cal T}^{(\nu,n)}_{ij}=\left(\begin{array}{ccc} \widetilde{\cal T}^{(\nu,n)}_{\bf RR} &~~~ \widetilde{\cal T}^{(\nu,n)}_{\bf RL} \\ \widetilde{\cal T}^{(\nu,n)}_{\bf LR} &~~~ \widetilde{\cal T}^{(\nu,n)}_{\bf LL}  \end{array}\right).~~~~~\eea
and the corresponding entries of the $\widetilde{\cal T}$ matrices for $p>>1$ limit are given by the following simplified expressions: 
\bea 
\widetilde{\cal T}^{(\nu)}_{\bf RR}&=&\widetilde{\cal T}^{(\nu)}_{\bf LL}=\left[\left(\cosh^2\alpha+\sinh^2\alpha~e^{-2i\pi\nu}\right)-\frac{1}{4}~\sinh 2\alpha~ e^{2p\pi}~e^{-i\pi\nu}\sec\pi\nu\right]\cos\pi\nu,\\
\widetilde{\cal T}^{(\nu)}_{\bf RL}&=&{\cal T}^{(\nu)}_{\bf LR}=i\left[\cosh^2\alpha+\sinh^2\alpha~e^{-2i\pi\nu}+\sinh 2\alpha~\cos\pi\nu~ e^{-i\pi\nu}\right]\frac{1}{2}~e^{\pi p},\\
\widetilde{\cal T}^{(\nu,n)}_{\bf RR}&=&\widetilde{\cal T}^{(\nu,n)}_{\bf LL}=\left[\left(\cosh^2\alpha+\sinh^2\alpha~e^{-2i\pi\nu}\right)-\frac{1}{4}~\sinh 2\alpha~ e^{2p_n\pi}~e^{-i\pi\nu}\sec\pi\nu\right]\cos\pi\nu,\\
\widetilde{\cal T}^{(\nu,n)}_{\bf RL}&=&\widetilde{\cal T}^{(\nu,n)}_{\bf LR}=i\left[\cosh^2\alpha+\sinh^2\alpha~e^{-2i\pi\nu}+\sinh 2\alpha~\cos\pi\nu~ e^{-i\pi\nu}\right]\frac{1}{2}~e^{\pi p_n}.~~~~~~~~~~~~~~~~\eea 
 For massless ($\nu=3/2$) axion, we get the following simplified expressions:
\bea 
              \tilde{m}_{ij}\approx 2~e^{i\theta}~e^{-2p\pi}~\widetilde{\cal T}^{(3/2)}_{ij}~{\rm sech}^2\alpha~~~~~ \\ \bar{\tilde{m}}_{ij,n}\approx 2~e^{i\theta}~ e^{-2p_n\pi}~\widetilde{\cal T}^{(3/2,n)}_{ij}~{\rm sech}^2\alpha~~~~~\eea
where the $\widetilde{\cal T}^{(3/2)}$ matrices (for $p>>1$)  are given by:
\bea ~~~~~~\widetilde{\cal T}^{(3/2)}_{ij}&=&\left(\begin{array}{ccc} \widetilde{\cal T}^{(3/2)}_{\bf RR} &~~~ \widetilde{\cal T}^{(3/2)}_{\bf RL} \\ \widetilde{\cal T}^{(3/2)}_{\bf LR} &~~~ \widetilde{\cal T}^{(3/2)}_{\bf LL}  \end{array}\right),~~~~~~~~~~\widetilde{\cal T}^{(3/2,n)}_{ij}=\left(\begin{array}{ccc} \widetilde{\cal T}^{(3/2,n)}_{\bf RR} &~~~ \widetilde{\cal T}^{(3/2,n)}_{\bf RL} \\ \widetilde{\cal T}^{(3/2,n)}_{\bf LR} &~~~ \widetilde{\cal T}^{(3/2,n)}_{\bf LL}  \end{array}\right).~~~~~\eea
and the corresponding entries of the $\widetilde{\cal T}^{(3/2)}$ matrices  are given by : 
\bea 
\widetilde{\cal T}^{(3/2)}_{\bf RR}&=&\widetilde{\cal T}^{(3/2)}_{\bf LL}=0,\\
\widetilde{\cal T}^{(3/2)}_{\bf RL}&=&{\cal T}^{(3/2)}_{\bf LR}=\frac{i}{2}~e^{\pi p},\\
\widetilde{\cal T}^{(3/2,n)}_{\bf RR}&=&\widetilde{\cal T}^{(3/2,n)}_{\bf LL}=0,\\
\widetilde{\cal T}^{(3/2,n)}_{\bf RL}&=&\widetilde{\cal T}^{(3/2,n)}_{\bf LR}=\frac{i}{2}~e^{\pi p_n}.~~~~~~~~~~~~~~~~\eea 
On the other hand, for small axion mass and for large wave number ($p<<1$) we have:       
 \bea 
              \tilde{m}_{ij}\approx e^{i\theta}~\frac{\sqrt{2}~e^{-p\pi}~\hat{\cal T}^{(\nu)}_{ij}}{\sqrt{\cos 2\pi \nu}\left(\cosh^2\alpha+\sinh^2\alpha~e^{-2\pi i\nu}\right)}~~~~~ \\ \bar{\tilde{m}}_{ij,n}\approx e^{i\theta}~ \frac{\sqrt{2}~e^{-p_n\pi}~\hat{\cal T}^{(\nu,n)}_{ij}}{\sqrt{\cos 2\pi \nu}\left(\cosh^2\alpha+\sinh^2\alpha~e^{-2\pi i\nu}\right)}~~~~~\eea
where  the $\hat{\cal T}$ matrices are defined as:
\bea ~~~~~~\hat{\cal T}^{(\nu)}_{ij}&=&\left(\begin{array}{ccc} \hat{\cal T}^{(\nu)}_{\bf RR} &~~~ \hat{\cal T}^{(\nu)}_{\bf RL} \\ \hat{\cal T}^{(\nu)}_{\bf LR} &~~~ \hat{\cal T}^{(\nu)}_{\bf LL}  \end{array}\right),~~~~~~~~~~\hat{\cal T}^{(\nu,n)}_{ij}=\left(\begin{array}{ccc} \hat{\cal T}^{(\nu,n)}_{\bf RR} &~~~ \hat{\cal T}^{(\nu,n)}_{\bf RL} \\ \hat{\cal T}^{(\nu,n)}_{\bf LR} &~~~ \hat{\cal T}^{(\nu,n)}_{\bf LL}  \end{array}\right)~~~~~\eea
and the corresponding entries of the $\hat{\cal T}$ matrices (for $p<<1$ ) are given by : 
\bea 
\hat{\cal T}^{(\nu)}_{\bf RR}&=&\hat{\cal T}^{(\nu)}_{\bf LL}=\left[\left(\cosh^2\alpha+\sinh^2\alpha~e^{-2i\pi\nu}\right)-\sinh 2\alpha~  \pi^2 p^2~e^{-i\pi\nu}\sec\pi\nu\right]\cos\pi\nu,\\
\hat{\cal T}^{(\nu)}_{\bf RL}&=&\hat{\cal T}^{(\nu)}_{\bf LR}=i\left[\cosh^2\alpha+\sinh^2\alpha~e^{-2i\pi\nu}+\sinh 2\alpha~\cos\pi\nu~ e^{-i\pi\nu}\right] \pi p,\\
\hat{\cal T}^{(\nu,n)}_{\bf RR}&=&\hat{\cal T}^{(\nu,n)}_{\bf LL}=\left[\left(\cosh^2\alpha+\sinh^2\alpha~e^{-2i\pi\nu}\right)-\sinh 2\alpha~ \pi^2 p^2_n~e^{-i\pi\nu}\sec\pi\nu\right]\cos\pi\nu,\\
\hat{\cal T}^{(\nu,n)}_{\bf RL}&=&\hat{\cal T}^{(\nu,n)}_{\bf LR}=i\left[\cosh^2\alpha+\sinh^2\alpha~e^{-2i\pi\nu}+\sinh 2\alpha~\cos\pi\nu~ e^{-i\pi\nu}\right] \pi p_n.~~~~~~~~~~~~~~~~\eea 
For the case of  massless ($\nu=3/2$) axion, we get the following simplified expressions:
\bea 
              \tilde{m}_{ij}&\approx& e^{i\theta}~\sqrt{2}~e^{-p\pi}~\hat{\cal T}^{(3/2)}_{ij}~~~~~ \\ \bar{\tilde{m}}_{ij,n}&\approx & e^{i\theta}~ \sqrt{2}~e^{-p_n\pi}~\hat{\cal T}^{(3/2,n)}_{ij}~~~~~\eea
with the $\hat{\cal T}$ matrices defined as:
\bea ~~~~~~\hat{\cal T}^{(3/2)}_{ij}&=&\left(\begin{array}{ccc} \hat{\cal T}^{(3/2)}_{\bf RR} &~~~ \hat{\cal T}^{(3/2)}_{\bf RL} \\ \hat{\cal T}^{(3/2)}_{\bf LR} &~~~ \hat{\cal T}^{(3/2)}_{\bf LL}  \end{array}\right),~~~~~~~~~~\hat{\cal T}^{(3/2,n)}_{ij}=\left(\begin{array}{ccc} \hat{\cal T}^{(3/2,n)}_{\bf RR} &~~~ \hat{\cal T}^{(3/2,n)}_{\bf RL} \\ \hat{\cal T}^{(3/2,n)}_{\bf LR} &~~~ \hat{\cal T}^{(3/2,n)}_{\bf LL}  \end{array}\right)~~~~~\eea
and the corresponding entries of the $\hat{\cal T}^{(3/2)}$ matrices (for $p<<1$ ) are given by: 
\bea 
\hat{\cal T}^{(3/2)}_{\bf RR}&=&\hat{\cal T}^{(3/2)}_{\bf LL}=0,\\
\hat{\cal T}^{(3/2)}_{\bf RL}&=&\hat{\cal T}^{(3/2)}_{\bf LR}=i \pi p,\\
\hat{\cal T}^{(3/2,n)}_{\bf RR}&=&\hat{\cal T}^{(3/2,n)}_{\bf LL}=0,\\
\hat{\cal T}^{(3/2,n)}_{\bf RL}&=&\hat{\cal T}^{(3/2,n)}_{\bf LR}=i \pi p_n.~~~~~~~~~~~~~~~~\eea 
 For further analysis, it is convenient to change over to a suitable basis by tracing  over all possible contributions from ${\bf R}$ and ${\bf L}$ region. To achieve this we  perform another Bogoliubov transformation by introducing new sets of operators :
    \bea 
    \label{kc1} \tilde{c}_{\bf R}&=& \tilde{u}~b_{\bf R}+\tilde{v}~b^{\dagger}_{\bf R},~~ \tilde{c}_{\bf L}= \bar{\tilde{u}}~b_{\bf L}+\bar{\tilde{v}}~b^{\dagger}_{\bf L},~~
       \label{kc3} \tilde{C}_{{\bf R},n}= \tilde{U}_n~b_{{\bf R},n}+\tilde{V}_n~b^{\dagger}_{{\bf R},n},~~ \tilde{C}_{{\bf L},n}= \bar{\tilde{U}}_n~b_{{\bf L},n}+\bar{\tilde{V}}_n~b^{\dagger}_{{\bf L},n},~~~~~~~~~~\eea   
      satisfying the following conditions:
         \bea  
            |\tilde{u}|^2-|\tilde{v}|^2&=& 1,~|\bar{\tilde{u}}|^2-|\bar{\tilde{v}}|^2= 1,~~  
                       |\tilde{U}_n|^2-|\tilde{V}_n|^2= 1,~~|\bar{\tilde{U}}_n|^2-|\bar{\tilde{V}}_n|^2= 1.\eea 
  Using these operators we write the $\alpha$-vacuum state in terms of new basis represented by the direct product of ${\bf R}^{'}$ and ${\bf L}^{'}$ vacuum state as:
  \bea |\alpha\rangle &=&\left[1-\left(|\gamma^{(\alpha)}_p|^2+\sum^{\infty}_{n=0}|\Gamma^{(\alpha)}_{p,n}|^2\right)\right]^{1/2}\exp\left(\gamma^{(\alpha)}_{p}~\tilde{c}^{\dagger}_{\bf R}~\tilde{c}^{\dagger}_{\bf L}+\sum^{\infty}_{n=0}\Gamma^{(\alpha)}_{p,n}~\tilde{C}^{\dagger}_{{\bf R},n}~\tilde{C}^{\dagger}_{{\bf L},n}\right)\left(|{\bf R}^{'}\rangle\otimes |{\bf L}^{'}\rangle\right)^{(\alpha)},~~~~~~~~~\eea  
      where $\gamma^{(\alpha)}_{p}$ and $\Gamma^{(\alpha)}_{p,n}$ are to be determined shortly. 
     We note that the the relationship between the new and the old basis is given by:
      \bea \left(|{\bf R}\rangle\otimes |{\bf L}\rangle\right)\rightarrow
      \left(|{\bf R}^{'}\rangle\otimes |{\bf L}^{'}\rangle\right)^{(\alpha)}&=&\left[1-\left(|\gamma^{(\alpha)}_p|^2+\sum^{\infty}_{n=0}|\Gamma^{(\alpha)}_{p,n}|^2\right)\right]^{-1/2}~\nonumber\\ &&\exp\left(-\gamma^{(\alpha)}_{p}~\tilde{c}^{\dagger}_{\bf R}~\tilde{c}^{\dagger}_{\bf L}-\sum^{\infty}_{n=0}\Gamma^{(\alpha)}_{p,n}~\tilde{C}^{\dagger}_{{\bf R},n}~\tilde{C}^{\dagger}_{{\bf L},n}\right)~\nonumber\\ &&\exp\left(\frac{1}{2}\sum_{i,j={\bf R},{\bf L}}m_{ij}~b^{\dagger}_{i}~b^{\dagger}_{j}+\frac{1}{2}\sum_{i,j={\bf R},{\bf L}}\sum^{\infty}_{n=0}\bar{m}_{ij,n}~\bar{b}^{\dagger}_{i,n}~\bar{b}^{\dagger}_{j,n}\right)\left(|{\bf R}\rangle\otimes |{\bf L}\rangle\right).~~~~~~~~\eea
      The commutation relations between  the creation and annihilation operators corresponding to the new sets of oscillators is taken as:
            \bea 
            \left[ \tilde{c}_i, \tilde{c}^{\dagger}_j\right]&=&\delta_{ij},~~\left[  \tilde{c}_i, \tilde{c}_j\right]=0=
            \left[  \tilde{c}^{\dagger}_i, \tilde{c}^{\dagger}_j\right],~~ 
                  \left[ \tilde{C}_{i,n},\tilde{C}^{\dagger}_{j,m}\right]=\delta_{ij}{\delta}_{nm},~~\left[ \tilde{C}_{i,n},\tilde{C}_{j,m}\right]= 0=
                  \left[ \tilde{C}^{\dagger}_{i,m}\tilde{C}^{\dagger}_{j,m}\right].~~~~~~~~~
                  \eea
                These operations act on the $\alpha$ vacuum state in the following way:
       \bea 
       \label{gv1} \tilde{c}_{\bf R}|\alpha\rangle &=&\gamma^{(\alpha)}_{p}~\tilde{c}^{\dagger}_{\bf L}|\alpha\rangle,~~
    \tilde{c}_{\bf R}|\alpha\rangle =\gamma^{(\alpha)}_{p}~\tilde{c}^{\dagger}_{\bf L}|\alpha\rangle,~~ 
   \tilde{C}_{{\bf R},n}|\alpha\rangle =\Gamma^{(\alpha)}_{p,n}~\tilde{C}^{\dagger}_{{\bf L},n}|\alpha\rangle,~~ \tilde{C}_{{\bf R},n}|\alpha\rangle =\Gamma^{(\alpha)}_{p,n}~\tilde{C}^{\dagger}_{{\bf L},n}|\alpha\rangle.~~~~~~~~\eea Further, one can express the new $c$ type annihilation operators in terms of the old $b$ type annihilation operators as: 
             \bea \tilde{c}_{J}&=& b_{I}\tilde{\cal G}^{I}_{J}=b_{I}\left(\begin{array}{ccc} \tilde{U}_q &~~~ \tilde{V}^{*}_q \\ \tilde{V}_q &~~~ \tilde{U}^{*}_q  \end{array}\right)
             ,~~~~~~~~~
             \tilde{C}_{J(n)}=\bar{b}_{J(n)}\left(\tilde{\cal G}_{(n)}\right)^{I}_{J}=\bar{b}_{J(n)}\left(\begin{array}{ccc} \bar{\tilde{U}}_{ q,n} &~~~ \bar{\tilde{V}}^{*}_{\sigma q,n} \\ \bar{\tilde{V}}_{ q,n} &~~~ \bar{\tilde{U}}^{*}_{ q,n}  \end{array}\right).\eea
                    Note that
                     $\tilde{U}_q \equiv {\rm \bf diag}\left(\tilde{u},\bar{\tilde{u}}\right)$,$\tilde{V}_q \equiv {\rm \bf diag}\left(\tilde{v},\bar{\tilde{v}}\right),$ $\bar{\tilde{U}}_{q,n} \equiv {\rm \bf diag}\left(\tilde{U}_n,\bar{\tilde{U}}_n\right),$ $\bar{\tilde{V}}_{q,n} \equiv {\rm \bf diag}\left(\tilde{V}_n,\bar{\tilde{V}}_n\right)$.
     From  Equations~(\ref{kc1})  and~(\ref{gv1}),  we obtain the following sets of homogeneous equations: 
  \bea 
   \underline{\textcolor{red}{\bf For~complementary~solution:}}~~~~~~~~~~~~~~~~~~~~~~~\nonumber\\ 
  \label{a1} \tilde{m}_{\bf RR}\tilde{u}+\tilde{v}-\gamma^{(\alpha)}_{p} \tilde{m}_{\bf RL}\bar{\tilde{v}}^{*}&=& 0,~~~\\
  \label{a2}  \tilde{m}_{\bf RR}\bar{\tilde{u}}+\bar{\tilde{v}}-\gamma^{(\alpha)}_{p} \tilde{m}_{\bf RL}\tilde{v}^{*}= 0,~~~~~~\\ 
  \label{a3}  \tilde{m}_{\bf RL}\tilde{u}-\gamma^{(\alpha)}_{p} \bar{\tilde{u}}^{*}-\gamma^{(\alpha)}_{p}\tilde{m}_{\bf RR}\bar{\tilde{v}}^{*}&=& 0,~~~\\
  \label{a4}  \tilde{m}_{\bf RL}\bar{\tilde{u}}-\gamma^{(\alpha)}_{p} \tilde{u}^{*}-\gamma^{(\alpha)}_{p}\tilde{m}_{\bf RR}\tilde{v}^{*}= 0,~~~~~~~~~~\\
     \underline{\textcolor{red}{\bf For~~particular~solution:}}~~~~~~~~~~~~~~~~~~~~~~~\nonumber\\ 
     \tilde{m}_{{\bf RR},n}\tilde{U}_n+\tilde{V}_n-\Gamma^{(\alpha)}_{p,n} \tilde{m}_{{\bf RL},n}\bar{\tilde{V}}^{*}_n&=& 0,~~~
        \tilde{m}_{{\bf RR},n}\bar{\tilde{U}}_n+\bar{\tilde{V}}_n-\Gamma^{(\alpha)}_{p,n} \tilde{m}_{{\bf RL},n}\tilde{V}^{*}_n= 0,~~~~~~\\ 
        \tilde{m}_{{\bf RL},n}\tilde{U}_n-\Gamma^{(\alpha)}_{p,n} \bar{\tilde{U}}^{*}_n-\Gamma^{(\alpha)}_{p,n} \tilde{m}_{{\bf RR},n}\bar{\tilde{V}}^{*}_n&=& 0,~~
        \tilde{m}_{{\bf RL},n}\bar{\tilde{U}}_n-\Gamma^{(\alpha)}_{p,n}  \tilde{U}^{*}_n-\Gamma^{(\alpha)}_{p,n} \tilde{m}_{{\bf RR},n}\tilde{V}^{*}_n= 0,~~~~~~~~~\eea
            Using the relations $\tilde{v}^{*}=\bar{\tilde{v}}, 
           \tilde{u}^{*}=\bar{\tilde{u}}$, $\tilde{V}^{*}_n=\bar{\tilde{V}}_n, 
           \tilde{U}^{*}_n=\bar{\tilde{U}}_n$, $|\tilde{u}|^2-|\tilde{v}|^2=1$ and $|\tilde{U}_n|^2-|\tilde{V}_n|^2=1$ the solutions of  these equations can be written as: 
         \bea
          \label{as1}\gamma^{(\alpha)}_{p}&=&\frac{1}{\sqrt{2}|\tilde{m}_{\bf RL}|}\left[\left(1+|\tilde{m}_{\bf RL}|^4+|\tilde{m}_{\bf RR}|^4\right.\right.\nonumber\\ && \left.\left.-2|\tilde{m}_{\bf RR}|^2-\tilde{m}^2_{\bf RR}(\tilde{m}^{*}_{\bf RL})^2-\tilde{m}^2_{\bf RL}(\tilde{m}^{*}_{\bf RR})^2\right) 
          \pm \left\{\left(-1-|\tilde{m}_{\bf RL}|^4-|\tilde{m}_{\bf RR}|^4\right.\right.\right.\nonumber\\ && \left.\left.\left.+2|\tilde{m}_{\bf RR}|^2+\tilde{m}^2_{\bf RR}(\tilde{m}^{*}_{\bf RL})^2+\tilde{m}^2_{\bf RL}(\tilde{m}^{*}_{\bf RR})^2\right)^2 
          -4|\tilde{m}_{\bf RL}|^4\right\}^{\frac{1}{2}}\right]^{\frac{1}{2}}\nonumber\\
          &\approx&i\frac{\sqrt{2} \left[\cosh^2\alpha+\sinh^2\alpha~e^{2i\pi\nu}+\sinh 2\alpha\cos \pi \nu~e^{i\pi\nu}\right]}{\left(\sqrt{\cosh 2\pi p +\cos 2\pi \nu}\pm\sqrt{\cosh 2\pi p +\cos 2\pi \nu+2}\right)\left(\cosh^2\alpha+\sinh^2\alpha~e^{-2\pi( p-i\nu)}\right)}\nonumber\\
          \underrightarrow{\alpha=0}~~~~~~~\gamma^{(0)}_{p}&=&\frac{1}{2m_{\bf RL}}\left[\left(1+m^2_{\bf RL}-m^2_{\bf RR}\right) 
          \pm \sqrt{\left(1+m^2_{\bf RL}-m^2_{\bf RR}\right)^2-4m^2_{\bf RL}}\right]\nonumber\\
          &\approx&i\frac{\sqrt{2}}{\sqrt{\cosh 2\pi p +\cos 2\pi \nu}\pm\sqrt{\cosh 2\pi p +\cos 2\pi \nu+2}},\eea\bea
           \label{as2}\Gamma^{(\alpha)}_{p,n}&=&\frac{1}{\sqrt{2}|\tilde{m}_{{\bf RL},n}|}\left[\left(1+|\tilde{m}_{{\bf RL},n}|^4+|\tilde{m}_{{\bf RR},n}|^4\right.\right.\nonumber\\ && \left.\left.-2|\tilde{m}_{{\bf RR},n}|^2-\tilde{m}^2_{{\bf RR},n}(\tilde{m}^{*}_{{\bf RL},n})^2-\tilde{m}^2_{{\bf RL},n}(\tilde{m}^{*}_{{\bf RR},n})^2\right) 
           \pm \left\{\left(-1-|\tilde{m}_{{\bf RL},n}|^4-|\tilde{m}_{{\bf RR},n}|^4\right.\right.\right.\nonumber\\ && \left.\left.\left.+2|\tilde{m}_{{\bf RR},n}|^2+\tilde{m}^2_{{\bf RR},n}(\tilde{m}^{*}_{{\bf RL},n})^2+\tilde{m}^2_{{\bf RL},n}(\tilde{m}^{*}_{{\bf RR},n})^2\right)^2 
           -4|\tilde{m}_{{\bf RL},n}|^4\right\}^{\frac{1}{2}}\right]^{\frac{1}{2}}\nonumber\\
           &\approx& i\frac{\sqrt{2} \left[\cosh^2\alpha+\sinh^2\alpha~e^{2i\pi\nu}+\sinh 2\alpha\cos \pi \nu~e^{i\pi\nu}\right]}{\left(\sqrt{\cosh 2\pi p_n +\cos 2\pi \nu}\pm\sqrt{\cosh 2\pi p_n +\cos 2\pi \nu+2}\right)\left(\cosh^2\alpha+\sinh^2\alpha~e^{-2\pi( p_n-i\nu)}\right)}\nonumber\\
            \underrightarrow{\alpha=0}~~~~~~~\Gamma^{(0)}_{p,n}&=&\frac{1}{2\bar{m}_{{\bf RL},n}}\left[\left(1+\bar{m}^2_{{\bf RL},n}-\bar{m}^2_{{\bf RR},n}\right)
            \pm \sqrt{\left(1+\bar{m}^2_{{\bf RL},n}-\bar{m}^2_{{\bf RR},n}\right)^2-4\bar{m}^2_{{\bf RL},n}}\right]\nonumber\\
           &\approx& i\frac{\sqrt{2}}{\sqrt{\cosh 2\pi p_n +\cos 2\pi \nu}\pm\sqrt{\cosh 2\pi p_n +\cos 2\pi \nu+2}},~~~~~~~~~~\eea         
           where the components $\tilde{m}_{\bf RR}=\tilde{m}_{\bf LL}$, $\tilde{m}_{\bf RL}=\tilde{m}_{\bf LR}$ and  $\tilde{m}_{{\bf RR},n}=\tilde{m}_{{\bf LL},n}$, $\tilde{m}_{{\bf RL},n}=\tilde{m}_{{\bf LR},n}$ are defined in equations (3.62--68) for general $\alpha$ vacua. Also the components without tilde symbol represent the contribution from $\alpha=0$, which is the Bunch Davies vacuum state.
           
           Further, for the massless ($\nu=3/2$) axion field we get the following simplified expressions:
            \bea
          \label{as1}\gamma^{(\alpha,3/2)}_{p}
          &\approx&i\frac{\sqrt{2} }{\left(\sqrt{\cosh 2\pi p -1}\pm\sqrt{\cosh 2\pi p +1}\right)\left(\cosh^2\alpha-\sinh^2\alpha~e^{-2\pi p}\right)}\nonumber\\
          \underrightarrow{\alpha=0}~~~~~~~\gamma^{(0,3/2)}_{p}
          &\approx&i\frac{\sqrt{2}}{\sqrt{\cosh 2\pi p -1}\pm\sqrt{\cosh 2\pi p +1}},\eea\bea
           \label{as2}\Gamma^{(\alpha)}_{p,n}
           &\approx& i\frac{\sqrt{2} }{\left(\sqrt{\cosh 2\pi p_n -1}\pm\sqrt{\cosh 2\pi p_n +1}\right)\left(\cosh^2\alpha-\sinh^2\alpha~e^{-2\pi p_n}\right)}\nonumber\\
            \underrightarrow{\alpha=0}~~~~~~~\Gamma^{(0)}_{p,n}
           &\approx& i\frac{\sqrt{2}}{\sqrt{\cosh 2\pi p_n -1}\pm\sqrt{\cosh 2\pi p_n +1}},~~~~~~~~~~\eea 
           
           In the large axion mass ($\nu^2<0$ where $\nu\rightarrow -i|\nu|$) limit the two solutions for the $\gamma^{(\alpha)}_p$ and $\Gamma^{(\alpha)}_{p,n}$for $\alpha$ vacuum are given by: 
\bea \gamma^{(\alpha)}_p&\approx&\frac{1}{2|\tilde{m}_{\bf RL}|}\left[\left(1+|\tilde{m}_{\bf RL}|^2-\tilde{m}^2_{\bf RR}\right) 
         \pm \sqrt{\left(1+|\tilde{m}_{\bf RL}|^2-\tilde{m}^2_{\bf RR}\right)^2-4|\tilde{m}_{\bf RL}|^2}\right].~~~~~~~~~~~\\
   \Gamma^{(\alpha)}_{p,n}&\approx&\frac{1}{2|\tilde{m}_{{\bf RL},n}|}\left[\left(1+|\tilde{m}_{{\bf RL},n}|^2-\tilde{m}^2_{{\bf RR},n}\right) 
         \pm \sqrt{\left(1+|\tilde{m}_{\bf RL}|^2-\tilde{m}^2_{\bf RR}\right)^2-4|\tilde{m}_{{\bf RL},n}|^2}\right]\eea      
 In this limit, we divide the total window of $p$ into two regions,  given by $0<p<|\nu|$ and $|\nu|<p<\Lambda_{\bf C}$. In these regions of interest, the two solutions for $\gamma^{(\alpha)}_p$ in presence of $\alpha$ vacuum can be approximately written as:
                     \bea
                                    \label{ty1}
                           \displaystyle |\gamma^{(\alpha)}_p| &\approx&\displaystyle\left\{\begin{array}{ll}
                          \displaystyle e^{\mp \pi |\nu|}\left(1+\tan\alpha\right)~~~~~~~~~~~~ &
                                                                                    \mbox{\small {\textcolor{red}{\bf for $0<p<|\nu|$}}}  
                                                                                   \\ 
                                   \displaystyle \frac{e^{\mp \pi p}\left(1+\tan\alpha\right)\left(1+\tan\alpha~e^{2\pi|\nu|}\right)}{\left(1+\tan^2\alpha~e^{-2\pi p}\right)} & \mbox{\small { \textcolor{red}{\bf for $|\nu|<p<\Lambda_{\bf C}/2\pi$}}}.~~~~~~~~
                                                                                             \end{array}
                                                                                   \right.\eea
                        and
                        \bea
                                       \label{ty2}
                              \displaystyle |\Gamma^{(\alpha)}_{p,n}| &=&\displaystyle\left\{\begin{array}{ll}
                             \displaystyle e^{\mp \pi |\nu|}\left(1+\tan\alpha\right)~~~~~~~~~~~~ &
                                                                                       \mbox{\small {\textcolor{red}{\bf for $0<p<|\nu|$}}}  
                                                                                      \\ 
                                      \displaystyle \frac{e^{\mp \pi p_n}\left(1+\tan\alpha\right)\left(1+\tan\alpha~e^{2\pi|\nu|}\right)}{\left(1+\tan^2\alpha~e^{-2\pi p_n}\right)} & \mbox{\small { \textcolor{red}{\bf for $|\nu|<p<\Lambda_{\bf C}/2\pi$}}}.~~~~~~~~
                                                                                                \end{array}
                                                                                      \right.\eea
   Further, in the limit $p>>1$ we get the following simplified results:
         \bea
          \label{as1}\gamma^{(\alpha)}_{p}&
          \approx&i\frac{2 \left[\cosh^2\alpha+\sinh^2\alpha~e^{2i\pi\nu}+\sinh 2\alpha\cos \pi \nu~e^{i\pi\nu}\right]{\rm sech}^2\alpha}{\left(\sqrt{|\cosh 2\pi p |}\pm\sqrt{|\cosh 2\pi p| +4}\right)}\nonumber\\
          \underrightarrow{\alpha=0}~~~~~~~\gamma^{(0)}_{p}
          &\approx&i\frac{2}{\sqrt{|\cosh 2\pi p |}\pm\sqrt{|\cosh 2\pi p|+4}},\eea\bea
           \label{as2}\Gamma^{(\alpha)}_{p,n}
           &\approx& i\frac{2 \left[\cosh^2\alpha+\sinh^2\alpha~e^{2i\pi\nu}+\sinh 2\alpha\cos \pi \nu~e^{i\pi\nu}\right]{\rm sech}^2\alpha}{\left(\sqrt{|\cosh 2\pi p_n|}\pm\sqrt{|\cosh 2\pi p_n| +4}\right)}\nonumber\\
          \underrightarrow{\alpha=0}~~~~~~~\Gamma^{(0)}_{p,n}
          &\approx&i\frac{2}{\sqrt{|\cosh 2\pi p_n |}\pm\sqrt{|\cosh 2\pi p_n|+4}},~~~~~~~~~~\eea 
 For massless ($\nu=3/2$) axion field this simplifies to :          
           \bea
          \label{as1}\gamma^{(\alpha,3/2)}_{p}&
          \approx&i\frac{2{\rm sech}^2\alpha}{\left(\sqrt{|\cosh 2\pi p |}\pm\sqrt{|\cosh 2\pi p| +4}\right)}\nonumber\\
          \underrightarrow{\alpha=0}~~~~~~~\gamma^{(0,3/2)}_{p}
          &\approx&i\frac{2}{\sqrt{|\cosh 2\pi p |}\pm\sqrt{|\cosh 2\pi p|+4}},\eea\bea
           \label{as2}\Gamma^{(\alpha,3/2)}_{p,n}
           &\approx& i\frac{2 {\rm sech}^2\alpha}{\left(\sqrt{|\cosh 2\pi p_n|}\pm\sqrt{|\cosh 2\pi p_n| +4}\right)}\nonumber\\
          \underrightarrow{\alpha=0}~~~~~~~\Gamma^{(0,3/2)}_{p,n}
          &\approx&i\frac{2}{\sqrt{|\cosh 2\pi p_n |}\pm\sqrt{|\cosh 2\pi p_n|+4}},~~~~~~~~~~\eea 
  On the other hand, in the limit $p<<1$ we get the following results:
         \bea
          \label{as1}\gamma^{(\alpha)}_{p}&
          \approx&i\frac{\sqrt{2} \left[\cosh^2\alpha+\sinh^2\alpha~e^{2i\pi\nu}+\sinh 2\alpha\cos \pi \nu~e^{i\pi\nu}\right]}{\left(\sqrt{\cos 2\pi \nu+1}\pm\sqrt{\cos 2\pi \nu+3}\right)\left(\cosh^2\alpha+\sinh^2\alpha~e^{2\pi  i\nu}\right)}\nonumber\\
          \underrightarrow{\alpha=0}~~~~~~~\gamma^{(0)}_{p}
          &\approx&i\frac{\sqrt{2}}{\sqrt{\cos 2\pi \nu+1}\pm\sqrt{\cos 2\pi \nu+3}},\eea\bea
           \label{as2}\Gamma^{(\alpha)}_{p,n}
           &\approx&i\frac{\sqrt{2} \left[\cosh^2\alpha+\sinh^2\alpha~e^{2i\pi\nu}+\sinh 2\alpha\cos \pi \nu~e^{i\pi\nu}\right]}{\left(\sqrt{\cos 2\pi \nu+1}\pm\sqrt{\cos 2\pi \nu+3}\right)\left(\cosh^2\alpha+\sinh^2\alpha~e^{2\pi  i\nu}\right)}\nonumber\\
            \underrightarrow{\alpha=0}~~~~~~~\Gamma^{(0)}_{p,n}&\approx& i\frac{\sqrt{2}}{\sqrt{\cos 2\pi \nu+1}\pm\sqrt{\cos 2\pi \nu+3}},~~~~~~~~~~\eea
which, for massless ($\nu=3/2$) axion field , simplifies to:                    
             \bea
          \label{as1}\gamma^{(\alpha,3/2)}_{p}&
          \approx&\pm i\frac{1}{\sqrt{2}}~~~
          \underrightarrow{\alpha=0}~~~~~~~\gamma^{(0,3/2)}_{p}
          \approx\pm  i\frac{1}{ \sqrt{2} },\eea\bea
           \label{as2}\Gamma^{(\alpha,3/2)}_{p,n}
           &\approx&\pm i\frac{1}{\sqrt{2} }~~~
            \underrightarrow{\alpha=0}~~~~~~~\Gamma^{(0,3/2)}_{p,n}\approx\pm i\frac{1}{ \sqrt{2}},~~~~~~~~~~\eea  
    and are very useful information for the computation of spectrum of vacuum fluctuation.

           Further, the Fourier mode of the total compact solution  in the region \textcolor{red}{\bf L} in case of $\alpha$ vacua can be re-expressed in terms of the oscillators defined in the new basis ($\tilde{c},\tilde{C}$) as well as the {\bf SO(1,3)} quantum numbers ($p,l,m$) as:
           \bea \widetilde{\phi_{{\bf L},plm}(t_{\bf L})}=\frac{H}{\sinh t_{\bf L}}\tilde{c}^{\bf T}_{\cal I}\tilde{\psi}^{\cal I}_{\bf T}=\frac{H}{\sinh t_{\bf L}}\left[\frac{1}{{\cal N}_{p}}\widetilde{\left(G^{-1}\right)^{I}_{J}}{\cal P}^{J}+\sum^{\infty}_{n=0}\frac{1}{{\cal N}_{p,(n)}}\widetilde{\left(G^{-1}_{(n)}\right)^{I}_{J}}{\cal P}^{J}_{(n)}\right], \eea
           where the total wave function $\tilde{\psi}^{\cal I}_{\bf T}$ is a column matrix and  for the complementary and particular part of the solution the inverse matrix $\widetilde{\left(G^{-1}\right)^{I}_{J}}$ and $\widetilde{\left(G^{-1}_{(n)}\right)^{I}_{J}}$ are defined as:
            \bea \widetilde{\left(G^{-1}\right)^{I}_{J}}&=&\left(\begin{array}{ccc} \tilde{\bar{u}}^{*} &~~~ -\tilde{\bar{v}}^{*}\\ -\tilde{\bar{v}}  &~~~ \tilde{\bar{u}}  \end{array}\right),~~~~~~  \widetilde{\left(G^{-1}_{(n)}\right)^{I}_{J}}=\left(\begin{array}{ccc} \tilde{\bar{U}}^{*}_{(n)} &~~~ -\tilde{\bar{V}}^{*}_{(n)}\\ -\tilde{\bar{V}}_{(n)}  &~~~ \tilde{\bar{U}}_{(n)}  \end{array}\right),~~~~~~
   \psi^{{\cal I},{\bf T}}=\left(\begin{array}{ccc} \psi^{{\bf L},{\bf T}}(t_{\bf L}) \\ \psi^{{\bf L}^{*},{\bf T}}(t_{\bf L})
   \end{array}\right).~~~~~~~~~~~~~\eea
 
 When we trace out the degrees of freedom over the right part of the Hilbert space, we obtain the following  reduced density matrix for the left part of the Hilbert space :
   \bea \label{ffz1}(\rho_{\bf L}(\alpha))_{p,l,m}&=&{\bf \rm Tr}_{\bf R}|\alpha\rangle \langle \alpha|,\eea
   where the $\alpha$ vacuum state is written in terms of $\tilde{c}$ type of oscillators as:
   \bea \label{ddqz1} |\alpha\rangle &\approx&\left[1-\left(|\gamma^{(\alpha)}_p|^2+\sum^{\infty}_{n=0}|\Gamma^{(\alpha)}_{p,n}|^2\right)\right]^{1/2}\exp\left[\gamma^{(\alpha)}_{p}~\tilde{c}^{\dagger}_{\bf R}~\tilde{c}^{\dagger}_{\bf L}+\sum^{\infty}_{n=0}\Gamma^{(\alpha)}_{p,n}~\tilde{C}^{\dagger}_{{\bf R},n}~\tilde{C}^{\dagger}_{{\bf L},n}\right]\left(|{\bf R}^{'}\rangle\otimes |{\bf L}^{'}\rangle\right)^{(\alpha)},~~~~~~~~~~\eea
    Substituting Eq~(\ref{ddqz1}) in Eq~(\ref{ffz1}), we get the expression for the reduced density matrix for the left part of the Hilbert space:
   \be \label{ff1x}(\rho_{\bf L}(\alpha))_{p,l,m}=\underbrace{\frac{\left(1-|\gamma^{(\alpha)}_{p}|^2\right)}{1+f^{(\alpha)}_{p}}\sum^{\infty}_{k=0}|\gamma^{(\alpha)}_{p}|^{2k}\widetilde{|k;p,l,m\rangle}\widetilde{\langle k;p,l,m|}}_{\textcolor{red}{\bf Complementary~part}} +\underbrace{\frac{(f^{(\alpha)}_{p})^2}{1+f^{(\alpha)}_p}\sum^{\infty}_{n=0}\sum^{\infty}_{r=0}|\Gamma^{(\alpha)}_{p,n}|^{2r}\widetilde{|n,r;p,l,m\rangle}\widetilde{\langle n,r;p,l,m|}}_{\textcolor{red}{\bf Particular~part}}.\ee
   where ${\it f}^{(\alpha)}_{p}$ is given by
   \bea {\it f}^{(\alpha)}_{p}&=&\left(\sum^{\infty}_{n=0}\frac{1}{1-|\Gamma^{(\alpha)}_{p,n}|^2}\right)^{-1},\eea 
   and the states $\widetilde{|k;p,l,m\rangle}$ and $\widetilde{|n,r;p,l,m\rangle}$ are expressed in terms of the new quantum state $|{\bf L}^{'}\rangle$ as:
   \bea \widetilde{|k;p,l,m\rangle}&=& \frac{1}{\sqrt{k!}}(\tilde{c}^{\dagger}_{\bf L})^{k}|{\bf L}^{'}\rangle,~~~~
   \widetilde{|n,r;p,l,m\rangle}= \frac{1}{\sqrt{r!}}(\tilde{C}^{\dagger}_{{\bf L},n})^{r}|{\bf L}^{'}\rangle.\eea           
Note that for $\alpha=0$, we get back the result obtained for Bunch Davies vacuum.     

 \subsubsection{Two point correlation function}
 \label{x2b}
 In this subsection we explicitly compute the two point correlation function and its significant role to obtain long range effect in the cosmological correlation using the  generalised $\alpha$ and Bunch Davies vacuum. For this purpose and using the expression for the reduced density matrix, derived in the previous subsection, we first compute the mean square quantum vacuum fluctuation, which is expressed for $\alpha$ vacua as:
 \bea\label{qqwfx} {{\rm Tr}_{\bf L}\left(\rho_{\bf L}(\alpha)\phi_{\bf L}(t_{\bf L})\phi^{\dagger}_{\bf L}(t_{\bf L})\right)}_{(\alpha)}&=&\exp\left(-2\alpha\right)\left[\underbrace{\left(1-|\gamma^{(\alpha)}_p|^2\right)\sum^{\infty}_{n=0}|\gamma^{(\alpha)}_p|^{2n}\widetilde{\langle n; p, l, m|} \phi_{\bf L}(t_{\bf L})\phi^{\dagger}_{\bf L}(t_{\bf L})\widetilde{|n;p,l,m\ \rangle}}_{\textcolor{red}{\bf Complementary ~part}} \right.\nonumber\\&&\left.~~~~~~+\underbrace{\frac{1}{\left(f^{(\alpha)}_{p}\right)^2}\sum^{\infty}_{r=0}\sum^{\infty}_{s=0}|\Gamma^{(\alpha)}_{p,r,s}|^{2r}\widetilde{\langle s,r; p, l, m|} \phi_{\bf L}(t_{\bf L})\phi^{\dagger}_{\bf L}(t_{\bf L})\widetilde{|s,r;p,l,m\ \rangle}}_{\textcolor{red}{\bf Particular~part}}\right].~~~~~~~~~~\eea   
 In the above,  we have used the shorthand notation $\phi_{\bf L}(t_{\bf L})=\phi_{{\bf L}plm}(t)$ for the field. Note that, setting $\alpha=0$ in Eq~(\ref{qqwfx}) we get the result for the Bunch Davies vacuum which is given by:
  \bea\label{qqwfxbd}{{\rm Tr}_{\bf L}\left(\rho_{\bf L}(\alpha)\phi_{\bf L}(t_{\bf L})\phi^{\dagger}_{\bf L}(t_{\bf L})\right)}_{(\bf BD)}&=&\underbrace{\left(1-|\gamma^{(0)}_p|^2\right)\sum^{\infty}_{n=0}|\gamma^{(0)}_p|^{2n}{\langle n; p, l, m|} \phi_{\bf L}(t_{\bf L})\phi^{\dagger}_{\bf L}(t_{\bf L}){|n;p,l,m\ \rangle}}_{\textcolor{red}{\bf Complementary ~part}} \nonumber\\&&~~~~~~+\underbrace{\frac{1}{\left(f^{(0)}_{p}\right)^2}\sum^{\infty}_{r=0}\sum^{\infty}_{s=0}|\Gamma^{(0)}_{p,r,s}|^{2r}{\langle s,r; p, l, m|} \phi_{\bf L}(t_{\bf L})\phi^{\dagger}_{\bf L}(t_{\bf L}){|s,r;p,l,m\ \rangle}}_{\textcolor{red}{\bf Particular~part}}.~~~~~~~~~~\eea 
  Here ${|s,r;p,l,m\ \rangle}$ is the Bunch Davies counterpart of the quantum state in the newly Bogoliubov transformed  basis and is obtained by simply setting $\alpha=0$ in the definition of the quantum state introduced in terms of the new oscillators. 
  
  The contributions from the complementary and the particular part, as appearing in the right hand side of Eq~(\ref{qqwfx}) for each $n$-particle state are found to be:
 \bea \label{kjv1}\widetilde{\langle n; p, l, m|} \phi_{\bf L}(t_{\bf L})\phi^{\dagger}_{\bf L}(t_{\bf L})\widetilde{|n;p,l,m\ \rangle}&=&\frac{H^2}{\sinh^2t_{\bf L}}\frac{1}{n!}\langle {\bf L}^{'}|(\tilde{c}_{\bf L})^{n}\left(\tilde{c}^{\bf T}_{\cal I}\tilde{\psi}^{\dagger {\cal I}}_{\bf T}\right)\left(\tilde{c}^{\bf T}_{\cal J}\tilde{\psi}^{\dagger {\cal J}}_{\bf T}\right)^{\dagger}(\tilde{c}^{\dagger}_{\bf L})^{n}|{\bf L}^{'}\rangle ~~~~~~\nonumber\\
 &=&\frac{H^2}{\sinh^2t_{\bf L}}\left(2n+1\right)|\tilde{\psi}^{\bf L}_{\bf T}|^2,\\
 \label{kjv3}\widetilde{\langle s,r; p, l, m|} \phi_{\bf L}(t_{\bf L})\phi^{\dagger}_{\bf L}(t_{\bf L})\widetilde{|s,r;p,l,m\ \rangle}&=&\frac{H^2}{\sinh^2t_{\bf L}}\frac{1}{r!}\langle {\bf L}^{'}|(\tilde{C}^{(s)}_{\bf L})^{r}\left(\tilde{c}^{\bf T}_{\cal I}\tilde{\psi}^{\dagger {\cal I}}_{\bf T}\right)\left(\tilde{c}^{\bf T}_{\cal J}\tilde{\psi}^{\dagger {\cal J}}_{\bf T}\right)^{\dagger}(\tilde{C}^{(s)\dagger}_{\bf L})^{r}|{\bf L}^{'}\rangle ~~~~~~\nonumber\\
 &=&\frac{H^2}{\sinh^2t_{\bf L}}\left(2r+1\right)|\tilde{\psi}^{\bf L}_{\bf T}|^2,\eea
 where $\tilde{\psi}^{\bf L}_{\bf T}$ is given by :
   \bea\label{ghv1} \tilde{\psi}^{\bf L}_{\bf T}&=&\left(\begin{array}{ccc} \tilde{\psi}^{\bf L}_{\bf T}(t) \\ \tilde{\psi}^{\bf L *}_{\bf T}(t)
    \end{array}\right)=\left(\begin{array}{ccc} {\cal E}_{\bf L}\widetilde{\cal P}^{\bf L}+ {\cal F}_{\bf L}\widetilde{\cal P}^{\bf L *}  \\ {\cal F}^{*}_{\bf L}\widetilde{\cal P}^{\bf L}+ {\cal E}^{*}_{\bf L}\widetilde{\cal P}^{\bf L *}
\end{array}\right)+\sum^{\infty}_{n=0}\left(\begin{array}{ccc} {\cal E}_{{\bf L},(n)}\widetilde{\cal P}^{\bf L}_{(n)}+ {\cal F}_{{\bf L},(n)}\widetilde{\cal P}^{\bf L *}_{(n)}  \\ {\cal F}^{ *}_{{\bf L},(n)}\widetilde{\cal P}^{\bf L}_{(n)}+ {\cal E}^{ *}_{{\bf L},(n)}\widetilde{\cal P}^{\bf L *}_{(n)}
\end{array}\right),\eea
with the entries of the column matrix for the complementary and particular integral part of the solution being:
\bea {\cal E}_{\bf L}&=& \frac{\bar{\tilde{u}}}{{\cal N}_c},\\ {\cal F}_{\bf L}&=& -\frac{\bar{\tilde{v}}}{{\cal N}_c},~~~~~~~\\{\cal E}_{{\bf L},(n)}&=& \frac{\bar{\tilde{U_n}}}{{\cal N}_{c,(n)}},\\{\cal F}_{{\bf L},(n)}&=&  -\frac{\bar{\tilde{V}}}{{\cal N}_{c,(n)}}.\eea
 The normalization constants ${\cal N}_c$ and ${\cal N}_{c,(n)}$ for the complementary part and particular integral part of the solution is defined as:
\bea {\cal N}_c&=& \sqrt{\frac{2}{\pi}}~e^{-\frac{\pi p}{2}}~\sqrt{\cosh 2\pi p+ cos 2\pi \nu},\\ 
{\cal N}_{c,(n)}&=& \sqrt{\frac{2}{\pi}}~e^{-\frac{\pi p_n}{2}}~\sqrt{\cosh 2\pi p_n+ cos 2\pi \nu}.\eea
 
 The expression for $(\bar{\tilde{u}},\bar{\tilde{v}})$ for complementary solution and $(\bar{\tilde{U}}_n,\bar{\tilde{V}}_n)$ for particular solution are given by the following expressions:
\bea &&\underline{\textcolor{red}{\bf For~complementary~part}:}\nonumber\\ \bar{\tilde{u}}&=&\frac{1-\gamma^{(\alpha)}_p \tilde{m}_{\bf LR}}{\sqrt{|1-\gamma^{(\alpha)}_p  \tilde{m}_{\bf LR}|^2-| \tilde{m}_{\bf RR}|^2}}~~~~~~~ \underrightarrow{\alpha=0}~~~~~~~\bar{u}=\frac{1-\gamma^{(0)}_p m_{\bf LR}}{\sqrt{|1-\gamma^{(0)}_p  m_{\bf LR}|^2-| m_{\bf RR}|^2}},~~~~~~\\
\bar{\tilde{v}}&=&\frac{ \tilde{m}_{\bf RR}}{\sqrt{|1-\gamma^{(\alpha)}_p  \tilde{m}_{\bf LR}|^2-| \tilde{m}_{\bf RR}|^2}}~~~~~~~ \underrightarrow{\alpha=0}~~~~~~\bar{\tilde{v}}=\frac{ m_{\bf RR}}{\sqrt{|1-\gamma^{(0)}_p m_{\bf LR}|^2-| m_{\bf RR}|^2}},~~~~~~\\
&&\underline{\textcolor{red}{\bf For~particular~part}:}\nonumber\\ \bar{\tilde{U}}_n&=&\frac{1-\Gamma^{(\alpha)}_{p,n} \tilde{m}_{\bf LR}}{\sqrt{|1-\Gamma^{(\alpha)}_{p,n} \tilde{m}_{\bf LR}|^2-| \tilde{m}_{\bf RR}|^2}}~~~~~~~ \underrightarrow{\alpha=0}~~~~~~\bar{U}_n=\frac{1-\Gamma^{(0)}_{p,n} m_{\bf LR}}{\sqrt{|1-\Gamma^{(0)}_{p,n} m_{\bf LR}|^2-| m_{\bf RR}|^2}},\\
\bar{\tilde{V}}_n&=&\frac{ \tilde{m}_{\bf LR}}{\sqrt{|1-\Gamma^{(\alpha)}_{p,n}  \tilde{m}_{\bf LR}|^2-| \tilde{m}_{\bf RR}|^2}}~~~~~~~ \underrightarrow{\alpha=0}~~~~~~\bar{V}_n=\frac{ m_{\bf LR}}{\sqrt{|1-\Gamma^{(0)}_{p,n}  m_{\bf LR}|^2-| m_{\bf RR}|^2}},\\ &&\nonumber\\
\nonumber &&\underbrace{\textcolor{red}{\bf Results ~for~generalised~\alpha~ vacua}}~~~~~~~~~~~~~\underbrace{\textcolor{red}{\bf Results ~for~Bunch~Davies~ vacuum}}.\eea
where the expression for $(\tilde{m}_{\bf LR},  \tilde{m}_{\bf RR})$ and $(\gamma^{(\alpha)}_{p},\Gamma^{(\alpha)}_{p,n})$ for the complementary and particular part of the solution are defined earlier in equations (3.62--68) and equations (3.119--120) respectively.
We have used  Eq~(\ref{a1}),   Eq~(\ref{a2}),  Eq~(\ref{a3}) and  Eq~(\ref{a4}) and also have imposed the normalization conditions, $|\bar{\tilde{u}}|^2-\bar{\tilde{v}}|^2=1$ and $|\bar{\tilde{u}}|^2-\bar{\tilde{v}}|^2=1$.  Note that the structural form of the equations for $\alpha=0$ corresponding to Bunch Davies vacuum is exactly same as that of $\alpha$ vacua. Only the significant changes appear  when we explicitly consider the  entries of $(m_{\bf LR},  m_{\bf RR})$ and $(\gamma_{p},\Gamma_{p,n})$ for the complementary and particular part of the solution.

Now, substituting Eq.~(\ref{kjv1}) and Eq.~(\ref{kjv3}) in Eq~(\ref{qqwfx}) we get the following simplified expression for the mean square quantum vacuum fluctuation for $\alpha$ vacua as:
\bea\label{qqwfxzx} {\rm Tr}_{\bf L}\left(\rho_{\bf L}(\alpha)\phi_{\bf L}(t_{\bf L})\phi^{\dagger}_{\bf L}(t_{\bf L})\right)_{(\alpha)}&=&\exp\left(-2\alpha\right)\left[\underbrace{\frac{H^2}{\sinh^2t_{\bf L}}|\tilde{\psi}^{\bf L}_{\bf T}|^2\left(1-|\gamma^{(\alpha)}_p|^2\right)\sum^{\infty}_{n=0}|\gamma^{(\alpha)}_p|^{2n}\left(2n+1\right)}_{\textcolor{red}{\bf Complementary ~part}} \right.\nonumber\\&&\left.~~~~~~+\underbrace{\frac{H^2}{\sinh^2t_{\bf L}}|\tilde{\psi}^{\bf L}_{\bf T}|^2\frac{1}{\left(f^{(\alpha)}_{p}\right)^2}\sum^{\infty}_{r=0}\sum^{\infty}_{s=0}|\Gamma^{(\alpha)}_{p,r,s}|^{2r}\left(2r+1\right)}_{\textcolor{red}{\bf Particular~part}}\right].~~~~~~~~~~\nonumber\\
&=&\frac{H^2}{\sinh^2t_{\bf L}}|\tilde{\psi}^{\bf L}_{\bf T}|^2\exp\left(-2\alpha\right)\left[\frac{1+|\gamma^{(\alpha)}_p|^2}{1-|\gamma^{(\alpha)}_p|^2}+\frac{1}{\left(f^{(\alpha)}_{p}\right)^2}\sum^{\infty}_{s=0}\frac{1+|\Gamma^{(\alpha)}_{p,s}|^2}{\left(1-|\Gamma^{(\alpha)}_{p,s}|^2\right)^2}\right].~~~~~~~~~~~~\eea  
 Setting $\alpha=0$ we get the expression for the Bunch Davies vacuum as :
\bea\label{qqwfxzx} {\rm Tr}_{\bf L}\left(\rho_{\bf L}(\alpha)\phi_{\bf L}(t_{\bf L})\phi^{\dagger}_{\bf L}(t_{\bf L})\right)_{(\bf BD)}&=&\underbrace{\frac{H^2}{\sinh^2t_{\bf L}}|{\psi}^{\bf L}_{\bf T}|^2\left(1-|\gamma^{(0)}_p|^2\right)\sum^{\infty}_{n=0}|\gamma^{(0)}_p|^{2n}\left(2n+1\right)}_{\textcolor{red}{\bf Complementary ~part}} \nonumber\\&&~~~~~~+\underbrace{\frac{H^2}{\sinh^2t_{\bf L}}|{\psi}^{\bf L}_{\bf T}|^2\frac{1}{\left(f^{(0)}_{p}\right)^2}\sum^{\infty}_{r=0}\sum^{\infty}_{s=0}|\Gamma^{(0)}_{p,r,s}|^{2r}\left(2r+1\right)}_{\textcolor{red}{\bf Particular~part}}.~~~~~~~~~~\nonumber\\
&=&\frac{H^2}{\sinh^2t_{\bf L}}|{\psi}^{\bf L}_{\bf T}|^2\left[\frac{1+|\gamma^{(0)}_p|^2}{1-|\gamma^{(0)}_p|^2}+\frac{1}{\left(f^{(0)}_{p}\right)^2}\sum^{\infty}_{s=0}\frac{1+|\Gamma^{(0)}_{p,s}|^2}{\left(1-|\Gamma^{(0)}_{p,s}|^2\right)^2}\right].~~~~~~~~~~~~\eea 
We note that, to derive this expression we have used the following identities: 
\bea \sum^{\infty}_{n=0}(2n+1)|\gamma^{(\alpha)}_p|^{2n}&=& \frac{1+|\gamma^{(\alpha)}_p|^2}{\left(1-\gamma^{(\alpha)}_p|^2\right)^2}~~\underrightarrow{\alpha=0}~~\sum^{\infty}_{n=0}(2n+1)|\gamma^{(0)}_p|^{2n}= \frac{1+|\gamma^{(0)}_p|^2}{\left(1-\gamma^{(0)}_p|^2\right)^2}, ~~~~~~~~~~\\
\sum^{\infty}_{s=0}\sum^{\infty}_{r=0}(2r+1)|\Gamma^{(\alpha)}_{p,r,s}|^{2r}&=& \sum^{\infty}_{s=0}\frac{1+|\Gamma^{(\alpha)}_{p,s}|^2}{\left(1-\Gamma^{(\alpha)}_{p,s}|^2\right)^2}~~\underrightarrow{\alpha=0}~~\sum^{\infty}_{s=0}\sum^{\infty}_{r=0}(2r+1)|\Gamma^{(0)}_{p,r,s}|^{2r}= \sum^{\infty}_{s=0}\frac{1+|\Gamma^{(0)}_{p,s}|^2}{\left(1-\Gamma^{(0)}_{p,s}|^2\right)^2}.~~~~~~~~~~~~\eea
The expression for $|\tilde{\psi}^{\bf L}_{\bf T}|^2$, now comes out to be:
\bea \label{opq1s} |\tilde{\psi}^{\bf L}_{\bf T}|^2=\left(\tilde{\psi}^{\bf L}_{\bf T}\right)^{\dagger}
\tilde{\psi}^{\bf L}_{\bf T}&=&\left[\left(|{\cal E}_{\bf L}|^2+|{\cal F}_{\bf L}|^2\right)\widetilde{\cal P}^{\bf L}\widetilde{\cal P}^{{\bf L}*}+{\cal E}_{\bf L}{\cal F}^{ *}_{\bf L}\left(\widetilde{\cal P}^{\bf L}\right)^2+{\cal E}^{ *}_{\bf L}{\cal F}_{\bf L}\left(\widetilde{\cal P}^{{\bf L}*}\right)^2\right.\nonumber\\ && \left.
~+\sum^{\infty}_{n=0}\left\{\left({\cal E}_{{\bf L}}{\cal E}^{*}_{{\bf L},(n)}+{\cal F}_{{\bf L}}{\cal F}^{*}_{{\bf L},(n)}\right)\widetilde{\cal P}^{\bf L}\widetilde{\cal P}^{{\bf L}*}_{(n)}\right.\right.\nonumber\\&&\left.\left.~~~~~~~~~~~~~~~~+\left({\cal E}_{{\bf L}}{\cal F}^{*}_{{\bf L},(n)}+{\cal E}_{{\bf L},(n)}{\cal F}^{*}_{{\bf L}}\right)\widetilde{\cal P}^{\bf L}\widetilde{\cal P}^{\bf L}_{(n)}\right.\right.\nonumber\\&&\left.\left.~~~~~~~~~~~~~~~~+\left({\cal E}^{*}_{{\bf L},(n)}{\cal F}_{{\bf L}}+{\cal E}^{*}_{{\bf L}}{\cal F}_{{\bf L},(n)}\right)\widetilde{\cal P}^{{\bf L}*}_{(n)}\widetilde{\cal P}^{{\bf L}*}\right\}\right.\nonumber\\ && \left.
~+\sum^{\infty}_{n=0}\sum^{\infty}_{m=0}\left\{\left({\cal E}_{{\bf L},(n)}{\cal E}^{*}_{{\bf L},(m)}+{\cal F}_{{\bf L},(n)}{\cal F}^{*}_{{\bf L},(m)}\right)\widetilde{\cal P}^{\bf L}_{(n)}\widetilde{\cal P}^{{\bf L}*}_{(m)}\right.\right.\nonumber\\&&\left.\left.~~~~~~~~~~~~~~~~+{\cal E}_{{\bf L},(n)}{\cal F}^{*}_{{\bf L},(m)}\widetilde{\cal P}^{\bf L}_{(n)}\widetilde{\cal P}^{\bf L}_{(m)}+{\cal E}^{*}_{{\bf L},(n)}{\cal F}_{{\bf L},(m)}\widetilde{\cal P}^{{\bf L}*}_{(n)}\widetilde{\cal P}^{{\bf L}*}_{(m)}\right\}\right]~~~~~~~~~~~\eea
Here also by fixing the parameter $\alpha=0$ one can get the expression for the square of the magnitude of the wave function for Bunch Davies vacuum in the newly defined Bogliubov transformed basis.

Using Eq~(\ref{opq1s}), the amplitude of the normalised power spectrum of axion from the generalised $\alpha$ vacua can be expressed in all time scales of region \textcolor{red}{\bf L} as:
\bea \label{po1xx} {\cal P}(p,\alpha,t_{\bf L})&=&\frac{p^3}{2\pi^2}~{\rm Tr}_{\bf L}\left(\rho_{\bf L}(\alpha)\phi_{\bf L}(t_{\bf L})\phi^{\dagger}_{\bf L}(t_{\bf L})\right)_{(\alpha)}\nonumber\\&=& \frac{p^3}{2\pi^2}~\frac{H^2}{\sinh^2 t_{\bf L}}|\tilde{\psi}^{\bf L}_{\bf T}|^2\exp\left(-2\alpha\right)\left[\frac{1+|\gamma^{(\alpha)}_p|^2}{1-|\gamma^{(\alpha)}_p|^2}+\frac{1}{\left(f^{(\alpha)}_{p}\right)^2}\sum^{\infty}_{s=0}\frac{1+|\Gamma^{(\alpha)}_{p,s}|^2}{\left(1-|\Gamma^{(\alpha)}_{p,s}|^2\right)^2}\right]\nonumber\\
&=&\frac{p^3}{2\pi^2}~\frac{H^2}{\sinh^2 t_{\bf L}}\exp\left(-2\alpha\right)\left[\frac{1+|\gamma^{(\alpha)}_p|^2}{1-|\gamma^{(\alpha)}_p|^2}+\frac{1}{\left(f^{(\alpha)}_{p}\right)^2}\sum^{\infty}_{s=0}\frac{1+|\Gamma^{(\alpha)}_{p,s}|^2}{\left(1-|\Gamma^{(\alpha)}_{p,s}|^2\right)^2}\right]\nonumber\\
&&~~~~~~~\left[\left(|{\cal E}_{\bf L}|^2+|{\cal F}_{\bf L}|^2\right)\widetilde{\cal P}^{\bf L}\widetilde{\cal P}^{{\bf L}*}+{\cal E}_{\bf L}{\cal F}^{ *}_{\bf L}\left(\widetilde{\cal P}^{\bf L}\right)^2+{\cal E}^{ *}_{\bf L}{\cal F}_{\bf L}\left(\widetilde{\cal P}^{{\bf L}*}\right)^2\right.\nonumber\\ && \left.
~+\sum^{\infty}_{n=0}\left\{\left({\cal E}_{{\bf L}}{\cal E}^{*}_{{\bf L},(n)}+{\cal F}_{{\bf L}}{\cal F}^{*}_{{\bf L},(n)}\right)\widetilde{\cal P}^{\bf L}\widetilde{\cal P}^{{\bf L}*}_{(n)}\right.\right.\nonumber\\&&\left.\left.~~~~~~~~~~~~~~~~+\left({\cal E}_{{\bf L}}{\cal F}^{*}_{{\bf L},(n)}+{\cal E}_{{\bf L},(n)}{\cal F}^{*}_{{\bf L}}\right)\widetilde{\cal P}^{\bf L}\widetilde{\cal P}^{\bf L}_{(n)}\right.\right.\nonumber\\&&\left.\left.~~~~~~~~~~~~~~~~+\left({\cal E}^{*}_{{\bf L},(n)}{\cal F}_{{\bf L}}+{\cal E}^{*}_{{\bf L}}{\cal F}_{{\bf L},(n)}\right)\widetilde{\cal P}^{{\bf L}*}_{(n)}\widetilde{\cal P}^{{\bf L}*}\right\}\right.\nonumber\\ && \left.
~+\sum^{\infty}_{n=0}\sum^{\infty}_{m=0}\left\{\left({\cal E}_{{\bf L},(n)}{\cal E}^{*}_{{\bf L},(m)}+{\cal F}_{{\bf L},(n)}{\cal F}^{*}_{{\bf L},(m)}\right)\widetilde{\cal P}^{\bf L}_{(n)}\widetilde{\cal P}^{{\bf L}*}_{(m)}\right.\right.\nonumber\\&&\left.\left.~~~~~~~~~~~~~~~~+{\cal E}_{{\bf L},(n)}{\cal F}^{*}_{{\bf L},(m)}\widetilde{\cal P}^{\bf L}_{(n)}\widetilde{\cal P}^{\bf L}_{(m)}+{\cal E}^{*}_{{\bf L},(n)}{\cal F}_{{\bf L},(m)}\widetilde{\cal P}^{{\bf L}*}_{(n)}\widetilde{\cal P}^{{\bf L}*}_{(m)}\right\}\right].~~~~~~~~~~~\eea
However, the above equation is very complicated to extract any physical information  for further cosmological predictions. For this reason we consider the superhorizon time scales ($t_{\bf L}>>1$) of region \textcolor{red}{\bf L}, in which the Legendre functions  appearing in the complementary part and the particular integral part of the time dependent solution can be approximated as the following simplified form:
\bea \left(\widetilde{\cal P}^{\bf L},\widetilde{\cal P}^{\bf L*}\right)&\equiv& P^{\pm ip}_{\nu-\frac{1}{2}}\left(\cosh t_{\bf L}\right)~\underrightarrow{t_{\bf L}>>1}~\frac{2^{\nu-\frac{1}{2}}\left(\cosh t_{\bf L}\right)^{\nu-\frac{1}{2}}\Gamma(\nu)}{\sqrt{\pi}\Gamma\left(\nu\mp ip +\frac{1}{2}\right)},\\
\left(\widetilde{\cal P}^{\bf L}_{(n)},\widetilde{\cal P}^{\bf L*}_{(n)}\right)&\equiv& P^{\pm ip_n}_{\nu-\frac{1}{2}}\left(\cosh t_{\bf L}\right)~\underrightarrow{t_{\bf L}>>1}~\frac{2^{\nu-\frac{1}{2}}\left(\cosh t_{\bf L}\right)^{\nu-\frac{1}{2}}\Gamma(\nu)}{\sqrt{\pi}\Gamma\left(\nu\mp ip_n +\frac{1}{2}\right)}.\eea
Consequently, in the superhorizon time scales ($t_{\bf L}>>1$) of region \textcolor{red}{\bf L} eqn~(\ref{po1xx}) can be  simplified for as:
\bea \label{df2xxx} |\tilde{\psi}^{\bf L}_{\bf T}|^2=\left(\tilde{\psi}^{\bf L}_{\bf T}\right)^{\dagger}
\tilde{\psi}^{\bf L}_{\bf T}&~\underrightarrow{t_{\bf L}>>1}~& \widetilde{{\cal Q}(p,\alpha,\nu)}\left(\cosh t_{\bf L}\right)^{2\nu-1}\eea
where the time independent function $\widetilde{{\cal Q}(p,\alpha,\nu)}$ for generalised $\alpha$ vacua is defined as:
\bea\widetilde{{\cal Q}(p,\alpha,\nu)}&=&\frac{2^{2\nu-1}\left(\Gamma(\nu)\right)^2}{\pi}\times \left[\frac{\left(|{\cal E}_{\bf L}|^2+|{\cal F}_{\bf L}|^2\right)}{|\Gamma\left(\nu+ip+\frac{1}{2}\right)|^2}+\frac{{\cal E}_{\bf L}{\cal F}^{ *}_{\bf L}}{\left(\Gamma\left(\nu-ip+\frac{1}{2}\right)\right)^2}+\frac{{\cal E}^{ *}_{\bf L}{\cal F}_{\bf L}}{\left(\Gamma\left(\nu+ip+\frac{1}{2}\right)\right)^2}\right.\nonumber\\ && \left.
~+\sum^{\infty}_{n=0}\left\{\frac{\left({\cal E}_{{\bf L}}{\cal E}^{*}_{{\bf L},(n)}+{\cal F}_{{\bf L}}{\cal F}^{*}_{{\bf L},(n)}\right)}{\Gamma\left(\nu-ip+\frac{1}{2}\right)\Gamma\left(\nu+ip_n+\frac{1}{2}\right)}\right.\right.\nonumber\\&&\left.\left.~~~~~~~~+\frac{\left({\cal E}_{{\bf L}}{\cal F}^{*}_{{\bf L},(n)}+{\cal E}_{{\bf L},(n)}{\cal F}^{*}_{{\bf L}}\right)}{\Gamma\left(\nu-ip+\frac{1}{2}\right)\Gamma\left(\nu-ip_n+\frac{1}{2}\right)}+\frac{\left({\cal E}^{*}_{{\bf L},(n)}{\cal F}_{{\bf L}}+{\cal E}^{*}_{{\bf L}}{\cal F}_{{\bf L},(n)}\right)}{\Gamma\left(\nu+ip+\frac{1}{2}\right)\Gamma\left(\nu+ip_n+\frac{1}{2}\right)}\right\}\right.\nonumber\\ && \left.
~+\sum^{\infty}_{n=0}\sum^{\infty}_{m=0}\left\{\frac{\left({\cal E}_{{\bf L},(n)}{\cal E}^{*}_{{\bf L},(m)}+{\cal F}_{{\bf L},(n)}{\cal F}^{*}_{{\bf L},(m)}\right)}{\Gamma\left(\nu-ip_n+\frac{1}{2}\right)\Gamma\left(\nu+ip_m+\frac{1}{2}\right)}\right.\right.\nonumber\\&&\left.\left.~~~+\frac{{\cal E}_{{\bf L},(n)}{\cal F}^{*}_{{\bf L},(m)}}{\Gamma\left(\nu-ip_n+\frac{1}{2}\right)\Gamma\left(\nu-ip_m+\frac{1}{2}\right)}+\frac{{\cal E}^{*}_{{\bf L},(n)}{\cal F}_{{\bf L},(m)}}{\Gamma\left(\nu+ip_n+\frac{1}{2}\right)\Gamma\left(\nu+ip_m+\frac{1}{2}\right)}\right\}\right].~~~~~~~~~\eea
As a result, in the superhorizon time scales ($t_{\bf L}>>1$) of region \textcolor{red}{\bf L} the amplitude of the normalised power spectrum of axion from generalised $\alpha$ vacua can be expressed as:
\bea \label{po2xxx} {\cal P}(p,\alpha,t_{\bf L})&=&\frac{p^3}{2\pi^2}~\frac{H^2}{\sinh^2 t_{\bf L}}|\tilde{\psi}^{\bf L}_{\bf T}|^2\exp\left(-2\alpha\right)\left[\frac{1+|\gamma^{(\alpha)}_p|^2}{1-|\gamma^{(\alpha)}_p|^2}+\frac{1}{\left(f^{(\alpha)}_{p}\right)^2}\sum^{\infty}_{s=0}\frac{1+|\Gamma^{(\alpha)}_{p,s}|^2}{\left(1-|\Gamma^{(\alpha)}_{p,s}|^2\right)^2}\right]\nonumber\\
\underrightarrow{t_{\bf L}>>1}&&\frac{p^3}{2\pi^2}~\frac{\left(\cosh t_{\bf L}\right)^{2\nu-1}}{\sinh^2 t_{\bf L}}~H^2\widetilde{{\cal Q}(p,\nu)}\exp\left(-2\alpha\right)\left[\frac{1+|\gamma^{(\alpha)}_p|^2}{1-|\gamma^{(\alpha)}_p|^2}+\frac{1}{\left(f^{(\alpha)}_{p}\right)^2}\sum^{\infty}_{s=0}\frac{1+|\Gamma^{(\alpha)}_{p,s}|^2}{\left(1-|\Gamma^{(\alpha)}_{p,s}|^2\right)^2}\right].~~~~~~~~~~~\eea
We note that in the superhorizon time scales ($t_{\bf L}>>1$) of region \textcolor{red}{\bf L} if we consider the massless case by fixing the mass parameter  $\nu=3/2$, then the time dependent contribution can be approximated as:
\bea \left(\frac{\left(\cosh t_{\bf L}\right)^{2\nu-1}}{\sinh^2 t_{\bf L}}\right)_{\nu=3/2}~\underrightarrow{t_{\bf L}>>1}~~~1.\eea 
From this we infer that for an arbitrary value of the parameter $\nu$ we can write:
\bea \left(\frac{\left(\cosh t_{\bf L}\right)^{2\nu-1}}{\sinh^2 t_{\bf L}}\right)~\underrightarrow{t_{\bf L}>>1}~~~\left(\cosh t_{\bf L}\right)^{2\nu-3}.\eea 
Consequently, in the super horizon time scales ($t_{\bf L}>>1$) of region \textcolor{red}{\bf L} considering the massless case ($\nu=3/2$) the amplitude of the normalised power spectrum of axion from generalised $\alpha$ vacua can be expressed as:
\bea \label{po3xxx} {\cal P}(p,\alpha,t_{\bf L})&=&\frac{p^3}{2\pi^2}~\frac{H^2}{\sinh^2 t_{\bf L}}|\tilde{\psi}^{\bf L}_{\bf T}|^2\exp\left(-2\alpha\right)\left[\frac{1+|\gamma^{(\alpha)}_p|^2}{1-|\gamma^{(\alpha)}_p|^2}+\frac{1}{\left(f^{(\alpha)}_{p}\right)^2}\sum^{\infty}_{s=0}\frac{1+|\Gamma^{(\alpha)}_{p,s}|^2}{\left(1-|\Gamma^{(\alpha)}_{p,s}|^2\right)^2}\right]\nonumber\\
~\underrightarrow{t_{\bf L}>>1,\nu=3/2}~&&\frac{p^3}{2\pi^2}~H^2\widetilde{{\cal Q}(p,\nu=3/2)}\exp\left(-2\alpha\right)\left[\frac{1+|\gamma^{(\alpha)}_p|^2}{1-|\gamma^{(\alpha)}_p|^2}+\frac{1}{\left(f^{(\alpha)}_{p}\right)^2}\sum^{\infty}_{s=0}\frac{1+|\Gamma^{(\alpha)}_{p,s}|^2}{\left(1-|\Gamma^{(\alpha)}_{p,s}|^2\right)^2}\right].~~~~~~~~~~~~\eea
Like the result in the case of field operator expansion method derived in the previous section, this result also implies that in the massless case ($\nu=3/2$) amplitude of the vacuum fluctuation gets frozen with respect to the time scale when the associated modes exit horizon.

Further to know the exact wave number dependence of the amplitude of the normalised power spectrum from generalised $\alpha$ vacua we need to know the behaviour of the power spectrum at very short wavelengths ($p,p_n>>1$). In this limit it is expected that the power spectrum of axion should match with the result obtained for spatially flat universe. In the short wave length approximation the time independent function $\widetilde{{\cal Q}(p>>1,\alpha,\nu)}$ for any arbitrary mass parameter $\nu$ can be expressed for generalised $\alpha$ vacua as:
\bea\widetilde{{\cal Q}(p>>1,\alpha,\nu)}&=&\frac{2^{2(\nu-1)}\left(\Gamma(\nu)\right)^2}{p^3\pi}\widetilde{{\cal G}(p>>1)}=\widetilde{{\cal M}(p,\nu)}~~~\forall \alpha,~~~~~~~~~~~\eea
where we have already defined the function $\widetilde{{\cal G}(p>>1)}$ in the earlier section.
Here for very large wave number $p,p_n>>1$ one can write, $\widetilde{{\cal G}(p>>1)}\sim 1+\cdots$, where all $\cdots$ are small correction terms. This also implies to the interesting fact that for large wavenumber limit and  for any values of the parameter $\alpha$, the time independent function ${{\cal Q}(p>>1,\alpha,\nu)}$ computed for generalised $\alpha$ vacua  exactly matches with the result obtained for Bunch Davies vacua in the earlier section i.e. $\widetilde{{\cal M}(p>>1,\nu)}$. This means that  the final result is independent of the choice of the parameter $\alpha$.

            \begin{figure*}[htb]
            \centering
           \subfigure[Large wave number dependence of RDM power spectrum for $\alpha=0$.]{
               \includegraphics[width=7.7cm,height=7.2cm] {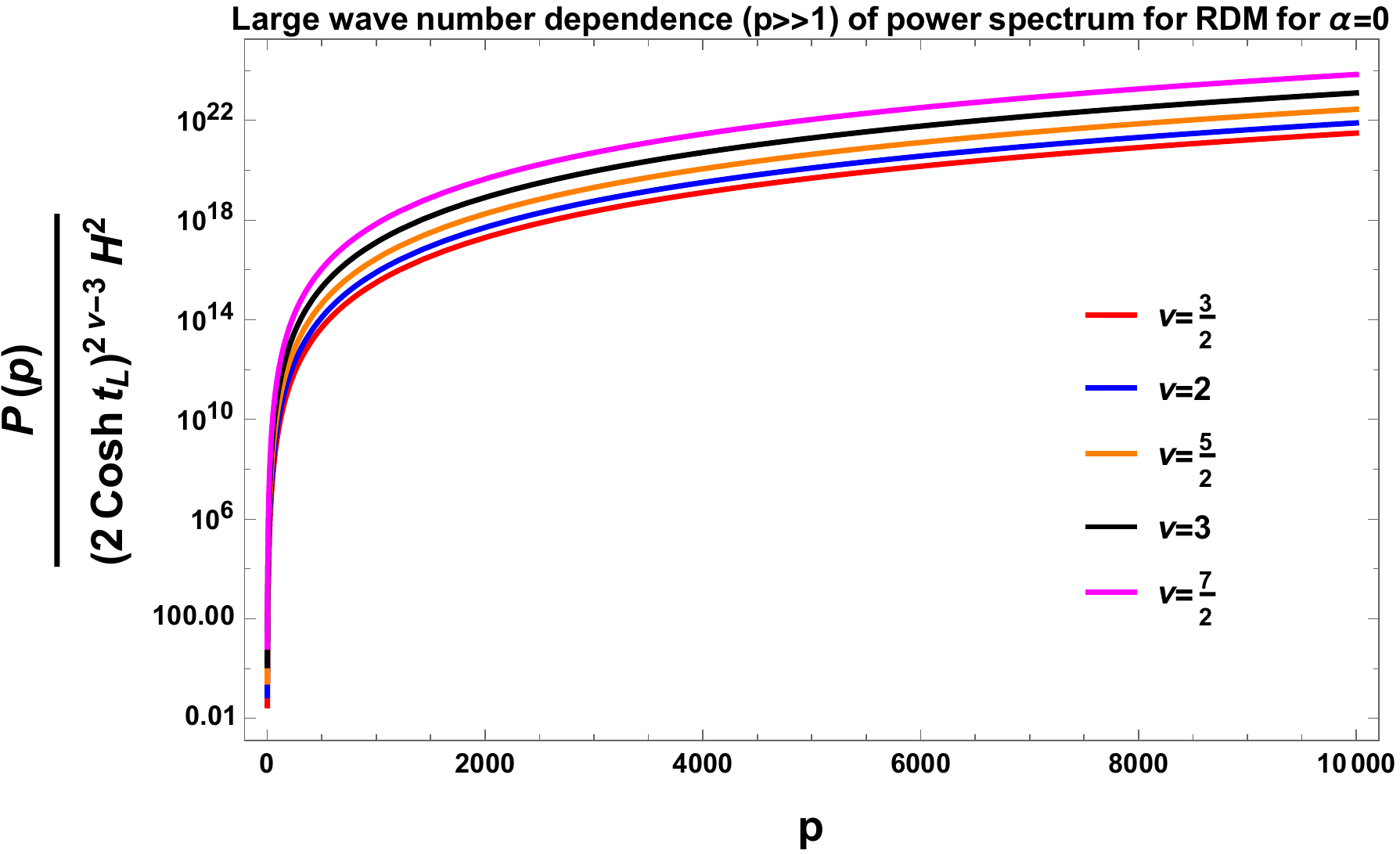}
               \label{fig3a}
            }
            \subfigure[Large wave number dependence of RDM power spectrum for $\alpha=0.1$.]{
                \includegraphics[width=7.7cm,height=7.2cm] {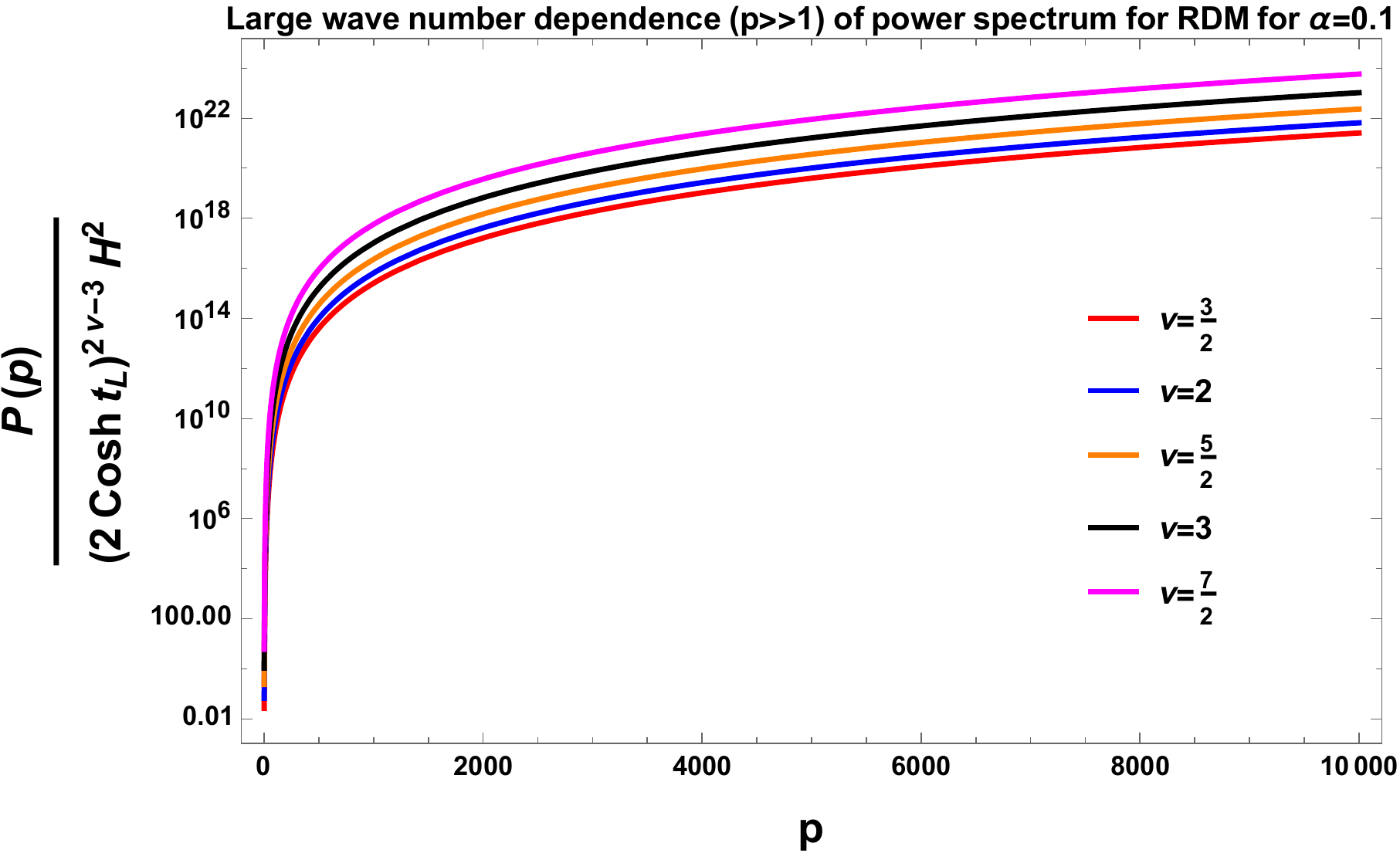}
               \label{fig3b}
              }
             \subfigure[Mass parameter dependence of RDM power spectrum in $p>>1$.]{
                    \includegraphics[width=10.7cm,height=7.5cm] {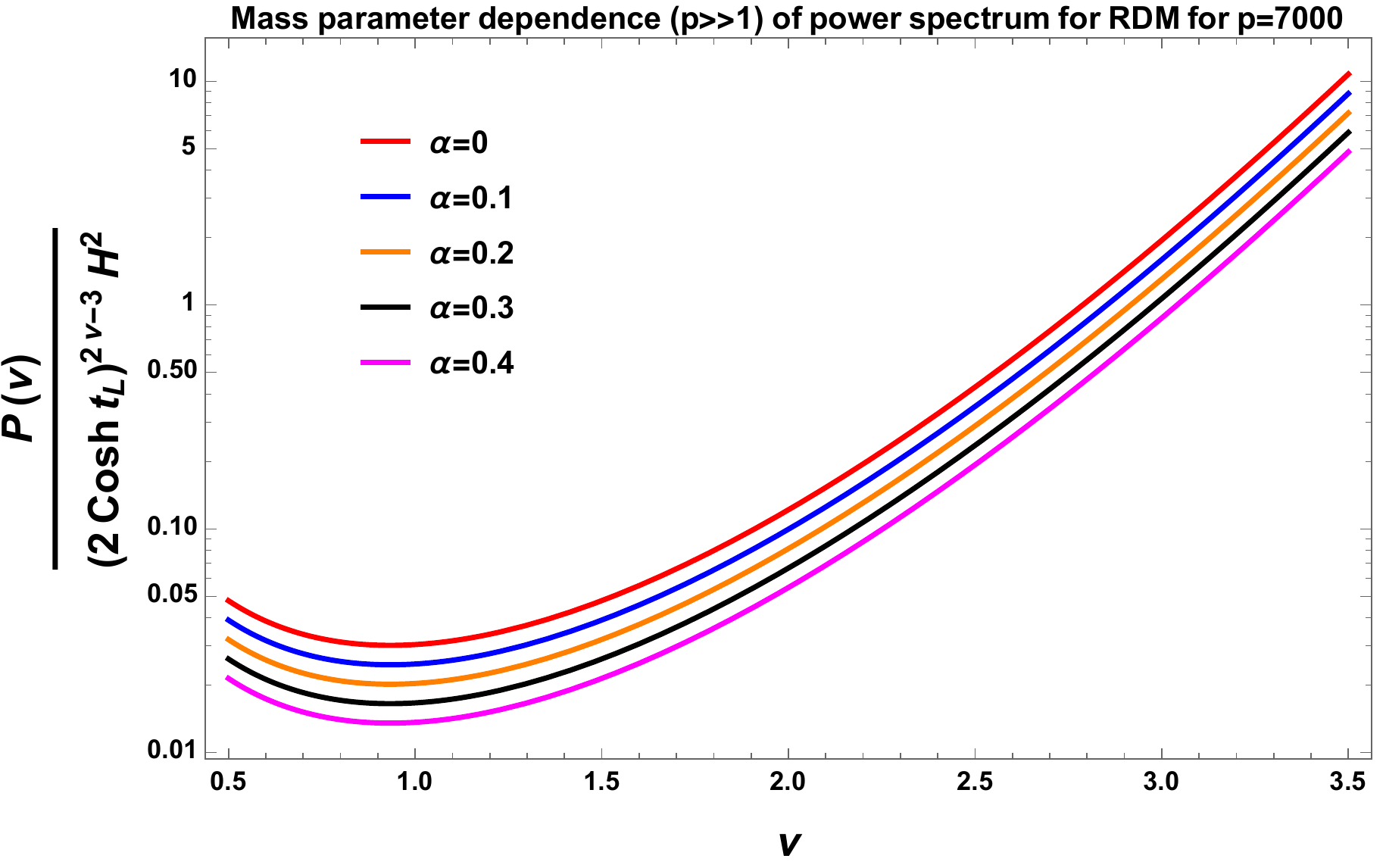}
                    \label{fig3c}   
           }
           \caption[Optional caption for list of figures]{Features of RDM power spectrum in large wave number region.} 
            \label{fig1x3}
            \end{figure*}

For the massless case ($\nu=3/2$) in the short wave length approximation, the time independent function $\widetilde{{\cal Q}(p>>1,\alpha,\nu=3/2)}$ can further be simplified to:
\bea\widetilde{{\cal Q}(p>>1,\alpha,\nu=3/2)}&=&\frac{\widetilde{{\cal G}(p>>1)}}{2p^3}=\widetilde{{\cal M}(p>>1,\nu=3/2)}~~~\forall\alpha.~~~~~~~~~~~\eea
Additionally, we note that the following important contribution appearing in the normalised power spectrum for axion can be simplified, in the large wave number limit, as: 
\bea \left[\frac{1+|\gamma^{(\alpha)}_p|^2}{1-|\gamma^{(\alpha)}_p|^2}+\frac{1}{\left(f^{(\alpha)}_{p}\right)^2}\sum^{\infty}_{s=0}\frac{1+|\Gamma^{(\alpha)}_{p,s}|^2}{\left(1-|\Gamma^{(\alpha)}_{p,s}|^2\right)^2}\right]&\stackrel{p>>1}{=}&\left[1+\underbrace{\left(\sum^{\infty}_{s=0}1\right)^{-1}}_{\bf =0}\right]~~~~~\forall \alpha.~~~~~~~~\eea
Finally, in the super horizon time scales ($t_{\bf L}>>1$) of region \textcolor{red}{\bf L}, the amplitude of the normalised power spectrum of axion, in the short wave length approximation, can be expressed as:
\bea \label{po2vccx} {\cal P}(p>>1,\alpha,t_{\bf L}>>1)&=&\frac{p^3}{2\pi^2}~\left(\cosh t_{\bf L}\right)^{2\nu-3}~\exp\left(-2\alpha\right)H^2\widetilde{{\cal Q}(p>>1,\alpha,\nu)}
\nonumber\\&=&\frac{p^3}{2\pi^2}~\left(\cosh t_{\bf L}\right)^{2\nu-3}~\exp\left(-2\alpha\right)H^2\widetilde{{\cal M}(p>>1,\nu)}\nonumber\\&=&\left(2\cosh t_{\bf L}\right)^{2\nu-3}~\left(\frac{H}{2\pi}\right)^2~\left(\frac{\Gamma(\nu)}{\Gamma\left(\frac{3}{2}\right)}\right)^2\widetilde{{\cal G}(p>>1)}.~~~~~~~~~~~\eea
 For the massless case ($\nu=3/2$), in the same scale and the same approximation, the above amplitude takes the form:
\bea \label{po3zcvc} {\cal P}(p>>1,\alpha,t_{\bf L}>>1)&=&\frac{p^3}{2\pi^2}~\exp\left(-2\alpha\right)H^2\widetilde{{\cal Q}(p>>1,\alpha,\nu=3/2)}\nonumber\\&=&\frac{p^3}{2\pi^2}~\exp\left(-2\alpha\right)H^2\widetilde{{\cal M}(p>>1,\nu=3/2)}\nonumber\\&=&\left(\frac{H}{2\pi}\right)^2~\exp\left(-2\alpha\right)\widetilde{{\cal G}(p>>1)}.~~~~~~~~~~~\eea
It is important to note that both of Eq~(\ref{po2vccx}) and Eq~(\ref{po3zcvc}) are valid after horizon exit. From the same results , we also observe that the normalised power spectrum  from generalised $\alpha$ vacua,in the leading order,  computed from reduced density matrix formalism is exactly same as that obtained in the previous sub-section, computed using field operator expansion method. 

For completeness, we present the result for the two point correlation function and the associated power spectrum for Bunch Davies vacuum  by fixing the parameter $\alpha=0$ in our previous equations and they can be expressed as:
 \bea \label{po2vvvccx} {\cal P}_{\bf BD}(p>>1,t_{\bf L}>>1)&=&\frac{p^3}{2\pi^2}~\left(\cosh t_{\bf L}\right)^{2\nu-3}~H^2\widetilde{{\cal Q}(p>>1,\alpha=0,\nu)}
\nonumber\\&=&\frac{p^3}{2\pi^2}~\left(\cosh t_{\bf L}\right)^{2\nu-3}~H^2\widetilde{{\cal M}(p>>1,\nu)}\nonumber\\&=&\left(2\cosh t_{\bf L}\right)^{2\nu-3}~\left(\frac{H}{2\pi}\right)^2~\left(\frac{\Gamma(\nu)}{\Gamma\left(\frac{3}{2}\right)}\right)^2\widetilde{{\cal G}(p>>1)}.~~~~~~~~~~~\eea
  For for the massless case ($\nu=3/2$) this can  be further simplified to:
\bea \label{po3zcvcccc} {\cal P}_{\bf BD}(p>>1,t_{\bf L}>>1)&=&\frac{p^3}{2\pi^2}~H^2\widetilde{{\cal Q}(p>>1,\alpha=0,\nu=3/2)}\nonumber\\&=&\frac{p^3}{2\pi^2}~H^2\widetilde{{\cal M}(p>>1,\nu=3/2)}\nonumber\\&=&\left(\frac{H}{2\pi}\right)^2~\widetilde{{\cal G}(p>>1)}.~~~~~~~~~~~\eea 
  
  In figure~(\ref{fig3a}) and figure~(\ref{fig3b}) we have shown the behaviour of the power spectrum of the mean square vacuum fluctuation computed from RDM formalism in the large wave number regime. We have considered $\alpha=0$ and $\alpha=0.1$  and  fixed values of the mass parameter $\nu$ respectively. Additionally, in figure~(\ref{fig3c}) we have depicted the behaviour of the power spectrum with respect to the mass parameter $\nu$ for fixed values of the parameter $\alpha=0,0.1,0.2,0.3,0.4$.  From the figures, we observe that the power spectrum shows two distinctive behaviour in $1/2<\nu<1$ and $\nu>1$ region. For $1/2<\nu<1$ region the amplitude of the power spectrum decrease to a certain value and just after $\nu=1$ it increases.  Also  note that in large wave number regime, the power spectrum obtained from RDM formalism behaves in the same as way as that obtained from FOE formalism in the previous section.

  On the other hand, to know the exact wave number dependence of the amplitude of the normalised power spectrum from generalised $\alpha$ vacua in the long wave length approximation, we need to know the behaviour of the power spectrum for $p,p_n<<1$. In this regime we expect that the power spectrum of axion should match with the result obtained for spatially flat universe. The time independent function $\widetilde{{\cal Q}(p<<1,\alpha,\nu)}$ for the mass parameter $\nu\neq 3/2$ can be expressed for generalised $\alpha$ vacua as:
\bea\widetilde{{\cal Q}(p<<1,\alpha,\nu)}&=&\frac{2^{2(\nu-1)}\left(\Gamma(\nu)\right)^2}{p^3\pi}\widetilde{{\cal G}(p<<1)}~~~\forall \alpha,~~~~~~~~~~~\eea
where the function $\widetilde{{\cal G}(p<<1)}$ is defined for $\nu \neq q/2$~\footnote{Here $q$ is any positive odd integer.} as:
\bea \widetilde{{\cal G}(p<<1)}&=&\frac{\pi p}{2|\cos \pi \nu|\left|\Gamma\left(\nu+ \frac{1}{2}\right)\right|^2}\frac{|1-\gamma^{(\alpha)}_p \tilde{m}_{\bf LR}|^2}{|1-\gamma^{(\alpha)}_p  \tilde{m}_{\bf LR}|^2-| \tilde{m}_{\bf RR}|^2}\nonumber\\ &&\times\left\{1+\frac{|\tilde{m}_{\bf RR}|^2+\left(1-\gamma^{(\alpha)}_p \tilde{m}_{\bf LR}\right)^{*}\tilde{m}_{\bf RR}+\left(1-\gamma^{(\alpha)}_p \tilde{m}_{\bf LR}\right)\tilde{m}^{*}_{\bf RR}}{|1-\gamma^{(\alpha)}_p \tilde{m}_{\bf LR}|^2}\right.\nonumber\\&&\left.~~~~+\sum^{\infty}_{n=0}\sqrt{\frac{p_n}{p}\frac{{|1-\gamma^{(\alpha)}_p  \tilde{m}_{\bf LR}|^2-| \tilde{m}_{\bf RR}|^2}}{{|1-\Gamma^{(\alpha)}_{p,n}  \tilde{m}_{{\bf LR},n}|^2-| \tilde{m}_{{\bf RR},n}|^2}}}\frac{1}{|1-\gamma^{(\alpha)}_p \tilde{m}_{\bf LR}|^2}\right.\nonumber\\&& \left.
~~~~~~~~~~\left[\left(1-\gamma^{(\alpha)}_p \tilde{m}_{\bf LR}\right)\left(1-\Gamma^{(\alpha)}_{p,n} \tilde{m}_{{\bf LR},n}\right)^{*}+\tilde{m}_{\bf RR}\tilde{m}^{*}_{{\bf RR},n}\right.\right.\nonumber\\&& \left.\left.~~~~~~~~~~+\left(1-\gamma^{(\alpha)}_p \tilde{m}_{\bf LR}\right)\tilde{m}^{*}_{{\bf RR},n}+\left(1-\Gamma^{(\alpha)}_{p,n} \tilde{m}_{{\bf LR},n}\right)\tilde{m}^{*}_{{\bf RR}}\right.\right.\nonumber\\&&\left.\left.~~~~~~~~~~~+\left(1-\gamma^{(\alpha)}_p \tilde{m}_{\bf LR}\right)^{*}\tilde{m}_{{\bf RR},n}+\left(1-\Gamma^{(\alpha)}_{p,n} \tilde{m}_{{\bf LR},n}\right)^{*}\tilde{m}_{{\bf RR}}\right]\right.\nonumber\\&&\left.~~~~+\sum^{\infty}_{n=0}\sum^{\infty}_{m=0}\sqrt{\frac{p_np_m}{p^2}\frac{{\left(|1-\gamma^{(\alpha)}_p  \tilde{m}_{\bf LR}|^2-| \tilde{m}_{\bf RR}|^2\right)^2}}{{\left(|1-\Gamma^{(\alpha)}_{p,n}  \tilde{m}_{{\bf LR},n}|^2-| \tilde{m}_{{\bf RR},n}|^2\right)\left(|1-\Gamma^{(\alpha)}_{p,m}  \tilde{m}_{{\bf LR},m}|^2-| \tilde{m}_{{\bf RR},m}|^2\right)}}}\right.\nonumber\\&& \left.
~~~~~~~~~~\frac{1}{|1-\gamma^{(\alpha)}_p \tilde{m}_{\bf LR}|^2}\left[\left(1-\Gamma^{(\alpha)}_{p,n} \tilde{m}_{{\bf LR},n}\right)\left(1-\Gamma^{(\alpha)}_{p,m} \tilde{m}_{{\bf LR},m}\right)^{*}+\tilde{m}_{{\bf RR},n}\tilde{m}^{*}_{{\bf RR},m}\right.\right.\nonumber\\&& \left.\left.~~~~~~~~~~+\left(1-\Gamma^{(\alpha)}_{p,n} \tilde{m}_{{\bf LR},n}\right)\tilde{m}^{*}_{{\bf RR},m}+\left(1-\Gamma^{(\alpha)}_{p,n} \tilde{m}_{{\bf LR},n}\right)\tilde{m}^{*}_{{\bf RR},m}\right.\right.\nonumber\\&&\left.\left.~~~~~~~~~~~+\left(1-\Gamma^{(\alpha)}_{p,n} \tilde{m}_{{\bf LR,n}}\right)^{*}\tilde{m}_{{\bf RR},m}+\left(1-\Gamma^{(\alpha)}_{p,n} \tilde{m}_{{\bf LR},n}\right)^{*}\tilde{m}_{{\bf RR},m}\right]\right\}\eea 
Here for very small wave number $p,p_n<<1$ one can write, $$\widetilde{{\cal G}(p<<1)}\sim \frac{\pi p }{2|\cos \pi \nu|\left|\Gamma\left(\nu+ \frac{1}{2}\right)\right|^2}\frac{|1-\gamma^{(\alpha)}_p \tilde{m}_{\bf LR}|^2}{|1-\gamma^{(\alpha)}_p  \tilde{m}_{\bf LR}|^2-| \tilde{m}_{\bf RR}|^2}\left[1+\cdots\right],$$ where all $\cdots$ are small correction terms.  For Bunch Davies vacuum once we fix $\alpha=0$, we find that the function $\widetilde{{\cal G}(p<<1)}$ only depends on the mass parameter $\nu$ for massive axion field.

On the contrary, for the case where $\nu=n/2$ (which also includes the massless situation $\nu=3/2$) the expression $\widetilde{{\cal G}(p<<1)}$ diverges due to  the overall factor $1/|\cos \pi \nu|$. But we can avoid such unwanted divergent contributions by rewriting all the expressions for $p,p_n<<1$ with $\nu=n/2$ that we have mentioned earlier. In such a situation for the massless case the time independent function$\widetilde{{\cal Q}(p<<1,\alpha,\nu=3/2)}$ can be further simplified as:
\bea\widetilde{{\cal Q}(p<<1,\alpha,\nu=3/2)}&=&\frac{\widetilde{{\cal G}(p<<1,\nu=3/2)}}{2p^3}~~~\forall\alpha,~~~~~~~~~~~\eea
where the function $\widetilde{{\cal G}(p<<1)}$ is defined for $\nu = 3/2$ as~\footnote{Here it is important to note the expression for the time dependent function $\widetilde{{\cal G}(p<<1)}$ for $\nu=q/2$ (where $q$ is any positive odd integer) in all cases are same. The only difference is appearing in the expression for the power spectrum. For $\nu=3/2$ case the power spectrum is scale invariant exactly. But for the  other values of $\nu=1/2,5/2,7/2,\cdots$ the power spectrum is not scale invariant and small deviation from the scale invariant feature can be observed easily.}:
\bea \widetilde{{\cal G}(p<<1,\nu=3/2)}&=&\frac{\pi }{2}\left\{1+\frac{\left(1\pm e^{i\theta}\pi p~e^{-p\pi}\right)}{{|1\pm e^{i\theta}\pi p~e^{-p\pi}|}}\sum^{\infty}_{n=0}\frac{\left(1\pm e^{-i\theta}\pi p_n~e^{-p_n\pi}\right)}{{|1\pm e^{i\theta}\pi p_n~e^{-p_n\pi}|}}\right.\nonumber\\&&\left.~~~~+\sum^{\infty}_{n=0}\sum^{\infty}_{m=0}\sqrt{\frac{\left(1\pm e^{i\theta}\pi p_n~e^{-p_n\pi}\right)}{|1\pm e^{i\theta}\pi p_n~e^{-p_n\pi}|}\frac{\left(1\pm e^{-i\theta}\pi p_m~e^{-p_m\pi}\right)}{|1\pm e^{i\theta}\pi p_m~e^{-p_m\pi}|}}\right\}\eea 
Here for very small wave number $p,p_n<<1$ with $\nu\neq 3/2$ and $\nu=3/2$ one can write, $$\widetilde{{\cal G}(p<<1)}\sim \frac{\pi }{2}\left[1+\cdots\right],$$ where all $\cdots$ are small correction terms.  For Bunch Davies vacuum we get the same result as the function $\widetilde{{\cal G}(p<<1)}$ for massless axion field ($\nu=3/2$) is independent of the parameter $\alpha$.
   
Moreover, it is important to note that the following contribution appearing in the normalised power spectrum for massive ($\nu\neq 3/2$) and massless ($\nu=3/2$) axion field can be simplified in the small wave number limit as:
\bea \left[\frac{1+|\gamma^{(\alpha)}_p|^2}{1-|\gamma^{(\alpha)}_p|^2}+\frac{1}{\left(f^{(\alpha)}_{p}\right)^2}\sum^{\infty}_{s=0}\frac{1+|\Gamma^{(\alpha)}_{p,s}|^2}{\left(1-|\Gamma^{(\alpha)}_{p,s}|^2\right)^2}\right]~~~~~~~~~~~~~~~~~~~~~~~~~~~~~~~~~~~~~~~~~~~~~~~~~~~~~\nonumber\\
\stackrel{p<<1}{\approx}\left[\frac{\frac{\left(\sqrt{\cos 2\pi \nu+1}\pm\sqrt{\cos 2\pi \nu+3}\right)^2\left(\cosh^2\alpha+\sinh^2\alpha~e^{2\pi  i\nu}\right)^2}{\left[\cosh^2\alpha+\sinh^2\alpha~e^{2i\pi\nu}+\sinh 2\alpha\cos \pi \nu~e^{i\pi\nu}\right]^2}+2 }{\frac{\left(\sqrt{\cos 2\pi \nu+1}\pm\sqrt{\cos 2\pi \nu+3}\right)^2\left(\cosh^2\alpha+\sinh^2\alpha~e^{2\pi  i\nu}\right)^2}{ \left[\cosh^2\alpha+\sinh^2\alpha~e^{2i\pi\nu}+\sinh 2\alpha\cos \pi \nu~e^{i\pi\nu}\right]^2}-2}~~~~~~~~~~~~~~~~~~~~~~~~\right.\nonumber \\ \left.~~~~~~~~~~~~~~~~~~~+\frac{1+\left|\frac{\sqrt{2} \left[\cosh^2\alpha+\sinh^2\alpha~e^{2i\pi\nu}+\sinh 2\alpha\cos \pi \nu~e^{i\pi\nu}\right]}{\left(\sqrt{\cos 2\pi \nu+1}\pm\sqrt{\cos 2\pi \nu+3}\right)\left(\cosh^2\alpha+\sinh^2\alpha~e^{2\pi  i\nu}\right)}\right|^2}{\left(1-\left|\frac{\sqrt{2} \left[\cosh^2\alpha+\sinh^2\alpha~e^{2i\pi\nu}+\sinh 2\alpha\cos \pi \nu~e^{i\pi\nu}\right]}{\left(\sqrt{\cos 2\pi \nu+1}\pm\sqrt{\cos 2\pi \nu+3}\right)\left(\cosh^2\alpha+\sinh^2\alpha~e^{2\pi  i\nu}\right)}\right|^2\right)^4}\underbrace{\left(\sum^{\infty}_{s=0} 1\right)^{-1}}_{\textcolor{red}{\bf =0}}\right]\nonumber\\
~~~~~~~~~~~~~=\left[\frac{\frac{\left(\sqrt{\cos 2\pi \nu+1}\pm\sqrt{\cos 2\pi \nu+3}\right)^2\left|\cosh^2\alpha+\sinh^2\alpha~e^{2\pi  i\nu}\right|^2}{\left|\cosh^2\alpha+\sinh^2\alpha~e^{2i\pi\nu}+\sinh 2\alpha\cos \pi \nu~e^{i\pi\nu}\right|^2}+2 }{\frac{\left(\sqrt{\cos 2\pi \nu+1}\pm\sqrt{\cos 2\pi \nu+3}\right)^2\left|\cosh^2\alpha+\sinh^2\alpha~e^{2\pi  i\nu}\right|^2}{ \left|\cosh^2\alpha+\sinh^2\alpha~e^{2i\pi\nu}+\sinh 2\alpha\cos \pi \nu~e^{i\pi\nu}\right|^2}-2}\right]~~~~~\forall \alpha~{\rm and}~ \nu\neq 3/2,~~~~~~~~\\
\left[\frac{1+|\gamma^{(\alpha,3/2)}_p|^2}{1-|\gamma^{(\alpha,3/2)}_p|^2}+\frac{1}{\left(f^{(\alpha,3/2)}_{p}\right)^2}\sum^{\infty}_{s=0}\frac{1+|\Gamma^{(\alpha,3/2)}_{p,s}|^2}{\left(1-|\Gamma^{(\alpha,3/2)}_{p,s}|^2\right)^2}\right]\stackrel{p<<1}{\approx}\left[1+\frac{1}{2}\underbrace{\left(\sum^{\infty}_{s=0} 1\right)^{-1}}_{\textcolor{red}{\bf =0}}\right]~~~~~~~~~\nonumber\\
= 1~~\forall \alpha~ {\rm and}~ \nu=3/2.~~~~~~~~~~~~~~\eea
Thus, in the superhorizon time scales ($t_{\bf L}>>1$) of region \textcolor{red}{\bf L} the amplitude of the normalised power spectrum of axion from generalised $\alpha$ vacua in the small wave number limit can be expressed as:
\bea \label{ph1} {\cal P}(p<<1,\alpha,t_{\bf L}>>1)&=&\frac{p^3}{2\pi^2}~\left(\cosh t_{\bf L}\right)^{2\nu-3}~\exp\left(-2\alpha\right)H^2\widetilde{{\cal Q}(p<<1,\alpha,\nu)}\nonumber\\
&&~~~~~~~~~~~~~\times \left[\frac{\frac{\left(\sqrt{\cos 2\pi \nu+1}\pm\sqrt{\cos 2\pi \nu+3}\right)^2\left|\cosh^2\alpha+\sinh^2\alpha~e^{2\pi  i\nu}\right|^2}{\left|\cosh^2\alpha+\sinh^2\alpha~e^{2i\pi\nu}+\sinh 2\alpha\cos \pi \nu~e^{i\pi\nu}\right|^2}+2 }{\frac{\left(\sqrt{\cos 2\pi \nu+1}\pm\sqrt{\cos 2\pi \nu+3}\right)^2\left|\cosh^2\alpha+\sinh^2\alpha~e^{2\pi  i\nu}\right|^2}{ \left|\cosh^2\alpha+\sinh^2\alpha~e^{2i\pi\nu}+\sinh 2\alpha\cos \pi \nu~e^{i\pi\nu}\right|^2}-2}\right]
\nonumber\\&=&\left(2\cosh t_{\bf L}\right)^{2\nu-3}~\left(\frac{H}{2\pi}\right)^2~\left(\frac{\Gamma(\nu)}{\Gamma\left(\frac{3}{2}\right)}\right)^2\widetilde{{\cal G}(p<<1)}\nonumber\\
&&~~~~~~~~~~~~~\times \left[\frac{\frac{\left(\sqrt{\cos 2\pi \nu+1}\pm\sqrt{\cos 2\pi \nu+3}\right)^2\left|\cosh^2\alpha+\sinh^2\alpha~e^{2\pi  i\nu}\right|^2}{\left|\cosh^2\alpha+\sinh^2\alpha~e^{2i\pi\nu}+\sinh 2\alpha\cos \pi \nu~e^{i\pi\nu}\right|^2}+2 }{\frac{\left(\sqrt{\cos 2\pi \nu+1}\pm\sqrt{\cos 2\pi \nu+3}\right)^2\left|\cosh^2\alpha+\sinh^2\alpha~e^{2\pi  i\nu}\right|^2}{ \left|\cosh^2\alpha+\sinh^2\alpha~e^{2i\pi\nu}+\sinh 2\alpha\cos \pi \nu~e^{i\pi\nu}\right|^2}-2}\right].~~~~~~~~~~~~~~~~~~\eea
 For the massless case ($\nu=3/2$) in the superhorizon time scales ($t_{\bf L}>>1$) of region \textcolor{red}{\bf L}, the amplitude of the normalised power spectrum of axion from generalised $\alpha$ vacua  in the small wave number limit can be simplified in the present context as:
\bea \label{az3} {\cal P}(p<<1,\alpha,t_{\bf L}>>1)&=&\frac{p^3}{2\pi^2}~\exp\left(-2\alpha\right)H^2\widetilde{{\cal Q}(p<<1,\alpha,\nu=3/2)}\nonumber\\&=&\left(\frac{H}{2\pi}\right)^2~\exp\left(-2\alpha\right)\widetilde{{\cal G}(p<<1,\nu=3/2)}.~~~~~~~~~~~\eea
            \begin{figure*}[htb]
            \centering
           \subfigure[Small wave number dependence of RDM power spectrum for $\alpha=0$ and $\nu=1,2,3,4,5$.]{
               \includegraphics[width=7.7cm,height=5.5cm] {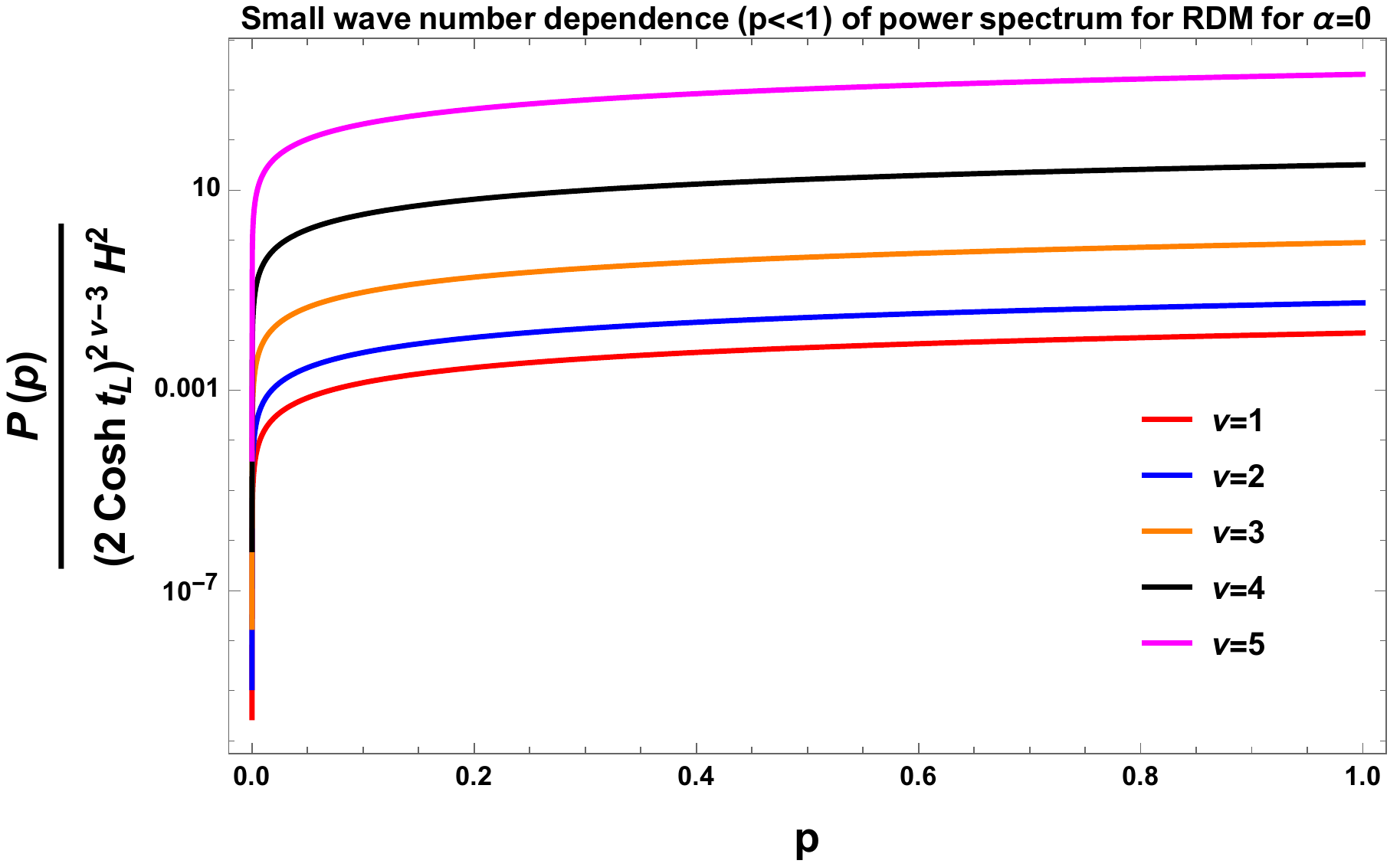}
               \label{fig4a}
            }
            \subfigure[Small wave number dependence of RDM power spectrum for $\alpha=0$ and $\nu=1/2,3/2,5/2,7/2,9/2$.]{
                \includegraphics[width=7.7cm,height=5.5cm] {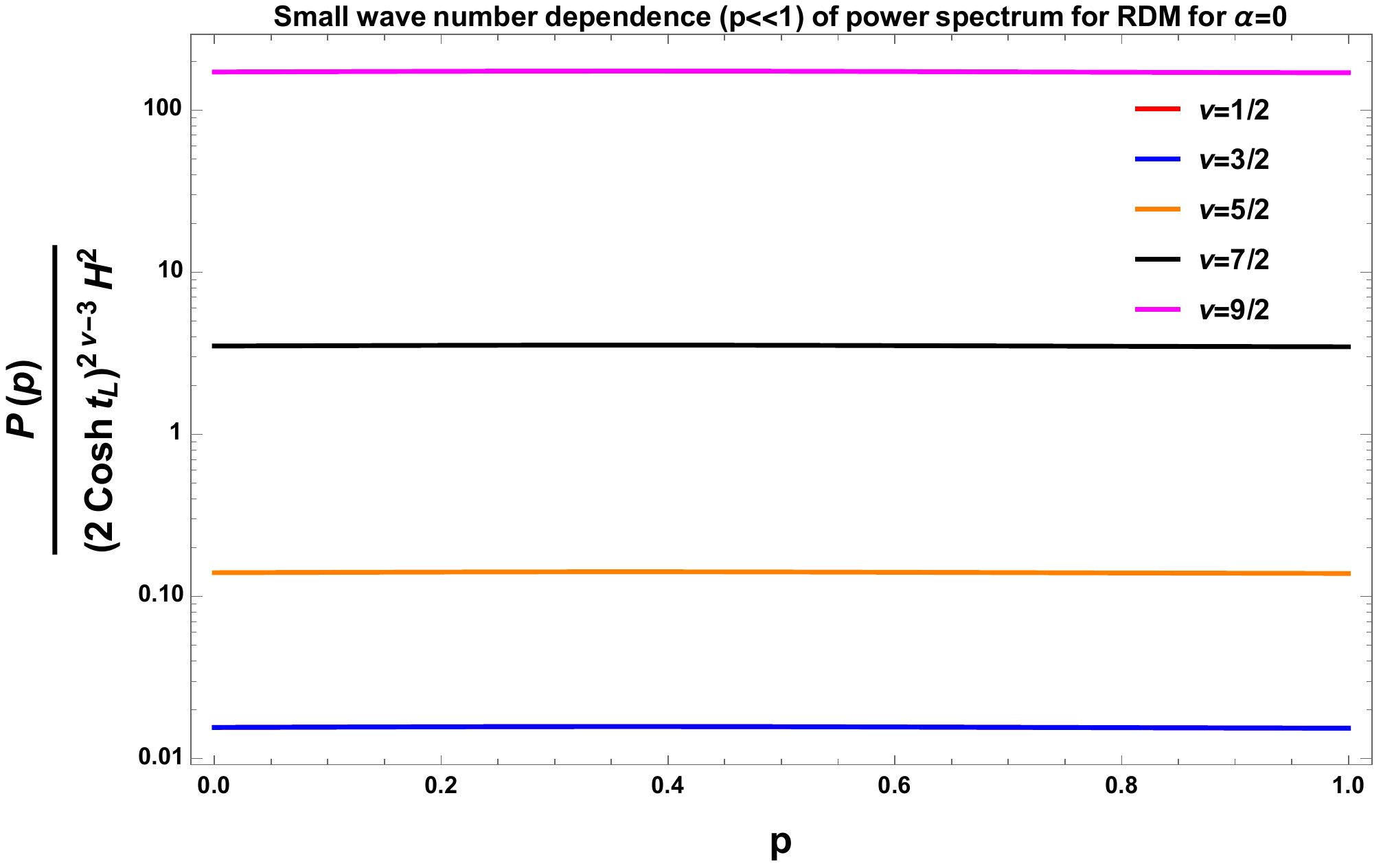}
               \label{fig4b}
              }
             \subfigure[Small wave number dependence of RDM power spectrum for $\alpha=0.1$ and $\nu=1,2,3,4,5$.]{
                    \includegraphics[width=7.7cm,height=5.5cm] {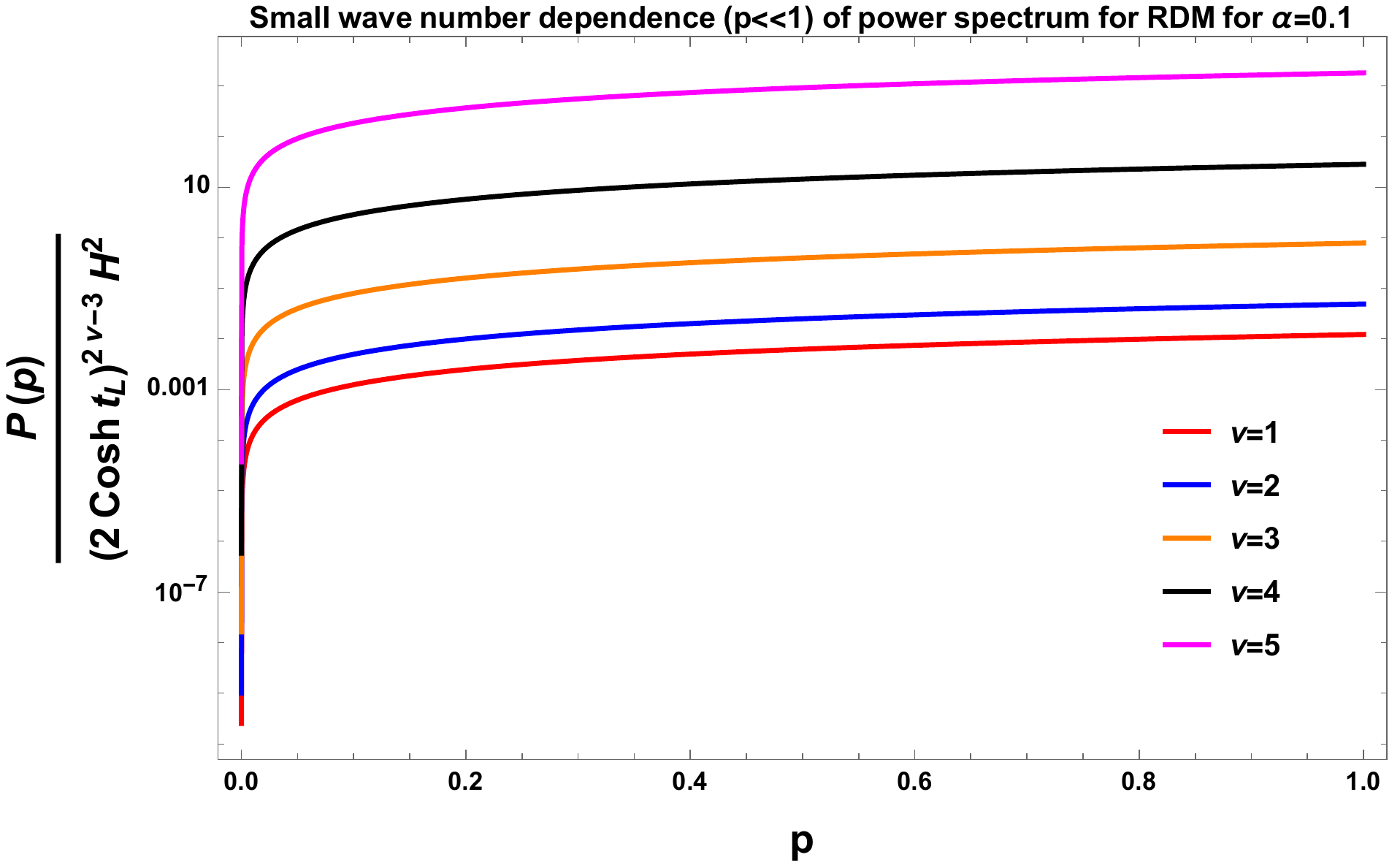}
                    \label{fig4c}   
           }
             \subfigure[Small wave number dependence of RDM power spectrum for $\alpha=0.1$ and $\nu=1/2,3/2,5/2,7/2,9/2$.]{
                    \includegraphics[width=7.7cm,height=5.5cm] {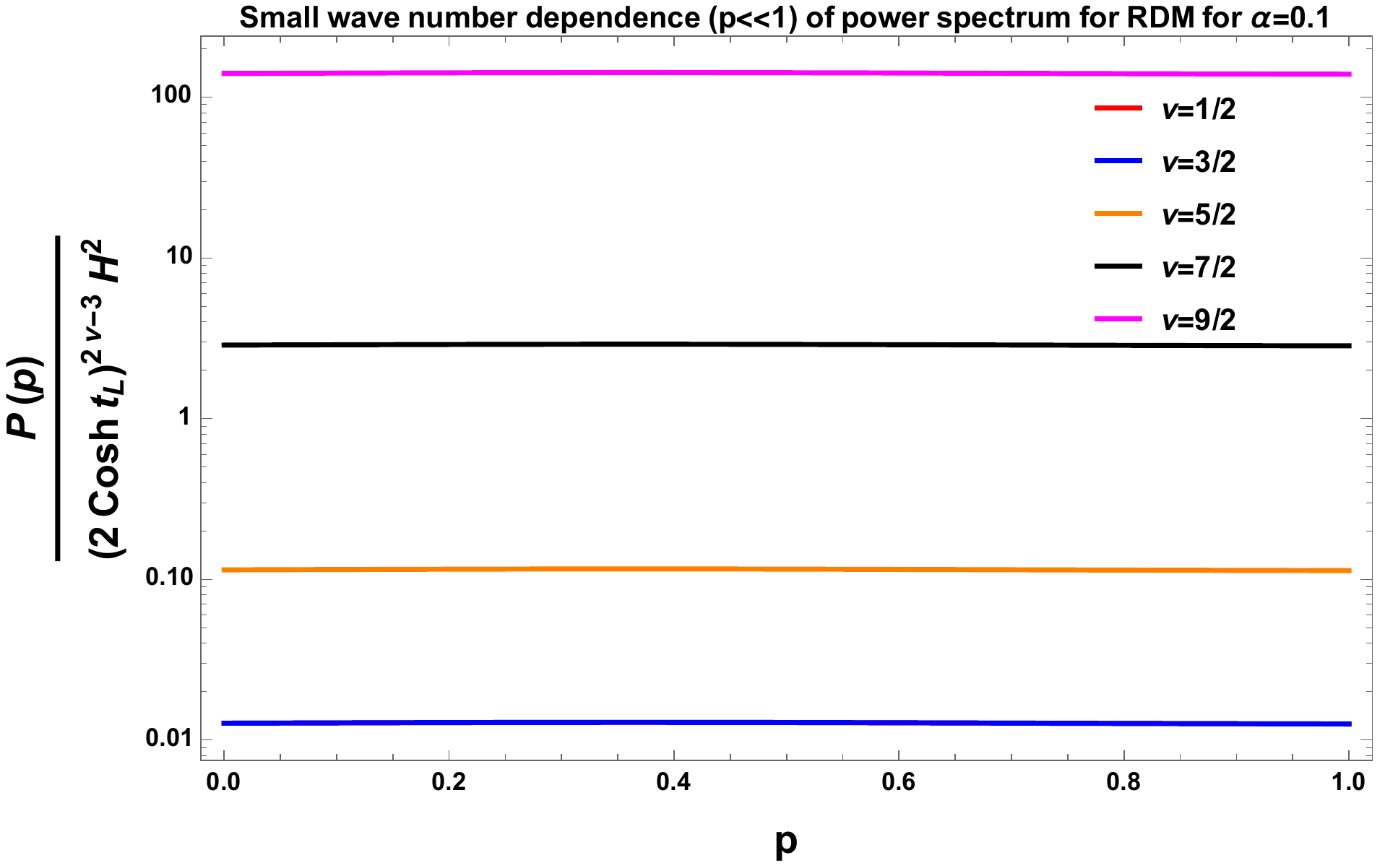}
                    \label{fig4d}   
           }
            \subfigure[Mass parameter dependence of RDM power spectrum in $p<<1$.]{
                    \includegraphics[width=10.7cm,height=5.5cm] {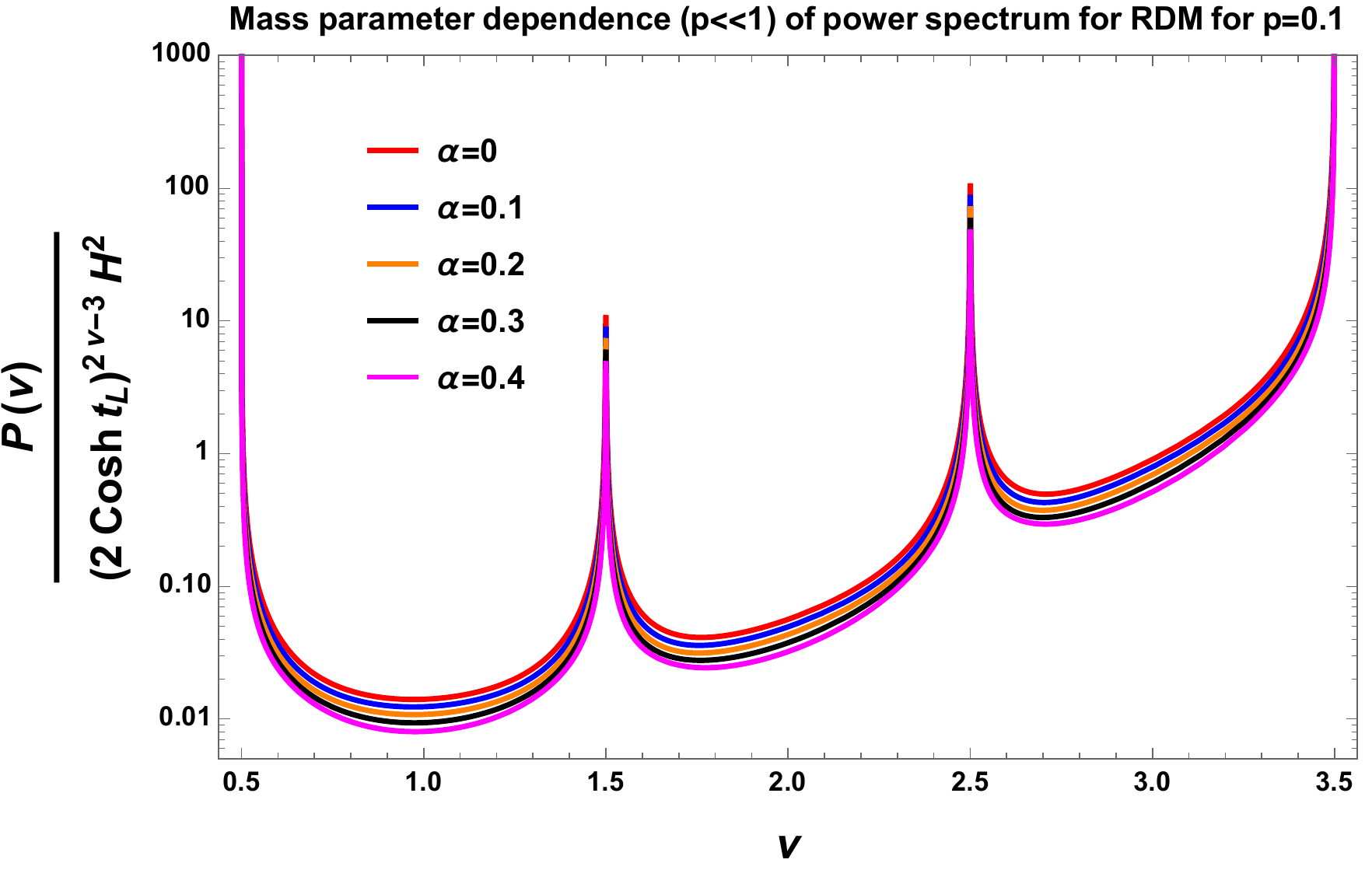}
                    \label{fig4e}   
           }
           \caption[Optional caption for list of figures]{Features of RDM power spectrum in small wave number region.} 
            \label{fig1x4}
            \end{figure*}
For Bunch Davies vacuum state  ( $\alpha=0$), the mean square vacuum fluctuation of axion can be expressed as:
 \bea \label{ph2} {\cal P}_{\bf BD}(p<<1,t_{\bf L}>>1)&=&\frac{p^3}{2\pi^2}~\left(\cosh t_{\bf L}\right)^{2\nu-3}~H^2\widetilde{{\cal Q}(p<<1,\alpha=0,\nu)}\nonumber\\
&&~~~~~~~~~~~~~\times \left[\frac{\left(\sqrt{\cos 2\pi \nu+1}\pm\sqrt{\cos 2\pi \nu+3}\right)^2+2 }{\left(\sqrt{\cos 2\pi \nu+1}\pm\sqrt{\cos 2\pi \nu+3}\right)^2-2}\right]
\nonumber\\&=&\left(2\cosh t_{\bf L}\right)^{2\nu-3}~\left(\frac{H}{2\pi}\right)^2~\left(\frac{\Gamma(\nu)}{\Gamma\left(\frac{3}{2}\right)}\right)^2\widetilde{{\cal G}(p<<1)}\nonumber\\
&&~~~~~~~~~~~~~\times \left[\frac{\left(\sqrt{\cos 2\pi \nu+1}\pm\sqrt{\cos 2\pi \nu+3}\right)^2+2 }{\left(\sqrt{\cos 2\pi \nu+1}\pm\sqrt{\cos 2\pi \nu+3}\right)^2-2}\right].~~~~~~~~~~~~~~~~~~\eea
  Also for the massless case ($\nu=3/2$) in the superhorizon time scales ($t_{\bf L}>>1$) of region \textcolor{red}{\bf L} the amplitude of the normalised power spectrum of axion from Bunch Davies vacuum in the small wave number limit can be simplified as:
\bea \label{az2} {\cal P}_{\bf BD}(p<<1,t_{\bf L}>>1)&=&\frac{p^3}{2\pi^2}~H^2\widetilde{{\cal Q}(p<<1,\alpha=0,\nu=3/2)}\nonumber\\&=&\left(\frac{H}{2\pi}\right)^2~\widetilde{{\cal G}(p<<1,\nu=3/2)}.~~~~~~~~~~~\eea 

 In figure~(\ref{fig4a}) and figure~(\ref{fig4c}) we have shown the behaviour of the power spectrum of the mean square vacuum fluctuation computed from RDM formalism in the small wave number regime for $\alpha=0$ and $\alpha=0.1$  and for fixed values of the mass parameter $\nu=1,2,3,3,4,5$ respectively. Moreover, in figure~(\ref{fig4e}) we have presented the behaviour of the power spectrum with respect to the mass parameter $\nu$ with fixed values of the parameter $\alpha=0,0.1,0.2,0.3,0.4$. For the mass parameter dependence here we get distinctive feature for RDM formalism compared to FOE formalism which we discussed in the last subsection and the NES formalism which we discuss in the next subsection.  From the plot, it is observed that for $\nu=1/2,3/2,5/2,7/2$ we get distinctive sharp peaks with constant and different magnitudes. On the other hand, in figure~(\ref{fig4b}) and figure~(\ref{fig4d}) we have shown the behaviour of the power spectrum in the small wave number regime for $\alpha=0$ and $\alpha=0.1$  with the fixed values of the mass parameter $\nu=1/2,3/2,5/2,7/2,9/2$. Here as the power spectrum is independent of the wave number, we get constant magnitude for different values of the mass parameter $\nu$.
  
\subsection{Quantum vacuum fluctuation with non entangled state (NES)}
\begin{figure*}[htb]
			    \centering
			    {
			        \includegraphics[width=18.5cm,height=12cm] {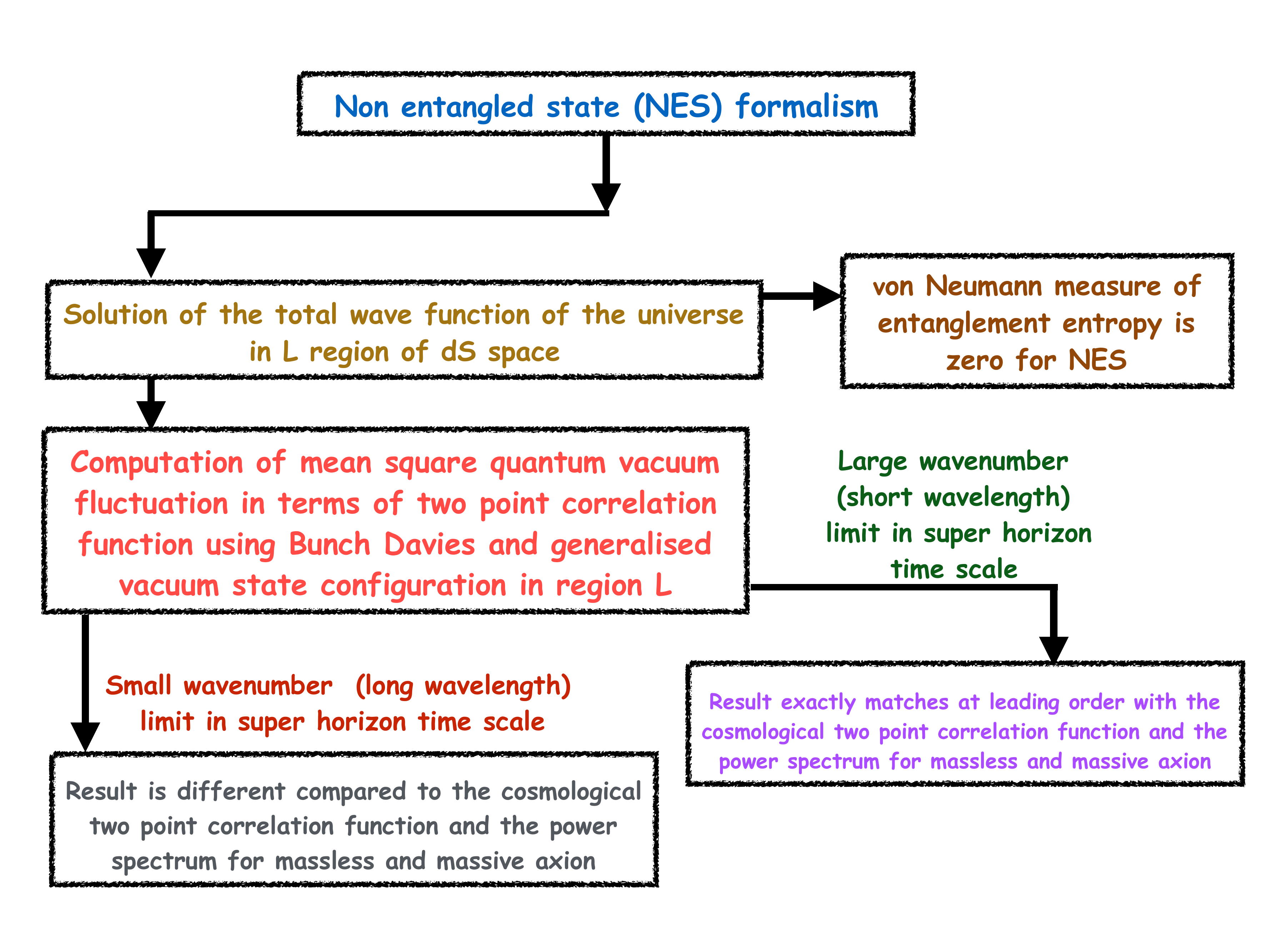}
			    }
			    \caption[Optional caption for list of figures]{Schematic diagram for the computation algorithm of NES formalism for non entangled quantum state of axion in de Sitter hyperbolic open chart. } 
			    \label{fzaanmvjx}
			    \end{figure*}

In this subsection, we describe the quantum vacuum fluctuation and its cosmological consequences using non entangled state (NES) formalism. In this formalism we assume that the wave function of the full de Sitter universe is described in the region \textcolor{red}{\bf L}. So we do not use anyt information from  the region \textcolor{red}{\bf R}.  In figure~(\ref{fzaanmvjx}) we have presented a schematic diagram for the computation algorithm of NES formalism for non entangled quantum state of axion in de Sitter hyperbolic open chart. 
\subsubsection{Non entangled state (NES) formalism}
\label{x3a}
In the region \textcolor{red}{\bf L} the total wave function of the universe is described by the non entangled state (NES) and for generalised $\alpha$ vacua it is given by:
\bea \tilde{\phi}^{\cal I}&=& \left(\begin{array}{ccc} \tilde{\cal \phi}^{L} \\ \tilde{\cal \phi}^{{L^*}}\\
   \end{array}\right)=\frac{1}{\tilde{\cal N}_b}\left(\begin{array}{ccc} \tilde{\cal P}^{L} \\ \tilde{\cal P}^{{L^*}}\\
   \end{array}\right)+\sum^{\infty}_{n=0}\frac{1}{\tilde{\cal N}_{b,(n)}}\left(\begin{array}{ccc} \tilde{\cal P}^{L,(n)} \\ \tilde{\cal P}^{{L^*,(n)}}\\
   \end{array}\right),\eea
   where the normalisation factors $\tilde{\cal N}_b$ and $\tilde{\cal N}_{b,(n)}$ are :
   \bea \tilde{\cal N}_b&=&\frac{\sqrt{2p}}{|\Gamma\left(1+ip\right)|},\\ 
   \tilde{\cal N}_{b,(n)}&=&\frac{\sqrt{2p_n}}{|\Gamma\left(1+ip_n\right)|}.\eea
   We can also express the total wave function of the universe in terms of the oscillator mode expansion as given by:
   \bea \tilde{\phi}^{L}(t_{\bf L})&=&\frac{H}{\sinh t_{\bf L}}\left[b_{\cal I}\tilde{\phi}^{\cal I}(t_{\bf L})+\sum^{\infty}_{n=0}b_{{\cal I},(n)}\tilde{\phi}^{\cal I}_{(n)}(t_{\bf L})\right].\eea 
   \subsubsection{Two point correlation function}
   \label{x3b}
  Using the above wave function we can further derive  the 
mean square vacuum fluctuation through the following two point correlation function :
\bea \langle {\bf L}|\tilde{\phi}^{L}_{plm}\tilde{\phi}^{\dagger L}_{p^{'}l^{'}m^{'}} |{\bf L}\rangle &=&\frac{H^2}{\sinh^2 t_{\bf L}}|\tilde{\phi}^{L}|^2\exp\left(-2\alpha\right)\delta(p-p^{'})\delta_{ll^{'}}\delta_{mm^{'}}\nonumber\\
&=& P(p,\alpha,t_{\bf L})\delta(p-p^{'})\delta_{ll^{'}}\delta_{mm^{'}},\eea
where $P(p,\alpha,t_{\bf L})$ is the power spectrum for non entangled state involving generalised $\alpha$ vacua. We can also define the normalised power spectrum for non entangled state as:
\bea {\cal P}(p,\alpha,t_{\bf L})&=&\frac{p^3}{2\pi^2}P(p,\alpha,t_{\bf L})=\frac{p^3}{2\pi^2}\frac{H^2}{\sinh^2 t_{\bf L}}|\tilde{\phi}^{L}|^2 \exp\left(-2\alpha\right).\eea
To quantify the normalised power spectrum for non entangled state, it is crcial to derive the expression for the square of the magnitude of the total wave function of the universe in the region \textcolor{red}{\bf L}, which is given by:
\bea\label{pog} |\tilde{\phi}^{L}|^2&=&\frac{1}{|\tilde{\cal N}_b|^2}\tilde{\cal P}^{L*}\tilde{\cal P}^{L}+\sum^{\infty}_{n=0}\frac{1}{{\cal N}_b{\cal N}^{*}_{b,(n)}}\left(\tilde{\cal P}^{L*}_{(n)}\tilde{\cal P}^{L}+\tilde{\cal P}^{L*}\tilde{\cal P}^{L}_{(n)}\right)+\sum^{\infty}_{n=0}\frac{1}{{\cal N}^{*}_b{\cal N}_{b,(n)}}\left(\tilde{\cal P}^{L*}_{(n)}\tilde{\cal P}^{L}+\tilde{\cal P}^{L*}\tilde{\cal P}^{L}_{(n)}\right)\nonumber\\
&&~~~~~~~~~~~~~~~~~~~~~~~~~~~~~~+\sum^{\infty}_{n=0}\sum^{\infty}_{m=0}\frac{1}{{\cal N}_{b,(m)}{\cal N}^{*}_{b,(n)}}\left(\tilde{\cal P}^{L*}_{(n)}\tilde{\cal P}^{L}_{(m)}+\tilde{\cal P}^{L*}_{(m)}\tilde{\cal P}^{L}_{(n)}\right).\eea
Further substituting the expressions for the normalisation factors, the above equation can be recast as:
\bea\label{pog1} |\tilde{\phi}^{L}|^2&=&\frac{1}{2p}|\Gamma(1+ip)|^2\tilde{\cal P}^{L*}\tilde{\cal P}^{L}+\sum^{\infty}_{n=0}\frac{1}{\sqrt{4pp_n}}|\Gamma(1+ip)| |\Gamma(1-ip_n)| \left(\tilde{\cal P}^{L*}_{(n)}\tilde{\cal P}^{L}+\tilde{\cal P}^{L*}\tilde{\cal P}^{L}_{(n)}\right)\nonumber\\
&&~~~~~~~~~~~~~~~~~~~~~~~~~~~~~~+\sum^{\infty}_{n=0}\frac{1}{4\sqrt{pp_n}}|\Gamma(1-ip)| |\Gamma(1+ip_n)| \left(\tilde{\cal P}^{L*}_{(n)}\tilde{\cal P}^{L}+\tilde{\cal P}^{L*}\tilde{\cal P}^{L}_{(n)}\right)\nonumber\\
&&~~~~~~~~~~~~~~~~~~~~~~~~~~~~~~+\sum^{\infty}_{n=0}\sum^{\infty}_{m=0}\frac{1}{\sqrt{4p_np_m}}|\Gamma(1-ip_n)| |\Gamma(1+ip_m)|\left(\tilde{\cal P}^{L*}_{(n)}\tilde{\cal P}^{L}_{(m)}+\tilde{\cal P}^{L*}_{(m)}\tilde{\cal P}^{L}_{(n)}\right).~~~~~~~~~~~~\eea
Consequently, the normalised power spectrum for non entangled state with generalised $\alpha$ vacua can be written as:
\bea\label{pog2} {\cal P}(p,\alpha,t_{\bf L})&=&\frac{p^3}{2\pi^2}\frac{H^2}{\sinh^2 t_{\bf L}}\left[\frac{1}{2p}|\Gamma(1+ip)|^2\tilde{\cal P}^{L*}\tilde{\cal P}^{L}+\sum^{\infty}_{n=0}\frac{1}{\sqrt{4pp_n}}|\Gamma(1+ip)| |\Gamma(1-ip_n)| \left(\tilde{\cal P}^{L*}_{(n)}\tilde{\cal P}^{L}+\tilde{\cal P}^{L*}\tilde{\cal P}^{L}_{(n)}\right)\right.\nonumber\\
&&\left.~~~~~~~~~~~~~~~~~~~~+\sum^{\infty}_{n=0}\frac{1}{4\sqrt{pp_n}}|\Gamma(1-ip)| |\Gamma(1+ip_n)| \left(\tilde{\cal P}^{L*}_{(n)}\tilde{\cal P}^{L}+\tilde{\cal P}^{L*}\tilde{\cal P}^{L}_{(n)}\right)\right.\nonumber\\
&&\left.~~~~~~~~~~~~~~~+\sum^{\infty}_{n=0}\sum^{\infty}_{m=0}\frac{1}{4\sqrt{p_np_m}}|\Gamma(1-ip_n)| |\Gamma(1+ip_m)|\left(\tilde{\cal P}^{L*}_{(n)}\tilde{\cal P}^{L}_{(m)}+\tilde{\cal P}^{L*}_{(m)}\tilde{\cal P}^{L}_{(n)}\right)\right].~~~~~~~~~~~~\eea
However,  to extract further physical information from Eqn~(\ref{po1xx}) for  cosmological predictions,  we consider the superhorizon time scales ($t_{\bf L}>>1$) of region \textcolor{red}{\bf L}. In  this limit, the Legendre functions as appearing in the complementary part and the particular integral part of the time dependent solution can be approximated to the following simplified form:
\bea \left(\widetilde{\cal P}^{\bf L},\widetilde{\cal P}^{\bf L*}\right)&\equiv& P^{\pm ip}_{\nu-\frac{1}{2}}\left(\cosh t_{\bf L}\right)~\underrightarrow{t_{\bf L}>>1}~\frac{2^{\nu-\frac{1}{2}}\left(\cosh t_{\bf L}\right)^{\nu-\frac{1}{2}}\Gamma(\nu)}{\sqrt{\pi}\Gamma\left(\nu\mp ip +\frac{1}{2}\right)},\\
\left(\widetilde{\cal P}^{\bf L}_{(n)},\widetilde{\cal P}^{\bf L*}_{(n)}\right)&\equiv& P^{\pm ip_n}_{\nu-\frac{1}{2}}\left(\cosh t_{\bf L}\right)~\underrightarrow{t_{\bf L}>>1}~\frac{2^{\nu-\frac{1}{2}}\left(\cosh t_{\bf L}\right)^{\nu-\frac{1}{2}}\Gamma(\nu)}{\sqrt{\pi}\Gamma\left(\nu\mp ip_n +\frac{1}{2}\right)}.\eea
Thus, in the superhorizon time scales ($t_{\bf L}>>1$) of region \textcolor{red}{\bf L},  eqn~(\ref{pog1}) can be further simplified as:
\bea \label{df2xxx} |\tilde{\phi}^{\bf L}|^2&~\underrightarrow{t_{\bf L}>>1}~& \widetilde{{\cal K}(p,\alpha,\nu)}\left(\cosh t_{\bf L}\right)^{2\nu-1}\eea
where the time independent function $\widetilde{{\cal K}(p,\alpha,\nu)}$ for generalised $\alpha$ vacua is defined as:
\bea\widetilde{{\cal K}(p,\alpha,\nu)}&=&\frac{2^{2\nu-1}\left(\Gamma(\nu)\right)^2}{\pi}\times \left[\frac{|\Gamma(1+ip)|^2}{2p|\Gamma\left(\nu+ip+\frac{1}{2}\right)|^2}\right.\nonumber\\ && \left.
~~~~~~+\sum^{\infty}_{n=0}\frac{|\Gamma(1-ip)| |\Gamma(1+ip_n)|+|\Gamma(1+ip)| |\Gamma(1-ip_n)|}{4\sqrt{pp_n}~\Gamma\left(\nu-ip+\frac{1}{2}\right)\Gamma\left(\nu+ip_n+\frac{1}{2}\right)}\right.\nonumber\\&&\left.~~~~~~~~+\sum^{\infty}_{n=0}\sum^{\infty}_{m=0}\frac{|\Gamma(1-ip_n)| |\Gamma(1+ip_m)|+|\Gamma(1+ip_n)| |\Gamma(1-ip_m)|}{4\sqrt{p_np_m}~\Gamma\left(\nu-ip_n+\frac{1}{2}\right)\Gamma\left(\nu+ip_m+\frac{1}{2}\right)}\right].~~~~~~~~~\eea

Also in the super horizon time scale ($t_{\bf L}>>1$) we get the following simplification in the normalised power spectrum for non entangled state :
\bea \label{po2xxxx1} {\cal P}(p,\alpha,t_{\bf L})&=&\frac{p^3}{2\pi^2}~\frac{H^2}{\sinh^2 t_{\bf L}} |\tilde{\phi}^{L}|^2\exp\left(-2\alpha\right)\nonumber\\
\underrightarrow{t_{\bf L}>>1}&&\frac{p^3}{2\pi^2}~\frac{\left(\cosh t_{\bf L}\right)^{2\nu-1}}{\sinh^2 t_{\bf L}}~H^2\widetilde{{\cal K}(p,\nu)}\exp\left(-2\alpha\right).~~~~~~~~~~~\eea
In this limit, for the massless case ( $\nu=3/2$), the time dependent contribution can be approximated into the following simplified form:
\bea \left(\frac{\left(\cosh t_{\bf L}\right)^{2\nu-1}}{\sinh^2 t_{\bf L}}\right)_{\nu=3/2}~\underrightarrow{t_{\bf L}>>1}~~~1.\eea 
This implies that for an arbitrary value of the parameter $\nu$ one can write:
\bea \left(\frac{\left(\cosh t_{\bf L}\right)^{2\nu-1}}{\sinh^2 t_{\bf L}}\right)~\underrightarrow{t_{\bf L}>>1}~~~\left(\cosh t_{\bf L}\right)^{2\nu-3}.\eea 
Consequently, in the superhorizon time scales ($t_{\bf L}>>1$) of region \textcolor{red}{\bf L} and for the massless case ($\nu=3/2$),  the amplitude of the normalised power spectrum can be expressed as:
\bea \label{po2xxxx1} {\cal P}(p,\alpha,t_{\bf L})&=&\frac{p^3}{2\pi^2}~\frac{H^2}{\sinh^2 t_{\bf L}} |\tilde{\phi}^{L}|^2\exp\left(-2\alpha\right)\nonumber\\
\underrightarrow{t_{\bf L}>>1,\nu=3/2}&&\frac{p^3}{2\pi^2}~H^2\widetilde{{\cal K}(p,\nu=3/2)}\exp\left(-2\alpha\right).~~~~~~~~~~~\eea
Like our  result derived in the previous section, this result also implies that for the massless case ($\nu=3/2$), the  amplitude of the vacuum fluctuation gets frozen with respect to the time scale when the associated modes exit horizon.

Further, to know the exact wavenumber dependence of the amplitude of the normalised power spectrum from generalised $\alpha$ vacua, we need to know the behaviour of the power spectrum at very short wavelengths ($p,p_n>>1$). In this limit, it is expected that the power spectrum of axion in the non entangled case should match with the result obtained for spatially flat universe. The time independent function $\widetilde{{\cal K}(p,\alpha,\nu)}$ in this limit and for arbitrary mass parameter $\nu$ can be expressed as:
\bea\widetilde{{\cal K}(p>>1,\alpha,\nu)}&=&\frac{2^{2(\nu-1)}\left(\Gamma(\nu)\right)^2}{p^3\pi}\widetilde{{\cal U}(p>>1)}~~~\forall \alpha,~~~~~~~~~~~\eea
where the function $\widetilde{{\cal U}(p>>1)}$ is defined as:
\bea \widetilde{{\cal U}(p>>1)}&=&\left[1+\underbrace{\sum^{\infty}_{n=0}\left(\frac{p}{p_n}\right)^{\frac{3}{2}}+\sum^{\infty}_{n=0}\sum^{\infty}_{m=0}\frac{p^3}{\left(p_np_m\right)^{\frac{3}{2}}}}_{\textcolor{red}{\bf Quantumm~correction~factor~for~axion~in~short~wave~length~limit}}\right].\eea 
Thus, for very large wave number ($p,p_n>>1$), we can write, $\widetilde{{\cal U}(p)}\sim 1+\cdots$, where all $\cdots$ are small correction terms. This also implies that for large wavenumber and  for any value of the mass parameter $\alpha$, the time independent function ${{\cal U}(p,\alpha,\nu)}$, computed with generalised $\alpha$ vacua,  matches with the result obtained for Bunch Davies vacua in the previous subsection at the leading order in $\widetilde{{\cal M}(p,\nu)}$.
            \begin{figure*}[htb]
            \centering
           \subfigure[Large wave number dependence of NES power spectrum for $\alpha=0$.]{
               \includegraphics[width=7.7cm,height=7.5cm] {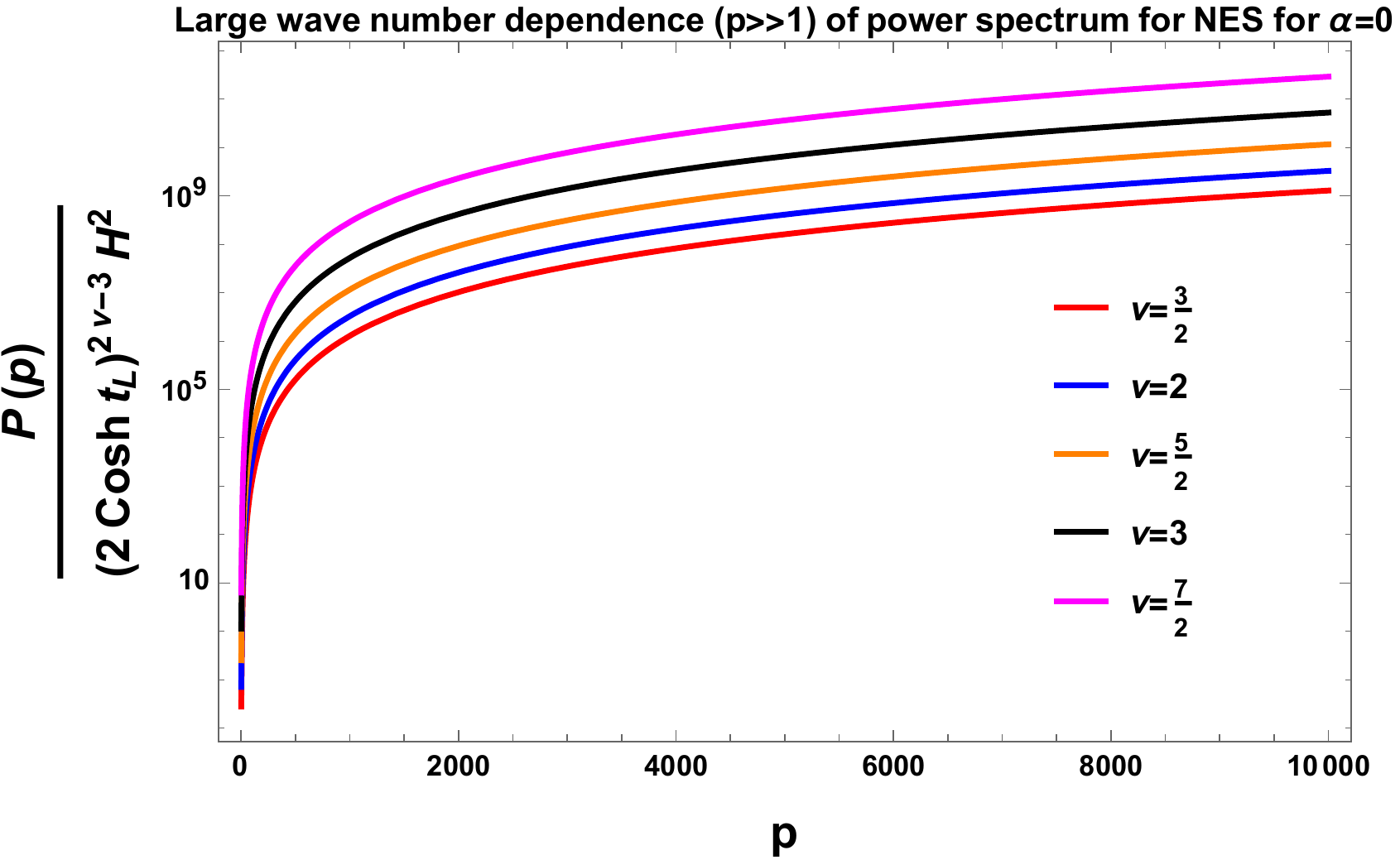}
               \label{fig5a}
            }
            \subfigure[Large wave number dependence of NES power spectrum for $\alpha=0.1$.]{
                \includegraphics[width=7.7cm,height=7.5cm] {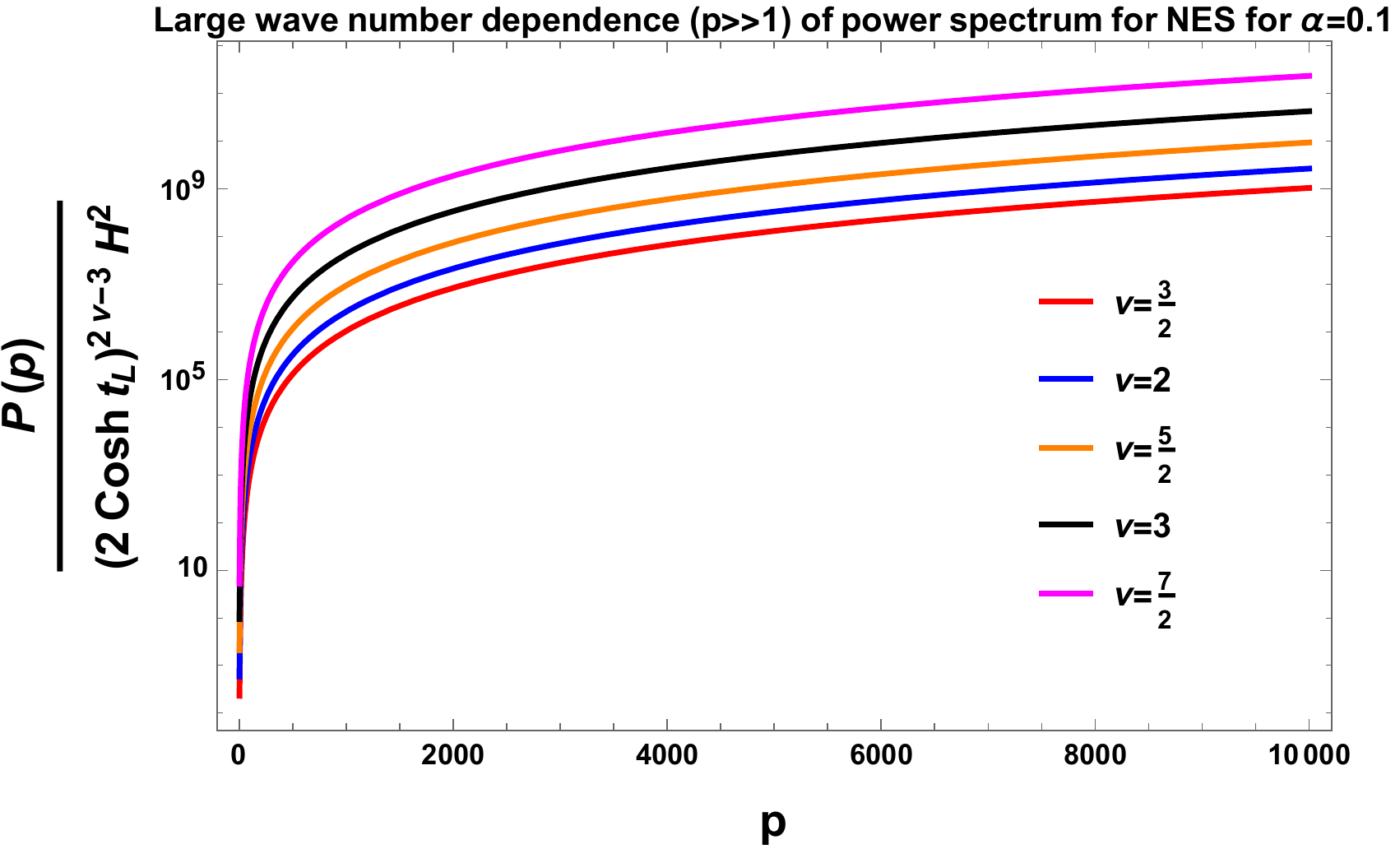}
               \label{fig5b}
              }
             \subfigure[Mass parameter dependence of NES power spectrum for $p>>1$.]{
                    \includegraphics[width=10.7cm,height=7.5cm] {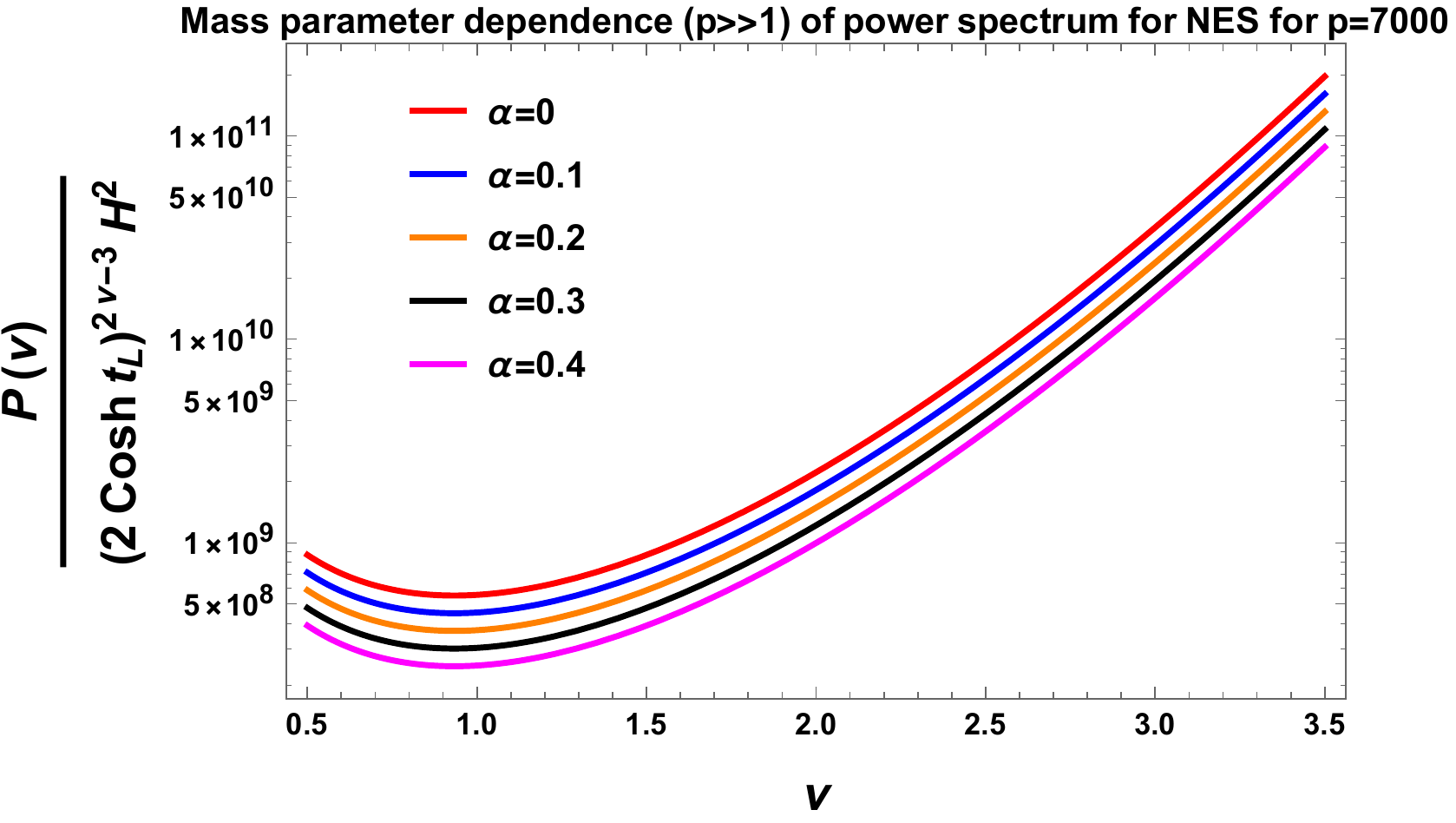}
                    \label{fig5c}   
           }
           \caption[Optional caption for list of figures]{Features of NES power spectrum in large wave number region.} 
            \label{fig1x5}
            \end{figure*}
Also for the massless case ($\nu=3/2$) the time independent function $\widetilde{{\cal K}(p,\alpha,\nu=3/2)}$ in the short wave length limit can further be simplified as:
\bea\widetilde{{\cal K}(p>>1,\alpha,\nu=3/2)}&=&\frac{\widetilde{{\cal U}(p>>1)}}{2p^3}~~~\forall\alpha.~~~~~~~~~~~\eea
Finally, in the superhorizon time scales ($t_{\bf L}>>1$) of region \textcolor{red}{\bf L} the amplitude of the normalised power spectrum of axion from generalised $\alpha$ vacua for non entangled state in short wave length limit can be expressed as:
\bea \label{po2vccxc1} {\cal P}(p>>1,\alpha,t_{\bf L}>>1)&=&\frac{p^3}{2\pi^2}~\left(\cosh t_{\bf L}\right)^{2\nu-3}~\exp\left(-2\alpha\right)H^2\widetilde{{\cal K}(p>>1,\alpha,\nu)}
\nonumber\\&=&\left(2\cosh t_{\bf L}\right)^{2\nu-3}~\left(\frac{H}{2\pi}\right)^2~\left(\frac{\Gamma(\nu)}{\Gamma\left(\frac{3}{2}\right)}\right)^2~\exp\left(-2\alpha\right)\widetilde{{\cal U}(p>>1)}.~~~~~~~~~~~\eea
For the massless case ($\nu=3/2$) in the superhorizon time scales ($t_{\bf L}>>1$) of region \textcolor{red}{\bf L}, the amplitude of the normalised power spectrum in short wave length limit can be simplified to:
\bea \label{po3zcvcc2} {\cal P}(p>>1,\alpha,t_{\bf L}>>1)&=&\frac{p^3}{2\pi^2}~\exp\left(-2\alpha\right)H^2\widetilde{{\cal K}(p>>1,\alpha,\nu=3/2)}\nonumber\\&=&\left(\frac{H}{2\pi}\right)^2~\exp\left(-2\alpha\right)\widetilde{{\cal U}(p>>1)}.~~~~~~~~~~~\eea
 Note that both the Eq~(\ref{po2vccxc1}) and Eq~(\ref{po3zcvcc2}) are valid after horizon exit. From these results we also observe that the power spectrum computed from non entangled state formalism is same, at the leading order approximation, as that computed from the  FOE and RDM formalism, computed in earlier subsections. This is true in the large wavenumber limit of superhorizon time scale in region \textcolor{red}{\bf L}.

The result for the two point correlation function and the associated power spectrum for Bunch Davies vacuum can be obtained by setting $\alpha=0$ in the above equation and is found to be:
 \bea \label{po2vvvccx} {\cal P}_{\bf BD}(p>>1,t_{\bf L}>>1)&=&\frac{p^3}{2\pi^2}~\left(\cosh t_{\bf L}\right)^{2\nu-3}~H^2\widetilde{{\cal K}(p>>1,\alpha=0,\nu)}
\nonumber\\&=&\left(2\cosh t_{\bf L}\right)^{2\nu-3}~\left(\frac{H}{2\pi}\right)^2~\left(\frac{\Gamma(\nu)}{\Gamma\left(\frac{3}{2}\right)}\right)^2\widetilde{{\cal U}(p>>1)}.~~~~~~~~~~~\eea
  For the massless case ($\nu=3/2$) it reduces to:
\bea \label{po3zcvcccc} {\cal P}_{\bf BD}(p>>1,t_{\bf L}>>1)&=&\frac{p^3}{2\pi^2}~H^2\widetilde{{\cal K}(p>>1,\alpha=0,\nu=3/2)}\nonumber\\&=&\left(\frac{H}{2\pi}\right)^2~\widetilde{{\cal U}(p>>1)}.~~~~~~~~~~~\eea 

 In figure~(\ref{fig5a}) and figure~(\ref{fig5b}) we have presented the behaviour of the power spectrum of the mean square vacuum fluctuation computed inNES formalism for the large wave number regime. This is shown for $\alpha=0$ and $\alpha=0.1$  and for fixed values of the mass parameter $\nu=3/2,2,5/2,3,7/2$ respectively. For both the values of $\alpha$, we get almost  similar behaviour. In figure~(\ref{fig5c}) we have shown the behaviour of the power spectrum with respect to the mass parameter $\nu$ with fixed values of the parameter $\alpha=0,0.1,0.2,0.3,0.4$. Here for $1/2<\nu<1$ region and $\nu>1$ region mass parameter dependence show two distinctive features. In $1/2<\nu<1$ region amplitude of the normalised power spectrum initially decrease and then just after $\nu=1$ the amplitude of the power spectrum increase. 

However, to examine the behaviour of the power spectrum in the long wavelength region and in the superhorizon time scale ($t_{\bf L}>>1$), we take the limit $p<<1$.  In the long wave length limit, the time independent function $\widetilde{{\cal K}(p,\alpha,\nu)}$ for any arbitrary mass parameter $\nu$ can be expressed (for $\alpha$ vacua) as:
\bea\widetilde{{\cal K}(p<<1,\alpha,\nu)}&=&\frac{2^{2(\nu-1)}\left(\Gamma(\nu)\right)^2}{p\pi}\widetilde{{\cal U}(p<<1)}~~~\forall \alpha,~~~~~~~~~~~\eea
where the function $\widetilde{{\cal U}(p<<1)}$ is given by:
\bea \widetilde{{\cal U}(p<<1)}&=&\left[1+\underbrace{\left(\frac{|\Gamma\left(\nu+\frac{1}{2}\right)|}{\Gamma\left(\nu+\frac{1}{2}\right)}\right)^2\left\{\sum^{\infty}_{n=0}\sqrt{\frac{p}{p_n}}+\sum^{\infty}_{n=0}\sum^{\infty}_{m=0}\frac{p}{\sqrt{p_np_m}}\right\}}_{\textcolor{red}{\bf Quantum ~correction ~factor ~for~axion~in ~long~wave~length~limit}}\right].\eea 
  For the massless case ($\nu=3/2$), this can be further simplified to:
\bea\widetilde{{\cal K}(p<<1,\alpha,\nu=3/2)}&=&\frac{\widetilde{{\cal U}(p<<1)}}{2p}~~~\forall\alpha.~~~~~~~~~~~\eea

		\begin{table*}
	\centering
	\footnotesize
	\begin{tabular}{|||c||c||c||c|||}
		\hline\hline
		\hline
		\textcolor{red}{\bf Feuatures} &\textcolor{red}{\bf FOE} &\textcolor{red}{\bf RDM}&\textcolor{red}{\bf NES}\\
		\hline\hline\hline
		\textcolor{blue}{\bf Wave}  &  Here we solve the  & Here we solve the & Here we only solve the
		\\
		\textcolor{blue}{\bf function}  &  wave function in {\bf L} region  & wave function in {\bf L} and {\bf R} region & wave function in {\bf L} region \\
		& of dS space.& of dS space.& of dS space.
		\\
		\hline\hline\hline
		\textcolor{blue}{\bf  Quantum }  & Here we deal with  &  Here we deal with& Here we deal with 
		\\
		\textcolor{blue}{\bf  state }&  entangled quantum state. &   mixed quantum state.  & non-entangled quantum state.
		\\
		\hline\hline\hline
		\textcolor{blue}{\bf  Quantum }  &  Power spectrum is & Power spectrum is & Power spectrum is
		\\
		\textcolor{blue}{\bf  number }& only dependent on {\bf SO(1,3)}    &  only dependent on {\bf SO(1,3)}     &only dependent on {\bf SO(1,3)}   
		\\
		\textcolor{blue}{\bf  dependence }& quantum number p  & quantum number p & quantum number p\\
		& and independent on l,m. & and independent on l,m. & and independent on l,m.
		\\
		\hline\hline\hline
		\textcolor{blue}{\bf  Time }  & Analysis is performed on  &  Analysis is performed on & Analysis is performed on
		\\
		\textcolor{blue}{\bf  scale }&  superhorizon  &  superhorizon  &superhorizon
		\\
		\textcolor{blue}{\bf  for computation }&  time scale.  &  time scale. &  time scale.
		\\
		\hline\hline\hline
		\textcolor{blue}{\bf  Power}  &  Leading order term  & Leading order term &Leading order term
		\\
		\textcolor{blue}{\bf  spectrum}  &  is $ \left(\frac{H}{2\pi}\right)^2~\exp(-2\alpha)$  &  is $ \left(\frac{H}{2\pi}\right)^2~\exp(-2\alpha)$  & is $ \left(\frac{H}{2\pi}\right)^2~\exp(-2\alpha)$ 
		\\
			\textcolor{blue}{\bf  spectrum}  &  and the next &  and the next &  and the next 
		\\
		\textcolor{blue}{\bf  at large}  &   order effects are different & order effects are different & order effects are different \\
	\textcolor{blue}{\bf  wave }	&  from RDM and NES &  from FOE and NES &  from FOE and RDM\\
	\textcolor{blue}{\bf  number}	& for massless axion ($\nu=3/2$). &  for massless axion ($\nu=3/2$).  &for massless axion ($\nu=3/2$).  \\
		\hline\hline\hline
		\textcolor{blue}{\bf  Power}  &  Leading order term  & Leading order term &Leading order term
		\\
		\textcolor{blue}{\bf  spectrum}  & is $\left(\frac{H}{2\pi}\right)^2~p^3~\exp(-2\alpha)$  & is $ \frac{H^2}{8\pi}~\exp(-2\alpha)$   &is $ \left(\frac{H}{2\pi}\right)^2~p^2~\exp(-2\alpha)$
		\\
		\textcolor{blue}{\bf  at small}  & and the next     &  and the next  & and the next 
		\\ 
		\textcolor{blue}{\bf  at small}  & order effects are different    &  order effects are different & order effects are different
		\\ 
		\textcolor{blue}{\bf  wave }  &   from RDM and NES & from FOE and NES &  from FOE and RDM\\
		\textcolor{blue}{\bf  number}	& for massless axion ($\nu=3/2$). &  for massless axion ($\nu=3/2$).  &for massless axion ($\nu=3/2$).  \\
		 \hline\hline\hline
	\end{tabular}
	 \caption{Comparison between FOE, RDM and NES formalism for $\alpha$ vacua.}\label{tabdfd}
	 \vspace{.2cm}
	\end{table*}

            \begin{figure*}[htb]
            \centering
           \subfigure[Small wave number dependence of NES power spectrum for $\alpha=0$.]{
               \includegraphics[width=7.7cm,height=7.5cm] {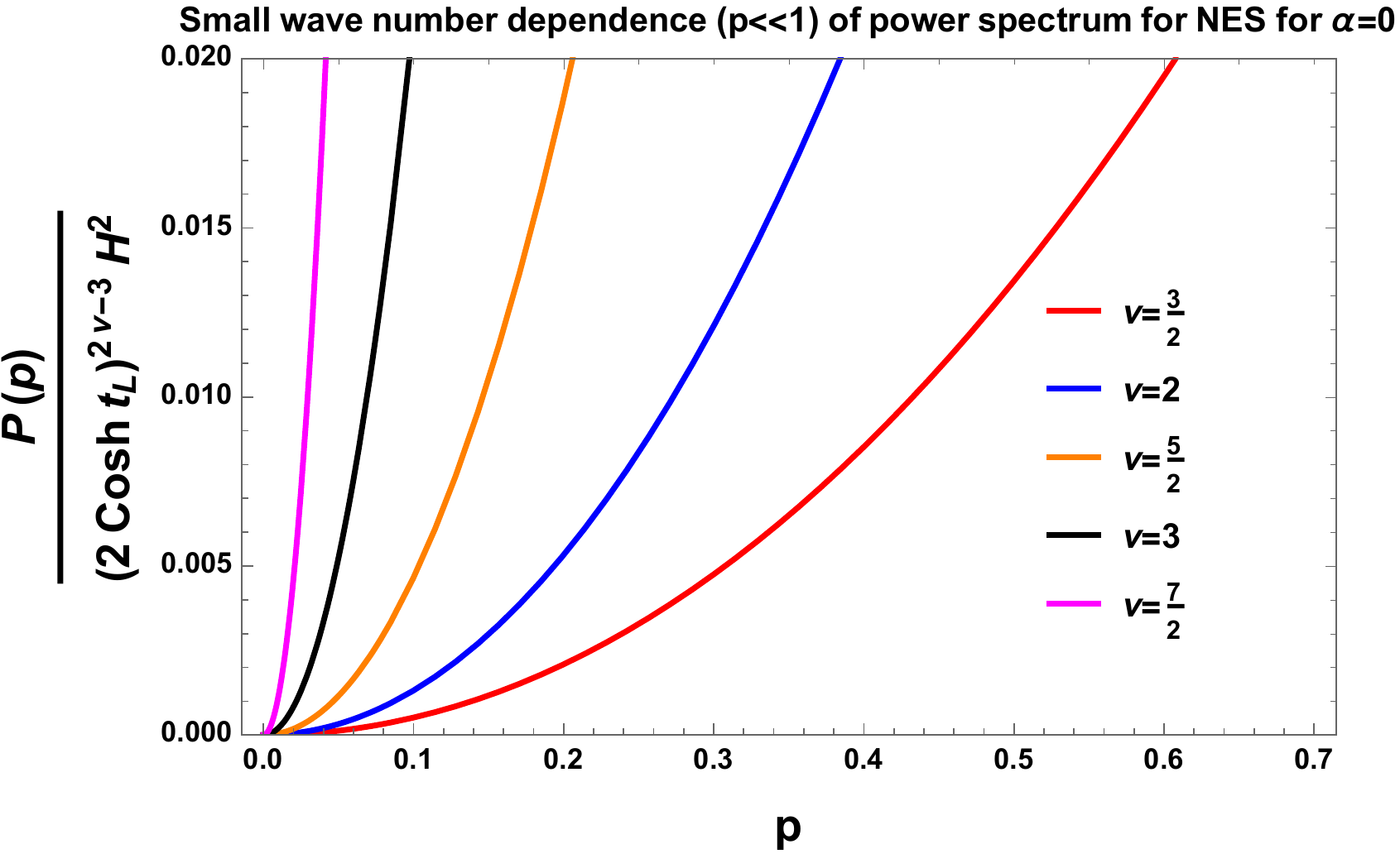}
               \label{fig6a}
            }
            \subfigure[Small wave number dependence of NES power spectrum for $\alpha=0.1$.]{
                \includegraphics[width=7.7cm,height=7.5cm] {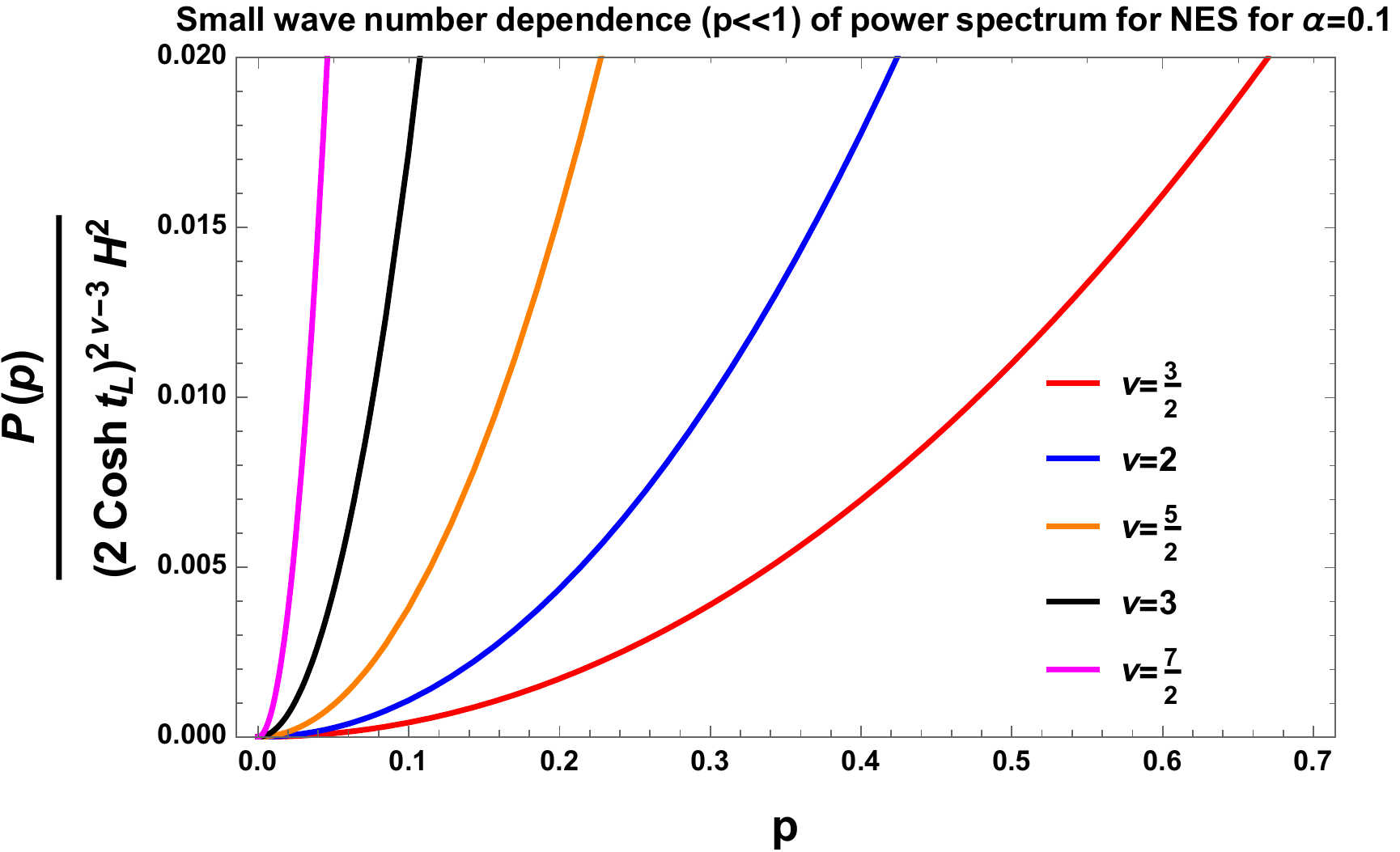}
               \label{fig6b}
              }
             \subfigure[Mass parameter dependence of NES power spectrum in $p<<1$.]{
                    \includegraphics[width=10.7cm,height=7.5cm] {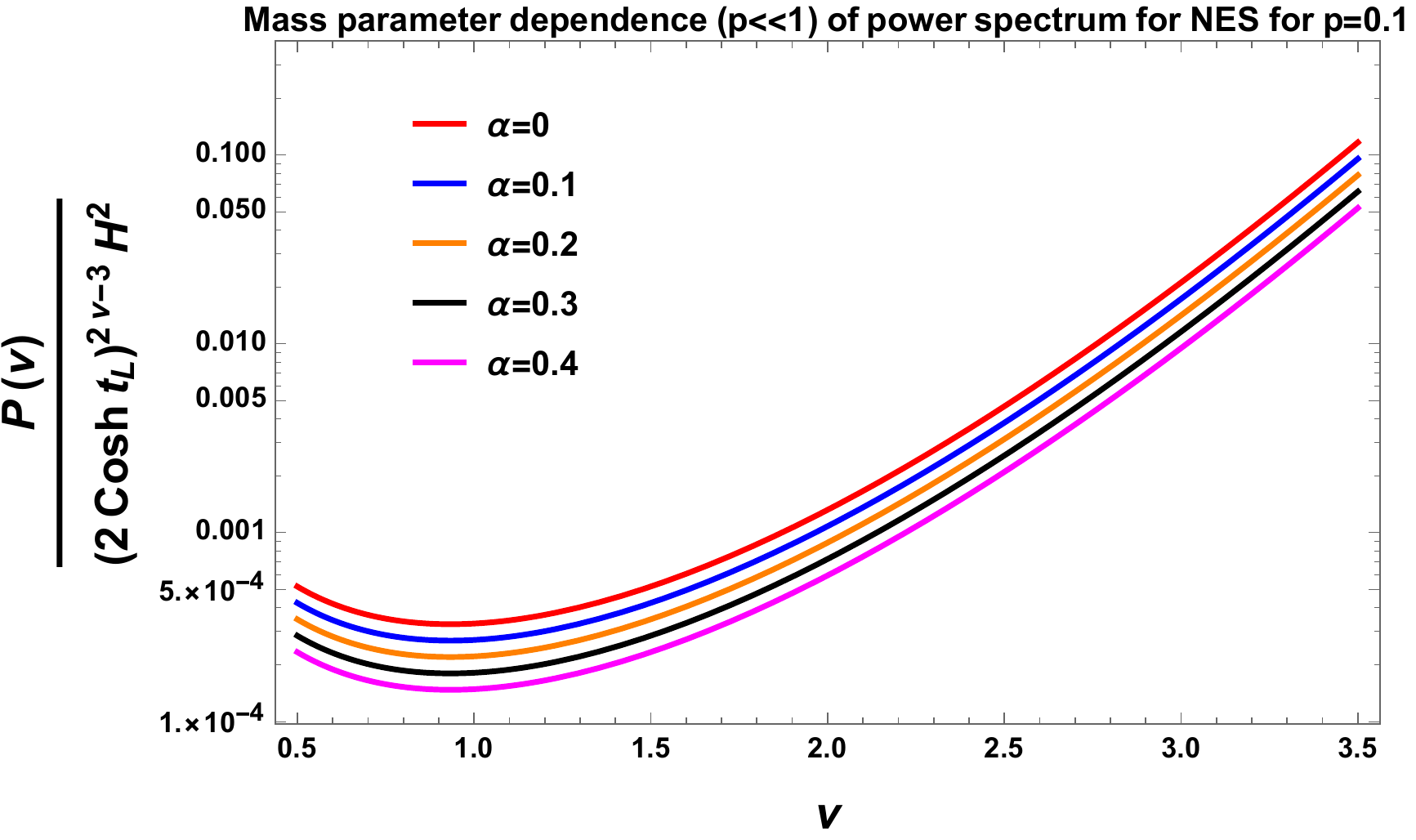}
                    \label{fig6c}   
           }
           \caption[Optional caption for list of figures]{Features of NES power spectrum in small wave number region.} 
            \label{fig1x6}
            \end{figure*}
Moreover, in the superhorizon time scales ($t_{\bf L}>>1$) of region \textcolor{red}{\bf L}, the amplitude of the normalised power spectrum ( for $\alpha$ vacua ) for non entangled state (in the long wave length limit)  can be expressed as:
\bea \label{po2vccxcx1cc} {\cal P}(p<<1,\alpha,t_{\bf L}>>1)&=&\frac{p^3}{2\pi^2}~\left(\cosh t_{\bf L}\right)^{2\nu-3}~\exp\left(-2\alpha\right)H^2\widetilde{{\cal K}(p<<1,\alpha,\nu)}
\nonumber\\&=&\left(2\cosh t_{\bf L}\right)^{2\nu-3}~\left(\frac{H}{2\pi}\right)^2~p^2~\exp\left(-2\alpha\right)~\left(\frac{\Gamma(\nu)}{\Gamma\left(\frac{3}{2}\right)}\right)^2\widetilde{{\cal U}(p<<1)}.~~~~~~~~~~~\eea
Also, for the massless case ($\nu=3/2$), this reduces to:
\bea \label{po3zcvcc2} {\cal P}(p<<1,\alpha,t_{\bf L}>>1)&=&\frac{p^3}{2\pi^2}~\exp\left(-2\alpha\right)H^2\widetilde{{\cal K}(p<<1,\alpha,\nu=3/2)}\nonumber\\&=&\left(\frac{H}{2\pi}\right)^2~p^2~\exp\left(-2\alpha\right)\widetilde{{\cal U}(p<<1)}.~~~~~~~~~~~\eea
The result for Bunch Davies vacuum is obtained by fixing $\alpha=0$ in above equation and is  expressed as:
 \bea \label{po2vvvccxcvbd} {\cal P}_{\bf BD}(p<<1,t_{\bf L}>>1)&=&\frac{p^3}{2\pi^2}~\left(\cosh t_{\bf L}\right)^{2\nu-3}~H^2~\widetilde{{\cal K}(p<<1,\alpha=0,\nu)}
\nonumber\\&=&\left(2\cosh t_{\bf L}\right)^{2\nu-3}~\left(\frac{H}{2\pi}\right)^2~p^2~\left(\frac{\Gamma(\nu)}{\Gamma\left(\frac{3}{2}\right)}\right)^2\widetilde{{\cal U}(p<<1)}~~~~~~~~~~~\eea
  which for the massless case ($\nu=3/2$)  reduces to :
\bea \label{po3zcvcccc} {\cal P}_{\bf BD}(p<<1,t_{\bf L}>>1)&=&\frac{p^3}{2\pi^2}~H^2\widetilde{{\cal K}(p<<1,\alpha=0,\nu=3/2)}\nonumber\\&=&\left(\frac{H}{2\pi}\right)^2~p^2~\widetilde{{\cal U}(p<<1)}.~~~~~~~~~~~\eea 

 In figure~(\ref{fig6a}) and figure~(\ref{fig6b}), we have shown the behaviour of the power spectrum of the mean square vacuum fluctuation in NES formalism in the small wave number regime for $\alpha=0$ and $\alpha=0.1$  with fixed values of the mass parameter $\nu=3/2,2,5/2,3,7/2$ respectively. Note that in both the cases we  find almost  similar behaviour. Also, in figure~(\ref{fig6c}) we have shown the behaviour of the power spectrum with respect to the mass parameter $\nu$ with fixed values of  $\alpha=0,0.1,0.2,0.3,0.4$. In this case we again observe two distinct regions of mass parameter dependence. 
 
 We have  explicitly presented the comparison among FOE, RDM and NES formalism for $\alpha$ vacua in table~(\ref{tabdfd}). The same table is valid for Bunch Davis vacuum when $\alpha = 0$. We have quoted the differences, among the findings from these formalism, for the primordial power spectrum from mean square vacuum fluctuation at large and small scales.

  \section{\textcolor{blue}{Summary}}
\label{x4}

To summarize, in this work, we have addressed the following issues:
\begin{itemize}
	\item  We have explicitly studied the power spectrum of mean squared vacuum fluctuation for axion field using the concept of quantum entanglement in de Sitter space. The effective action for the axion field, used here, has its origin from Type IIB String theory compactified to four dimensions. . For our analysis, we have chosen two initial vacuum states i.e. Bunch Davies and a generalised class of $\alpha$ vacua. The power spectrum of mean squared vacuum fluctuation is computed using three distinctive formalisms: (1) Field operator expansion (FOE), (2) Reduced density matrix (RDM) and (3) Non entangled state (NES). In all three cases, the  computation has been done  starting with two open charts in hyperbolic manifold of de Sitter space  consisting of two regions: \textcolor{red}{\bf L} and \textcolor{red}{\bf R}. Though the starting point is same,  the construction of these three formalisms  are different from each other and have their own physical significance. Each of the formalism has been discussed in text of the papers and some details of approximations for them are presented in the appendix. Similarities and differences from each other are presented in a table.

\item In case of FOE  formalism we  solve for the wave function  in the region \textcolor{red}{\bf L} and using this solution we compute the general expression for the mean square vacuum fluctuation and its quantum correction in terms of two point correlation function.  The result is evaluated at all momentum scales. We considered two limiting approximation in the characteristic momentum scales, i.e. large wave number (small wave length in which the corresponding scale is smaller than the curvature radius of the de Sitter hyperbolic open chart) regime and small wave number (long wave length  in which the corresponding scale is larger than the curvature radius of the de Sitter hyperbolic open chart) regime. We have observed distinctive features in the  power spectrum of of mean squared vacuum fluctuation in these two different regimes. In the large wave number (small wave length) regime we  found that the leading order result for the  power spectrum is  consistent with the known result for observed cosmological correlation function  in the super horizon time scale. The  correction to the leading order result that we computed for the power spectrum can be interpreted as the sub-leading effect in the observed cosmological power spectrum. This is a strong information from the perspective of cosmological observation since such  effects, possibly due to quantum entanglement of states, can play a big role  to break the degeneracy of the observed cosmological power spectrum in the small wave length regime. On the other hand, in the long wave length regime we found that the power spectrum follows completely different momentum dependence in the super horizon time scale. Since in this regime and in this time scale, at present, we lack adequate observational data on power spectrum we are unable to comment on our result with observation.  But  our result for the power spectrum in long wave length limit and super horizon time scale  can be used as a theoretical probe to study the physical implications and its observational cosmological consequences in near future. Our result also implies that the mean square vacuum fluctuation for axion field, in super horizon time scale, gets enhanced in long wave length regime and freezes in the small wave length regime. We also observe  that for a massive axion, the power spectrum is nearly scale invariant in all momentum scales. On the other hand, for massless axion we observe exact scale invariance only in large wave number (small wave length) regime and for the  Bunch Davies initial quantum state. For generalised $\alpha$ initial state, we find slight modification in the corresponding power spectrum of the mean square vacuum fluctuation. The modification factor is proportional to $\exp(-2\alpha)$ which is valid for all values of the parameter $\alpha$. It also implies that for large value of the parameter $\alpha$ we get additional exponential suppression  for the power spectrum. This information  can be used to distinguish between the role of Bunch Davies vacuum ($\alpha=0$) and any  $\alpha$ vacua quantum initial state during analysis of observational data. 

	\item In  RDM formalism, the wave function for the axion field is solved in \textcolor{red}{\bf L} and \textcolor{red}{\bf R} regions  of the de Sitter open chart. This solution has been used to compute the mean square vacuum fluctuation and its quantum correction for both Bunch Davies and $\alpha$ vacuum state. Corresponding results are evaluated at all momentum scales by partially tracing out all the information from the region \textcolor{red}{\bf R}. Like in the case of FOE, we considered the small and large wavelength approximations in the characteristic momentum scales and  found distinct features in the corresponding power spectrum. In the small wave length regime  again the leading order result, in super horizon time scales matched with known result (same as FOE). However, the sub-leading order result for the power spectrum is different from the result obtained from FOE formalism which distinguishes the two approaches.  Moreover,  in the long wave length regime  the power spectrum has  completely different momentum dependence compared to FOE formalism. We also notice that the enhancement of mean square vacuum fluctuation for axion field, in long wave length regime, is different (slower) in nature compared to FOE formalism but the freezing in short wavelength regime is of same nature. The observation on scale invariance of power spectrum in this formalism remains similar to that in FOE formalism.
	
	\item In  the last formalism i.e.NES, the wave function of axion field is solved in the region \textcolor{red}{\bf L} of the de Sitter hyperbolic open chart. With the help of this solution, t we computed  the mean square vacuum fluctuation using Bunch Davies and $\alpha$ vacuum state configuration.  The corresponding result is evaluated at all momentum scales. Like the previous two cases, here also we reverted to two limiting approximations  i.e. large wave number (small wave length ) regime and small wave number (long wave length) regime. We again observed  distinctive behaviour in the power spectrum  in these two different regimes. In the large wave number (small wave length) regime, the leading order result for power spectrum matches with the known result for observed cosmological correlation function just as the cases of FOE and RDM formalism. However, the sub-leading order result s completely different FOE as well as RDM formalism. Thus, it is the sub-leading  terms which distinguish these formalisms from each other and they can be confronted with future observational data. On the other hand, in the small wave number (long wave length) regime, even the leading order result for the power spectrum differs, in momentum dependence, compared to  the result obtained from FOE and RDM formalism. Also the nature of enhancement of the mean square vacuum fluctuation in NES formalism is found to be different from that in FOE and RDM formalism  but the nature of freezing and the observation on scale invariance of power spectrum remains same in all the three cases.

	\item For completeness, we discuss the actual reason for the results obtained for the power spectra from quantum entangled state as appearing in FOE formalism and the mixed state which is used to construct the RDM formalism. To do so, we consider two subsystems, $L$ and $R$ using which one can construct the quantum mechanical state vector of axion field as $|\Psi\rangle_{\textcolor{red}{\bf axion}}$. In our computation, these subsystems are defined in the region \textcolor{red}{\bf L} and \textcolor{red}{\bf R} respectively in the de Sitter hyperbolic open chart.  Now using this state vector of axion field we can define the density matrix as :
	\be \rho_{\textcolor{red}{\bf axion}}=|\Psi_{\textcolor{red}{\bf axion}}\rangle\langle \Psi_{\textcolor{red}{\bf axion}}|,\ee
in both the subsystems, $L$ and $R$ for FOE and RDM formalism and only  the system $L$ for NES formalism. Using this density matrix we can  express the expectation value  (for the total system) of a quantum mechanical operator $\widetilde{\cal O}_{\textcolor{red}{\bf axion}}$,  applicable for FOE and RDM formalism, as:
	\bea\label{simp} {\bf Tr}\left(\rho_{\textcolor{red}{\bf axion}}\widetilde{\cal O}_{\textcolor{red}{\bf axion}}\right)&=&\sum_{\textcolor{red}{\bf L}}\sum_{\textcolor{red}{\bf R}} \langle \textcolor{red}{\bf L},\textcolor{red}{\bf R}|\Psi_{\textcolor{red}{\bf axion}}\rangle\langle \Psi_{\textcolor{red}{\bf axion}}| \widetilde{\cal O}_{\textcolor{red}{\bf axion}}|  \textcolor{red}{\bf L},\textcolor{red}{\bf R}\rangle \nonumber\\&\equiv& \langle \Psi_{\textcolor{red}{\bf axion}}| \widetilde{\cal O}_{\textcolor{red}{\bf axion}}|\Psi_{\textcolor{red}{\bf axion}}\rangle\nonumber\\&\equiv& \langle \widetilde{\cal O}_{\textcolor{red}{\bf axion}} \rangle.~~~~~~~\eea
	This is  an important observation as it is related to the measurement and quantification of any physical cosmological observable in the quantum regime. 
	But in the case of NES formalism one can rewrite Eq~(\ref{simp}) as :
		\bea\label{simps} {\bf Tr}\left(\rho_{\textcolor{red}{\bf axion}}\widetilde{\cal O}_{\textcolor{red}{\bf axion}}\right)&=&\sum_{\textcolor{red}{\bf L}}\sum_{\textcolor{red}{\bf R}} \langle \textcolor{red}{\bf L},\textcolor{red}{\bf R}|\Psi_{\textcolor{red}{\bf axion}}\rangle\langle \Psi_{\textcolor{red}{\bf axion}}| \widetilde{\cal O}_{\textcolor{red}{\bf axion}}| \textcolor{red}{\bf L},\textcolor{red}{\bf R}\rangle \nonumber\\&=& \sum_{\textcolor{red}{\bf L}}\sum_{\textcolor{red}{\bf R}} \sum_{\textcolor{red}{\bf L^{'}}}\sum_{\textcolor{red}{\bf R^{'}}}\langle \textcolor{red}{\bf L},\textcolor{red}{\bf R}|\Psi_{\textcolor{red}{\bf axion}}\rangle\langle \Psi_{\textcolor{red}{\bf axion}}| \textcolor{red}{\bf L^{'}},\textcolor{red}{\bf R^{'}}\rangle\langle \textcolor{red}{\bf L^{'}},\textcolor{red}{\bf R^{'}}|\widetilde{\cal O}^{\textcolor{red}{\bf L}}_{\textcolor{red}{\bf axion}}| \textcolor{red}{\bf L},\textcolor{red}{\bf R}\rangle \nonumber\\&=& \sum_{\textcolor{red}{\bf L}}\sum_{\textcolor{red}{\bf R}} \sum_{\textcolor{red}{\bf L^{'}}}\sum_{\textcolor{red}{\bf R^{'}}}\langle \textcolor{red}{\bf L},\textcolor{red}{\bf R}|\Psi_{\textcolor{red}{\bf axion}}\rangle\langle \Psi_{\textcolor{red}{\bf axion}}| \textcolor{red}{\bf L^{'}},\textcolor{red}{\bf R^{'}}\rangle\langle \textcolor{red}{\bf L^{'}}|\widetilde{\cal O}^{\textcolor{red}{\bf L}}_{\textcolor{red}{\bf axion}}| \textcolor{red}{\bf L}\rangle\delta_{\textcolor{red}{\bf R}\textcolor{red}{\bf R^{'}}}\nonumber\\
		\nonumber\\&=& \sum_{\textcolor{red}{\bf L}}\sum_{\textcolor{red}{\bf R}} \sum_{\textcolor{red}{\bf L^{'}}}\langle \textcolor{red}{\bf L},\textcolor{red}{\bf R}|\Psi_{\textcolor{red}{\bf axion}}\rangle\langle \Psi_{\textcolor{red}{\bf axion}}| \textcolor{red}{\bf L^{'}},\textcolor{red}{\bf R^{'}}\rangle\langle \textcolor{red}{\bf L^{'}}|\widetilde{\cal O}^{\textcolor{red}{\bf L}}_{\textcolor{red}{\bf axion}}| \textcolor{red}{\bf L}\rangle\nonumber\\
		&=&{\bf Tr}\left(\rho^{\textcolor{red}{\bf L}}_{\textcolor{red}{\bf axion}}\widetilde{\cal O}^{\textcolor{red}{\bf L}}_{\textcolor{red}{\bf axion}}\right),~~~~~~~\eea
		where the operator $\widetilde{\cal O}^{\textcolor{red}{\bf L}}_{\textcolor{red}{\bf axion}}$ solely in the region \textcolor{red}{\bf L} is defined by the following expression for NES formalism:
		\bea \langle \textcolor{red}{\bf L^{'}},\textcolor{red}{\bf R^{'}}|\widetilde{\cal O}^{\textcolor{red}{\bf L}}_{\textcolor{red}{\bf axion}}| \textcolor{red}{\bf L},\textcolor{red}{\bf R}\rangle &=& \langle \textcolor{red}{\bf L^{'}}|\widetilde{\cal O}^{\textcolor{red}{\bf L}}_{\textcolor{red}{\bf axion}}| \textcolor{red}{\bf L}\rangle\langle \textcolor{red}{\bf R^{'}}|\textcolor{red}{\bf R}\rangle \nonumber\\
		&=& \langle \textcolor{red}{\bf L^{'}}|\widetilde{\cal O}^{\textcolor{red}{\bf L}}_{\textcolor{red}{\bf axion}}| \textcolor{red}{\bf L}\rangle \delta_{\textcolor{red}{\bf R}\textcolor{red}{\bf R^{'}}}.\eea
		Also in NES formalism the density matrix $\rho^{\textcolor{red}{\bf L}}_{\textcolor{red}{\bf axion}}$ for the region \textcolor{red}{\bf L} is described by the following expression:
	\bea \rho^{\textcolor{red}{\bf L}}_{\textcolor{red}{\bf axion}}&=& {\bf Tr}_{\textcolor{red}{\bf R}}\rho_{\textcolor{red}{\bf axion}}\nonumber\\
	&=& \sum_{\textcolor{red}{\bf L}} \sum_{\textcolor{red}{\bf L^{'}}}|\textcolor{red}{\bf L}\rangle \left(\sum_{\textcolor{red}{\bf R}}\langle \textcolor{red}{\bf L},\textcolor{red}{\bf R}|\Psi_{\textcolor{red}{\bf axion}}\rangle\langle \Psi_{\textcolor{red}{\bf axion}}|\textcolor{red}{\bf L^{'}},\textcolor{red}{\bf R^{'}}\rangle\right)\langle \textcolor{red}{\bf L^{'}}|\nonumber\\
	&=& \sum_{\textcolor{red}{\bf L}} \sum_{\textcolor{red}{\bf L^{'}}}|\textcolor{red}{\bf L}\rangle \left(\sum_{\textcolor{red}{\bf R}} \Psi_{\textcolor{red}{\bf axion}}(\textcolor{red}{\bf L},\textcolor{red}{\bf R}) \Psi^{*}_{\textcolor{red}{\bf axion}}(\textcolor{red}{\bf L^{'}},\textcolor{red}{\bf R^{'}})\right)\langle \textcolor{red}{\bf L^{'}}|.\eea
	This implies that in NES formalism, the physical operator is solely described by the information from the region \textcolor{red}{\bf L} and consequently the expectation value of such operator satisfy the following condition:
\bea\label{simpqz} \langle \widetilde{O}_{\textcolor{red}{\bf axion}}\rangle={\bf Tr}\left(\rho_{\textcolor{red}{\bf axion}}\widetilde{\cal O}_{\textcolor{red}{\bf axion}}\right)&=&
		{\bf Tr}\left(\rho^{\textcolor{red}{\bf L}}_{\textcolor{red}{\bf axion}}\widetilde{\cal O}^{\textcolor{red}{\bf L}}_{\textcolor{red}{\bf axion}}\right)=\langle \widetilde{O}^{\textcolor{red}{\bf L}}_{\textcolor{red}{\bf axion}}\rangle.\eea
	 The above analysis can help us to explain the differences between the power spectra of mean square vacuum fluctuation obtained from FOE, RDM and NES formalism on large scale (or small wave number or large wave length regime). It clearly points towards the fact that in FOE and RDM formalism the creation and annihilation operators for axion field includes new set of creation and annihilation operators coming from the Bogoliubov transformation from one quantum basis to the other. This means that the field operator in the FOE formalism also involves these extra creation and annihilation operators even if the computation is being performed on a particularly specified temporal slice defined in the region \textcolor{red}{\bf L} of the Hilbert space. On the other hand, after applying the partial trace over the degrees of freedom from the region \textcolor{red}{\bf R}, the mixed quantum state, using which we formulate the RDM formalism, is prepared by the creation and annihilation operators in the region \textcolor{red}{\bf L} of the Hilbert space.  Thus, in RDM formalism, the field operator is only defined in the region \textcolor{red}{\bf L} and not in the region \textcolor{red}{\bf R} of the Hilbert space. This implies  that the field operator defined before partially tracing over the  degrees of freedom from region \textcolor{red}{\bf R} for FOE formalism is different from the field operator in region \textcolor{red}{\bf L} used in RDM formalism since for this case we have performed the partial trace over the degrees of freedom in region \textcolor{red}{\bf R}. Thus, any general quantum mechanical operator defined in the framework of FOE is not same as that of RDM formalism. 
		
Before we conclude, we point out that apart from the quantification of the mean square vacuum fluctuation in the formalisms we discussed here, we have also computed the entanglement entropy  using von Neumann measure and the Renyi entropy in our previous work \cite{Choudhury:2017bou, Choudhury:2017qyl}.

	\section*{\textcolor{blue}{Acknowledgments}}
	SC would like to thank Quantum Gravity and Unified Theory and Theoretical Cosmology Group, Max Planck Institute for Gravitational Physics, Albert Einstein Institute for providing the Post-Doctoral
	Research Fellowship. SC take this opportunity to thank sincerely to  Jean-Luc Lehners, Shiraz Minwalla and Varun Sahni for their constant support
	and inspiration. SC also thank the organisers of Advanced String School 2017, ST$^{4}$ 2017 and Kavli Asian Winter School on Strings, Particles and Cosmology 2018, Summer School on Cosmology 2018, ICTP, Trieste, 15 th Marcel Grossman  Meeting, Rome
	for providing the local hospitality during the work. SC also thank ICTP, Trieste, La Sapienza University, Rome, DTP, TIFR, Mumbai, ICTS, TIFR, Bengaluru, IOP, CMI, SINP and 
IACS for
	providing the academic visit during the work. SP acknowledges the J. C. Bose National Fellowship for support of his research. Last but not the least, we would like to acknowledge our
	debt to the people of India for their generous and steady support for research in natural sciences, especially
	for theoretical high energy physics, string theory and cosmology.

\appendix

\vspace{0.2cm}
\section{\textcolor{blue}{Quantum correction to the power spectrum  in FOE formalism}}
At  the superhorizon time scales ($t_{\bf L}>>1$) of region \textcolor{red}{\bf L} one can write the amplitude of the FOE power spectrum as:
\bea \label{df2} \sum_{\sigma=\pm 1}|\widetilde{\chi^{\sigma}}|^2=\sum_{\sigma=\pm 1}\left(\widetilde{\chi^{\sigma}}\right)^{\dagger}\widetilde{\chi^{\sigma}}&~\underrightarrow{t_{\bf L}>>1}~& \widetilde{{\cal M}(p,\nu)}\left(\cosh t_{\bf L}\right)^{2\nu-1}\eea
where the time independent function $\widetilde{{\cal M}(p,\nu)}$ is defined as:
\bea\widetilde{{\cal M}(p,\nu)}&=&\frac{2^{2\nu-1}\left(\Gamma(\nu)\right)^2}{\pi}\times \sum_{\sigma=\pm 1}\left[\frac{\left(|{\cal A}^{\sigma}_{\bf L}|^2+|{\cal B}^{\sigma}_{\bf L}|^2\right)}{\left|\Gamma\left(\nu+ ip +\frac{1}{2}\right)\right|^2}+\frac{{\cal A}^{\sigma}_{\bf L}{\cal B}^{\sigma *}_{\bf L}}{\left(\Gamma\left(\nu- ip +\frac{1}{2}\right)\right)^2}+\frac{{\cal A}^{\sigma *}_{\bf L}{\cal B}^{\sigma}_{\bf L}}{\left(\Gamma\left(\nu+ ip +\frac{1}{2}\right)\right)^2}\right.\nonumber\\ && \left.
~+\sum^{\infty}_{n=0}\left\{\frac{\left({\cal A}^{\sigma}_{{\bf L}}{\cal A}^{\sigma*}_{{\bf L},(n)}+{\cal B}^{\sigma}_{{\bf L}}{\cal B}^{\sigma*}_{{\bf L},(n)}\right)}{\Gamma\left(\nu- ip +\frac{1}{2}\right)\Gamma\left(\nu+ ip_n +\frac{1}{2}\right)}+\frac{\left({\cal A}^{\sigma}_{{\bf L}}{\cal B}^{\sigma*}_{{\bf L},(n)}+{\cal A}^{\sigma}_{{\bf L},(n)}{\cal B}^{\sigma*}_{{\bf L}}\right)}{\Gamma\left(\nu- ip +\frac{1}{2}\right)\Gamma\left(\nu- ip_n +\frac{1}{2}\right)}\right.\right.\nonumber\\&&\left.\left.~~~~~~~~~~~~~~~~~~~~~~~~~~~~~~~~~~~~~~~~~~~~~~~~+\frac{\left({\cal A}^{\sigma*}_{{\bf L},(n)}{\cal B}^{\sigma}_{{\bf L}}+{\cal A}^{\sigma*}_{{\bf L}}{\cal B}^{\sigma}_{{\bf L},(n)}\right)}{\Gamma\left(\nu+ ip_n +\frac{1}{2}\right)\Gamma\left(\nu+ ip +\frac{1}{2}\right)}\right\}\right.\nonumber\\ && \left.
~+\sum^{\infty}_{n=0}\sum^{\infty}_{m=0}\left\{\frac{\left({\cal A}^{\sigma}_{{\bf L},(n)}{\cal A}^{\sigma*}_{{\bf L},(m)}+{\cal B}^{\sigma}_{{\bf L},(n)}{\cal B}^{\sigma*}_{{\bf L},(m)}\right)}{\Gamma\left(\nu- ip_n +\frac{1}{2}\right)\Gamma\left(\nu+ ip_m +\frac{1}{2}\right)}\right.\right.\nonumber\\&&\left.\left.~~~~~~~~~~~~~~~~~+\frac{{\cal A}^{\sigma}_{{\bf L},(n)}{\cal B}^{\sigma*}_{{\bf L},(m)}}{\Gamma\left(\nu- ip_n +\frac{1}{2}\right)\Gamma\left(\nu- ip_m +\frac{1}{2}\right)}+\frac{{\cal A}^{\sigma*}_{{\bf L},(n)}{\cal B}^{\sigma}_{{\bf L},(m)}}{\Gamma\left(\nu+ ip_n +\frac{1}{2}\right)\Gamma\left(\nu+ ip_m +\frac{1}{2}\right)}\right\}\right].~~~~~~\eea
\subsection{For large wave number}

Further to know the exact wave number dependence of the amplitude of the normalized power spectrum from Bunch Davies vacuum we need to know the behaviour of the power spectrum at very short wavelengths ($p,p_n>>1$). After taking this limit it is expected that the power spectrum of axion match with the result obtained for spatially flat universe. In general for an arbitrary value of the mass parameter $\nu$, we get the following approximated contributions in the short wavelength limit ($p,p_n>>1$), which are explicitly appearing in the expression for the amplitude of the normalized power spectrum from Bunch Davies vacuum:
\bea \sum_{\sigma=\pm 1}\frac{|{\cal A}^{\sigma}_{\bf L}|^2}{\left|\Gamma\left(\nu+ ip +\frac{1}{2}\right)\right|^2}
&\stackrel{p>>1}{\approx}&\frac{\pi e^{-\pi p}}{2p^4\left|\Gamma\left(ip\right)\right|^2},\\
\sum_{\sigma=\pm 1}\frac{|{\cal B}^{\sigma}_{\bf L}|^2}{\left|\Gamma\left(\nu+ ip +\frac{1}{2}\right)\right|^2}
&\stackrel{p>>1}{\approx}&\frac{\pi e^{-5\pi p}}{2p^4\left|\Gamma\left(ip\right)\right|^2},\\
\sum_{\sigma=\pm 1}\frac{{\cal A}^{\sigma}_{\bf L}{\cal B}^{\sigma *}_{\bf L}}{\left(\Gamma\left(\nu- ip +\frac{1}{2}\right)\right)^2}
&\stackrel{p>>1}{\approx}&\frac{\pi e^{-3\pi p}}{2p^4\left|\Gamma\left(ip\right)\right|^2},\\
\sum_{\sigma=\pm 1}\frac{{\cal A}^{\sigma*}_{\bf L}{\cal B}^{\sigma}_{\bf L}}{\left(\Gamma\left(\nu+ ip +\frac{1}{2}\right)\right)^2}
&\stackrel{p>>1}{\approx}&\frac{\pi e^{-3\pi p}}{2p^4\left|\Gamma\left(ip\right)\right|^2}\\
\sum_{\sigma=\pm 1}\frac{{\cal A}^{\sigma}_{\bf L}{\cal A}^{\sigma*}_{{\bf L},(n)}}{\Gamma\left(\nu- ip +\frac{1}{2}\right)\Gamma\left(\nu+ ip_n +\frac{1}{2}\right)}
&\stackrel{p,p_n>>1}{\approx}&\frac{\pi e^{-\pi (p+p_n)/2}}{2p^2p^2_n\left|\Gamma\left(ip\right)\right|\left|\Gamma\left(ip_n\right)\right|}\\
\sum_{\sigma=\pm 1}\frac{{\cal B}^{\sigma}_{\bf L}{\cal B}^{\sigma*}_{{\bf L},(n)}}{\Gamma\left(\nu- ip +\frac{1}{2}\right)\Gamma\left(\nu+ ip_n +\frac{1}{2}\right)}
&\stackrel{p,p_n>>1}{\approx}&\frac{\pi e^{-5\pi (p+p_n)/2}}{2p^2p^2_n\left|\Gamma\left(ip\right)\right|\left|\Gamma\left(ip_n\right)\right|}\\
\sum_{\sigma=\pm 1}\frac{{\cal A}^{\sigma}_{\bf L}{\cal B}^{\sigma*}_{{\bf L},(n)}}{\Gamma\left(\nu- ip +\frac{1}{2}\right)\Gamma\left(\nu- ip_n +\frac{1}{2}\right)}
&\stackrel{p,p_n>>1}{\approx}&\frac{\pi e^{-\pi (p+5p_n)/2}}{2p^2p^2_n \left|\Gamma\left(ip\right)\right|\left|\Gamma\left(ip_n\right)\right|}\\
\sum_{\sigma=\pm 1}\frac{{\cal A}^{\sigma}_{{\bf L},(n)}{\cal B}^{\sigma*}_{{\bf L}}}{\Gamma\left(\nu- ip +\frac{1}{2}\right)\Gamma\left(\nu- ip_n +\frac{1}{2}\right)}
&\stackrel{p,p_n>>1}{\approx}&\frac{\pi e^{-\pi (5p+p_n)/2}}{2p^2p^2_n\left|\Gamma\left(ip\right)\right|\left|\Gamma\left(ip_n\right)\right|}\eea\bea
\sum_{\sigma=\pm 1}\frac{{\cal A}^{\sigma*}_{{\bf L},(n)}{\cal B}^{\sigma}_{\bf L}}{\Gamma\left(\nu+ ip +\frac{1}{2}\right)\Gamma\left(\nu+ ip_n +\frac{1}{2}\right)}
&\stackrel{p,p_n>>1}{\approx}&\frac{\pi e^{-\pi (5p+p_n)/2}}{2p^2p^2_n\left|\Gamma\left(ip\right)\right|\left|\Gamma\left(ip_n\right)\right|}\\
\sum_{\sigma=\pm 1}\frac{{\cal A}^{\sigma*}_{{\bf L}}{\cal B}^{\sigma}_{{\bf L},(n)}}{\Gamma\left(\nu+ ip +\frac{1}{2}\right)\Gamma\left(\nu+ ip_n +\frac{1}{2}\right)}
&\stackrel{p,p_n>>1}{\approx}&\frac{\pi e^{-\pi (p+5p_n)/2}}{2p^2p^2_n\left|\Gamma\left(ip\right)\right|\left|\Gamma\left(ip_n\right)\right|}\\
\sum_{\sigma=\pm 1}\frac{{\cal A}^{\sigma}_{{\bf L},(n)}{\cal A}^{\sigma*}_{{\bf L},(m)}}{\Gamma\left(\nu- ip_n +\frac{1}{2}\right)\Gamma\left(\nu+ ip_m +\frac{1}{2}\right)}
&\stackrel{p_n, p_m>>1}{\approx}&\frac{\pi e^{-\pi(p_n+p_m)/2}}{2p^2_n p^2_m\left|\Gamma\left(ip_m\right)\right|\left|\Gamma\left(ip_n\right)\right|}\\
\sum_{\sigma=\pm 1}\frac{{\cal B}^{\sigma}_{{\bf L},(n)}{\cal B}^{\sigma*}_{{\bf L},(m)}}{\Gamma\left(\nu- ip_n +\frac{1}{2}\right)\Gamma\left(\nu+ ip_m +\frac{1}{2}\right)}
&\stackrel{p_n, p_m>>1}{\approx}&\frac{\pi e^{-5\pi(p_n+p_m)/2}}{2p^2_n p^2_m\left|\Gamma\left(ip_m\right)\right|\left|\Gamma\left(ip_n\right)\right|}\\
\sum_{\sigma=\pm 1}\frac{{\cal A}^{\sigma}_{{\bf L},(n)}{\cal B}^{\sigma*}_{{\bf L},(m)}}{\Gamma\left(\nu- ip_n +\frac{1}{2}\right)\Gamma\left(\nu- ip_m +\frac{1}{2}\right)}
&\stackrel{p_n, p_m>>1}{\approx}&\frac{\pi e^{-3\pi(p_n+p_m)/2}}{2p^2_n p^2_m\left|\Gamma\left(ip_m\right)\right|\left|\Gamma\left(ip_n\right)\right|}\\
\sum_{\sigma=\pm 1}\frac{{\cal A}^{\sigma*}_{{\bf L},(n)}{\cal B}^{\sigma}_{{\bf L},(m)}}{\Gamma\left(\nu+ ip_n +\frac{1}{2}\right)\Gamma\left(\nu+ ip_m +\frac{1}{2}\right)}
&\stackrel{p_n, p_m>>1}{\approx}&\frac{\pi e^{-3\pi(p_n+p_m)/2}}{2p^2_n p^2_m\left|\Gamma\left(ip_m\right)\right|\left|\Gamma\left(ip_n\right)\right|}\eea
Further, we apply Stirling's formula to approximate Gamma functions for large wavenumbers $p,p_n>>1$ to simplify the expression for the power spectrum:
\bea \Gamma(ip)&\sim & \sqrt{2\pi}~(ip)^{ip-\frac{1}{2}}e^{-ip}\left(1+\frac{1}{12ip}-\frac{1}{288p^2}+\cdots\right),\\
\Gamma(ip_n)&\sim & \sqrt{2\pi}~(ip_n)^{ip_n-\frac{1}{2}}e^{-ip_n}\left(1+\frac{1}{12ip_n}-\frac{1}{288p^2_n}+\cdots\right).\eea
Consequently, we get the following simplified expressions in large wavenumber ($p,p_n>>1$) limit:
\bea \sum_{\sigma=\pm 1}\frac{|{\cal A}^{\sigma}_{\bf L}|^2}{\left|\Gamma\left(\nu+ ip +\frac{1}{2}\right)\right|^2}
&\sim&\frac{1}{2p^3\left(1+\frac{1}{82944p^4}\right)},\\
\sum_{\sigma=\pm 1}\frac{|{\cal B}^{\sigma}_{\bf L}|^2}{\left|\Gamma\left(\nu+ ip +\frac{1}{2}\right)\right|^2}
&\sim &\frac{e^{-4\pi p}}{2p^3\left(1+\frac{1}{82944p^4}\right)},\\
\sum_{\sigma=\pm 1}\frac{{\cal A}^{\sigma}_{\bf L}{\cal B}^{\sigma *}_{\bf L}}{\left(\Gamma\left(\nu- ip +\frac{1}{2}\right)\right)^2}
&\sim &\frac{e^{-2\pi p}}{2p^3\left(1+\frac{1}{82944p^4}\right)},\\
\sum_{\sigma=\pm 1}\frac{{\cal A}^{\sigma*}_{\bf L}{\cal B}^{\sigma}_{\bf L}}{\left(\Gamma\left(\nu+ ip +\frac{1}{2}\right)\right)^2}
&\sim&\frac{e^{-2\pi p}}{2p^3\left(1+\frac{1}{82944p^4}\right)}\\ 
\sum_{\sigma=\pm 1}\frac{{\cal A}^{\sigma}_{\bf L}{\cal A}^{\sigma*}_{{\bf L},(n)}}{\Gamma\left(\nu- ip +\frac{1}{2}\right)\Gamma\left(\nu+ ip_n +\frac{1}{2}\right)}
&\sim &\frac{1}{2p^{3/2} p^{3/2}_n \sqrt{1+\frac{1}{82944p^4}}\sqrt{1+\frac{1}{82944p^4_n}}}\\
\sum_{\sigma=\pm 1}\frac{{\cal B}^{\sigma}_{\bf L}{\cal B}^{\sigma*}_{{\bf L},(n)}}{\Gamma\left(\nu- ip +\frac{1}{2}\right)\Gamma\left(\nu+ ip_n +\frac{1}{2}\right)}
&\sim&\frac{e^{-2\pi (p+p_n)}}{2p^{3/2} p^{3/2}_n\sqrt{1+\frac{1}{82944p^4}}\sqrt{1+\frac{1}{82944p^4_n}}}\\
\sum_{\sigma=\pm 1}\frac{{\cal A}^{\sigma}_{\bf L}{\cal B}^{\sigma*}_{{\bf L},(n)}}{\Gamma\left(\nu- ip +\frac{1}{2}\right)\Gamma\left(\nu- ip_n +\frac{1}{2}\right)}
&\sim&\frac{e^{-2\pi p_n}}{2p^{3/2} p^{3/2}_n\sqrt{1+\frac{1}{82944p^4}}\sqrt{1+\frac{1}{82944p^4_n}}}\eea\bea
\sum_{\sigma=\pm 1}\frac{{\cal A}^{\sigma}_{{\bf L},(n)}{\cal B}^{\sigma*}_{{\bf L}}}{\Gamma\left(\nu- ip +\frac{1}{2}\right)\Gamma\left(\nu- ip_n +\frac{1}{2}\right)}
&\sim&\frac{e^{-2\pi p}}{2p^{3/2} p^{3/2}_n\sqrt{1+\frac{1}{82944p^4}}\sqrt{1+\frac{1}{82944p^4_n}}}\\
\sum_{\sigma=\pm 1}\frac{{\cal A}^{\sigma*}_{{\bf L},(n)}{\cal B}^{\sigma}_{\bf L}}{\Gamma\left(\nu+ ip +\frac{1}{2}\right)\Gamma\left(\nu+ ip_n +\frac{1}{2}\right)}
&\sim&\frac{e^{-2\pi p}}{2p^{3/2} p^{3/2}_n\sqrt{1+\frac{1}{82944p^4}}\sqrt{1+\frac{1}{82944p^4_n}}}\\
\sum_{\sigma=\pm 1}\frac{{\cal A}^{\sigma*}_{{\bf L}}{\cal B}^{\sigma}_{{\bf L},(n)}}{\Gamma\left(\nu+ ip +\frac{1}{2}\right)\Gamma\left(\nu+ ip_n +\frac{1}{2}\right)}
&\sim &\frac{e^{-2\pi p_n}}{2p^{3/2} p^{3/2}_n\sqrt{1+\frac{1}{82944p^4}}\sqrt{1+\frac{1}{82944p^4_n}}}\\
\sum_{\sigma=\pm 1}\frac{{\cal A}^{\sigma}_{{\bf L},(n)}{\cal A}^{\sigma*}_{{\bf L},(m)}}{\Gamma\left(\nu- ip_n +\frac{1}{2}\right)\Gamma\left(\nu+ ip_m +\frac{1}{2}\right)}
&\sim &\frac{1}{2p^{3/2}_m p^{3/2}_n\sqrt{1+\frac{1}{82944p^4_m}}\sqrt{1+\frac{1}{82944p^4_n}}}\\
\sum_{\sigma=\pm 1}\frac{{\cal B}^{\sigma}_{{\bf L},(n)}{\cal B}^{\sigma*}_{{\bf L},(m)}}{\Gamma\left(\nu- ip_n +\frac{1}{2}\right)\Gamma\left(\nu+ ip_m +\frac{1}{2}\right)}
&\sim &\frac{e^{-2\pi(p_n+p_m)}}{2p^{3/2}_m p^{3/2}_n\sqrt{1+\frac{1}{82944p^4_m}}\sqrt{1+\frac{1}{82944p^4_n}}}\\
\sum_{\sigma=\pm 1}\frac{{\cal A}^{\sigma}_{{\bf L},(n)}{\cal B}^{\sigma*}_{{\bf L},(m)}}{\Gamma\left(\nu- ip_n +\frac{1}{2}\right)\Gamma\left(\nu- ip_m +\frac{1}{2}\right)}
&\sim &\frac{e^{-\pi(p_n+p_m)}}{2p^{3/2}_m p^{3/2}_n\sqrt{1+\frac{1}{82944p^4_m}}\sqrt{1+\frac{1}{82944p^4_n}}}\\
\sum_{\sigma=\pm 1}\frac{{\cal A}^{\sigma*}_{{\bf L},(n)}{\cal B}^{\sigma}_{{\bf L},(m)}}{\Gamma\left(\nu+ ip_n +\frac{1}{2}\right)\Gamma\left(\nu+ ip_m +\frac{1}{2}\right)}
&\sim &\frac{e^{-\pi(p_n+p_m)}}{2p^{3/2}_m p^{3/2}_n\sqrt{1+\frac{1}{82944p^4_m}}\sqrt{1+\frac{1}{82944p^4_n}}}\eea
As a result, in the short wave length approximation the time independent function $\widetilde{{\cal M}(p>>1,\nu)}$ for any arbitrary mass parameter $\nu$ can be expressed as:
\bea\widetilde{{\cal M}(p>>1,\nu)}&=&\frac{2^{2(\nu-1)}\left(\Gamma(\nu)\right)^2}{p^3\pi}\widetilde{{\cal G}(p>>1)},~~~~~~~~~~~\eea
where we define a new function $\widetilde{{\cal G}(p>>1)}$ in the short wave length limit as given by:
\bea\widetilde{{\cal G}(p)}&=&\frac{1}{\left(1+\frac{1}{82944p^4}\right)}\times \nonumber\\
&&\left[\left(1+e^{-2\pi p}\right)^2+\sum^{\infty}_{n=0}\left(\frac{p}{p_n}\right)^{\frac{3}{2}}\frac{\sqrt{1+\frac{1}{82944p^4}}}{\sqrt{1+\frac{1}{82944p^4_n}}}\left(1+2\left(e^{-2\pi p}+e^{-2\pi p_n}\right)+e^{-2\pi (p+p_n)}\right)\right.\nonumber\\ && \left.
~+\sum^{\infty}_{n=0}\sum^{\infty}_{m=0}\frac{p^3}{\left(p_n p_m\right)^{3/2}}\frac{\left(1+\frac{1}{82944p^4}\right)}{\sqrt{1+\frac{1}{82944p^4_n}}\sqrt{1+\frac{1}{82944p^4_m}}}\left(1+e^{-\pi (p_m+p_n)}\right)^2\right],~~~~~~~~~~~\eea
\subsection{For small wave number}
Similarly to know the exact wavenumber dependence of the amplitude of the normalised power spectrum from Bunch Davies vacuum in the long wavelength limit we need to know the behaviour of the power spectrum for $p,p_n<<1$. In this limit it is expected that the power spectrum of axion should match with the result obtained for spatially flat universe. In general for an arbitrary value of the mass parameter $\nu$, we get the following approximated contributions in the long wavelength limit ($p,p_n<<1$), which are explicitly appearing in the expression for the amplitude of the normalised power spectrum from Bunch Davies vacuum:
\bea \sum_{\sigma=\pm 1}\frac{|{\cal A}^{\sigma}_{\bf L}|^2}{\left|\Gamma\left(\nu+ ip +\frac{1}{2}\right)\right|^2}
&\stackrel{p<<1}{\approx}&\frac{\pi}{4\left|\Gamma\left(\nu+\frac{1}{2}\right)\right|^2},\\
\sum_{\sigma=\pm 1}\frac{|{\cal B}^{\sigma}_{\bf L}|^2}{\left|\Gamma\left(\nu+ ip +\frac{1}{2}\right)\right|^2}
&\stackrel{p<<1}{\approx}&\frac{\pi}{4\left|\Gamma\left(\nu+\frac{1}{2}\right)\right|^2},\\
\sum_{\sigma=\pm 1}\frac{{\cal A}^{\sigma}_{\bf L}{\cal B}^{\sigma *}_{\bf L}}{\left(\Gamma\left(\nu- ip +\frac{1}{2}\right)\right)^2}
&\stackrel{p<<1}{\approx}&\frac{\pi}{4\left(\Gamma\left(\nu +\frac{1}{2}\right)\right)^2},\\
\sum_{\sigma=\pm 1}\frac{{\cal A}^{\sigma*}_{\bf L}{\cal B}^{\sigma}_{\bf L}}{\left(\Gamma\left(\nu+ ip +\frac{1}{2}\right)\right)^2}
&\stackrel{p<<1}{\approx}&\frac{\pi}{4\left(\Gamma\left(\nu +\frac{1}{2}\right)\right)^2}\eea\bea
\sum_{\sigma=\pm 1}\frac{{\cal A}^{\sigma}_{\bf L}{\cal A}^{\sigma*}_{{\bf L},(n)}}{\Gamma\left(\nu- ip +\frac{1}{2}\right)\Gamma\left(\nu+ ip_n +\frac{1}{2}\right)}
&\stackrel{p,p_n<<1}{\approx}&\frac{\pi e^{-\pi(p+p_n)}}{2\left(\Gamma\left(\nu +\frac{1}{2}\right)\right)^2}\\
\sum_{\sigma=\pm 1}\frac{{\cal B}^{\sigma}_{\bf L}{\cal B}^{\sigma*}_{{\bf L},(n)}}{\Gamma\left(\nu- ip +\frac{1}{2}\right)\Gamma\left(\nu+ ip_n +\frac{1}{2}\right)}
&\stackrel{p,p_n<<1}{\approx}&\frac{\pi e^{-\pi(p+p_n)}}{2\left(\Gamma\left(\nu +\frac{1}{2}\right)\right)^2}~~~~~\\
\sum_{\sigma=\pm 1}\frac{{\cal A}^{\sigma}_{\bf L}{\cal B}^{\sigma*}_{{\bf L},(n)}}{\Gamma\left(\nu- ip +\frac{1}{2}\right)\Gamma\left(\nu- ip_n +\frac{1}{2}\right)}
&\stackrel{p,p_n<<1}{\approx}&\frac{\pi e^{-\pi(p+p_n)}}{2\left(\Gamma\left(\nu +\frac{1}{2}\right)\right)^2}\\
\sum_{\sigma=\pm 1}\frac{{\cal A}^{\sigma}_{{\bf L},(n)}{\cal B}^{\sigma*}_{{\bf L}}}{\Gamma\left(\nu- ip +\frac{1}{2}\right)\Gamma\left(\nu- ip_n +\frac{1}{2}\right)}
&\stackrel{p,p_n<<1}{\approx}&\frac{\pi e^{-\pi(p+p_n)}}{2\left(\Gamma\left(\nu +\frac{1}{2}\right)\right)^2}\\
\sum_{\sigma=\pm 1}\frac{{\cal A}^{\sigma*}_{{\bf L},(n)}{\cal B}^{\sigma}_{\bf L}}{\Gamma\left(\nu+ ip +\frac{1}{2}\right)\Gamma\left(\nu+ ip_n +\frac{1}{2}\right)}
&\stackrel{p,p_n<<1}{\approx}&\frac{\pi e^{-\pi(p+p_n)}}{2\left(\Gamma\left(\nu +\frac{1}{2}\right)\right)^2}\\
\sum_{\sigma=\pm 1}\frac{{\cal A}^{\sigma*}_{{\bf L}}{\cal B}^{\sigma}_{{\bf L},(n)}}{\Gamma\left(\nu+ ip +\frac{1}{2}\right)\Gamma\left(\nu+ ip_n +\frac{1}{2}\right)}
&\stackrel{p,p_n<<1}{\approx}&\frac{\pi e^{-\pi(p+p_n)}}{2\left(\Gamma\left(\nu +\frac{1}{2}\right)\right)^2}\\
\sum_{\sigma=\pm 1}\frac{{\cal A}^{\sigma}_{{\bf L},(n)}{\cal A}^{\sigma*}_{{\bf L},(m)}}{\Gamma\left(\nu- ip_n +\frac{1}{2}\right)\Gamma\left(\nu+ ip_m +\frac{1}{2}\right)}
&\stackrel{p_n, p_m<<1}{\approx}&\frac{\pi e^{-\pi(p_n+p_m)}}{2\left(\Gamma\left(\nu +\frac{1}{2}\right)\right)^2}\\
\sum_{\sigma=\pm 1}\frac{{\cal B}^{\sigma}_{{\bf L},(n)}{\cal B}^{\sigma*}_{{\bf L},(m)}}{\Gamma\left(\nu- ip_n +\frac{1}{2}\right)\Gamma\left(\nu+ ip_m +\frac{1}{2}\right)}
&\stackrel{p_n, p_m<<1}{\approx}&\frac{\pi e^{-\pi(p_n+p_m)}}{2\left(\Gamma\left(\nu +\frac{1}{2}\right)\right)^2}\\
\sum_{\sigma=\pm 1}\frac{{\cal A}^{\sigma}_{{\bf L},(n)}{\cal B}^{\sigma*}_{{\bf L},(m)}}{\Gamma\left(\nu- ip_n +\frac{1}{2}\right)\Gamma\left(\nu- ip_m +\frac{1}{2}\right)}
&\stackrel{p_n, p_m<<1}{\approx}&\frac{\pi e^{-\pi(p_n+p_m)}}{2\left(\Gamma\left(\nu +\frac{1}{2}\right)\right)^2}\\
\sum_{\sigma=\pm 1}\frac{{\cal A}^{\sigma*}_{{\bf L},(n)}{\cal B}^{\sigma}_{{\bf L},(m)}}{\Gamma\left(\nu+ ip_n +\frac{1}{2}\right)\Gamma\left(\nu+ ip_m +\frac{1}{2}\right)}
&\stackrel{p_n, p_m<<1}{\approx}&\frac{\pi e^{-\pi(p_n+p_m)}}{2\left(\Gamma\left(\nu +\frac{1}{2}\right)\right)^2}.\eea
As a result, the time independent function $\widetilde{{\cal M}(p<<1,\nu)}$ for any arbitrary mass parameter $\nu$ can be expressed as:
\bea\widetilde{{\cal M}(p<<1,\nu)}&=&\frac{2^{2(\nu-1)}\left(\Gamma(\nu)\right)^2}{\pi}\widetilde{{\cal G}(p<<1)},~~~~~~~~~~~\eea
where we define a new function $\widetilde{{\cal G}(p<<1)}$ in the long wave length limit as given by:
\bea\widetilde{{\cal G}(p<<1)}&=&\frac{\pi}{|\Gamma\left(\nu+\frac{1}{2}\right)|^2}\left[1+\frac{|\Gamma\left(\nu+\frac{1}{2}\right)|^2}{\left(\Gamma\left(\nu+\frac{1}{2}\right)\right)^2}\left\{1+3e^{-\pi p}\sum^{\infty}_{n=0}e^{-\pi p_n}+2\sum^{\infty}_{n=0}\sum^{\infty}_{m=0}e^{-\pi (p_n+p_m)}\right\}\right].~~~~~~~~~~~\eea

\section{\textcolor{blue}{Quantum correction to the power spectrum  in RDM formalism}}
At  the super horizon time scales ($t_{\bf L}>>1$) of region \textcolor{red}{\bf L} one can write the amplitude of the RDM power spectrum as:
\bea \label{df2xxxss} |\tilde{\psi}^{\bf L}_{\bf T}|^2=\left(\tilde{\psi}^{\bf L}_{\bf T}\right)^{\dagger}
\tilde{\psi}^{\bf L}_{\bf T}&~\underrightarrow{t_{\bf L}>>1}~& \widetilde{{\cal Q}(p,\alpha,\nu)}\left(\cosh t_{\bf L}\right)^{2\nu-1}\eea
where the time independent function $\widetilde{{\cal Q}(p,\alpha,\nu)}$ for generalised $\alpha$ vacua is defined as:
\bea\widetilde{{\cal Q}(p,\alpha,\nu)}&=&\frac{2^{2\nu-1}\left(\Gamma(\nu)\right)^2}{\pi}\times \left[\frac{\left(|{\cal E}_{\bf L}|^2+|{\cal F}_{\bf L}|^2\right)}{|\Gamma\left(\nu+ip+\frac{1}{2}\right)|^2}+\frac{{\cal E}_{\bf L}{\cal F}^{ *}_{\bf L}}{\left(\Gamma\left(\nu-ip+\frac{1}{2}\right)\right)^2}+\frac{{\cal E}^{ *}_{\bf L}{\cal F}_{\bf L}}{\left(\Gamma\left(\nu+ip+\frac{1}{2}\right)\right)^2}\right.\nonumber\\ && \left.
~+\sum^{\infty}_{n=0}\left\{\frac{\left({\cal E}_{{\bf L}}{\cal E}^{*}_{{\bf L},(n)}+{\cal F}_{{\bf L}}{\cal F}^{*}_{{\bf L},(n)}\right)}{\Gamma\left(\nu-ip+\frac{1}{2}\right)\Gamma\left(\nu+ip_n+\frac{1}{2}\right)}\right.\right.\nonumber\\&&\left.\left.~~~~~~~~+\frac{\left({\cal E}_{{\bf L}}{\cal F}^{*}_{{\bf L},(n)}+{\cal E}_{{\bf L},(n)}{\cal F}^{*}_{{\bf L}}\right)}{\Gamma\left(\nu-ip+\frac{1}{2}\right)\Gamma\left(\nu-ip_n+\frac{1}{2}\right)}+\frac{\left({\cal E}^{*}_{{\bf L},(n)}{\cal F}_{{\bf L}}+{\cal E}^{*}_{{\bf L}}{\cal F}_{{\bf L},(n)}\right)}{\Gamma\left(\nu+ip+\frac{1}{2}\right)\Gamma\left(\nu+ip_n+\frac{1}{2}\right)}\right\}\right.\nonumber\\ && \left.
~+\sum^{\infty}_{n=0}\sum^{\infty}_{m=0}\left\{\frac{\left({\cal E}_{{\bf L},(n)}{\cal E}^{*}_{{\bf L},(m)}+{\cal F}_{{\bf L},(n)}{\cal F}^{*}_{{\bf L},(m)}\right)}{\Gamma\left(\nu-ip_n+\frac{1}{2}\right)\Gamma\left(\nu+ip_m+\frac{1}{2}\right)}\right.\right.\nonumber\\&&\left.\left.~~~+\frac{{\cal E}_{{\bf L},(n)}{\cal F}^{*}_{{\bf L},(m)}}{\Gamma\left(\nu-ip_n+\frac{1}{2}\right)\Gamma\left(\nu-ip_m+\frac{1}{2}\right)}+\frac{{\cal E}^{*}_{{\bf L},(n)}{\cal F}_{{\bf L},(m)}}{\Gamma\left(\nu+ip_n+\frac{1}{2}\right)\Gamma\left(\nu+ip_m+\frac{1}{2}\right)}\right\}\right].~~~~~~~~~\eea

\subsection{For large wave number}

Further to know the exact wave number dependence of the amplitude of the normalised power spectrum from generalised $\alpha$ vacua we need to know the behaviour of the power spectrum at very short wavelengths ($p,p_n>>1$). After taking this limit it is expected that the power spectrum of axion should match with the result obtained for spatially flat universe. In general for an arbitrary value of the mass parameter $\nu$, we get the following approximated contributions in the short wavelength limit ($p,p_n>>1$), which are explicitly appearing in the expression for the amplitude of the normalised power spectrum from generalised $\alpha$ vacua:
\bea \frac{|{\cal E}^{}_{\bf L}|^2}{\left|\Gamma\left(\nu+ ip +\frac{1}{2}\right)\right|^2}
&\stackrel{p>>1}{\approx}&\frac{\pi e^{-\pi p}}{2p^4\left|\Gamma\left(ip\right)\right|^2},\\
\frac{|{\cal F}^{}_{\bf L}|^2}{\left|\Gamma\left(\nu+ ip +\frac{1}{2}\right)\right|^2}
&\stackrel{p>>1}{\approx}&\frac{\pi e^{-5\pi p}}{2p^4\left|\Gamma\left(ip\right)\right|^2},\\
\frac{{\cal E}^{}_{\bf L}{\cal F}^{ *}_{\bf L}}{\left(\Gamma\left(\nu- ip +\frac{1}{2}\right)\right)^2}
&\stackrel{p>>1}{\approx}&\frac{\pi e^{-3\pi p}}{2p^4\left|\Gamma\left(ip\right)\right|^2},\\
\frac{{\cal E}^{*}_{\bf L}{\cal F}^{}_{\bf L}}{\left(\Gamma\left(\nu+ ip +\frac{1}{2}\right)\right)^2}
&\stackrel{p>>1}{\approx}&\frac{\pi e^{-3\pi p}}{2p^4\left|\Gamma\left(ip\right)\right|^2}\\
\frac{{\cal E}^{}_{\bf L}{\cal E}^{*}_{{\bf L},(n)}}{\Gamma\left(\nu- ip +\frac{1}{2}\right)\Gamma\left(\nu+ ip_n +\frac{1}{2}\right)}
&\stackrel{p,p_n>>1}{\approx}&\frac{\pi e^{-\pi (p+p_n)/2}}{2p^2p^2_n\left|\Gamma\left(ip\right)\right|\left|\Gamma\left(ip_n\right)\right|}\\
\frac{{\cal F}^{}_{\bf L}{\cal F}^{*}_{{\bf L},(n)}}{\Gamma\left(\nu- ip +\frac{1}{2}\right)\Gamma\left(\nu+ ip_n +\frac{1}{2}\right)}
&\stackrel{p,p_n>>1}{\approx}&\frac{\pi e^{-5\pi (p+p_n)/2}}{2p^2p^2_n\left|\Gamma\left(ip\right)\right|\left|\Gamma\left(ip_n\right)\right|}\eea\bea
\frac{{\cal E}^{}_{\bf L}{\cal F}^{*}_{{\bf L},(n)}}{\Gamma\left(\nu- ip +\frac{1}{2}\right)\Gamma\left(\nu- ip_n +\frac{1}{2}\right)}
&\stackrel{p,p_n>>1}{\approx}&\frac{\pi e^{-\pi (p+5p_n)/2}}{2p^2p^2_n \left|\Gamma\left(ip\right)\right|\left|\Gamma\left(ip_n\right)\right|}\\
\frac{{\cal E}^{}_{{\bf L},(n)}{\cal F}^{*}_{{\bf L}}}{\Gamma\left(\nu- ip +\frac{1}{2}\right)\Gamma\left(\nu- ip_n +\frac{1}{2}\right)}
&\stackrel{p,p_n>>1}{\approx}&\frac{\pi e^{-\pi (5p+p_n)/2}}{2p^2p^2_n\left|\Gamma\left(ip\right)\right|\left|\Gamma\left(ip_n\right)\right|}\\
\frac{{\cal E}^{*}_{{\bf L},(n)}{\cal F}^{}_{\bf L}}{\Gamma\left(\nu+ ip +\frac{1}{2}\right)\Gamma\left(\nu+ ip_n +\frac{1}{2}\right)}
&\stackrel{p,p_n>>1}{\approx}&\frac{\pi e^{-\pi (5p+p_n)/2}}{2p^2p^2_n\left|\Gamma\left(ip\right)\right|\left|\Gamma\left(ip_n\right)\right|}\\
\frac{{\cal E}^{*}_{{\bf L}}{\cal F}^{}_{{\bf L},(n)}}{\Gamma\left(\nu+ ip +\frac{1}{2}\right)\Gamma\left(\nu+ ip_n +\frac{1}{2}\right)}
&\stackrel{p,p_n>>1}{\approx}&\frac{\pi e^{-\pi (p+5p_n)/2}}{2p^2p^2_n\left|\Gamma\left(ip\right)\right|\left|\Gamma\left(ip_n\right)\right|}\\
\frac{{\cal E}^{}_{{\bf L},(n)}{\cal E}^{*}_{{\bf L},(m)}}{\Gamma\left(\nu- ip_n +\frac{1}{2}\right)\Gamma\left(\nu+ ip_m +\frac{1}{2}\right)}
&\stackrel{p_n, p_m>>1}{\approx}&\frac{\pi e^{-\pi(p_n+p_m)/2}}{2p^2_n p^2_m\left|\Gamma\left(ip_m\right)\right|\left|\Gamma\left(ip_n\right)\right|}\\
\frac{{\cal F}^{}_{{\bf L},(n)}{\cal F}^{*}_{{\bf L},(m)}}{\Gamma\left(\nu- ip_n +\frac{1}{2}\right)\Gamma\left(\nu+ ip_m +\frac{1}{2}\right)}
&\stackrel{p_n, p_m>>1}{\approx}&\frac{\pi e^{-5\pi(p_n+p_m)/2}}{2p^2_n p^2_m\left|\Gamma\left(ip_m\right)\right|\left|\Gamma\left(ip_n\right)\right|}\\
\frac{{\cal E}^{}_{{\bf L},(n)}{\cal F}^{*}_{{\bf L},(m)}}{\Gamma\left(\nu- ip_n +\frac{1}{2}\right)\Gamma\left(\nu- ip_m +\frac{1}{2}\right)}
&\stackrel{p_n, p_m>>1}{\approx}&\frac{\pi e^{-3\pi(p_n+p_m)/2}}{2p^2_n p^2_m\left|\Gamma\left(ip_m\right)\right|\left|\Gamma\left(ip_n\right)\right|}\\
\frac{{\cal E}^{\sigma*}_{{\bf L},(n)}{\cal F}^{}_{{\bf L},(m)}}{\Gamma\left(\nu+ ip_n +\frac{1}{2}\right)\Gamma\left(\nu+ ip_m +\frac{1}{2}\right)}
&\stackrel{p_n, p_m>>1}{\approx}&\frac{\pi e^{-3\pi(p_n+p_m)/2}}{2p^2_n p^2_m\left|\Gamma\left(ip_m\right)\right|\left|\Gamma\left(ip_n\right)\right|}\eea
Further, we apply Stirling's formula to approximate Gamma functions for large wavenumbers $p,p_n>>1$ to simplify the expression for the power spectrum:
\bea \Gamma(ip)&\sim & \sqrt{2\pi}~(ip)^{ip-\frac{1}{2}}e^{-ip}\left(1+\frac{1}{12ip}-\frac{1}{288p^2}+\cdots\right),\\
\Gamma(ip_n)&\sim & \sqrt{2\pi}~(ip_n)^{ip_n-\frac{1}{2}}e^{-ip_n}\left(1+\frac{1}{12ip_n}-\frac{1}{288p^2_n}+\cdots\right).\eea
Consequently, we get the following simplified expressions for large wavenumber $p,p_n>>1$ limit in the case of generalised $\alpha$ vacua:
\bea \frac{|{\cal E}^{}_{\bf L}|^2}{\left|\Gamma\left(\nu+ ip +\frac{1}{2}\right)\right|^2}
&\sim&\frac{1}{2p^3\left(1+\frac{1}{82944p^4}\right)},\\
\frac{|{\cal F}^{}_{\bf L}|^2}{\left|\Gamma\left(\nu+ ip +\frac{1}{2}\right)\right|^2}
&\sim &\frac{e^{-4\pi p}}{2p^3\left(1+\frac{1}{82944p^4}\right)},\\
\frac{{\cal E}^{}_{\bf L}{\cal F}^{ *}_{\bf L}}{\left(\Gamma\left(\nu- ip +\frac{1}{2}\right)\right)^2}
&\sim &\frac{e^{-2\pi p}}{2p^3\left(1+\frac{1}{82944p^4}\right)},\\
\frac{{\cal E}^{\sigma*}_{\bf L}{\cal F}^{\sigma}_{\bf L}}{\left(\Gamma\left(\nu+ ip +\frac{1}{2}\right)\right)^2}
&\sim&\frac{e^{-2\pi p}}{2p^3\left(1+\frac{1}{82944p^4}\right)}\\
\frac{{\cal E}^{}_{\bf L}{\cal E}^{*}_{{\bf L},(n)}}{\Gamma\left(\nu- ip +\frac{1}{2}\right)\Gamma\left(\nu+ ip_n +\frac{1}{2}\right)}
&\sim &\frac{1}{2p^{3/2} p^{3/2}_n \sqrt{1+\frac{1}{82944p^4}}\sqrt{1+\frac{1}{82944p^4_n}}}\eea\bea
\frac{{\cal F}^{}_{\bf L}{\cal F}^{*}_{{\bf L},(n)}}{\Gamma\left(\nu- ip +\frac{1}{2}\right)\Gamma\left(\nu+ ip_n +\frac{1}{2}\right)}
&\sim&\frac{e^{-2\pi (p+p_n)}}{2p^{3/2} p^{3/2}_n\sqrt{1+\frac{1}{82944p^4}}\sqrt{1+\frac{1}{82944p^4_n}}}\\
\frac{{\cal E}^{}_{\bf L}{\cal F}^{*}_{{\bf L},(n)}}{\Gamma\left(\nu- ip +\frac{1}{2}\right)\Gamma\left(\nu- ip_n +\frac{1}{2}\right)}
&\sim&\frac{e^{-2\pi p_n}}{2p^{3/2} p^{3/2}_n\sqrt{1+\frac{1}{82944p^4}}\sqrt{1+\frac{1}{82944p^4_n}}}\\
\frac{{\cal E}^{}_{{\bf L},(n)}{\cal F}^{*}_{{\bf L}}}{\Gamma\left(\nu- ip +\frac{1}{2}\right)\Gamma\left(\nu- ip_n +\frac{1}{2}\right)}
&\sim&\frac{e^{-2\pi p}}{2p^{3/2} p^{3/2}_n\sqrt{1+\frac{1}{82944p^4}}\sqrt{1+\frac{1}{82944p^4_n}}}\\
\frac{{\cal E}^{*}_{{\bf L},(n)}{\cal F}^{}_{\bf L}}{\Gamma\left(\nu+ ip +\frac{1}{2}\right)\Gamma\left(\nu+ ip_n +\frac{1}{2}\right)}
&\sim&\frac{e^{-2\pi p}}{2p^{3/2} p^{3/2}_n\sqrt{1+\frac{1}{82944p^4}}\sqrt{1+\frac{1}{82944p^4_n}}}\\
\frac{{\cal E}^{*}_{{\bf L}}{\cal F}^{}_{{\bf L},(n)}}{\Gamma\left(\nu+ ip +\frac{1}{2}\right)\Gamma\left(\nu+ ip_n +\frac{1}{2}\right)}
&\sim &\frac{e^{-2\pi p_n}}{2p^{3/2} p^{3/2}_n\sqrt{1+\frac{1}{82944p^4}}\sqrt{1+\frac{1}{82944p^4_n}}}\\
\frac{{\cal E}^{\sigma}_{{\bf L},(n)}{\cal E}^{*}_{{\bf L},(m)}}{\Gamma\left(\nu- ip_n +\frac{1}{2}\right)\Gamma\left(\nu+ ip_m +\frac{1}{2}\right)}
&\sim &\frac{1}{2p^{3/2}_m p^{3/2}_n\sqrt{1+\frac{1}{82944p^4_m}}\sqrt{1+\frac{1}{82944p^4_n}}}\\
\frac{{\cal F}^{}_{{\bf L},(n)}{\cal F}^{*}_{{\bf L},(m)}}{\Gamma\left(\nu- ip_n +\frac{1}{2}\right)\Gamma\left(\nu+ ip_m +\frac{1}{2}\right)}
&\sim &\frac{e^{-2\pi(p_n+p_m)}}{2p^{3/2}_m p^{3/2}_n\sqrt{1+\frac{1}{82944p^4_m}}\sqrt{1+\frac{1}{82944p^4_n}}}\\
\frac{{\cal E}^{}_{{\bf L},(n)}{\cal F}^{*}_{{\bf L},(m)}}{\Gamma\left(\nu- ip_n +\frac{1}{2}\right)\Gamma\left(\nu- ip_m +\frac{1}{2}\right)}
&\sim &\frac{e^{-\pi(p_n+p_m)}}{2p^{3/2}_m p^{3/2}_n\sqrt{1+\frac{1}{82944p^4_m}}\sqrt{1+\frac{1}{82944p^4_n}}}\\
\frac{{\cal E}^{*}_{{\bf L},(n)}{\cal F}^{}_{{\bf L},(m)}}{\Gamma\left(\nu+ ip_n +\frac{1}{2}\right)\Gamma\left(\nu+ ip_m +\frac{1}{2}\right)}
&\sim &\frac{e^{-\pi(p_n+p_m)}}{2p^{3/2}_m p^{3/2}_n\sqrt{1+\frac{1}{82944p^4_m}}\sqrt{1+\frac{1}{82944p^4_n}}}\eea

As a result, in the short wave length approximation the time independent function $\widetilde{{\cal Q}(p>>1,\alpha,\nu)}$ for any arbitrary mass parameter $\nu$ can be expressed for generalised $\alpha$ vacua as:
\bea\widetilde{{\cal Q}(p>>1,\alpha,\nu)}&=&\frac{2^{2(\nu-1)}\left(\Gamma(\nu)\right)^2}{p^3\pi}\widetilde{{\cal G}(p>>1)}=\widetilde{{\cal M}(p,\nu)}~~~\forall \alpha,~~~~~~~~~~~\eea
where we have already defined the function $\widetilde{{\cal G}(p>>1)}$ in the earlier section of the Appendix.

\subsection{For small wave number}
Similarly to know the exact wave number dependence of the amplitude of the normalised power spectrum from generalised $\alpha$ vacua in the long wave length approximation we need to know the behaviour of the power spectrum at $p,p_n<<1$. After taking this limit it is expected that the power spectrum of axion should match with the result obtained for spatially flat universe. In general for an arbitrary value of the mass parameter $\nu$, we get the following approximated contributions in the in the long wave length approximation, which are explicitly appearing in the expression for the amplitude of the normalised power spectrum from generalised $\alpha$ vacua:
\bea \frac{|{\cal E}^{}_{\bf L}|^2}{\left|\Gamma\left(\nu+ ip +\frac{1}{2}\right)\right|^2}
&\stackrel{p<<1}{\approx}&\frac{\pi p}{2|\cos \pi \nu|\left|\Gamma\left(\nu+ \frac{1}{2}\right)\right|^2}\frac{|1-\gamma^{(\alpha)}_p \tilde{m}_{\bf LR}|^2}{|1-\gamma^{(\alpha)}_p  \tilde{m}_{\bf LR}|^2-| \tilde{m}_{\bf RR}|^2}~~~~~~~~~~~\\
\frac{|{\cal F}^{}_{\bf L}|^2}{\left|\Gamma\left(\nu+ ip +\frac{1}{2}\right)\right|^2}
&\stackrel{p<<1}{\approx}&\frac{\pi p}{2|\cos \pi \nu|\left|\Gamma\left(\nu+ \frac{1}{2}\right)\right|^2}\frac{| \tilde{m}_{\bf RR}|^2}{|1-\gamma^{(\alpha)}_p  \tilde{m}_{\bf LR}|^2-| \tilde{m}_{\bf RR}|^2}~~~~~~~~~~~\\
\frac{{\cal E}^{}_{\bf L}{\cal F}^{ *}_{\bf L}}{\left(\Gamma\left(\nu- ip +\frac{1}{2}\right)\right)^2}
&\stackrel{p<<1}{\approx}&\frac{\pi p}{2|\cos \pi \nu|\left(\Gamma\left(\nu+ \frac{1}{2}\right)\right)^2}\frac{\left(1-\gamma^{(\alpha)}_p \tilde{m}_{\bf LR}\right)\tilde{m}^{*}_{\bf RR}}{|1-\gamma^{(\alpha)}_p  \tilde{m}_{\bf LR}|^2-| \tilde{m}_{\bf RR}|^2}~~~~~~~~~~~\\
\frac{{\cal E}^{*}_{\bf L}{\cal F}^{}_{\bf L}}{\left(\Gamma\left(\nu+ ip +\frac{1}{2}\right)\right)^2}
&\stackrel{p<<1}{\approx}&\frac{\pi p}{2|\cos \pi \nu|\left(\Gamma\left(\nu+ \frac{1}{2}\right)\right)^2}\frac{\left(1-\gamma^{(\alpha)}_p \tilde{m}_{\bf LR}\right)^{*}\tilde{m}_{\bf RR}}{|1-\gamma^{(\alpha)}_p  \tilde{m}_{\bf LR}|^2-| \tilde{m}_{\bf RR}|^2}~~~~~~~~~~~\\
\frac{{\cal E}^{}_{\bf L}{\cal E}^{*}_{{\bf L},(n)}}{\Gamma\left(\nu- ip +\frac{1}{2}\right)\Gamma\left(\nu+ ip_n +\frac{1}{2}\right)}
&\stackrel{p,p_n<<1}{\approx}&\frac{\pi\sqrt{pp_n} }{2|\cos \pi \nu|\left(\Gamma\left(\nu+ \frac{1}{2}\right)\right)^2}\frac{\left(1-\gamma^{(\alpha)}_p \tilde{m}_{\bf LR}\right)}{\sqrt{|1-\gamma^{(\alpha)}_p  \tilde{m}_{\bf LR}|^2-| \tilde{m}_{\bf RR}|^2}}\nonumber\\
&&~~~~~~\times \frac{\left(1-\Gamma^{(\alpha)}_{p,n} \tilde{m}_{{\bf LR},n}\right)^{*}}{\sqrt{|1-\Gamma^{(\alpha)}_{p,n}  \tilde{m}_{{\bf LR},n}|^2-| \tilde{m}_{{\bf RR},n}|^2}}~~~~~~~~~~~\\
\frac{{\cal F}^{}_{\bf L}{\cal F}^{*}_{{\bf L},(n)}}{\Gamma\left(\nu- ip +\frac{1}{2}\right)\Gamma\left(\nu+ ip_n +\frac{1}{2}\right)}
&\stackrel{p,p_n<<1}{\approx}&\frac{\pi \sqrt{pp_n}}{2|\cos \pi \nu|\left(\Gamma\left(\nu+ \frac{1}{2}\right)\right)^2}\frac{\tilde{m}_{\bf RR}}{\sqrt{|1-\gamma^{(\alpha)}_p  \tilde{m}_{\bf LR}|^2-| \tilde{m}_{\bf RR}|^2}}\nonumber\\
&&~~~~~~\times \frac{\tilde{m}^{*}_{{\bf RR},n}}{\sqrt{|1-\Gamma^{(\alpha)}_{p,n}  \tilde{m}_{{\bf LR},n}|^2-| \tilde{m}_{{\bf RR},n}|^2}}~~~~~~~~~~~\\
\frac{{\cal E}^{}_{\bf L}{\cal F}^{*}_{{\bf L},(n)}}{\Gamma\left(\nu- ip +\frac{1}{2}\right)\Gamma\left(\nu- ip_n +\frac{1}{2}\right)}
&\stackrel{p,p_n<<1}{\approx}&\frac{\pi \sqrt{pp_n}}{2|\cos \pi \nu|\left(\Gamma\left(\nu+ \frac{1}{2}\right)\right)^2}\frac{\left(1-\gamma^{(\alpha)}_p \tilde{m}_{\bf LR}\right)}{\sqrt{|1-\gamma^{(\alpha)}_p  \tilde{m}_{\bf LR}|^2-| \tilde{m}_{\bf RR}|^2}}\nonumber\\
&&~~~~~~\times \frac{\tilde{m}^{*}_{{\bf RR},n}}{\sqrt{|1-\Gamma^{(\alpha)}_{p,n}  \tilde{m}_{{\bf LR},n}|^2-| \tilde{m}_{{\bf RR},n}|^2}}~~~~~~~~~~~\\
\frac{{\cal E}^{}_{{\bf L},(n)}{\cal F}^{*}_{{\bf L}}}{\Gamma\left(\nu- ip +\frac{1}{2}\right)\Gamma\left(\nu- ip_n +\frac{1}{2}\right)}
&\stackrel{p,p_n<<1}{\approx}&\frac{\pi \sqrt{pp_n}}{2|\cos \pi \nu|\left(\Gamma\left(\nu+ \frac{1}{2}\right)\right)^2}\frac{\left(1-\Gamma^{(\alpha)}_{p,n} \tilde{m}_{{\bf LR},n}\right)}{\sqrt{|1-\Gamma^{(\alpha)}_{p,n}  \tilde{m}_{{\bf LR},n}|^2-| \tilde{m}_{{\bf RR},n}|^2}}\nonumber\\
&&~~~~~~\times \frac{\tilde{m}^{*}_{{\bf RR}}}{\sqrt{|1-\gamma^{(\alpha)}_{p}  \tilde{m}_{{\bf LR}}|^2-| \tilde{m}_{{\bf RR}}|^2}}~~~~~~~~~~~\\
\frac{{\cal E}^{*}_{{\bf L},(n)}{\cal F}^{}_{\bf L}}{\Gamma\left(\nu+ ip +\frac{1}{2}\right)\Gamma\left(\nu+ ip_n +\frac{1}{2}\right)}
&\stackrel{p,p_n<<1}{\approx}&\frac{\pi \sqrt{pp_n}}{2|\cos \pi \nu|\left(\Gamma\left(\nu+ \frac{1}{2}\right)\right)^2}\frac{\left(1-\Gamma^{(\alpha)}_{p,n} \tilde{m}_{{\bf LR},n}\right)^{*}}{\sqrt{|1-\Gamma^{(\alpha)}_{p,n}  \tilde{m}_{{\bf LR},n}|^2-| \tilde{m}_{{\bf RR},n}|^2}}\nonumber\\
&&~~~~~~\times \frac{\tilde{m}_{{\bf RR}}}{\sqrt{|1-\gamma^{(\alpha)}_{p}  \tilde{m}_{{\bf LR}}|^2-| \tilde{m}_{{\bf RR}}|^2}}~~~~~~~~~~~\\
\frac{{\cal E}^{*}_{{\bf L}}{\cal F}^{}_{{\bf L},(n)}}{\Gamma\left(\nu+ ip +\frac{1}{2}\right)\Gamma\left(\nu+ ip_n +\frac{1}{2}\right)}
&\stackrel{p,p_n<<1}{\approx}&\frac{\pi \sqrt{pp_n} }{2|\cos \pi \nu|\left(\Gamma\left(\nu+ \frac{1}{2}\right)\right)^2}\frac{\left(1-\gamma^{(\alpha)}_{p} \tilde{m}_{{\bf LR}}\right)^{*}}{\sqrt{|1-\gamma^{(\alpha)}_{p}  \tilde{m}_{{\bf LR}}|^2-| \tilde{m}_{\bf RR}|^2}}\nonumber\\
&&~~~~~~\times \frac{\tilde{m}_{{\bf RR},n}}{\sqrt{|1-\Gamma^{(\alpha)}_{p,n}  \tilde{m}_{{\bf LR},n}|^2-| \tilde{m}_{{\bf RR},n}|^2}}~~~~~~~~~~~\eea\bea
\frac{{\cal E}^{}_{{\bf L},(n)}{\cal E}^{*}_{{\bf L},(m)}}{\Gamma\left(\nu- ip_n +\frac{1}{2}\right)\Gamma\left(\nu+ ip_m +\frac{1}{2}\right)}
&\stackrel{p_n, p_m<<1}{\approx}&\frac{\pi \sqrt{p_np_m} }{2|\cos \pi \nu|\left(\Gamma\left(\nu+ \frac{1}{2}\right)\right)^2}\frac{\left(1-\Gamma^{(\alpha)}_{p,n} \tilde{m}_{{\bf LR},n}\right)}{\sqrt{|1-\Gamma^{(\alpha)}_{p,n}  \tilde{m}_{{\bf LR},n}|^2-| \tilde{m}_{{\bf RR},n}|^2}}\nonumber\\
&&~~~~~~\times \frac{\left(1-\Gamma^{(\alpha)}_{p,m} \tilde{m}_{{\bf LR},m}\right)^{*}}{\sqrt{|1-\Gamma^{(\alpha)}_{p,m}  \tilde{m}_{{\bf LR},m}|^2-| \tilde{m}_{{\bf RR},m}|^2}}~~~~~~~~~~~\\
\frac{{\cal F}^{}_{{\bf L},(n)}{\cal F}^{*}_{{\bf L},(m)}}{\Gamma\left(\nu- ip_n +\frac{1}{2}\right)\Gamma\left(\nu+ ip_m +\frac{1}{2}\right)}
&\stackrel{p_n, p_m<<1}{\approx}&\frac{\pi  \sqrt{p_np_m}}{2|\cos \pi \nu|\left(\Gamma\left(\nu+ \frac{1}{2}\right)\right)^2}\frac{\tilde{m}_{{\bf RR},n}}{\sqrt{|1-\Gamma^{(\alpha)}_{p,n}  \tilde{m}_{{\bf LR},n}|^2-| \tilde{m}_{{\bf RR},n}|^2}}\nonumber\\
&&~~~~~~\times \frac{\tilde{m}^{*}_{{\bf RR},m}}{\sqrt{|1-\Gamma^{(\alpha)}_{p,m}  \tilde{m}_{{\bf LR},m}|^2-| \tilde{m}_{{\bf RR},m}|^2}}~~~~~~~~~~~\\
\frac{{\cal E}^{}_{{\bf L},(n)}{\cal F}^{*}_{{\bf L},(m)}}{\Gamma\left(\nu- ip_n +\frac{1}{2}\right)\Gamma\left(\nu- ip_m +\frac{1}{2}\right)}
&\stackrel{p_n, p_m<<1}{\approx}&\frac{\pi \sqrt{p_np_m}}{2|\cos \pi \nu|\left(\Gamma\left(\nu+ \frac{1}{2}\right)\right)^2}\frac{\left(1-\Gamma^{(\alpha)}_{p,n} \tilde{m}_{{\bf LR},n}\right)}{\sqrt{|1-\Gamma^{(\alpha)}_{p,n}  \tilde{m}_{{\bf LR},n}|^2-| \tilde{m}_{{\bf RR},n}|^2}}\nonumber\\
&&~~~~~~\times \frac{\tilde{m}^{*}_{{\bf RR},m}}{\sqrt{|1-\Gamma^{(\alpha)}_{p,m}  \tilde{m}_{{\bf LR},m}|^2-| \tilde{m}_{{\bf RR},m}|^2}}~~~~~~~~~~~\\
\frac{{\cal E}^{\sigma*}_{{\bf L},(n)}{\cal F}^{}_{{\bf L},(m)}}{\Gamma\left(\nu+ ip_n +\frac{1}{2}\right)\Gamma\left(\nu+ ip_m +\frac{1}{2}\right)}
&\stackrel{p_n, p_m<<1}{\approx}&\frac{\pi \sqrt{p_np_m}}{2|\cos \pi \nu|\left(\Gamma\left(\nu+ \frac{1}{2}\right)\right)^2}\frac{\left(1-\Gamma^{(\alpha)}_{p,n} \tilde{m}_{{\bf LR},n}\right)^{*}}{\sqrt{|1-\Gamma^{(\alpha)}_{p,n}  \tilde{m}_{{\bf LR},n}|^2-| \tilde{m}_{{\bf RR},n}|^2}}\nonumber\\
&&~~~~~~\times \frac{\tilde{m}_{{\bf RR},m}}{\sqrt{|1-\Gamma^{(\alpha)}_{p,m}  \tilde{m}_{{\bf LR}}|^2-| \tilde{m}_{{\bf RR},m}|^2}}~~~~~~~~~~~\eea
where all the entries of the right hand side of the above expressions for $p,p_n<<1$ are explicitly computed earlier in this paper.

As a result, the time independent function $\widetilde{{\cal Q}(p<<1,\alpha,\nu)}$ for the mass parameter $\nu\neq q/2$ (where $q$ is any half integer) can be expressed for generalised $\alpha$ vacua as:
\bea\widetilde{{\cal Q}(p<<1,\alpha,\nu)}&=&\frac{2^{2(\nu-1)}\left(\Gamma(\nu)\right)^2}{p^3\pi}\widetilde{{\cal G}(p<<1)}~~~\forall \alpha,~~~~~~~~~~~\eea
where the function $\widetilde{{\cal G}(p<<1)}$ is defined for $\nu \neq 3/2$ as:
\bea \widetilde{{\cal G}(p<<1)}&=&\frac{\pi p}{2|\cos \pi \nu|\left|\Gamma\left(\nu+ \frac{1}{2}\right)\right|^2}\frac{|1-\gamma^{(\alpha)}_p \tilde{m}_{\bf LR}|^2}{|1-\gamma^{(\alpha)}_p  \tilde{m}_{\bf LR}|^2-| \tilde{m}_{\bf RR}|^2}\nonumber\eea
\bea &&\times\left\{1+\frac{|\tilde{m}_{\bf RR}|^2+\left(1-\gamma^{(\alpha)}_p \tilde{m}_{\bf LR}\right)^{*}\tilde{m}_{\bf RR}+\left(1-\gamma^{(\alpha)}_p \tilde{m}_{\bf LR}\right)\tilde{m}^{*}_{\bf RR}}{|1-\gamma^{(\alpha)}_p \tilde{m}_{\bf LR}|^2}\right.\nonumber\\&&\left.~~~~+\sum^{\infty}_{n=0}\sqrt{\frac{p_n}{p}\frac{{|1-\gamma^{(\alpha)}_p  \tilde{m}_{\bf LR}|^2-| \tilde{m}_{\bf RR}|^2}}{{|1-\Gamma^{(\alpha)}_{p,n}  \tilde{m}_{{\bf LR},n}|^2-| \tilde{m}_{{\bf RR},n}|^2}}}\frac{1}{|1-\gamma^{(\alpha)}_p \tilde{m}_{\bf LR}|^2}\right.\nonumber\\&& \left.
~~~~~~~~~~\left[\left(1-\gamma^{(\alpha)}_p \tilde{m}_{\bf LR}\right)\left(1-\Gamma^{(\alpha)}_{p,n} \tilde{m}_{{\bf LR},n}\right)^{*}+\tilde{m}_{\bf RR}\tilde{m}^{*}_{{\bf RR},n}\right.\right.\nonumber\\&& \left.\left.~~~~~~~~~~+\left(1-\gamma^{(\alpha)}_p \tilde{m}_{\bf LR}\right)\tilde{m}^{*}_{{\bf RR},n}+\left(1-\Gamma^{(\alpha)}_{p,n} \tilde{m}_{{\bf LR},n}\right)\tilde{m}^{*}_{{\bf RR}}\right.\right.\nonumber\\&&\left.\left.~~~~~~~~~~~+\left(1-\gamma^{(\alpha)}_p \tilde{m}_{\bf LR}\right)^{*}\tilde{m}_{{\bf RR},n}+\left(1-\Gamma^{(\alpha)}_{p,n} \tilde{m}_{{\bf LR},n}\right)^{*}\tilde{m}_{{\bf RR}}\right]\right.\nonumber\\&&\left.~~~~+\sum^{\infty}_{n=0}\sum^{\infty}_{m=0}\sqrt{\frac{p_np_m}{p^2}\frac{{\left(|1-\gamma^{(\alpha)}_p  \tilde{m}_{\bf LR}|^2-| \tilde{m}_{\bf RR}|^2\right)^2}}{{\left(|1-\Gamma^{(\alpha)}_{p,n}  \tilde{m}_{{\bf LR},n}|^2-| \tilde{m}_{{\bf RR},n}|^2\right)\left(|1-\Gamma^{(\alpha)}_{p,m}  \tilde{m}_{{\bf LR},m}|^2-| \tilde{m}_{{\bf RR},m}|^2\right)}}}\right.\nonumber\\&& \left.
~~~~~~~~~~\frac{1}{|1-\gamma^{(\alpha)}_p \tilde{m}_{\bf LR}|^2}\left[\left(1-\Gamma^{(\alpha)}_{p,n} \tilde{m}_{{\bf LR},n}\right)\left(1-\Gamma^{(\alpha)}_{p,m} \tilde{m}_{{\bf LR},m}\right)^{*}+\tilde{m}_{{\bf RR},n}\tilde{m}^{*}_{{\bf RR},m}\right.\right.\nonumber\\&& \left.\left.~~~~~~~~~~+\left(1-\Gamma^{(\alpha)}_{p,n} \tilde{m}_{{\bf LR},n}\right)\tilde{m}^{*}_{{\bf RR},m}+\left(1-\Gamma^{(\alpha)}_{p,n} \tilde{m}_{{\bf LR},n}\right)\tilde{m}^{*}_{{\bf RR},m}\right.\right.\nonumber\\&&\left.\left.~~~~~~~~~~~+\left(1-\Gamma^{(\alpha)}_{p,n} \tilde{m}_{{\bf LR,n}}\right)^{*}\tilde{m}_{{\bf RR},m}+\left(1-\Gamma^{(\alpha)}_{p,n} \tilde{m}_{{\bf LR},n}\right)^{*}\tilde{m}_{{\bf RR},m}\right]\right\}\eea 

On the other hand,  if we set $\nu=q/2$ (including the massless case for $\nu=3/2$) in the previous expressions obtained for general $\nu$ then due to the presence of the overall factor $1/|\cos \pi \nu|$ the final expression for the power spectrum in small wave number limit diverges. This is very obvious from the obtained expressions but one can be able to avoid such  unwanted divergent contributions very easily.  To serve this purpose let us rewrite all the expressions for $p,p_n<<1$ with $\nu=q/2$ that we have mentioned earlier:
\bea \frac{|{\cal E}^{}_{\bf L}|^2}{\left|\Gamma\left(\nu+ ip +\frac{1}{2}\right)\right|^2}
&\stackrel{p<<1}{\approx}&\frac{\pi }{2}~~~~~~~~~~~\\
\frac{|{\cal F}^{}_{\bf L}|^2}{\left|\Gamma\left(\nu+ ip +\frac{1}{2}\right)\right|^2}
&\stackrel{p<<1}{\approx}&0~~~~~~~~~~~\\
\frac{{\cal E}^{}_{\bf L}{\cal F}^{ *}_{\bf L}}{\left(\Gamma\left(\nu- ip +\frac{1}{2}\right)\right)^2}
&\stackrel{p<<1}{\approx}&0~~~~~~~~~~~\\
\frac{{\cal E}^{*}_{\bf L}{\cal F}^{}_{\bf L}}{\left(\Gamma\left(\nu+ ip +\frac{1}{2}\right)\right)^2}
&\stackrel{p<<1}{\approx}&0~~~~~~~~~~~\\
\frac{{\cal E}^{}_{\bf L}{\cal E}^{*}_{{\bf L},(n)}}{\Gamma\left(\nu- ip +\frac{1}{2}\right)\Gamma\left(\nu+ ip_n +\frac{1}{2}\right)}
&\stackrel{p,p_n<<1}{\approx}&\frac{\pi }{2}\frac{\left(1\pm \pi p~e^{-p\pi}~e^{i\theta}\right)}{|1\pm \pi p~e^{-p\pi}~e^{i\theta}|}\frac{\left(1\pm \pi p_n~e^{-p_n\pi}e^{-i\theta}\right)}{|1\pm \pi p_n~e^{-p_n\pi}~e^{i\theta|}}~~~~~~~~~~~\\
\frac{{\cal F}^{}_{\bf L}{\cal F}^{*}_{{\bf L},(n)}}{\Gamma\left(\nu- ip +\frac{1}{2}\right)\Gamma\left(\nu+ ip_n +\frac{1}{2}\right)}
&\stackrel{p,p_n<<1}{\approx}&0~~~~~~~~~~~\\
\frac{{\cal E}^{}_{\bf L}{\cal F}^{*}_{{\bf L},(n)}}{\Gamma\left(\nu- ip +\frac{1}{2}\right)\Gamma\left(\nu- ip_n +\frac{1}{2}\right)}
&\stackrel{p,p_n<<1}{\approx}&0~~~~~~~~~~~\eea\bea
\frac{{\cal E}^{}_{{\bf L},(n)}{\cal F}^{*}_{{\bf L}}}{\Gamma\left(\nu- ip +\frac{1}{2}\right)\Gamma\left(\nu- ip_n +\frac{1}{2}\right)}
&\stackrel{p,p_n<<1}{\approx}&0~~~~~~~~~~~\\
\frac{{\cal E}^{*}_{{\bf L},(n)}{\cal F}^{}_{\bf L}}{\Gamma\left(\nu+ ip +\frac{1}{2}\right)\Gamma\left(\nu+ ip_n +\frac{1}{2}\right)}
&\stackrel{p,p_n<<1}{\approx}&0~~~~~~~~~~~\\
\frac{{\cal E}^{*}_{{\bf L}}{\cal F}^{}_{{\bf L},(n)}}{\Gamma\left(\nu+ ip +\frac{1}{2}\right)\Gamma\left(\nu+ ip_n +\frac{1}{2}\right)}
&\stackrel{p,p_n<<1}{\approx}&0~~~~~~~~~~~\\
\frac{{\cal E}^{}_{{\bf L},(n)}{\cal E}^{*}_{{\bf L},(m)}}{\Gamma\left(\nu- ip_n +\frac{1}{2}\right)\Gamma\left(\nu+ ip_m +\frac{1}{2}\right)}
&\stackrel{p_n, p_m<<1}{\approx}&\frac{\pi }{2}\frac{\left(1\pm \pi p_n~e^{-p_n\pi} ~e^{i\theta}\right)}{|1\pm \pi p_n~e^{-p_n\pi} ~e^{i\theta}|}\frac{\left(1\pm \pi p_m~e^{-p_m\pi}~e^{i\theta}\right)}{|1\pm \pi p_m~e^{-p_m\pi}~e^{i\theta}|}~~~~~~~~~~~\\
\frac{{\cal F}^{}_{{\bf L},(n)}{\cal F}^{*}_{{\bf L},(m)}}{\Gamma\left(\nu- ip_n +\frac{1}{2}\right)\Gamma\left(\nu+ ip_m +\frac{1}{2}\right)}
&\stackrel{p_n, p_m<<1}{\approx}&0~~~~~~~~~~~\\
\frac{{\cal E}^{}_{{\bf L},(n)}{\cal F}^{*}_{{\bf L},(m)}}{\Gamma\left(\nu- ip_n +\frac{1}{2}\right)\Gamma\left(\nu- ip_m +\frac{1}{2}\right)}
&\stackrel{p_n, p_m<<1}{\approx}&0~~~~~~~~~~~\\
\frac{{\cal E}^{\sigma*}_{{\bf L},(n)}{\cal F}^{}_{{\bf L},(m)}}{\Gamma\left(\nu+ ip_n +\frac{1}{2}\right)\Gamma\left(\nu+ ip_m +\frac{1}{2}\right)}
&\stackrel{p_n, p_m<<1}{\approx}&0~~~~~~~~~~~\eea
Also for the massless case ($\nu=3/2$) the time independent function $\widetilde{{\cal Q}(p<<1,\alpha,\nu=3/2)}$ can be further simplified as:
\bea\widetilde{{\cal Q}(p<<1,\alpha,\nu=3/2)}&=&\frac{\widetilde{{\cal G}(p<<1,\nu=3/2)}}{2p^3}~~~\forall\alpha,~~~~~~~~~~~\eea
where the function $\widetilde{{\cal G}(p<<1)}$ is defined for $\nu = 3/2$ as:
\bea \widetilde{{\cal G}(p<<1,\nu=3/2)}&=&\frac{\pi }{2}\left\{1+\frac{\left(1\pm e^{i\theta}\pi p~e^{-p\pi}\right)}{{|1\pm e^{i\theta}\pi p~e^{-p\pi}|}}\sum^{\infty}_{n=0}\frac{\left(1\pm e^{-i\theta}\pi p_n~e^{-p_n\pi}\right)}{{|1\pm e^{i\theta}\pi p_n~e^{-p_n\pi}|}}\right.\nonumber\\&&\left.~~~~+\sum^{\infty}_{n=0}\sum^{\infty}_{m=0}\sqrt{\frac{\left(1\pm e^{i\theta}\pi p_n~e^{-p_n\pi}\right)}{|1\pm e^{i\theta}\pi p_n~e^{-p_n\pi}|}\frac{\left(1\pm e^{-i\theta}\pi p_m~e^{-p_m\pi}\right)}{|1\pm e^{i\theta}\pi p_m~e^{-p_m\pi}|}}\right\}\eea

\section{\textcolor{blue}{Quantum correction to the power spectrum  in NES formalism}}
At the superhorizon time scales ($t_{\bf L}>>1$) of region \textcolor{red}{\bf L}  the amplitude of the NES power spectrum can be expressed as:
\bea \label{df2xxxcv} |\tilde{\phi}^{\bf L}|^2&~\underrightarrow{t_{\bf L}>>1}~& \widetilde{{\cal K}(p,\alpha,\nu)}\left(\cosh t_{\bf L}\right)^{2\nu-1}\eea
where the time independent function $\widetilde{{\cal K}(p,\alpha,\nu)}$ for generalised $\alpha$ vacua is defined as:
\bea\widetilde{{\cal K}(p,\alpha,\nu)}&=&\frac{2^{2\nu-1}\left(\Gamma(\nu)\right)^2}{\pi}\times \left[\frac{|\Gamma(1+ip)|^2}{2p|\Gamma\left(\nu+ip+\frac{1}{2}\right)|^2}\right.\nonumber\\ && \left.
~~~~~~+\sum^{\infty}_{n=0}\frac{|\Gamma(1-ip)| |\Gamma(1+ip_n)|+|\Gamma(1+ip)| |\Gamma(1-ip_n)|}{4\sqrt{pp_n}~\Gamma\left(\nu-ip+\frac{1}{2}\right)\Gamma\left(\nu+ip_n+\frac{1}{2}\right)}\right.\nonumber\\&&\left.~~~~~~~~+\sum^{\infty}_{n=0}\sum^{\infty}_{m=0}\frac{|\Gamma(1-ip_n)| |\Gamma(1+ip_m)|+|\Gamma(1+ip_n)| |\Gamma(1-ip_m)|}{4\sqrt{p_np_m}~\Gamma\left(\nu-ip_n+\frac{1}{2}\right)\Gamma\left(\nu+ip_m+\frac{1}{2}\right)}\right].~~~~~~~~~\eea

\subsection{For large wave number}

Further, to know the exact wave number dependence of the amplitude of the normalised power spectrum from generalised $\alpha$ vacua we need to know the behaviour of the power spectrum at very short wavelengths ($p,p_n>>1$). After taking this limit it is expected that the power spectrum of axion in the non entangled case should match with the result obtained for spatially flat universe. In general for an arbitrary value of the mass parameter $\nu$, we get the following approximated contributions in the short wavelength limit ($p,p_n>>1$), which are explicitly appearing in the expression for the amplitude of the normalised power spectrum from generalised $\alpha$ vacua:
\bea \frac{|\Gamma(1+ip)|^2}{2p|\Gamma\left(\nu+ip+\frac{1}{2}\right)|^2}&\stackrel{p>>1}{\approx}&\frac{1}{2p^3}, \\ 
\frac{|\Gamma(1-ip)| |\Gamma(1+ip_n)|+|\Gamma(1+ip)| |\Gamma(1-ip_n)|}{4\sqrt{pp_n}~\Gamma\left(\nu-ip+\frac{1}{2}\right)\Gamma\left(\nu+ip_n+\frac{1}{2}\right)}&\stackrel{p,p_n>>1}{\approx}&\frac{1}{2(pp_n)^{\frac{3}{2}}}\\  
\frac{|\Gamma(1-ip_n)| |\Gamma(1+ip_m)|+|\Gamma(1+ip_n)| |\Gamma(1-ip_m)|}{4\sqrt{p_np_m}~\Gamma\left(\nu-ip_n+\frac{1}{2}\right)\Gamma\left(\nu+ip_m+\frac{1}{2}\right)}&\stackrel{p_n,p_m>>1}{\approx}&\frac{1}{2(p_np_m)^{\frac{3}{2}}}.\eea
As a result, the time independent function $\widetilde{{\cal K}(p,\alpha,\nu)}$ in the short wave length limit for any arbitrary mass parameter $\nu$ can be expressed for generalised $\alpha$ vacua as:
\bea\widetilde{{\cal K}(p>>1,\alpha,\nu)}&=&\frac{2^{2(\nu-1)}\left(\Gamma(\nu)\right)^2}{p^3\pi}\widetilde{{\cal U}(p>>1)}~~~\forall \alpha,~~~~~~~~~~~\eea
where the function $\widetilde{{\cal U}(p>>1)}$ is defined as:
\bea \widetilde{{\cal U}(p>>1)}&=&\left[1+\sum^{\infty}_{n=0}\left(\frac{p}{p_n}\right)^{\frac{3}{2}}+\sum^{\infty}_{n=0}\sum^{\infty}_{m=0}\frac{p^3}{\left(p_np_m\right)^{\frac{3}{2}}}\right].\eea 
Here for very large wave number $p,p_n>>1$ one can write, $\widetilde{{\cal U}(p)}\sim 1+\cdots$, where all $\cdots$ are small correction terms. This also implies to the nice fact that for large wave number limit for any values of the parameter $\alpha$ the time independent function ${{\cal U}(p,\alpha,\nu)}$ computed for generalised $\alpha$ vacua is exactly matches with the result obtained for Bunch Davies vacua in the earlier section at the leading order in $\widetilde{{\cal M}(p,\nu)}$.

Also for the massless case ($\nu=3/2$) the time independent function $\widetilde{{\cal K}(p,\alpha,\nu=3/2)}$ in the short wave length limit can be further simplified as:
\bea\widetilde{{\cal K}(p>>1,\alpha,\nu=3/2)}&=&\frac{\widetilde{{\cal U}(p>>1)}}{2p^3}~~~\forall\alpha.~~~~~~~~~~~\eea
\subsection{For small wave number}

Similarly to see the behaviour of the power spectrum in the long wavelength region in the super horizon time scale ($t_{\bf L}>>1$) we take the limit $p<<1$ and further expand the expression for the power spectrum in $p$.  In general for an arbitrary value of the mass parameter $\nu$, we get the following approximated contributions in the long wavelength limit ($p,p_n<<1$), which are explicitly appearing in the expression for the amplitude of the normalised power spectrum from generalised $\alpha$ vacua:
\bea \frac{|\Gamma(1+ip)|^2}{2p|\Gamma\left(\nu+ip+\frac{1}{2}\right)|^2}&\stackrel{p<<1}{\approx}&\frac{1}{2p|\Gamma\left(\nu+\frac{1}{2}\right)|^2}, \\ 
\frac{|\Gamma(1-ip)| |\Gamma(1+ip_n)|+|\Gamma(1+ip)| |\Gamma(1-ip_n)|}{4\sqrt{pp_n}~\Gamma\left(\nu-ip+\frac{1}{2}\right)\Gamma\left(\nu+ip_n+\frac{1}{2}\right)}&\stackrel{p,p_n<<1}{\approx}&\frac{1}{2\sqrt{pp_n}\left(\Gamma\left(\nu+\frac{1}{2}\right)\right)^2}\\  
\frac{|\Gamma(1-ip_n)| |\Gamma(1+ip_m)|+|\Gamma(1+ip_n)| |\Gamma(1-ip_m)|}{4\sqrt{p_np_m}~\Gamma\left(\nu-ip_n+\frac{1}{2}\right)\Gamma\left(\nu+ip_m+\frac{1}{2}\right)}&\stackrel{p_n,p_m<<1}{\approx}&\frac{1}{2\sqrt{p_np_m}\left(\Gamma\left(\nu+\frac{1}{2}\right)\right)^2}.\eea
As a result, in the long wave length limit the time independent function $\widetilde{{\cal K}(p,\alpha,\nu)}$ for any arbitrary mass parameter $\nu$ can be expressed for generalised $\alpha$ vacua as:
\bea\widetilde{{\cal K}(p<<1,\alpha,\nu)}&=&\frac{2^{2(\nu-1)}\left(\Gamma(\nu)\right)^2}{p\pi}\widetilde{{\cal U}(p<<1)}~~~\forall \alpha,~~~~~~~~~~~\eea
where the function $\widetilde{{\cal U}(p<<1)}$ is defined in the long wave length limit as:
\bea \widetilde{{\cal U}(p<<1)}&=&\left[1+\left(\frac{|\Gamma\left(\nu+\frac{1}{2}\right)|}{\Gamma\left(\nu+\frac{1}{2}\right)}\right)^2\left\{\sum^{\infty}_{n=0}\sqrt{\frac{p}{p_n}}+\sum^{\infty}_{n=0}\sum^{\infty}_{m=0}\frac{p}{\sqrt{p_np_m}}\right\}\right].\eea 
  Also for the massless case ($\nu=3/2$) the time independent function $\widetilde{{\cal K}(p,\alpha,\nu=3/2)}$ can be further simplified as:
\bea\widetilde{{\cal K}(p<<1,\alpha,\nu=3/2)}&=&\frac{\widetilde{{\cal U}(p<<1)}}{2p}~~~\forall\alpha.~~~~~~~~~~~\eea

\end{itemize}


\begin{thebibliography}{}

	
	\bibitem{Horodecki:2009zz}
	\textsf{R.~Horodecki, P.~Horodecki, M.~Horodecki and K.~Horodecki },
	\textcolor{red}{\textsf{Quantum entanglement,}}
	\textcolor{violet}{{Rev.\ Mod.\ Phys.\  {\bf 81} (2009) 865
			[quant-ph/0702225].}}
			
			\bibitem{MartinMartinez:2012sg}
	\textsf{E.~Martin-Martinez and N.~C.~Menicucci},
	\textcolor{red}{\textsf{Cosmological quantum entanglement,}}
	\textcolor{violet}{{Class.\ Quant.\ Grav.\  {\bf 29} (2012) 224003
			[arXiv:1204.4918 [gr-qc]].  }}   
	
	\bibitem{Nambu:2008my}
	\textsf{Y.~Nambu},
	\textcolor{red}{\textsf{Entanglement of Quantum Fluctuations in the Inflationary Universe,}}
	\textcolor{violet}{{Phys.\ Rev.\ D {\bf 78} (2008) 044023
			[arXiv:0805.1471 [gr-qc]].  }}     
			
				
	\bibitem{Bell:1964kc}
	\textsf{J.~S.~Bell},
	\textcolor{red}{\textsf{On the Einstein-Podolsky-Rosen paradox,}}
	\textcolor{violet}{{Physics {\bf 1} (1964) 195.  }} 
			
			
\bibitem{Coleman:1980aw}
  \textsf{S.~R.~Coleman and F.~De Luccia,}
  \textcolor{red}{\textsf{``Gravitational Effects on and of Vacuum Decay,''}}
  \textcolor{violet}{Phys.\ Rev.\ D {\bf 21} (1980) 3305.}
  
  \bibitem{Garriga:2012qp}
  \textsf{ J.~Garriga, S.~Kanno, M.~Sasaki, J.~Soda and A.~Vilenkin},
 \textcolor{red}{\textsf{``Observer dependence of bubble nucleation and Schwinger pair production,''}}
   \textcolor{violet}{JCAP {\bf 1212} (2012) 006 
  [arXiv:1208.1335 [hep-th]]}.
  
  \bibitem{Garriga:2013pga}
  \textsf{ J.~Garriga, S.~Kanno and T.~Tanaka},
   \textcolor{red}{\textsf{``Rest frame of bubble nucleation,''}}
  \textcolor{violet}{JCAP {\bf 1306} (2013) 034
  [arXiv:1304.6681 [hep-th]]}.

	\bibitem{Frob:2014zka}
  \textsf{M.~B.~Fröb, J.~Garriga, S.~Kanno, M.~Sasaki, J.~Soda, T.~Tanaka and A.~Vilenkin},
 \textcolor{red}{\textsf{``Schwinger effect in de Sitter space,''}}
  \textcolor{red}{JCAP {\bf 1404} (2014) 009
  [arXiv:1401.4137 [hep-th]]};
	\textsf{W.~Fischler, P.~H.~Nguyen, J.~F.~Pedraza and W.~Tangarife},
	\textcolor{red}{\textsf{Holographic Schwinger effect in de Sitter space,}}
	\textcolor{violet}{{Phys.\ Rev.\ D {\bf 91} (2015) no.8,  086015 [arXiv:1411.1787 [hep-th]].  }}

	
	\bibitem{Laflorencie:2015eck}
	\textsf{N.~Laflorencie},
	\textcolor{red}{\textsf{Quantum entanglement in condensed matter systems,}}
	\textcolor{violet}{{Phys.\ Rept.\  {\bf 646} (2016) 1
			[arXiv:1512.03388 [cond-mat.str-el]].}}  
  
  
  \bibitem{Ryu:2006bv}
   \textsf{S.~Ryu and T.~Takayanagi},
  \textcolor{red}{\textsf{``Holographic derivation of entanglement entropy from AdS/CFT,''}}
   \textcolor{red}{Phys.\ Rev.\ Lett.\  {\bf 96} (2006) 181602
  [hep-th/0603001]}.
  
  
  \bibitem{Takayanagi:2012kg}
  \textsf{T.~Takayanagi},
  \textcolor{red}{\textsf{``Entanglement Entropy from a Holographic Viewpoint,''}}
    \textcolor{red}{Class.\ Quant.\ Grav.\  {\bf 29} (2012) 153001
  [arXiv:1204.2450 [gr-qc]]};
	\textsf{S.~Ryu and T.~Takayanagi},
	\textcolor{red}{\textsf{Holographic derivation of entanglement entropy from AdS/CFT,}}
	\textcolor{violet}{{Phys.\ Rev.\ Lett.\  {\bf 96} (2006) 181602 [hep-th/0603001].  }} ;
	\textsf{S.~Ryu and T.~Takayanagi},
	\textcolor{red}{\textsf{Aspects of Holographic Entanglement Entropy,}}
	\textcolor{violet}{{JHEP {\bf 0608} (2006) 045 [hep-th/0605073].  }};
	\textsf{T.~Nishioka, S.~Ryu and T.~Takayanagi},
	\textcolor{red}{\textsf{Holographic Entanglement Entropy: An Overview,}}
	\textcolor{violet}{{J.\ Phys.\ A {\bf 42} (2009) 504008 [arXiv:0905.0932 [hep-th]].  }} ;
	\textsf{M.~Rangamani and T.~Takayanagi},
	\textcolor{red}{\textsf{Holographic Entanglement Entropy,}}
	\textcolor{violet}{{Lect.\ Notes Phys.\  {\bf 931} (2017) [arXiv:1609.01287 [hep-th]].  }} ;
	\textsf{V.~E.~Hubeny, M.~Rangamani and T.~Takayanagi},
	\textcolor{red}{\textsf{A Covariant holographic entanglement entropy proposal,}}
	\textcolor{violet}{{JHEP {\bf 0707} (2007) 062 [arXiv:0705.0016 [hep-th]].  }}  
  
  \bibitem{Maldacena:2012xp}
   \textsf{J.~Maldacena and G.~L.~Pimentel},
   \textcolor{red}{\textsf{``Entanglement entropy in de Sitter space,''}}
  \textcolor{red}{JHEP {\bf 1302} (2013) 038
  [arXiv:1210.7244 [hep-th]]}.
  
  
  \bibitem{Kanno:2014lma}
    \textsf{S.~Kanno, J.~Murugan, J.~P.~Shock and J.~Soda},
   \textcolor{red}{\textsf{``Entanglement entropy of $\alpha$-vacua in de Sitter space,''}}
  \textcolor{red}{JHEP {\bf 1407} (2014) 072
  [arXiv:1404.6815 [hep-th]]}.
  
  
	\bibitem{Allen:1985ux}
	\textsf{B.~Allen},
	\textcolor{red}{\textsf{Vacuum States in de Sitter Space,}}
	\textcolor{violet}{{Phys.\ Rev.\ D {\bf 32} (1985) 3136}};
	\textsf{K.~Goldstein and D.~A.~Lowe},
	\textcolor{red}{\textsf{A Note on alpha vacua and interacting field theory in de Sitter space,}}
	\textcolor{violet}{{Nucl.\ Phys.\ B {\bf 669} (2003) 325
			[hep-th/0302050]}};
	\textsf{J.~de Boer, V.~Jejjala and D.~Minic},
	\textcolor{red}{\textsf{Alpha-states in de Sitter space,}}
	\textcolor{violet}{{Phys.\ Rev.\ D {\bf 71} (2005) 044013
			[hep-th/0406217]}};
	\textsf{R.~Brunetti, K.~Fredenhagen and S.~Hollands},
	\textcolor{red}{\textsf{A Remark on alpha vacua for quantum field theories on de Sitter space,}}
	\textcolor{violet}{{JHEP {\bf 0505} (2005) 063
			[hep-th/0503022].}}
  
  \bibitem{Choudhury:2017bou}
    \textsf{S.~Choudhury and S.~Panda},
   \textcolor{red}{\textsf{``Entangled de Sitter from stringy axionic Bell pair I: an analysis using Bunch–Davies vacuum,''}}
  \textcolor{red}{Eur.\ Phys.\ J.\ C {\bf 78} (2018) no.1,  52
  [arXiv:1708.02265 [hep-th]]}.
  
  \bibitem{Choudhury:2017qyl}
   \textsf{S.~Choudhury and S.~Panda},
  \textcolor{red}{\textsf{``Quantum entanglement in de Sitter space from Stringy Axion: An analysis using $\alpha$ vacua,''}}
   \textcolor{red}{arXiv:1712.08299 [hep-th].}
   
   \bibitem{Capolupo:2019peg}
 \textsf{A.~Capolupo, G.~Lambiase, A.~Quaranta and S.~M.~Giampaolo},
  \textcolor{red}{\textsf{Probing axion mediated fermion–fermion interaction by means of entanglement,}}
 \textcolor{red}{Phys. Lett. B \textbf{804} (2020), 135407
[arXiv:1910.01533 [hep-ph]].}

\bibitem{Patrascu:2018sia}
 \textsf{A.~T.~Patrascu},
  \textcolor{red}{\textsf{Axion mass and quantum information,}}
\textcolor{red}{Phys. Lett. B \textbf{786} (2018), 1-4
[arXiv:1804.10522 [hep-th]].}
   
   \bibitem{Maldacena:2015bha}
	\textsf{J.~Maldacena},
	\textcolor{red}{\textsf{A model with cosmological Bell inequalities,}}
	\textcolor{violet}{{Fortsch.\ Phys.\  {\bf 64} (2016) 10
			[arXiv:1508.01082 [hep-th]].  }} 
	
	\bibitem{Choudhury:2016cso}
	\textsf{S.~Choudhury, S.~Panda and R.~Singh},
	\textcolor{red}{\textsf{Bell violation in the Sky,}}
	\textcolor{violet}{{Eur.\ Phys.\ J.\ C {\bf 77} (2017) no.2,  60
			[arXiv:1607.00237 [hep-th]].  }}   
	
	\bibitem{Choudhury:2016pfr}
	\textsf{S.~Choudhury, S.~Panda and R.~Singh},
	\textcolor{red}{\textsf{Bell violation in primordial cosmology,}}
	\textcolor{violet}{{Universe {\bf 3} (2017) no.1,  13
			[arXiv:1612.09445 [hep-th]].  }}   
			
			
	
\bibitem{Kanno:2017dci}
	\textsf{S.~Kanno and J.~Soda},
	\textcolor{red}{\textsf{Infinite violation of Bell inequalities in inflation,}}
	\textcolor{violet}{{arXiv:1705.06199 [hep-th].  }}   
	
	
  
  \bibitem{Panda:2010uq}
   \textsf{S.~Panda, Y.~Sumitomo and S.~P.~Trivedi},
   \textcolor{red}{\textsf{``Axions as Quintessence in String Theory,''}}
  \textcolor{red}{Phys.\ Rev.\ D {\bf 83} (2011) 083506
  [arXiv:1011.5877 [hep-th]]};
\textsf{L.~McAllister, E.~Silverstein and A.~Westphal},
	\textcolor{red}{\textsf{Gravity Waves and Linear Inflation from Axion Monodromy,}}
	\textcolor{violet}{{Phys.\ Rev.\ D {\bf 82} (2010) 046003 [arXiv:0808.0706 [hep-th]] }};
	\textsf{ E.~Silverstein and A.~Westphal},
	\textcolor{red}{\textsf{Monodromy in the CMB: Gravity Waves and String Inflation,}}
	\textcolor{violet}{{Phys.\ Rev.\ D {\bf 78} (2008) 106003 [arXiv:0803.3085 [hep-th]]}};
\textsf{ L.~McAllister, E.~Silverstein, A.~Westphal and T.~Wrase},
	\textcolor{red}{\textsf{The Powers of Monodromy,}}
	\textcolor{violet}{{JHEP {\bf 1409} (2014) 123 [arXiv:1405.3652 [hep-th]].  }}                          
	
  
 
	
	\bibitem{Kanno:2014ifa}
	\textsf{S.~Kanno},
	\textcolor{red}{\textsf{Impact of quantum entanglement on spectrum of cosmological fluctuations,}}
	\textcolor{violet}{{JCAP {\bf 1407} (2014) 029 [arXiv:1405.7793 [hep-th]].  }} 
	
	
\bibitem{Kolevatov:2017dze}
 \textsf{ R.~Kolevatov, S.~Mironov, V.~Rubakov, N.~Sukhov and V.~Volkova},
  \textcolor{red}{\textsf{Perturbations in generalized Galileon theories,}}
  \textcolor{violet}{{Phys.\ Rev.\ D {\bf 96} (2017) no.12,  125012
  [arXiv:1708.04262 [hep-th]]}},
	  \textsf{ M.~Libanov, S.~Mironov and V.~Rubakov},
  \textcolor{red}{\textsf{Generalized Galileons: instabilities of bouncing and Genesis cosmologies and modified Genesis,}}
   \textcolor{violet}{{JCAP {\bf 1608} (2016) 037
  [arXiv:1605.05992 [hep-th]]}}, 
	 \textsf{ M.~Libanov, V.~Rubakov and G.~Rubtsov},
  \textcolor{red}{\textsf{Towards conformal cosmology,}}
  \textcolor{violet}{{JETP Lett.\  {\bf 102} (2015) no.8,  561
   [Pisma Zh.\ Eksp.\ Teor.\ Fiz.\  {\bf 102} (2015) no.8,  630]
  [arXiv:1508.07728 [hep-th]]}},
	 \textsf{ M.~Libanov and V.~Rubakov},
   \textcolor{red}{\textsf{Conformal Universe as false vacuum decay,}}
    \textcolor{violet}{{Phys.\ Rev.\ D {\bf 91} (2015) no.10,  103515
  [arXiv:1502.05897 [hep-th]]}},
	 \textsf{M.~Libanov, V.~Rubakov and S.~Sibiryakov},
  \textcolor{red}{\textsf{TOn holography for (pseudo-)conformal cosmology,}}
   \textcolor{violet}{{Phys.\ Lett.\ B {\bf 741} (2015) 239
  [arXiv:1409.4363 [hep-th]]}},
  \textsf{V.~A.~Rubakov},
   \textcolor{red}{\textsf{Cosmology,}}
  \textcolor{violet}{{arXiv:1504.03587 [astro-ph.CO]}},
	\textsf{S.~A.~Mironov, S.~R.~Ramazanov and V.~A.~Rubakov},
   \textcolor{red}{\textsf{Effect of intermediate Minkowskian evolution on CMB bispectrum,}}
   \textcolor{violet}{{JCAP {\bf 1404} (2014) 015
  [arXiv:1312.7808 [astro-ph.CO]]}},
	 \textsf{M.~Osipov and V.~Rubakov},
   \textcolor{red}{\textsf{Galileon bounce after ekpyrotic contraction,}}
  \textcolor{violet}{{JCAP {\bf 1311} (2013) 031
  [arXiv:1303.1221 [hep-th]]}},
\textsf{M.~V.~Libanov and V.~A.~Rubakov},
  \textcolor{red}{\textsf{Cosmological density perturbations in a conformal scalar field theory,}}
   \textcolor{violet}{{Theor.\ Math.\ Phys.\  {\bf 170} (2012) 151
   [Teor.\ Mat.\ Fiz.\  {\bf 170} (2012) 188]}},
   \textsf{M.~Libanov, S.~Mironov and V.~Rubakov},
  \textcolor{red}{\textsf{Non-Gaussianity of scalar perturbations generated by conformal mechanisms,}}
   \textcolor{violet}{{Phys.\ Rev.\ D {\bf 84} (2011) 083502
  [arXiv:1105.6230 [astro-ph.CO]]}},
	\textsf{M.~Libanov, S.~Ramazanov and V.~Rubakov},
    \textcolor{red}{\textsf{Scalar perturbations in conformal rolling scenario with intermediate stage,}}
  \textcolor{violet}{{JCAP {\bf 1106} (2011) 010
  [arXiv:1102.1390 [hep-th]]}},
   \textsf{M.~Libanov and V.~Rubakov},
   \textcolor{red}{\textsf{Cosmological density perturbations from conformal scalar field: infrared properties and statistical anisotropy,}}
   \textcolor{violet}{{JCAP {\bf 1011} (2010) 045
  [arXiv:1007.4949 [hep-th]]}},
    \textsf{M.~Osipov and V.~Rubakov},
  \textcolor{red}{\textsf{Scalar tilt from broken conformal invariance,}}
   \textcolor{violet}{{JETP Lett.\  {\bf 93} (2011) 52
  [arXiv:1007.3417 [hep-th]]}},
   \textsf{V.~A.~Rubakov},
   \textcolor{red}{\textsf{Harrison-Zeldovich spectrum from conformal invariance,}}
   \textcolor{violet}{{JCAP {\bf 0909} (2009) 030
  [arXiv:0906.3693 [hep-th]]}},
   \textsf{M.~V.~Libanov and V.~A.~Rubakov},
\textcolor{red}{\textsf{Lorentz-violating brane worlds and cosmological perturbations,}}
  \textcolor{violet}{{Phys.\ Rev.\ D {\bf 72} (2005) 123503
  [hep-ph/0509148]}},
   \textsf{V.~A.~Rubakov},
  \textcolor{red}{\textsf{Relaxation of the cosmological constant at inflation?,}}
  \textcolor{violet}{{Phys.\ Rev.\ D {\bf 61} (2000) 061501
  [hep-ph/9911305]}}.
  
  
  
  \bibitem{Kopeikin:2001uk}
   \textsf{S.~M.~Kopeikin, J.~Ramirez, B.~Mashhoon and M.~V.~Sazhin},
    \textcolor{red}{\textsf{Cosmological perturbations: A New gauge invariant approach,}}
   \textcolor{violet}{{Phys.\ Lett.\ A {\bf 292} (2001) 173
  [gr-qc/0106064]}},
  \textsf{V.~A.~Rubakov, M.~V.~Sazhin and A.~V.~Veryaskin},
  \textcolor{red}{\textsf{Graviton Creation in the Inflationary Universe and the Grand Unification Scale,}}
  \textcolor{violet}{{Phys.\ Lett.\  {\bf 115B} (1982) 189}},
  
  
  
  
	\bibitem{Choudhury:2015hvr}
	\textsf{S.~Choudhury and S.~Panda},
	\textcolor{red}{\textsf{COSMOS-$e^{'}$-GTachyon from string theory,}}
	\textcolor{violet}{{Eur.\ Phys.\ J.\ C {\bf 76} (2016) no.5,  278 [arXiv:1511.05734 [hep-th]]  }}, \textsf{S.~Choudhury},
	\textcolor{red}{\textsf{COSMOS-$e^{'}$- soft Higgsotic attractors,}}
	\textcolor{violet}{{Eur.\ Phys.\ J.\ C {\bf 77} (2017) no.7,  469 [arXiv:1703.01750 [hep-th]]  }},
	\textsf{S.~Choudhury and S.~Pal},
	\textcolor{red}{\textsf{Primordial non-Gaussian features from DBI Galileon inflation,}}
	\textcolor{violet}{{Eur.\ Phys.\ J.\ C {\bf 75} (2015) no.6,  241 [arXiv:1210.4478 [hep-th]]  }},
	\textsf{S.~Choudhury and S.~Pal},
	\textcolor{red}{\textsf{DBI Galileon inflation in background SUGRA,}}
	\textcolor{violet}{{Nucl.\ Phys.\ B {\bf 874} (2013) 85 [arXiv:1208.4433 [hep-th]]  }}, 
	\textsf{S.~Choudhury and S.~Pal},
	\textcolor{red}{\textsf{Fourth level MSSM inflation from new flat directions,}}
	\textcolor{violet}{{JCAP {\bf 1204} (2012) 018 [arXiv:1111.3441 [hep-ph]]  }},
	\textsf{S.~Choudhury and S.~Pal},
	\textcolor{red}{\textsf{Brane inflation in background supergravity,}}
	\textcolor{violet}{{Phys.\ Rev.\ D {\bf 85} (2012) 043529 [arXiv:1102.4206 [hep-th]]  }}, 
	\textsf{S.~Choudhury},
	\textcolor{red}{\textsf{Can Effective Field Theory of inflation generate large tensor-to-scalar ratio within Randall–Sundrum single braneworld?,}}
	\textcolor{violet}{{Nucl.\ Phys.\ B {\bf 894} (2015) 29 [arXiv:1406.7618 [hep-th]]  }}, 
	\textsf{S.~Choudhury, B.~K.~Pal, B.~Basu and P.~Bandyopadhyay},
	\textcolor{red}{\textsf{Quantum Gravity Effect in Torsion Driven Inflation and CP violation,}}
	\textcolor{violet}{{JHEP {\bf 1510} (2015) 194 [arXiv:1409.6036 [hep-th]]  }}, 
	\textsf{S.~Choudhury},
	\textcolor{red}{\textsf{Reconstructing inflationary paradigm within Effective Field Theory framework,}}
	\textcolor{violet}{{Phys.\ Dark Univ.\  {\bf 11} (2016) 16 [arXiv:1508.00269 [astro-ph.CO]]  }},
	\textsf{S.~Choudhury and A.~Mazumdar},
	\textcolor{red}{\textsf{An accurate bound on tensor-to-scalar ratio and the scale of inflation,}}
	\textcolor{violet}{{Nucl.\ Phys.\ B {\bf 882} (2014) 386 [arXiv:1306.4496 [hep-ph]]  }}, 
	\textsf{S.~Choudhury and A.~Mazumdar},
	\textcolor{red}{\textsf{Primordial blackholes and gravitational waves for an inflection-point model of inflation,}}
	\textcolor{violet}{{Phys.\ Lett.\ B {\bf 733} (2014) 270 [arXiv:1307.5119 [astro-ph.CO]]  }},
	\textsf{S.~Choudhury and A.~Mazumdar},
	\textcolor{red}{\textsf{Reconstructing inflationary potential from BICEP2 and running of tensor modes,}}
	\textcolor{violet}{{arXiv:1403.5549 [hep-th]  }},
	\textsf{S.~Choudhury, A.~Mazumdar and E.~Pukartas},
	\textcolor{red}{\textsf{Constraining ${\cal N}=1$ supergravity inflationary framework with non-minimal Kähler operators,}}
	\textcolor{violet}{{JHEP {\bf 1404} (2014) 077 [arXiv:1402.1227 [hep-th]]  }},
	\textsf{S.~Choudhury},
	\textcolor{red}{\textsf{Constraining ${\cal N}=1$ supergravity inflation with non-minimal Kaehler operators using $\delta$N formalism,}}
	\textcolor{violet}{{JHEP {\bf 1404} (2014) 105 [arXiv:1402.1251 [hep-th]]  }},
	\textsf{S.~Choudhury, A.~Mazumdar and S.~Pal},
	\textcolor{red}{\textsf{Low \& High scale MSSM inflation, gravitational waves and constraints from Planck,}}
	\textcolor{violet}{{JCAP {\bf 1307} (2013) 041 [arXiv:1305.6398 [hep-ph]]  }},
	\textsf{S.~Choudhury},
  \textcolor{red}{\textsf{The Cosmological OTOC: Formulating new cosmological micro-canonical correlation functions for random chaotic fluctuations in Out-of-Equilibrium Quantum Statistical Field Theory,}}
  \textcolor{violet}{{arXiv:2005.11750 [hep-th]}},
 \textsf{ S.~Banerjee, S.~Choudhury, S.~Chowdhury, R.~N.~Das, N.~Gupta, S.~Panda and A.~Swain},
   \textcolor{red}{\textsf{Indirect detection of Cosmological Constant from large $N$ entangled open quantum system,}}
 \textcolor{violet}{{ arXiv:2004.13058 [hep-th]}},
	\textsf{ S.~Akhtar, S.~Choudhury, S.~Chowdhury, D.~Goswami, S.~Panda and A.~Swain},
   \textcolor{red}{\textsf{Open Quantum Cosmology: A study of two body quantum entanglement in static patch of De Sitter space,}}
 \textcolor{violet}{{  arXiv:1908.09929 [hep-th]}},
	 \textsf{H.~Bohra, S.~Choudhury, P.~Chauhan, A.~Mukherjee, P.~Narayan, S.~Panda and A.~Swain},
   \textcolor{red}{\textsf{Relating the curvature of De Sitter Universe to Open Quantum Lamb Shift Spectroscopy,}}
  \textcolor{violet}{{ arXiv:1905.07403 [physics.gen-ph]}},
	\textsf{S.~Choudhury and A.~Mukherjee},
  \textcolor{red}{\textsf{Quantum randomness in the Sky,}}
   \textcolor{violet}{{Eur.\ Phys.\ J.\ C {\bf 79} (2019) no.7,  554
  [arXiv:1812.04107 [physics.gen-ph]]}},
	\textsf{S.~Choudhury, A.~Mukherjee, P.~Chauhan and S.~Bhattacherjee},
  \textcolor{red}{\textsf{Quantum Out-of-Equilibrium Cosmology,}}
   \textcolor{violet}{{Eur.\ Phys.\ J.\ C {\bf 79} (2019) no.4,  320
  [arXiv:1809.02732 [hep-th]]}},
	 \textsf{S.~Choudhury},
    \textcolor{red}{\textsf{Quantum Field Theory approaches to Early Universe Cosmology,}}
      \textcolor{violet}{{LAP Lambert Academic Publishing (10 July 2018), ISBN-13: 978-6139840908}}.
	
	\bibitem{Maharana:1997cz}
	\textsf{J.~Maharana, S.~Mukherji and S.~Panda},
	\textcolor{red}{\textsf{Notes on axion, inflation and graceful exit in stringy cosmology,}}
	\textcolor{violet}{{Mod.\ Phys.\ Lett.\ A {\bf 12} (1997) 447
			[hep-th/9701115]  }},  \textsf{A.~Mazumdar, S.~Panda and A.~Perez-Lorenzana},
	\textcolor{red}{\textsf{Assisted inflation via tachyon condensation,}}
	\textcolor{violet}{{Nucl.\ Phys.\ B {\bf 614} (2001) 101 [hep-ph/0107058]  }},   
	\textsf{D.~Choudhury, D.~Ghoshal, D.~P.~Jatkar and S.~Panda},  \textcolor{red}{\textsf{Hybrid inflation and brane - anti-brane system,}} \textcolor{violet}{{JCAP {\bf 0307} (2003) 009 [hep-th/0305104]  }},
	\textsf{D.~Choudhury, D.~Ghoshal, D.~P.~Jatkar and S.~Panda},  \textcolor{red}{\textsf{On the cosmological relevance of the tachyon,}} \textcolor{violet}{{Phys.\ Lett.\ B {\bf 544} (2002) 231 [hep-th/0204204]  }},
	\textsf{P.~Chingangbam, S.~Panda and A.~Deshamukhya},  \textcolor{red}{\textsf{Non-minimally coupled tachyonic inflation in warped string background,}} \textcolor{violet}{{JHEP {\bf 0502} (2005) 052 [hep-th/0411210]  }},
	\textsf{A.~Deshamukhya and S.~Panda},  \textcolor{red}{\textsf{Warm tachyonic inflation in warped background,}} \textcolor{violet}{{Int.\ J.\ Mod.\ Phys.\ D {\bf 18} (2009) 2093
			[arXiv:0901.0471 [hep-th]] }}, \textsf{P.~Vargas Moniz, S.~Panda and J.~Ward},  \textcolor{red}{\textsf{Higher order corrections to Heterotic M-theory inflation,}} \textcolor{violet}{{Class.\ Quant.\ Grav.\  {\bf 26} (2009) 245003
			[arXiv:0907.0711 [astro-ph.CO]] }}, \textsf{A.~Ali, A.~Deshamukhya, S.~Panda and M.~Sami},  \textcolor{red}{\textsf{Inflation with improved D3-brane potential and the fine tunings associated with the model,}} \textcolor{violet}{{Eur.\ Phys.\ J.\ C {\bf 71} (2011) 1672
			[arXiv:1010.1407 [hep-th]] }}, \textsf{A.~Bhattacharjee, A.~Deshamukhya and S.~Panda},  \textcolor{red}{\textsf{A note on low energy effective theory of chromo-natural inflation in the light of BICEP2 results,}} \textcolor{violet}{{Mod.\ Phys.\ Lett.\ A {\bf 30} (2015) no.11,  1550040
			[arXiv:1406.5858 [astro-ph.CO]] }},
	\textsf{S.~Panda, M.~Sami and S.~Tsujikawa},  \textcolor{red}{\textsf{Inflation and dark energy arising from geometrical tachyons,}} \textcolor{violet}{{Phys.\ Rev.\ D {\bf 73} (2006) 023515 [hep-th/0510112]  }},
	\textsf{S.~Panda, M.~Sami, S.~Tsujikawa and J.~Ward},  \textcolor{red}{\textsf{Inflation from D3-brane motion in the background of D5-branes,}} \textcolor{violet}{{Phys.\ Rev.\ D {\bf 73} (2006) 083512 [hep-th/0601037]  }},
	\textsf{S.~Panda, M.~Sami and S.~Tsujikawa},  \textcolor{red}{\textsf{Prospects of inflation in delicate D-brane cosmology,}} \textcolor{violet}{{Phys.\ Rev.\ D {\bf 76} (2007) 103512 [arXiv:0707.2848 [hep-th]].  }}		
	
	\bibitem{Baumann:2009ds}       
	\textsf{D.~Baumann},
	\textcolor{red}{\textsf{TASI lectures on Inflation 2009,}}
	\textcolor{violet}{{arXiv:0907.5424 [hep-th]}}, 
	\textsf{D.~Baumann, A.~Dymarsky, I.~R.~Klebanov and L.~McAllister},
	\textcolor{red}{\textsf{Towards an Explicit Model of D-brane Inflation,}}
	\textcolor{violet}{{JCAP {\bf 0801} (2008) 024 [arXiv:0706.0360 [hep-th]]}},                   
	\textsf{D.~Baumann and L.~McAllister},
	\textcolor{red}{\textsf{Advances in Inflation in String Theory,}}
	\textcolor{violet}{{Ann.\ Rev.\ Nucl.\ Part.\ Sci.\  {\bf 59} (2009) 67 [arXiv:0901.0265 [hep-th]]}},    
	\textsf{V.~Assassi, D.~Baumann and D.~Green},
	\textcolor{red}{\textsf{Symmetries and Loops in Inflation,}}
	\textcolor{violet}{{JHEP {\bf 1302} (2013) 151 [arXiv:1210.7792 [hep-th]]}},
	\textsf{D.~Baumann and L.~McAllister},
	\textcolor{red}{\textsf{Inflation and String Theory,}}
	\textcolor{violet}{{arXiv:1404.2601 [hep-th]}},
	\textsf{D.~Baumann, A.~Dymarsky, S.~Kachru, I.~R.~Klebanov and L.~McAllister},
	\textcolor{red}{\textsf{Holographic Systematics of D-brane Inflation,}}
	\textcolor{violet}{{JHEP {\bf 0903} (2009) 093 [arXiv:0808.2811 [hep-th]]}}, \textsf{H.~V.~Peiris, D.~Baumann, B.~Friedman and A.~Cooray},
	\textcolor{red}{\textsf{Phenomenology of D-Brane Inflation with General Speed of Sound,}}
	\textcolor{violet}{{Phys.\ Rev.\ D {\bf 76} (2007) 103517 [arXiv:0706.1240 [astro-ph]].}}
	                                                                        
	
	
	
	
	\bibitem{Svrcek:2006yi}
	\textsf{P.~Svrcek and E.~Witten},
	\textcolor{red}{\textsf{Axions In String Theory,}}
	\textcolor{violet}{{JHEP {\bf 0606} (2006) 051 [hep-th/0605206].  }}                                                  
	
	


\end{thebibliography}
\end{document}